\documentclass[useAMS,usenatbib]{mn2e}  
\usepackage {graphicx,subfig,aas_macros,float,amsmath}
\newcommand{\equ}[1]{eq.~(\ref{eq:#1})}
\newcommand{\equs}[1]{eqs.~(\ref{eq:#1})}
\newcommand{\equm}[1]{(\ref{eq:#1})}
\newcommand{\Equ}[1]{Eq.~(\ref{eq:#1})}

\newcommand{\equnp}[1]{eq.~\ref{eq:#1}}
\newcommand{\equsnp}[1]{eqs.~\ref{eq:#1}}
\newcommand{\equmnp}[1]{\ref{eq:#1}}
\newcommand{\se}[1]{\S\ref{sec:#1}}
\newcommand{\fig}[1]{Fig.~\ref{fig:#1}}
\newcommand{\figs}[1]{Figs.~\ref{fig:#1}}
\newcommand{\figss}[1]{\ref{fig:#1}}
\newcommand{\Fig}[1]{Figure~\ref{fig:#1}}

\newcommand{\tab}[1]{Table~\ref{tab:#1}}
\newcommand{\be}{\begin{equation}}
\newcommand{\ee}{\end{equation}}
\newcommand{\bea}{\begin{eqnarray}}
\newcommand{\eea}{\end{eqnarray}}

\newcommand{\no}{\noindent}

\newcommand{\msun}{{\rm M}_\odot}
\newcommand{\Msun}{M_\odot}

\newcommand{\ifm}[1]{\relax\ifmmode#1\else$\mathsurround=0pt #1$\fi}
\newcommand{\kms}{\ifmmode\,{\rm km}\,{\rm s}^{-1}\else km$\,$s$^{-1}$\fi}

\newcommand{\K}{\,{\rm K}}

\newcommand{\ltsima}{$\; \buildrel < \over \sim \;$}
\newcommand{\lsim}{\lower.5ex\hbox{\ltsima}}
\newcommand{\gtsima}{$\; \buildrel > \over \sim \;$}
\newcommand{\gsim}{\lower.5ex\hbox{\gtsima}}

\def\la{\langle}

\def\sy{\,M_\odot\, {\rm yr}^{-1}}

\def\M*{M_{\rm *}}
\def\Mv{M_{\rm v}}
\def\Rv{R_{\rm v}}
\def\Vv{V_{\rm v}}
\def\tv{t_{\rm v}}

\def\Tv{T_{\rm v}}
\def\Tb{T_{\rm b}}
\def\Ts{T_{\rm s}}
\def\Rs{R_{\rm s}}
\def\rhob{\rho_{\rm b}}
\def\rhos{\rho_{\rm s}}
\def\rhobs{\rho_{\rm b,s}}
\def\cb{c_{\rm b}}
\def\cs{c_{\rm s}}
\def\cbs{c_{\rm b,s}}
\def\Vb{V_{\rm b}}
\def\Vs{V_{\rm s}}

\def\tkh{t_{\rm KH}}
\def\tsc{t_{\rm sc}}
\def\Mb{M_{\rm b}}
\def\Ms{M_{\rm s}}

\def\hb{h_{\rm b}}
\def\hs{h_{\rm s}}
\def\Ev{E_{\rm v}}
\def\dmb{\rm dm_b}
\def\dms{\rm dm_s}
\def\dV{\rm dV}

\def\Pi{\varpi_{_{\rm I}}}

\usepackage{color}

\begin{document} 

\large 

\title[3d KHI in Cold Streams]
{Instability of Supersonic Cold Streams Feeding Galaxies III: Kelvin-Helmholtz Instability in Three Dimensions}

\author[Mandelker et al.] 
{\parbox[t]{\textwidth} 
{ 
Nir Mandelker$^{1,2}$\thanks{E-mail: nir.mandelker@yale.edu },
Daisuke Nagai$^{1,3}$,
Han Aung$^3$,
Avishai Dekel$^4$,
Dan Padnos$^4$,
Yuval Birnboim$^4$
} 
\\ \\ 
$^1$Department of Astronomy, Yale University, PO Box 208101, New Haven, CT, USA;\\
$^2$Heidelberger Institut f{\"u}r Theoretische Studien, Schloss-Wolfsbrunnenweg 35, 69118 Heidelberg, Germany;\\ 
$^3$Department of Physics, Yale University, New Haven, CT 06520, USA;\\
$^4$Centre for Astrophysics and Planetary Science, Racah Institute of Physics, The Hebrew University, Jerusalem 91904, Israel}
\date{} 
 
\pagerange{\pageref{firstpage}--\pageref{lastpage}} \pubyear{0000} 
 
\maketitle 
 
\label{firstpage} 
 
\begin{abstract}
We study the effects of Kelvin-Helmholtz instability (KHI) on the cold streams that feed high-redshift galaxies through their hot haloes, generalizing our earlier analyses of a 2D slab to a 3D cylinder, but still limiting our analysis to the adiabatic case with no gravity. We combine analytic modeling and numerical simulations in the linear and non-linear regimes. For subsonic or transonic streams with respect to the halo sound speed, the instability in 3D is qualitatively similar to 2D, but progresses at a faster pace. For supersonic streams, the instability grows much faster in 3D and can be qualitatively different due to azimuthal modes, which introduce a strong dependence on the initial width of the stream-background interface. Using analytic toy models and approximations supported by high-resolution simulations, we apply our idealized hydrodynamical analysis to the astrophysical scenario. The upper limit for the radius of a stream that disintegrates prior to reaching the central galaxy is $\sim 70\%$ larger than the 2D estimate; it is in the range $0.5-5\%$ of the halo virial radius, decreasing with increasing stream density and velocity. Stream disruption generates a turbulent mixing zone around the stream with velocities at the level of $\sim 20\%$ of the initial stream velocity. KHI can cause significant stream deceleration and energy dissipation in 3d, contrary to 2D estimates. For typical streams, up to $10-50\%$ of the gravitational energy gained by inflow down the dark-matter halo potential can be dissipated, capable of powering Lyman-alpha blobs if most of it is dissipated into radiation.
\end{abstract} 
 

%%%%%%%%%%%%%%%%%%%%%%%%%%%%% 
\section{Introduction}
\label{sec:intro}
\smallskip
Dark matter haloes with virial masses $\Mv\gsim 10^{12}\msun$ are predicted to contain hot gas at the 
virial temperature, $\Tv \gsim 10^6\K$, with cooling times exceeding the Hubble time \citep{Rees77,
White78,bd03,db06,Fielding17}. However, during the peak phase of star- and galaxy-formation at redshifts 
$z=1-4$, massive galaxies of $\sim 10^{11}\msun$ in baryons, which are predicted to reside in such haloes, 
exhibit star-formation rates (SFRs) of order $100\sy$ \citep{Genzel06,Forster06,Elmegreen07,Genzel08,Stark08}, 
only a factor of $\sim 2$ below the theoretical cosmological gas accretion rate 
\citep{Dekel09,Dekel13}\footnote{Note that this does not imply that the stellar-to-halo mass ratio is 
only a factor of $\sim 2$ below the cosmic baryon fraction, see \citet{DM14}}. This implies that accreted 
gas must efficiently cool, or never be heated, and penetrate down to the central galaxy. 

\smallskip
According to the developing theoretical picture of galaxy formation, these massive galaxies 
reside at the nodes of the cosmic web, and are penetrated by cosmic filaments of dark matter 
\citep{Bond96,Springel05}. Gas flowing along these filaments is significantly denser 
than the halo gas, allowing it to cool rapidly \citep{db06}. These ``cold streams'' are expected 
to penetrate through the hot circumgalactic medium (CGM) onto the central galaxy while retaining 
a temperature of $\Ts\gsim 10^4\K$, set by the drop in the cooling rate below this temperature 
\citep{SD93}, and by photo-heating by the UV background, though the interiors of streams 
are expected to be at least partly self-shielded \citep{Goerdt10,FG10}.
%According to the standard $\Lambda {\rm CDM}$ cosmological model, the most massive dark matter halos at any epoch lie at the intersections of cosmic web filaments \citep{Zeldovich70,Bond96,Springel05}. At redshift $z=1-4$, when star-formation is at its peak and most of the mass is assembled into galaxies \citep{Madau98,hopkins06,Madau14}, these include halos with a virial mass exceeding $\Mv\sim 10^{12}\Msun$, which contain hot gas at the virial temperature, $\Tv \gsim 10^6\K$ \citep{bd03,db06}. However, gas flowing along the filaments that feed such halos is significantly denser than the halo gas, allowing it to cool rapidly and preventing the formation of a stable accretion shock within the filaments. Instead, these gas streams are expected to penetrate through the hot circumgalactic medium (CGM) onto the central galaxy, all the while remaining cold and dense \citep{db06,Dekel09}, with a typical temeperture of $\Ts\gsim 10^4\K$, set by the steep drop in the cooling rate below that temperature \citep{SD93}.

\smallskip
The above theoretical picture is supported by cosmological simulations 
\citep{Keres05,Ocvirk08,Dekel09,CDB,FG11,vdv11}, which show cold streams 
with diameters of a few to ten percent of the halo virial radius penetrating 
deep into the haloes of massive star-forming galaxies (SFGs). These streams 
supply gas to the haloes at rates of $\sim 100\sy$, comparable to both the 
theoretical cosmological gas accretion rate and the observed SFR in SFGs. 
This implies that cold streams must carry a significant fracton of the 
cosmological gas accretion rate onto the central galaxy \citep{Dekel09,Dekel13}.
%The gas accretion rate via cold streams in simulations is on the order of $\sim 100\sy$, comparable to both the theoretical cosmological gas accretion rate \citep{Dekel09,Dekel13} and the observed star-formation rate (SFR) inSFGs \citep{Genzel06,Forster06,Elmegreen07,Genzel08,Stark08}. This implies that cold streams must carry a significant fracton of the cosmological gas accretion rate onto the central galaxy \citep{Dekel09,Dekel13}. 
% Additionally, the streams are thought to play a key role in the buildup of angular momentum in disk galaxies \citep{Pichon11,Kimm11,Stewart11,Stewart13,Codis12,Danovich12,Danovich15}. 

\smallskip
In cosmological simulations, the streams maintain roughly constant inflow velocities 
as they travel from the outer halo to the central galaxy \citep{Dekel09,Goerdt15a}, 
rather than accelerating in the halo gravitational potential. This indicates that a 
dissipation process acts upon the streams in the CGM, though its source is yet to be 
identified. As the cold streams are likely dense enough to be self-shielded 
from the UV background \citep{Goerdt10,FG10}, they consist mostly of neutral Hydrogen and 
the associated energy loss may be observed as Lyman-$\alpha$ emission \citep{Dijkstra09,Goerdt10,
FG10}, possibly accounting for observed Lyman-$\alpha$ ``blobs" at $z>2$ \citep{Steidel00,
Matsuda06,Matsuda11}. Recent observations have revealed massive extended cold components in 
the CGM of high-redshift galaxies, whose spatial and kinematic properties are consistent with 
predictions for cold streams \citep{Cantalupo14,Martin14a,Martin14b,Borisova16,Fumagalli17,
Leclercq17,Arrigoni18}.The cold streams may also be visible in Lyman-$\alpha$ absorption, 
possibly accounting for several observed systems \citep{Fumagalli11,Goerdt12,vdv12,Bouche13,
Bouche16,Prochaska14}. 
%The simulated streams maintain roughly constant inflow velocities as they travel from the outer halo to the central galaxy \citep{Dekel09,Goerdt15a}, instead of accelerating as expected due to the halo gravitational potential. This indicates that a dissipation process, as yet unidentified, acts upon the streams along the way. The associated energy loss may be observed as Lyman-$\alpha$ emission \citep{Dijkstra09,Goerdt10,FG10}, possibly accounting for Lyman-$\alpha$ ``blobs" observed at $z>2$ \citep{Steidel00,Matsuda06,Matsuda11}. The cold streams, consisting of mostly neutral Hydrogen, may also be visible in Lyman-$\alpha$ absorption, possibly accounting for several observed systems \citep{Fumagalli11,Goerdt12,vdv12,Bouche13,Bouche16,Prochaska14}. Recent observations have revealed massive extended cold components in the CGM of high-redshift SFGs, with spatial and kinematic properties consistent with predictions for cold streams \citep{Cantalupo14,Martin14a,Martin14b,Borisova16,Fumagalli17,Leclercq17,Arrigoni18}.

\smallskip
While there is growing evidence that cold streams play an important role in galaxy 
formation at high redshift, their evolution in the CGM is still a matter of debate. 
In particular, it remains unclear whether the streams indeed penetrate all the way 
to the central galaxy or whether they dissolve or fragment along the way, what fraction 
of their energy is dissipated in the halo and whether this dissipation is observable, 
and what the net effect of all this is on the gas that eventually joins the galaxy. 
Cosmological simulations used to study these issues typically reach a resolution of 
one hundred to a few hundred pc within streams in the outer halo, comparable to the 
stream width. Hydrodynamic and other instabilities at smaller scales are thus not captured 
properly\footnote{Global stream properties such as their radius and mean density may be 
resolved.}, rendering current cosmological simulations ill-suited to investigate the detailed 
evolution of cold streams. This may be the cause of apparent contradictions between properties 
of cold streams predicted by different simulations. For example, simulations using the moving 
mesh code \texttt{AREPO} \citep{Springel10,Vogelsberger12} suggest that streams heat up and 
dissolve at $\gsim 0.5\Rv$ \citep{Nelson13}, contrary to comparable Eulerian AMR \citep{CDB,
Danovich15} and Lagrangian SPH \citep{Keres05,FG10} simulations, where the streams remain cold 
and collimated outside of an interaction region at $\sim 0.25\Rv$. The interpretation of these 
results is uncertain due to insufficient resolution \citep[see also][]{Nelson16}, motivating a 
more careful study of cold stream evolution in the CGM. 

\smallskip
As an alternative to cosmological simulations, in this series of papers we use analytic 
models and idealized simulations, progressively increasing the complexity of our analysis.
% by adding fundamental physical processes one-by-one. 
In two previous papers, \citet{M16}, hereafter M16, and \citet{P18}, hereafter P18, 
we studied the effects of Kelvin-Helmholtz Instability (KHI) on the evolution of cold 
streams. We found that for a reasonable range of stream density, velocity and radius, 
KHI was expected to become highly nonlinear within a virial crossing time, with the 
number of e-foldings of growth experienced by a linear perturbation ranging from $\sim 0.1-100$ 
(M16). A detailed analysis of the nonlinear evolution of KHI in two dimensions revealed 
that sufficiently narrow streams should dissintegrate in the CGM prior to reaching the 
central galaxy (P18). The condition for breakup ranged from $\Rs\lsim 0.003\Rv$ to $\Rs\lsim 0.03\Rv$, 
where $\Rs$ is the stream radius and $\Rv$ is the halo virial radius, with denser, faster 
streams having smaller critical radii for disintegration. However, due to the large stream 
inertia, KHI was found to have only a small effect on the stream inflow rate and a small 
contribution to heating and subsequent Lyman-$\alpha$ cooling emission. 

\smallskip
In this paper, we extend the study of the nonlinear evolution of KHI in cold streams to three 
dimensional cylinders, using both analytic models and numerical simulations. As described in 
detail in \se{theory}, the two dimensional analysis presented in P18 is limited in a number 
of ways, both quantitative and qualitative. Indeed, KHI is known to evolve more rapidly and 
more violently in three dimensions \citep{Bassett95,Bodo98,Xu00}, thus motivating our current 
analysis. 

\smallskip
While several previous works have used numerical simulations to study the nonlinear 
evolution of KHI in cylindrical geometry \citep[e.g.][]{Hardee95,Bassett95,Bodo98,Freund00,Bogey11}, 
almost all of them focussed on light or equidense jets, with $\delta=\rhos/\rhob\le 1$, and none 
of them explored the regime $\delta>10$, relevant for cosmic cold streams. Furthermore, many of 
these studies focussed on spatial, rather than temporal stability analysis (see \se{theory}), and 
are thus not precisely equivalent to our study. A notable exception is \citet{Bodo98} who studied 
the temporal stability of a 3d cylindrical jet with $\delta=10$ and compared it to that of a 2d 
slab. However, only one such simulation was presented and no attempt was made to estimate how 
properties such as stream deceleration or disruption might scale with Mach number or density 
contrast. Furthermore, the resolution in our simulations is much higher than those of \citet{Bodo98}, 
reaching up to 5 times as many grid cells per stream diameter. Our work offers the first comprehensive 
study of temporal nonlinear growth of KHI in dense cylindrical streams, and the first to focus on 
deceleration due to KHI in these systems, providing estimates for the relevant timescales as a 
function of Mach number and density contrast.

\smallskip
The remainder of this paper is organized as follows. In \se{theory} we summarize the theoretical 
understanding of the evolution of KHI in the linear and nonlinear regimes, in 2d and in 3d. In 
\se{methods} we discuss the numerical simulations used to study KHI in 3d cylinders and the techniques 
used for their analysis. In \se{results} we present the results of our numerical analysis and compare 
these to our analytic predictions. In \se{application} we apply the results of our idealized models 
to the astrophysical scenario of cold streams in hot halos. We obtain estimates of the potential 
fragmentation, reduction of inflow rates, and Lyman-$\alpha$ emission due to KHI in cold streams. 
Readers only interested in the astrophysical implications of our analysis rather than the detailed 
hydrodynamics can skip directly to this section without loss of clarity. In \se{phys} we discuss 
the potential effects of additional physics not included in our current analysis, and outline future 
work. We summarize our conclusions in \se{conc}.

%%%%%%%%%%%%%%%%%%%%%%%%%%%%%%%%%%%%%%%%%%%%%%%%  
\section{Analytic Theory}
\label{sec:theory}
\smallskip
In this section, we review the linear theory of KHI in 2d slabs and 3d cylinders, 
and the nonlinear growth of KHI in 2d slabs. We limit this discussion to the 
elements necessary for understanding our current analysis, and refer interested 
readers to M16 and P18 for further details and additional references. We then discuss 
our expectations for the nonlinear behaviour of KHI in 3d cylinders, which will be 
tested with simulations in \se{results}. 

%-------------------------------------------
\subsection{Linear Analysis}
\label{sec:theory_linear}

\smallskip
We consider the case of a cold, dense stream confined in a hot, dilute background, with no 
radiative cooling or gravity. The fluids are characterized by their respective densities and 
speeds of sound, $\rhobs$ and $\cbs$, and are assumed to have an ideal gas equation of state 
with adiabatic index $\gamma=5/3$. We assume that the fluids are in pressure equilibrium. We 
adopt a reference frame where the background is initially stationary, $\Vb=0$, while the stream 
has velocity $\Vs=V$ parallel to the stream-background interface. The instability is dominated 
by two dimensionless parameters: the density contrast, $\delta \equiv \rhos / \rhob$, and the Mach 
number with respect to the background speed of sound, $\Mb \equiv V / \cb$. Due to pressure equilibrium, 
the temperatures and speeds of sound satisfy $\Tb/\Ts=\delta$ and $\cb/\cs=\delta^{1/2}$. The Mach 
number with respect to the stream speed of sound satisfies $\Ms \equiv V/\cs=\delta^{1/2}\Mb$.

\smallskip
We limit our discussion to temporal stability analysis\footnote{Generally, there are two 
approaches to linear stability analysis; \emph{temporal} and \emph{spatial}. In the former, the 
wavenumber $k$ is real while the frequency $\omega$ is complex. This represents seeding the entire 
system with a spatially-oscillating perturbation and studying its \emph{temporal growth}. In the 
latter, $\omega$ is real while $k$ is complex, which represents seeding a temporally-oscillating  
perturbation at the stream origin and studying its downstream \emph{spatial growth}.}, finding 
the growth rates, $\omega_{\rm I}$, as a function of wavenumber, $k=2\pi/\lambda$, for all 
unstable eigenmodes of the system. In the linear regime, unstable eigenmodes grow exponentially 
with time as $\propto {\rm exp}(t/t_{\rm KH})$, where $t_{\rm KH}=\omega_{\rm I}^{-1}$ is the 
\textit{Kelvin-Helmholtz time} of the associated eigenmode. 
%Three variants of the problem are considered:
%\begin{itemize}
%	\item A \emph{planar sheet}, where two semi-infinite fluids are initially separated by a single planar interface at $x=0$. 
%	\item A \emph{planar slab}, where the stream fluid is initially confined to a slab of finite thickness, $-\Rs<x<\Rs$, surrounded by the background fluid. 
%	\item A \emph{cylindrical stream}, where the stream fluid is initially confined to a cylinder of finite radius, $r<\Rs$, surrounded by the background fluid.
%\end{itemize}

\smallskip
The simplest variant of the problem is a \emph{planar sheet}, where two semi-infinite fluids are 
initially separated by a single planar interface at $x=0$. The planar sheet admits unstable eigenmodes 
that spatially decay exponentially away from the initial interface and are therefore called \emph{surface 
modes}. Instability occurs if and only if the Mach number is below a critical value, 
\be 
\label{eq:Mcrit}
\Mb < M_{\rm crit} = \left(1 + \delta^{-1/3}\right)^{3/2}.
\ee
{\no}If \equ{Mcrit} is satisfied, perturbations at all wavelength are unstable with 
$t_{\rm KH}\propto \lambda/V$, while the proportionality constant depends on $\Mb$ 
and $\delta$ (M16).

\smallskip
Two additional, more complicated, variants of the problem are a \emph{planar slab}, 
where the stream fluid is initially confined to a slab of finite thickness, $-\Rs<x<\Rs$, 
surrounded by the background fluid, and a \emph{cylindrical stream}, where the stream 
fluid is initially confined to a cylinder of finite radius, $r<\Rs$, surrounded by the 
background fluid. Both of these also admit surface mode solutions, which converge 
to the same dispersion relation as in the planar sheet in the incompressible ($\Mb\ll 1$), 
short wavelength ($\lambda\lsim\Rs$) limit. Each unstable mode is characterized by a symmetry-order, 
$m$. $m=0$ corresponds to axisymmetric perturbations, called \emph{pinch-modes} or \textit{P-modes}. 
$m=1$ corresponds to antisymmetric perturbations, called \emph{sinusoidal-modes} or \textit{S-modes} 
in the slab, and \emph{helical-modes} in the cylinder. In slab geometry, these are the only two 
symmetry modes. However, a cylinder admits infinitely many symmetry modes, collectively referred 
to as \textit{fluting-modes}. These have $m = 2,3,...$ corresponding to the number of azimuthal 
nodes on the circumference of the cylinder, and azimuthal wavelengths $\lambda_{\varphi}=2\pi\Rs/m$.

\smallskip
When $\Mb>M_{\rm crit}$, surface modes stabilize. However, another class of unstable solutions, called 
\emph{body modes}, are excited. 
%As the Mach number is increased, surface modes stabilize and are replaced by another class of unstable solutions, called \emph{body modes}.
These are associated with waves reverberating between the stream boundaries, forming a 
pattern of nodes inside the stream that resembles standing waves propagating through a waveguide. Body 
modes are unstable if and only if %A necessary condition for body modes to be unstable is 
\be 
\label{eq:Mtot}
M_{\rm tot} \equiv \frac{V}{\cs+\cb} = \frac{\sqrt{\delta}}{1+\sqrt{\delta}}\Mb > 1,
\ee
{\no}which is roughly the opposite of \equ{Mcrit}. The system is therefore \textit{always} unstable, with 
the $(\Mb,\delta)$ parameter space divided into a surface-mode-dominated region and a body-mode-dominated 
region, and a narrow range of parameters allowing coexistence (M16). 

\smallskip
At shorter and shorter wavelengths, an ever-increasing number of unstable body modes appear, 
characterized by the number of transverse nodes in the perturbed variables within the stream. These 
form a discrete set with different frequencies, $\{\omega_{\rm m,n}(k)\}$. For each symmetry-oder $m$, 
the $n=0$ mode is called the \textit{fundamental $m$ mode}, while modes with $n\ge 1$ are referred to 
as \textit{reflected modes}. The \emph{effective} Kelvin-Helmholtz time at a given wavelength is determined 
by the mode with the largest growth rate at that wavelength, the \textit{fastest growing mode}. At short 
wavelengths, $t_{\rm KH}\sim \tsc/{\rm ln}(\beta \Rs/\lambda)$, with $\tsc=2\Rs/\cs$ the stream sound 
crossing time, and $\beta$ a function of $\Mb$ and $\delta$ (M16). While shorter wavelength perturbations 
have  
%an approximation for the complex frequency of the fastest growing mode in both the slab and cylinder is
%\be 
%\label{eq:lin_body}
%\omega \simeq \frac{\sqrt{\delta}}{1+\sqrt{\delta}}kV + i\tsc^{-1}{\rm ln}\left(4M_{\rm tot}\frac{\sqrt{\delta}}{1+\sqrt{\delta}}k\Rs\right),
%\ee
%{\no}where $\tsc=2\Rs/\cs$ is the stream sound crossing time. While shorter wavelength perturbations have 
larger growth rates for both surface and body modes, the dependence on $\lambda$ is weaker for body 
modes, logarithmic rather than linear. In general, the growth rates of body modes are smaller than 
those of surface modes, while for both surface and body modes the instability is attenuated as either 
the Mach number or the density contrast are increased. 

%-------------------------------------------
\subsection{Nonlinear Evolution of Surface Modes}

%-------------------------------------------
\subsubsection{Surface Modes at $\Mb<M_{\rm crit}$}
\label{sec:theory_surface}

\smallskip
We begin by considering 2d slab geometry. Each interface of a slab behaves as a 
\textit{vortex sheet}, with the vorticity perpendicular to the plane. The nonlinear 
behaviour is dominated by vortex mergers, resulting in self-similar growth of the 
shear layer separating the fluids. The width of the shear layer, $h$, evolves as 
\be 
\label{eq:shear_growth}
\frac{h}{\Rs}=\frac{\alpha V t}{\Rs} = \frac{\alpha t}{t_{\rm s}},
\ee
{\no}where $t_{\rm s}=\Rs/V$ is the characteristic time for surface mode evolution, and 
$\alpha$ is a dimensionless growth rate that depends primarily on $M_{\rm tot}$, and is 
typically in the range $\alpha\sim 0.05-0.25$. This behaviour is independent of the initial 
perturbations. An empirical fit to $\alpha$ was proposed by \citet{Dimotakis91}, 
\be 
\label{eq:alpha_fit}
\alpha \simeq 0.21\times \left[0.8{\rm exp}\left(-3 M_{\rm tot}^2\right)+0.2\right].
\ee

\smallskip
The centres of the largest eddies in the shear layer, with sizes of order the shear layer 
thickness $h$, move downstream at the \textit{convection velocity}, 
\be 
\label{eq:convective}
V_{\rm c}=\frac{\sqrt{\delta}}{1+\sqrt{\delta}}V,
\ee
{\no}which can be derived by assuming that in between each pair of eddies there is a 
stagnation point, where the ram pressure from both fluids must be equal \citep{Coles85,
Dimotakis86}. In 2d slabs this corresponds to the center of mass velocity in the shear 
layer (P18). Combining this result with conservation of mass and momentum of material 
entering the shear layer as it expands yields the \textit{entrainment ratio}, the ratio 
of shear layer penetration into the stream to penetration into the background (P18),
\be 
\label{eq:Ev}
E_{\rm v}\equiv \frac{\hs}{\hb} = \delta^{-1/2}.
\ee
{\no}Combining \equs{shear_growth} and \equm{Ev} with $h=\hs+\hb$ yields 
\be 
\label{eq:hs_growth}
\frac{\hs}{\Rs} = \frac{1}{1+\sqrt{\delta}}\frac{\alpha t}{t_{\rm s}},
\ee
\be 
\label{eq:hb_growth}
\frac{\hb}{\Rs} = \frac{\sqrt{\delta}}{1+\sqrt{\delta}}\frac{\alpha t}{t_{\rm s}},
\ee
{\no}Stream disruption occurs when $\hs=\Rs$ so the shear layer encompasses the 
entire stream. This occurs at time (P18)
\be 
\label{eq:tau_diss_2d}
t_{\rm dis,\,surface} = \frac{1+\sqrt{\delta}}{\alpha}t_{\rm s}.
\ee

\smallskip
As the shear layer expands into the background, the initial momentum of the stream is 
distributed over more and more material. The deceleration of the stream thus occurs 
over a characteristic timescale (P18)
\be 
\label{eq:tau_surface_2d}
\tau_{\rm surface,\,2d} = \frac{\delta+\sqrt{\delta}}{\alpha}t_{\rm s} = \sqrt{\delta} ~ t_{\rm dis,\,surface}.
\ee
{\no}This marks the time when the amount of background mass swept up by the shear layer 
is equal to the initial stream mass, and thus corresponds to the time when the stream 
velocity is reduced to half its initial value. This is longer than the disruption timescale 
for any $\delta>1$.

\smallskip
To develop an analogous description of the nonlinear evolution of surface modes in 3d 
cylinders, we model the cylindrical interface between the stream and the background as 
a \textit{vortex ring}, where the vorticity is concentrated entirely in the azimuthal 
direction\footnote{This is clearly true in the unperturbed initial conditions, where 
the only non-zero gradient in the fluid velocity is $\partial v_{\rm z}/\partial r$. 
At $t>0$, the growing perturbations induce motions in the azimuthal direction as well, 
leading to vorticity in all directions. However, these are confined to small scales while 
on large scales the vortex ring structure is preserved (see \fig{vorticity_panel_M1D1})}. 
This is supported by simulations as discussed in \se{surface}. In this model, the fluid 
motion on large scales remains confined to the $z-r$ plane at each azimuthal angle $\varphi$. 
This implies that any cross-section through the stream along its axis (at a constant $\varphi$) 
will appear identical to a planar slab, growing by vortex mergers in the $z-r$ plane. We thus 
predict that \equs{shear_growth}-\equm{tau_diss_2d} will hold for shear layers in cylindrical 
streams as well.

\smallskip
A qualitative difference between shear layer growth in 2d and 3d arrises due to the nature of the energy 
cascade. In 2d, there is only an inverse cascade to larger scales, so the largest eddies remain coherent 
and grow larger as they merge, with $\alpha$ roughly constant throughout the evolution. In 3d, the inverse 
cascade coexists with a direct cascade to smaller scales which breaks up large eddies, generates 
turbulence, and enhances mixing. This may cause the shear layer growth rate to decline, as energy is 
transferred from the largest scales which drive the growth to small scales which drive turbulence and 
generate heat through dissipation. 

\smallskip
Deceleration is expected to occur faster in 3d cylinders than in 2d slabs, because the 
shear layer will sweep up mass at a higher rate as it expands into the background. The 
penetration depth of the shear layer into the background when it has swept up a mass equal 
to the initial stream mass is given by 
\be 
\label{eq:mom_loss_3d}
\rhob \pi [(\Rs+\hb)^2-\Rs^2] L = \rhos \pi \Rs^2 L.
\ee
{\no}Combined with \equ{hb_growth} this yields the expected deceleration timescale for a 3d cylinder, 
\be 
\label{eq:tau_surface_3d}
\begin{array}{c}
\tau_{\rm surface,\,3d} = \dfrac{\left(1+\sqrt{\delta}\right)\left(\sqrt{1+\delta}-1\right)}{\alpha\sqrt{\delta}}t_{\rm s} \\
= \dfrac{\sqrt{1+\delta}-1}{\delta}\tau_{\rm surface,\,2d}.
\end{array}
\ee
{\no}For $\delta=1$, $10$, and $100$, we have $\tau_{\rm surface,\,3d}/\tau_{\rm surface,\,2d}\simeq 0.41$, 
$0.23$, and $0.09$ respectively. Furthermore, while for 2d slabs the deceleration timescale is always longer 
than the disruption timescale, for 3d cylinders we have $\tau_{\rm surface,\,3d}/t_{\rm dis,\,surface}\simeq 0.41$, 
$0.73$ and $0.91$ for $\delta=1$, $10$, and $100$. Stream deceleration is thus predicted to be much more 
significant in 3d cylinders than 2d slabs.

%-------------------------------------------
\subsubsection{Surface Modes at $\Mb>M_{\rm crit}$}
\label{sec:theory_surface_3d}
\smallskip
The largest qualitative difference between 2d and 3d systems is that \textit{the strict 
separation between a surface-mode- and a body-mode-dominated regime is an accurate description 
only in 2d}. In 3d, unstable surface-modes exist at $\Mb>M_{\rm crit}$ as well, associated 
with large values of the azimuthal wave number, $m$. This is because the Mach number determining 
surface mode stability in \equ{Mcrit} corresponds to the velocity component parallel to the 
perturbation wave-vector, $V_{\rm k} = \vec{V}\cdot \hat{k}$ (M16). In 2d systems, $V_{\rm k}=V$ 
by definition. However, in cylindrical geometry, modes with $m>1$ have components in the $\hat{\varphi}$ 
direction, while the velocity is in the $\hat{z}$ direction. As a result, the effective Mach number 
associated with an azimuthal mode number $m$ is reduced by a factor $[1+(m/K)^2]^{-1/2}$, 
where $K=k\Rs$, and surface modes are unstable for
\be 
\label{eq:mcrit}
m>m_{\rm crit}=K[(\Mb/M_{\rm crit})^2-1]^{1/2}.
\ee
{\no}For example, for wavelengths $\lambda\sim 2\Rs$ and $\Mb\sim 1.5 M_{\rm crit}$ (reasonable 
for cold streams), \equ{mcrit} predicts that surface modes will be unstable for $m\gsim 4$. Note 
that $m_{\rm crit}$ depends on both $\Mb$ and $\delta$ through $M_{\rm crit}$, as well as on $\lambda/\Rs$ 
through $K$, making the overall behaviour very complicated. Regardless of the value of $m$, we expect 
unstable surface modes to behave according to the description in \se{theory_surface}. 

\begin{table} 
	\centering
	\caption{
	Critical wavelengths and corresponding linear growth rates for stream disruption 
	due to body modes, for the $(\Mb,\delta)$ parameters presented in this \se{body}. The 
	columns show, from left to right, the Mach number of the stream with respect to the 
	background sound speed, $\Mb$, the density contrast between the stream and the background, 
	$\delta$, the critical wavelength for stream disruption due to body modes in units of the 
	stream radius, $\lambda_{\rm crit}/\Rs$, for 2d slabs and 3d cylinders, and the corresponding 
	linear growth times in units of the sound crossing time, $\tkh/\tsc$, for 2d slabs and 3d 
	cylinders. This growth time can be used in \equ{tNL} to estimate the time for stream disruption 
	due to body modes. For 2d slabs we have $\tkh\sim\tsc$ while for 3d cylinders we have 
	$\tkh\sim 0.5\tsc$}
	\label{tab:body2}
	\begin{tabular}{cccccc}
		\hline
		$\Mb$ & $\delta$ & $\lambda_{\rm crit,2d}/\Rs$ & $\lambda_{\rm crit,3d}/\Rs$ & $t_{\rm KH,2d}/\tsc$ & $t_{\rm KH,3d}/\tsc$ \\
		\hline
		5.0 & 1   & 17 & 13  & 1.17 & 0.75 \\
		2.5 & 5   & 14 & 9   & 0.91 & 0.55 \\
		2.0 & 10  & 11 & 8.5 & 0.86 & 0.54 \\
		2.5 & 20  & 12 & 11  & 0.93 & 0.53 \\
		2.0 & 100 & 13 & 11  & 0.86 & 0.51 \\
		\hline
	\end{tabular}
\end{table}

%-------------------------------------------
\subsection{Nonlinear Evolution of Body Modes}
\label{sec:theory_body}
\smallskip
%By their very nature, body modes are related to the global scales of the entire stream, namely its radius, $\Rs$, and sound crossing time, $\tsc$. The existence of these preferred scales means that the evolution of body modes depends on the initial conditions, as opposed to the scale free nature of surface modes. 
Nonlinear evolution of body modes occurs through global deformation of the stream. Locally, deformation 
is measured by the radial displacement of a Lagrangian fluid element, $\xi$. For a given eigenmode in 
the linear regime, it can be shown that inside the stream\footnote{See Appendix C in P18 for the derivation 
in the slab case. The derivation in the cylinder case is analogous, and follows from section 2.4 in M16.} 
\be 
\label{eq:xi_fourier}
\xi_{\rm m,n}=H_0\frac{\mathcal{I}_{\rm m}(q_{\rm s;\,m,n} r)}{\mathcal{I}_{\rm m}(q_{\rm s;\,m,n} \Rs)}{\rm exp}[i(kz+m\varphi-\omega_{\rm m,n} t)],
\ee
{\no}where $H_0$ is the maximal displacement of the stream-background interface at $t=0$, and 
\be 
\label{eq:qs}
q_{\rm s;\,m,n} = k\left[1-\left(\dfrac{\omega_{\rm m,n}-kV}{k\cs}\right)^2\right]^{1/2}.
\ee
%{\no}For a cylinder, $m$ is any non-negative integer and $\varphi\in [0,2\pi)$, while for a slab $r=|x|$, 
{\no}For a slab\footnote{\Equ{xi_fourier} is valid for a slab as well as a cylinder by taking $r=|x|$, 
and $\varphi=0$ or $\pi$ for $x>0$ or $x<0$ respectively.}, $\mathcal{I}_{\rm 0}={\rm cosh}$ 
and $\mathcal{I}_{\rm 1}={\rm sinh}$, while for a cylinder, $\mathcal{I}_{\rm m}$ is the $m$-th 
order modified Bessel function of the first kind. Note that $\xi$ is complex, so the physical 
displacement of the fluid is given by the real part of \equ{xi_fourier}. 

\smallskip
Due to the non-trivial form of $\xi$, as it grows in amplitude two fluid elements will eventually 
cross somewhere inside the stream. This fluid crossing will lead to shocks, and marks the transition 
to the nonlinear regime, where the eigenmode in question ceases to grow exponentially and temporarily 
saturates. While shorter wavelength perturbations grow more rapidly in the linear regime, %(\equnp{lin_body}), 
they transition to non-linearity and saturate at smaller amplitudes \citep[][P18]{Hardee95,Hardee97}. This 
can be intuitively understood by realising that in the linear regime, $\xi\propto u_{\rm r}/\omega$, where 
$u_{\rm r}$ is the perturbation in the radial velocity. Transition to non-linearity occurs when $u_{\rm r}\sim \cs$, 
and therefore $\xi\propto \omega^{-1}$, which increases towards longer wavelengths. In addition, eigenmodes 
with larger $n$ tend to saturate at smaller amplitudes \citep[][P18]{Hardee95,Hardee97,Bodo98}, because 
they have more nodes across the stream diameter, leaving less room for a fluid element to move before it 
crosses such a node. 
%In almost all cases, it is the fundamental mode, $n=0$, which achieves the largest amplitude in the linear regime (P18, appendix D). 
For similar reasons, modes with larger $m$ saturate at smaller amplitudes \citep{Hardee95}. 
%However, in most cases, it is the helical mode, $m=1$, that saturates at the largest amplitude, while the pinch mode, $m=0$, is secondary. Modes with $m\ge 2$ reach smaller and smaller amplitudes as $m$ is increased. 

\smallskip
For a given eigenmode and wavelength, we define the maximal displacement of 
the stream-background interface at the time when fluid crossing first occurs 
inside the stream as $H_{\rm NL}$. 
%Note that this does not depend on the initial amplitude of the perturbation. 
Since perturbations grow exponentially in the linear regime, the time at which 
this occurs is given by 
\be 
\label{eq:tNL}
t_{\rm NL} = t_{\rm KH}{\rm ln}(H_{\rm NL}/H_0),
\ee
{\no}where $H_0$ is the initial displacement at $t=0$. A mode is disruptive to the stream if it has 
$H_{\rm NL}\ge \Rs$. Consequently, the mode expected to ultimately break the stream, hereafter 
the \textit{critical mode}, is the fastest growing among those with $H_{\rm NL}\ge \Rs$. For slabs 
spanning a wide range of $(\Mb,\delta)$ values typical of cold streams, the critical mode is the 
fundamental ($n=0$) S-mode with a wavelength of $\lambda\sim 10-20\Rs$ and $t_{\rm KH}\simeq \tsc$ (P18 
and \tab{body2}). 
At $t\gsim t_{\rm NL}$, the slab becomes dominated by a large scale sinusoidal perturbation whose amplitude 
reaches $\sim 5-10\Rs$ within $1-2\tsc$, effectively disrupting the stream. The timescale for stream disruption 
due to body modes is thus 
\be 
\label{eq:tau_break_body}
t_{\rm dis,\,body,\,2d} \simeq \tsc\left[1+{\rm ln}(\Rs/H_0)\right].
\ee
{\no}Note that this depends on the initial displacement amplitude of the stream-background interface, $H_0$, 
in stark contrast to the corresponding expression for surface modes, \equ{tau_diss_2d}, which depended only 
on the unperturbed initial conditions, $(\Mb,\delta,\Rs)$.

\begin{figure}
\includegraphics[trim={0.4cm 0.0cm 1.3cm 0.0cm}, clip, width =0.47 \textwidth]{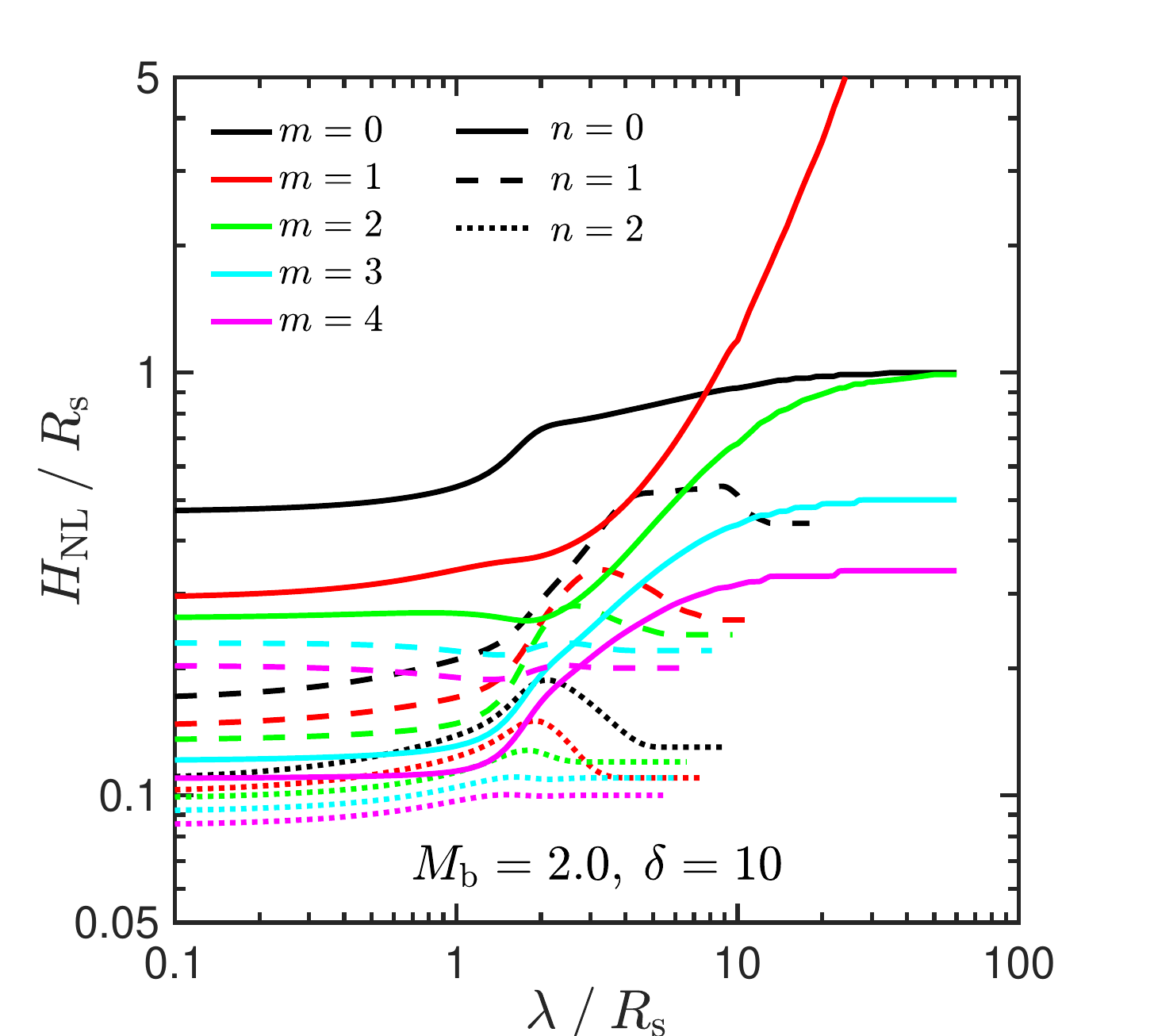}
\caption{
Transition to nonlinearity of different body modes as a function of wavelength. Shown is 
$H_{\rm NL}$, the maximal displacement of the stream-background interface at the time when fluid 
crossing first occurs inside the stream, as a function of the perturbation wavelength (both normalized 
to the stream radius, $\Rs$) for the case $(\Mb,\delta)=(2,10)$. Different coloured lines correspond 
to different azimuthal symmetries, $m=0-4$. Different line types correspond to different radial modes, 
solid lines for the fundamental mode ($n=0$), dashed and dotted lines for the first and second reflected 
modes ($n=1$ and $2$) respectively. In general, $H_{\rm NL}$ decreases as either $n$ or $m$ are increased. 
A mode is considered disruptive to the stream if it reaches $H_{\rm NL}\ge \Rs$. Only the $(m,n)=(1,0);\,(0,0);\,(2,0)$ 
modes fulfil this criterion, at wavelengths of $\lambda/\Rs\gsim 8,\,20$, and $40$ respectively. 
The corresponding KH times for these modes at these wavelengths are $\tkh\sim 0.53,\,3.85$, and 
$1.76\tsc$. Therefore, if all modes begin with similar amplitudes, the $(1,0)$ mode will reach $H_{\rm NL}=\Rs$ 
first, making it the critical mode. Similar results are found for all $(\Mb,\delta)$ values relevant 
for cold streams. 
}
\label{fig:HNL} 
\end{figure}

\smallskip
Numerical evaluation of $H_{\rm NL}$ and $t_{\rm NL}$ for 3d cylinders spanning a wide range 
of $(\Mb,\delta)$ values typical of cold streams reveals that the critical mode is always the 
fundamental helical mode, $(m,n)=(1,0)$ similar to the 2d slab, though at slightly shorter 
wavelengths, $\lambda\sim 8-16\Rs$. An example of this is shown in \fig{HNL}. The KH times of 
the critical perturbations in 3d turn out to always be roughly a factor of 2 shorter than in 
2d, namely $t_{\rm KH}\sim 0.5\tsc$ (\tab{body2}). \textit{This leads to faster stream 
disruption in 3d compared to 2d,} on a timescale 
\be 
\label{eq:tau_break_body_3d}
t_{\rm dis,\,body,\,3d} \simeq \tsc\left[1+0.5~{\rm ln}(\Rs/H_0)\right].
\ee

\smallskip
At wavelengths $\sim 10\Rs$, close to the critical wavelength, \equ{mcrit} predicts that surface modes 
are unstable for $m\ge 1$ if $\Mb/M_{\rm crit}\lsim 2$, which is almost certainly the case for cold streams. 
\textit{The evolution at $t>t_{\rm NL}$ may thus be qualitatively different in 3d than 2d, characterised 
by much more efficient mixing.}

\smallskip
Regarding stream deceleration, at $t<t_{\rm NL}$, the stream velocity remains roughly constant. 
At $t>t_{\rm NL}$, the critical perturbation bends the stream into a sinusoidal shape, effectively 
driving a piston through the background medium at every crest of the sinusoid. This produces a 
periodic pattern of weak shocks propagating away from the stream at approximately the speed of 
sound, $\cb$, and fascilitating the transfer of momentum from the stream to the background. The 
characteristic timescale for this process is the time it takes such an outward propagating wave 
to encounter a mass of background fluid equal to the initial stream mass. For 2d slabs this means 
$\cb\tau_{\rm body}\rhob=\Rs\rhos$, so 
\be 
\label{eq:tau_body_2d}
\tau_{\rm body,\,2d} = \delta \frac{\Rs}{\cb} = \frac{\sqrt{\delta}}{2}\tsc.
\ee
{\no}For 3d cylinders this means $[(\Rs+\cb\tau_{\rm body})^2-\Rs^2]\rhob=\Rs^2\rhos$, so 
\be 
\label{eq:tau_body_3d}
\tau_{\rm body,\,3d} = \left[\sqrt{1+\delta}-1\right]\frac{\Rs}{\cb} = \frac{\sqrt{1+\delta}-1}{2\sqrt{\delta}}\tsc.
\ee
For $\delta=1,\,10,\,100$, $\tau_{\rm body,\,3d}/\tau_{\rm body,\,2d}\simeq 0.41,\,0.23,\,0.09$ respectively, 
the same as for surface modes, since in both cases the difference between $\tau_{\rm 3d}$ and $\tau_{\rm 2d}$ 
arrises from a purely geometrical effect. Furthermore, while $\tau_{\rm body,\,2d}>\tsc$ for $\delta>4$ and 
reaches $4\tsc$ for $\delta=100$, $\tau_{\rm body,\,3d}<\tsc$ for any $\delta$, with an assymptotic value of 
$0.5\tsc$ for $\delta\rightarrow \infty$. It is also worth noting that in both 2d and 3d,
\be 
\label{eq:tau_ratio}
\frac{\tau_{\rm surface}}{\tau_{\rm body}} = \frac{1+\delta^{-1/2}}{\alpha \Mb}.
\ee
{\no}We conclude that stream deceleration due to body modes is much more rapid in 3d than in 2d, especially 
for dense streams, similar to the conclusion reached for surface modes. 

\smallskip
The deceleration process described above is not very efficient. Slabs initially loose roughly $12\%$ of 
their velocity per $\tau_{\rm body,\,2d}$ at $t\gsim t_{\rm NL}$, but once the stream breaks this decreases 
to $\sim 1.6\%$ per $\tau_{\rm body,\,2d}$, before the velocity has reached half its initial value (P18). 
The origin of these efficiencies is unclear, and we cannot predict from first principles 
what they might be for 3d cylinders. However, so long as they are not significantly smaller the stream should 
reach half its initial velocity before the deceleration rate decreases, due to the much smaller ratio of 
$\tau_{\rm body,\,3d}$ to $\tsc$. This will further enhance the effective deceleration rate due to body modes 
in 3d compared to 2d. 

%%%%%%%%%%%%%%%%%%%%%%%%%%%%%%%%%%%%%%%%%%%%%%%%  
\section{Numerical Methods}
\label{sec:methods}
\smallskip
In this section we describe the details of our simulation code and setup, as well 
as our analysis method. In \se{methods-comp} we compare our setup to that used in P18.

\subsection{Hydrodynamic Code}
\label{sec:methods-cod}
\smallskip
We use the Eulerian AMR code \texttt{RAMSES} \citep{Teyssier02}, with a piecewise-linear 
reconstruction using the MonCen slope limiter \citep{vanLeer77} and an HLLC approximate 
Riemann solver \citep{Toro94}, identical to M16 and P18. 

%-------------------------------------------
\subsection{Unperturbed Initial Conditions}
\label{sec:methods-unperturbed}
\smallskip
One of our main goals is to compare the non-linear evolution of KHI in a two dimensional slab 
and a three dimensional cylinder under identical initial conditions. For slabs, the simulation 
domain is a square of side $L=1$, representing the $xz$ plane, extending from $-0.5$ to $0.5$ 
in both the $x$ and $z$ directions. The slab is centered at $x=0$, such that the stream fluid 
occupies the region $|x|<\Rs$, while the background fluid fills the rest of the domain. Analogously, 
for cylinders, the simulation domain is a cube of side $L=1$, extending from $-0.5$ to $0.5$ in 
all directions. The cylinder axis is placed along the $z$ axis, at $r=0$.
%, where $r=(x^2+y^2)^{1/2}$ is the cylindrical radial coordinate in the $xy$ plane. 
The stream fluid occupies the region $r<\Rs$ while the background fluid occupies the rest of the 
domain. We hereafter use standard cylindrical coordinates, $(r,\varphi,z)$, when discussing both 
3d and 2d simulations, with the convention that for 2d simulations, $r=|x|$ and $\varphi=0$ or 
$\pi$ for $x>0$ and $x<0$ respectively. 

\smallskip
We set the stream radius to $\Rs=1/32$ in all simulations. Both fluids are 
ideal gasses with adiabatic index $\gamma=5/3$, and initial uniform pressure $P_0=1$. The background 
is initialized with density $\rhob=1$ and velocity $\vec{v}_{\rm b}=0$. The stream is initialized 
with density $\rhos=\delta$ and velocity $\vec{v}_{\rm s}=V \hat{z}=\Mb\cb \hat{z}$, where $\cb=\sqrt{5/3}$ 
is the background sound speed in simulation units. Since the stream and the background are initially 
in pressure equilibrium, the sound speed in the stream is $\cs=\delta^{-1/2}\cb$.

\smallskip
In the setup described above, the density and velocity are discontinuous at the interface between 
the stream and the background. This generates numerical perturbations at the grid scale, which grow 
faster than the intended perturbations in the linear regime, and may dominate the instability at late 
times depending on their amplitude. Furthermore, since smaller scales grow more rapidly in the linear 
regime, these numerical perturbations become more severe as the resolution is increased, preventing 
convergence of the solution \citep{Robertson10}. This is remedied by smoothing the unperturbed density 
and velocity profiles around the interfaces using the ramp function proposed by \citet{Robertson10}, 
which was also used in M16 and P18, 
\be 
\label{eq:ramp1}
f(r)=f_{\rm b} + \left(f_{\rm s}-f_{\rm b}\right)\times \mathcal{R}(r),
\ee 
\be 
\label{eq:ramp2}
\mathcal{R}(r)=\frac{1}{2}\left[1-{\rm tanh}\left(\frac{r-\Rs}{\sigma}\right)\right],
\ee
{\no}where $f$ stands for either $\rho$ of $\vec{v}$. This yields $f=f_{\rm s}$ inside 
the stream, at $r<\Rs$, while $f=f_{\rm b}$ in the background. The parameter $\sigma$ 
determines the width of the transition zone. The function $\mathcal{R}(r)$ transitions 
from $0.05$ to $0.95$ over a full width of $\sim 3\sigma$ in $(r-\Rs)$. 
For the surface mode simulations presented in \se{surface} we adopt $\sigma=\Rs/32$. 
For the body mode simulations presented in \se{body} we use values of $\sigma$ ranging 
from $\Rs/32$ to $\Rs/8$ (\tab{body}). 

%-------------------------------------------
\subsection{Boundary Conditions}
\label{sec:methods-boundary}
\smallskip
We use periodic boundary conditions at $z=\pm 0.5$, and outflow boundary 
conditions at $x=\pm 0.5$ and $y=\pm 0.5$ (for 3d simulations), such that 
gas crossing the boundary is lost from the simulation domain. The boundary 
conditions at $x,\,y=\pm 0.5$ may affect the interface region once a 
sound crossing time from the interface to the boundary has elapsed. For an 
interface at $r=\Rs$, the minimal time for this interaction to occur (assuming 
shocks in the background dissolve into sound waves quickly) is 
$T_{\rm box}=(L-2\Rs)/\cb \sim  0.73$ in simulation units for $\Rs=1/32$. 
All of our simulations were run for between 10-20 stream sound crossing 
times, $T_{\rm end}=10-20\tsc=(10-20)\Rs/\cs\sim (0.48-0.96) \delta^{1/2}$. 
For $\delta=1$, this is typically less than $T_{\rm box}$, so our 
results are not influenced by the outflow boundary conditions. However, for 
$\delta=100$ we have $T_{\rm end}\sim (6.6-13.2) T_{\rm box}$. While we 
do not explicitly test the influence of the boundary conditions in our 3d 
simulations, we showed in P18 that this is negligible in 2d simulations with 
comparable ratios of $T_{\rm end}$ to $T_{\rm box}$. 

%-------------------------------------------
\subsection{Computational Grid}
\label{sec:methods-grid}
\smallskip
We used a statically refined grid in all runs, with the resolution gradually decreasing 
away from the stream axis. For most of our runs, the highest resolution region was 
${\rm max}(|x|,|y|)<3\Rs$, with cell size $\Delta=2^{-11}$. For $\Rs=1/32$ this corresponds 
to $\Delta=\Rs/64$, or 128 cells per stream diameter. The cell size increases by a factor 
of 2 every $3\Rs$ in the $x$ and $y$-directions, up to a maximal cell size of $2^{-7}$. The 
resolution is uniform along the $z$ direction, parallel to the stream axis. 

\smallskip
Overall, our results are converged in terms of the computational grid. We have tested the 
dependence of our results to increasing or decreasing the resolution by a factor of 2, such 
that the cell sizes range from $2^{-12}-2^{-8}$ or $2^{-10}-2^{-6}$ respectively. We have also 
tested the effect of changing the refinement intervals from $3\Rs$ to $1.5\Rs$. These results 
are presented in Appendix \se{convergence}. 

%-------------------------------------------
\subsection{Perturbations} 
\label{sec:methods-pert}
\smallskip
We initialize nearly all of our simulations with a random realization of periodic 
perturbations in the radial component of the velocity, $v_{\rm r}$. In practice, 
we initialize the following perturbations in the Cartesian components of the velocity, 
\be 
\label{eq:pertx}
\begin{array}{c}
v_{\rm x}^{\rm pert}(r,\varphi ,z) = \sum_{j=1}^{N_{\rm pert}} v_{\rm 0,j}~{\rm cos}\left(k_{\rm j}z+m_{\rm j}\varphi + \phi_{\rm j}\right)\\
\\
\times {\rm exp}\left[-\dfrac{(r-\Rs)^2}{2\sigma_{\rm pert}^2}\right]{\rm cos}\left(\varphi\right)
\end{array},
\ee
\be 
\label{eq:perty}
\begin{array}{c}
v_{\rm y}^{\rm pert}(r,\varphi ,z) = \sum_{j=1}^{N_{\rm pert}} v_{\rm 0,j}~{\rm cos}\left(k_{\rm j}z+m_{\rm j}\varphi + \phi_{\rm j}\right)\\
\\
\times {\rm exp}\left[-\dfrac{(r-\Rs)^2}{2\sigma_{\rm pert}^2}\right]{\rm sin}\left(\varphi\right)
\end{array}.
\ee
{\no}In 3d, this results in a perturbation to $v_{\rm r}=v_{\rm x}{\rm cos}(\varphi)+v_{\rm y}{\rm sin}(\varphi)$. 
In 2d, since $\varphi$ is always $0$ or $\pi$, $v_{\rm y}^{\rm pert}=0$ everywhere, while each 
individual mode, $j$, in $v_{\rm x}^{\rm pert}$ contributes to both of the slab interfaces. 
As discussed below, in one simulation we initilized perturbations in the stream-background interface 
rather than the radial velocity. This has the form 
\be 
\label{eq:pertr}
r_{\rm s} = \Rs\left[1 + \sum_{j=1}^{N_{\rm pert}} h_{\rm 0,j}~{\rm cos}\left(k_{\rm j}z+m_{\rm j}\varphi + \phi_{\rm j}\right)\right].
\ee

\smallskip
The velocity perturbations are localized on the stream-background interface, with a penetration depth set by the parameter 
$\sigma_{\rm pert}$. In all runs with $\sigma=\Rs/32$ in \equ{ramp2} we used $\sigma_{\rm pert}=\Rs/16$, 
while in the runs with $\sigma=\Rs/8$ presented in \se{body} we used $\sigma_{\rm pert}=\Rs/8$. However, we experimented 
with both $\sigma_{\rm pert}=\Rs/16$ and $\Rs/8$ and found no noticeable difference in our results. 

\smallskip
To comply with periodic boundary conditions, all wavelengths were harmonics of the box length, $k_{\rm j}=2\pi n_{\rm j}$ 
where $n_{\rm j}$ is an integer, corresponding to a wavelength $\lambda_{\rm j}=1/n_{\rm j}$. In each simulation, we 
include all wavenumbers in the range $n_{\rm j}=16-64$, corresponding to all available wavelengths in the range 
$\Rs/2 - 2\Rs$. As discussed below, in one simulation we expanded the wavenumber range to $n_{\rm j}=2-64$, 
corresponding to all available wavelengths in the range $\Rs/2 - 16\Rs$.

\smallskip
Each perturbation mode is also assigned a symmetry mode, represented by the index $m_{\rm j}$ in \equs{pertx} and 
\equm{perty}, and discussed in \se{theory}. In order to initialize self-consistent perturbations in 2d and 3d, in 
nearly all simulations we only consider $m=0,1$, as these are the only symmetry modes available in 2d. For each 
wavenumber $k_{\rm j}$ we include both an $m=0$ mode and an $m=1$ mode. This results in a total of 
$N_{\rm pert}=2\times 49=98$ modes per simulation. As discussed below, we also performed one 3d simulation where we 
initialized the full range $m=0-4$ for each wavenumber $k_{\rm j}$, resulting in $N_{\rm pert}=5\times 49=245$ modes. 

\smallskip
Each mode is then given a random phase $\phi_{\rm j} \in [0,2\pi)$. The stochastic variability from changing the 
random phases was extremely small, and is discussed in Appendix \se{convergence}. The amplitude of each mode, 
$v_{\rm 0,j}$ was identical, resulting in a white noise specturm. We set the normalization such that the rms 
amplitude was $0.01\cs$. In the one case where the stream-background interface was perturbed, the rms amplitude 
was set to $0.1\Rs$.

\smallskip
In Appendix \se{linear_sims} we demonstrate that our numerical setup properly captures the behaviour of KHI in 
cylindrical geometry in the linear regime, both in terms of the linear growth rates, and the convergence to 
eigenmodes. This serves both as a validation of our code and numerical setup, as well as a test of the predictions 
presented in M16.

%-------------------------------------------
\subsection{Tracing the Two Fluids} 
\label{sec:methods-trace}
\smallskip
In order to track the expansion of the stream into the background and the mixing of the two fluids, our 
simulations include a passive scalar field, denoted by $\psi(r, \varphi , z, t)$. The passive scalar is 
initialized such that $\psi=1$ in the stream and $\psi=0$ in the background. Since this field is advected with 
the flow, it serves as a Lagrangian tracer for the fluid in the simulation (which is Eulerian). An element 
characterized by passive scalar value $\psi$, density $\rho$ and volume $\dV$, contains a mass of stream and 
background fluid given by 
\be 
\label{eq:mass-stream-background}
\dms = \psi \rho \dV \qquad \text{and} \qquad \dmb = (1-\psi) \rho \dV.
\ee

\smallskip
Following P18, we use the passive scalar to define the edges of the perturbed region around the initial interface. 
The volume-weighted average radial profile of the passive scalar in 3d simulations is given by 
\be 
\label{eq:volume-averaged-colour-3d}
\overline{\psi}_{\rm 3d}(r,t) = \frac{\int_{-L/2}^{L/2}\int_{0}^{2\pi} \psi_{(r,\varphi ,z,t)} r~{\rm d\varphi \,dz}}{2\pi r L},
\ee
{\no}while in 2d simulations it is given by 
\be 
\label{eq:volume-averaged-colour-2d}
\overline{\psi}_{\rm 2d}(r,t) = \frac{\int_{-L/2}^{L/2} \left[\psi_{(x,z,t)}+\psi_{(-x,z,t)}\right]~{\rm dz}}{2L},
\ee
{\no}where as before $r=|x|$. We hereafter omit the subscript 2d or 3d and simply use $\overline{\psi}(r,t)$ 
with the dimensionality being clear from the context.

\begin{table} 
	\centering
	\caption{
	Parameters of simulations studying the nonlinear evolution of surface modes. 
	The columns show, from left to right, the Mach number of the stream with respect to the 
	background sound speed, $\Mb$, the density contrast between the stream 	and the background, 
	$\delta$, the Mach number of the stream with respect to the sum of the two sound speeds, 
	$M_{\rm tot}$ (\equnp{Mtot}), the number of cells per stream radius, $\Rs/\Delta$ where 
	$\Delta$ is the cell size in the highest resolution region, and the refinement scheme, 
	where the cell size is increased by a factor of 2 every $\Delta_{\rm ref}$ in the $x$ and 
	$y$-directions. The high resolution region is thus ${\rm max}(|x|,|y|)<\Delta_{\rm ref}$. 
	The smoothing width parameter in \equ{ramp2} was $\sigma=\Rs/32$ in all cases. The first 
	eight rows, above the horizontal line, represent our fiducial runs presented in this section. 
	The following five rows represent tests to check convergence with respect to resolution and 
	the refinement strategy, and are presented in Appendix \se{convergence}.}
	\label{tab:surface}
	\begin{tabular}{ccccc}
		\hline
		$\Mb$ &$\delta$ &$M_{\rm tot}$ &$\Rs/\Delta$ &$\Delta_{\rm ref}/\Rs$ \\
		\hline
		0.1 & 1    & 0.05 & 64 & 3.0 \\
		0.5 & 1    & 0.25 & 64 & 3.0 \\
		1.0 & 1    & 0.50 & 64 & 3.0 \\
		1.5 & 1    & 0.75 & 64 & 3.0 \\
		0.5 & 10   & 0.38 & 64 & 3.0 \\
		1.0 & 10   & 0.76 & 64 & 3.0 \\
		0.5 & 100  & 0.45 & 64 & 3.0 \\
		1.0 & 100  & 0.91 & 64 & 3.0 \\
		\hline
		1.0 & 1    & 0.50 & 32  & 1.5 \\
		1.0 & 1    & 0.50 & 64  & 1.5 \\
		1.0 & 1    & 0.50 & 128 & 1.5 \\
		1.0 & 10   & 0.76 & 32  & 3.0 \\
		1.0 & 100  & 0.91 & 32  & 3.0 \\	  	 	  	
		\hline
	\end{tabular}
\end{table}

\begin{figure*}
\begin{center}
\includegraphics[trim={0.0cm 1.238cm 3.3cm 0}, clip, width =0.393 \textwidth]{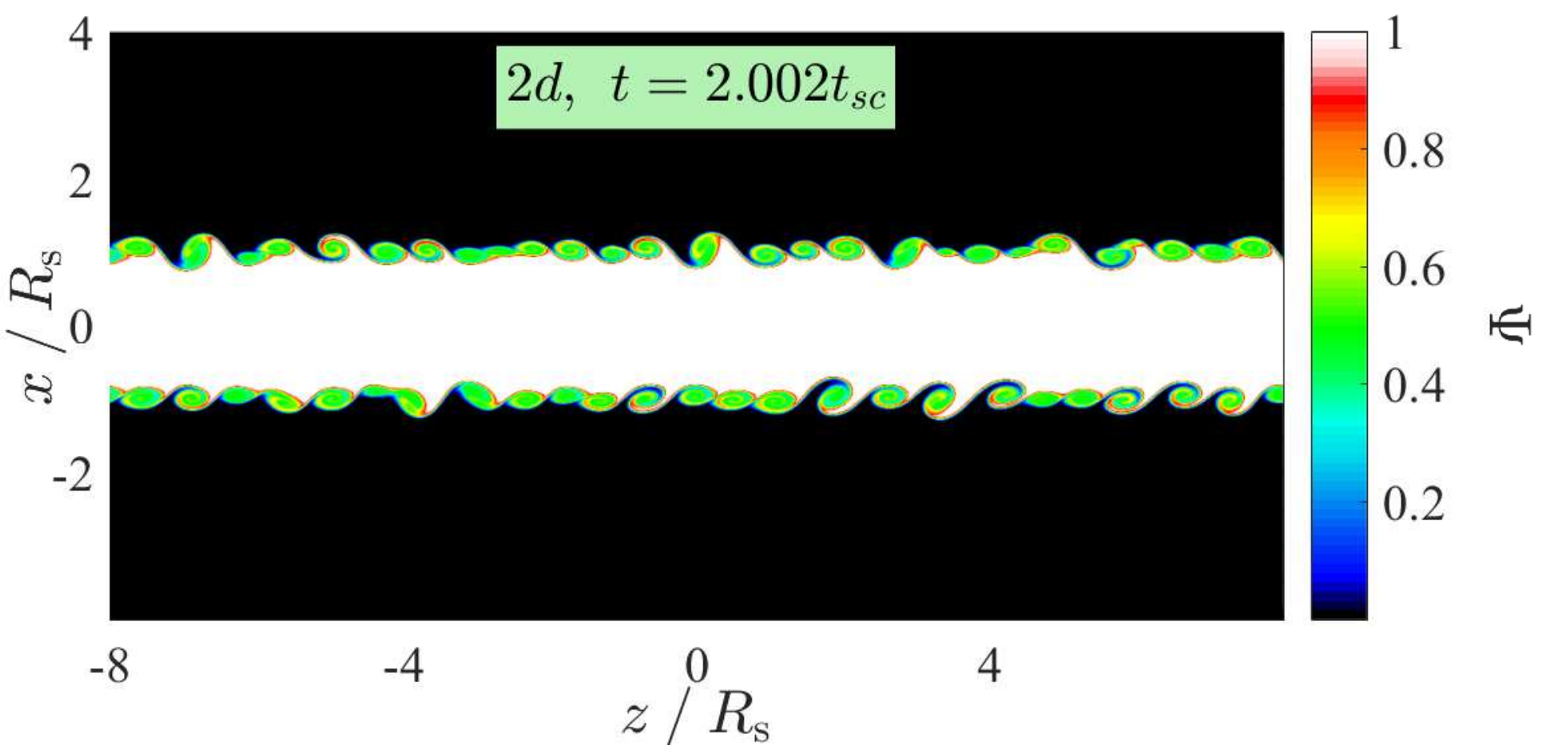}
\hspace{-0.3cm}
\includegraphics[trim={1.3cm 1.238cm 3.3cm 0}, clip, width =0.363 \textwidth]{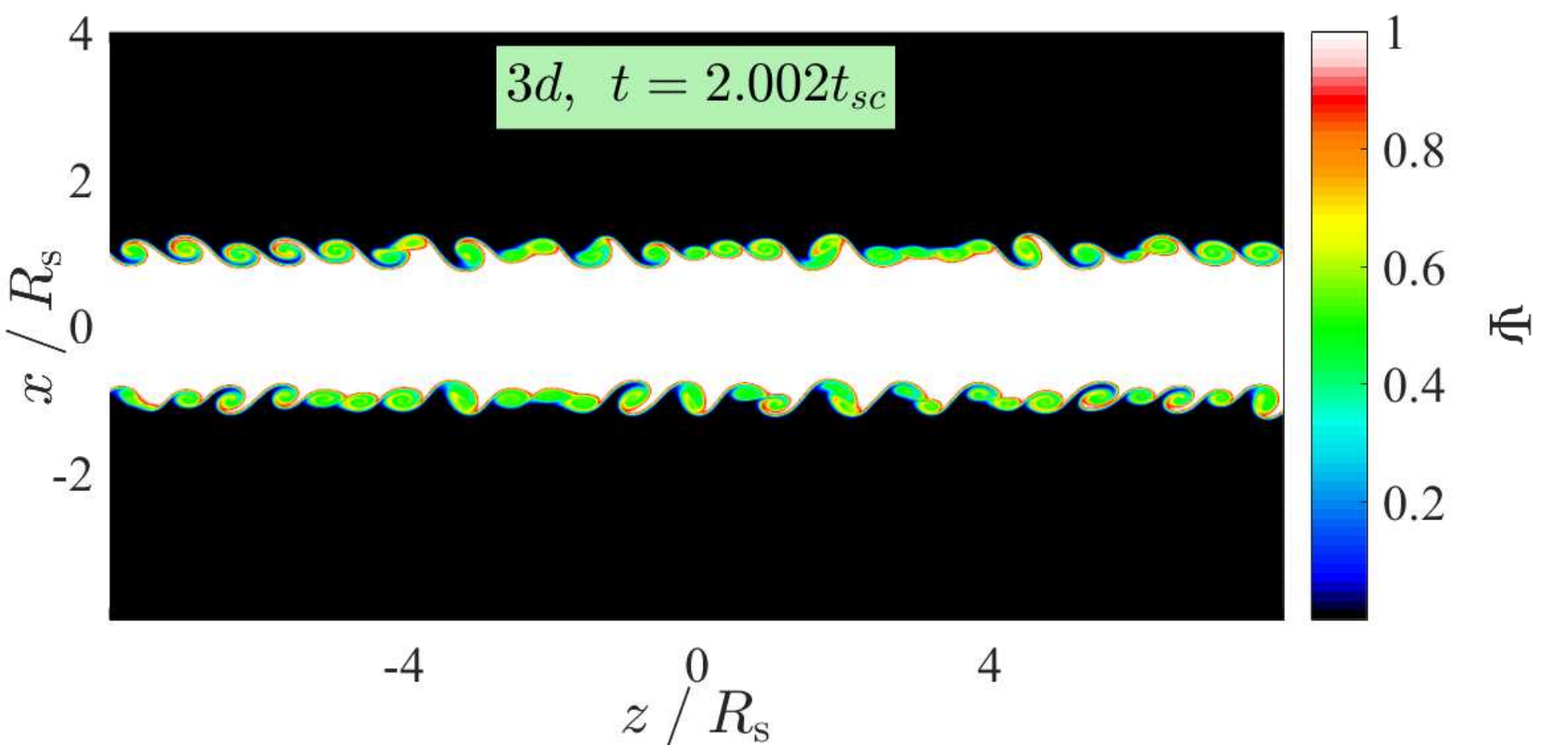}
\hspace{-0.29cm}
\includegraphics[trim={5.1cm 1.238cm 4.05cm 0}, clip, width =0.257 \textwidth]{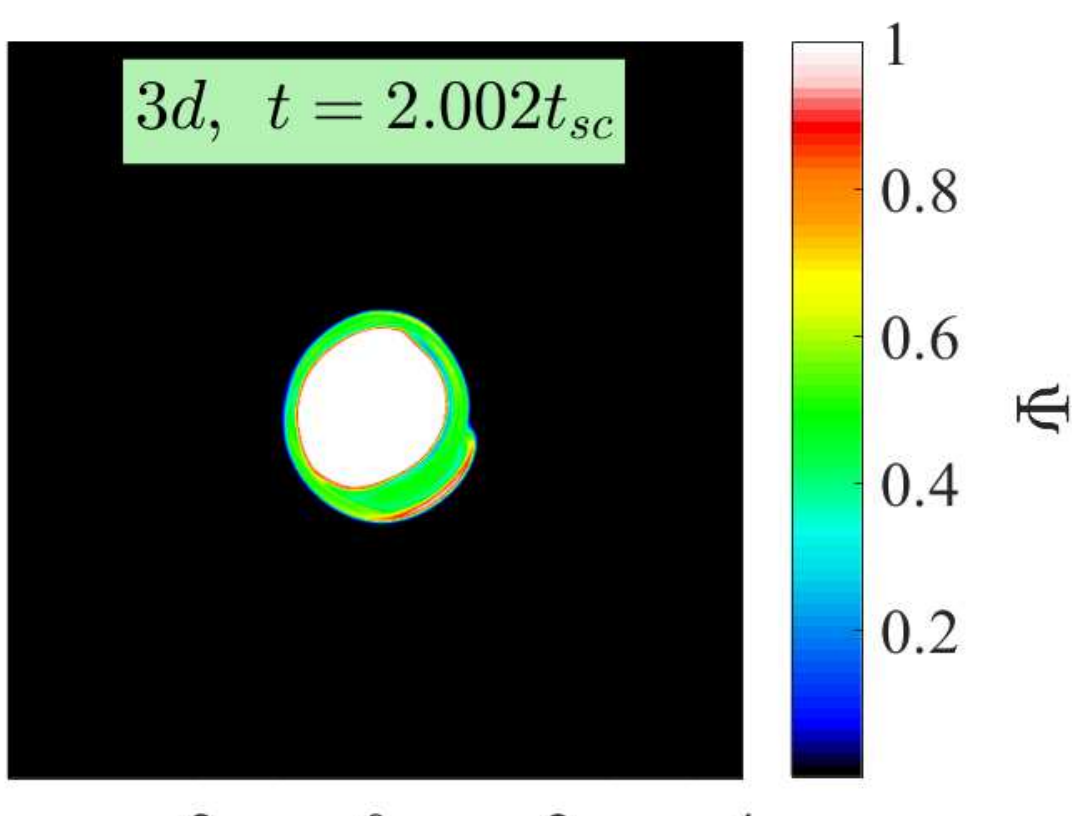}\\
\vspace{-0.09cm}
\includegraphics[trim={0.0cm 1.238cm 3.3cm 0.22cm}, clip, width =0.393 \textwidth]{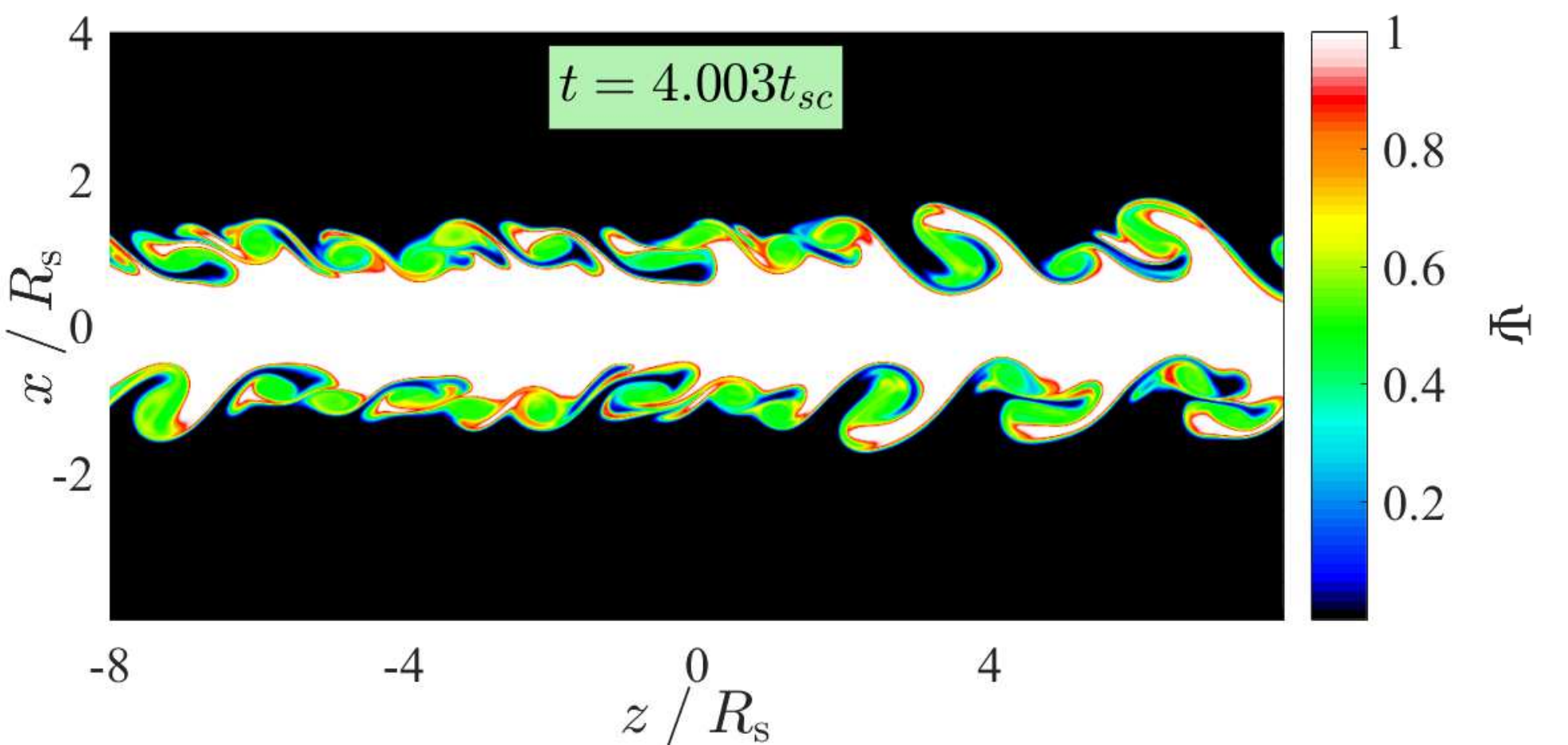}
\hspace{-0.3cm}
\includegraphics[trim={1.3cm 1.238cm 3.3cm 0.22cm}, clip, width =0.363 \textwidth]{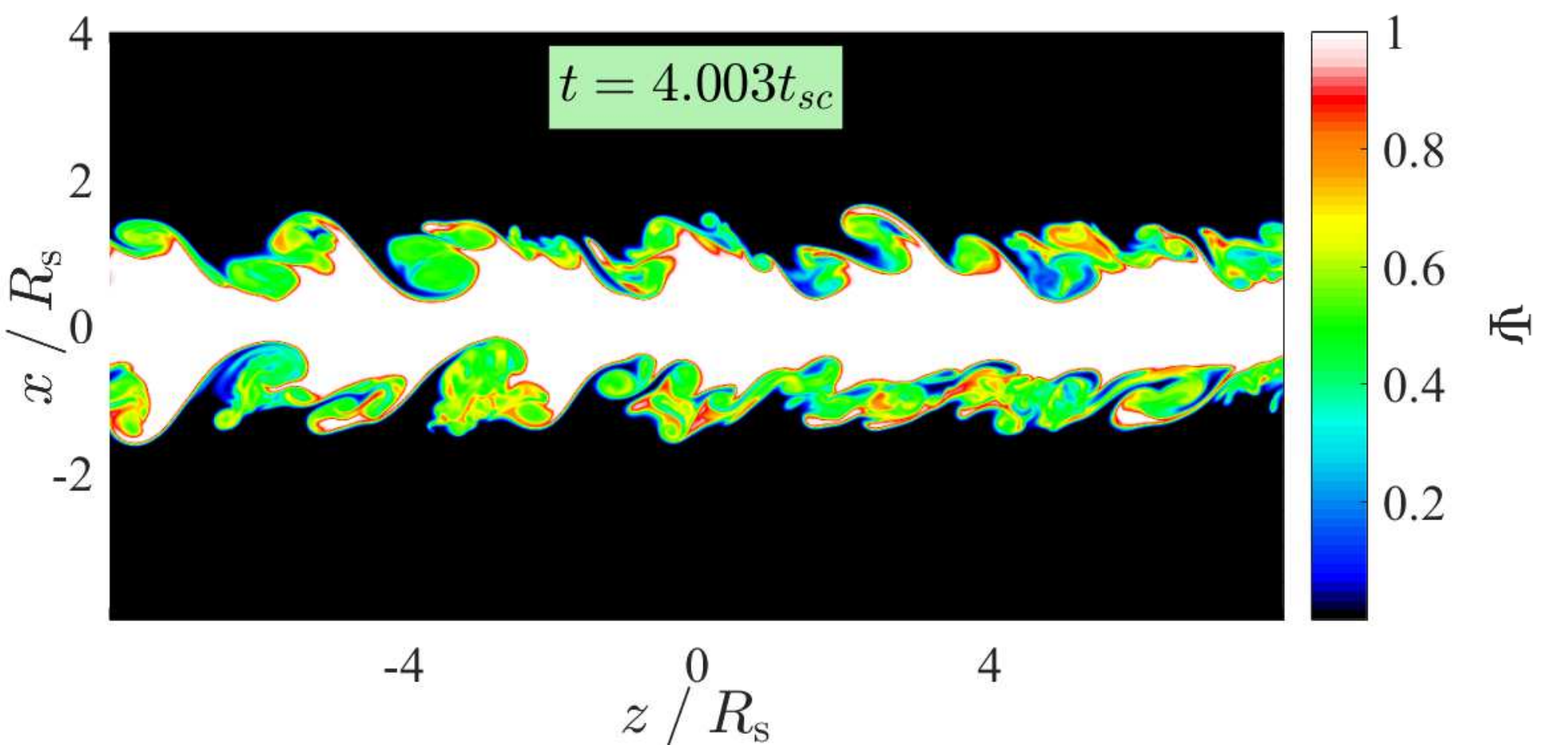}
\hspace{-0.29cm}
\includegraphics[trim={5.1cm 1.238cm 4.05cm 0.22cm}, clip, width =0.257 \textwidth]{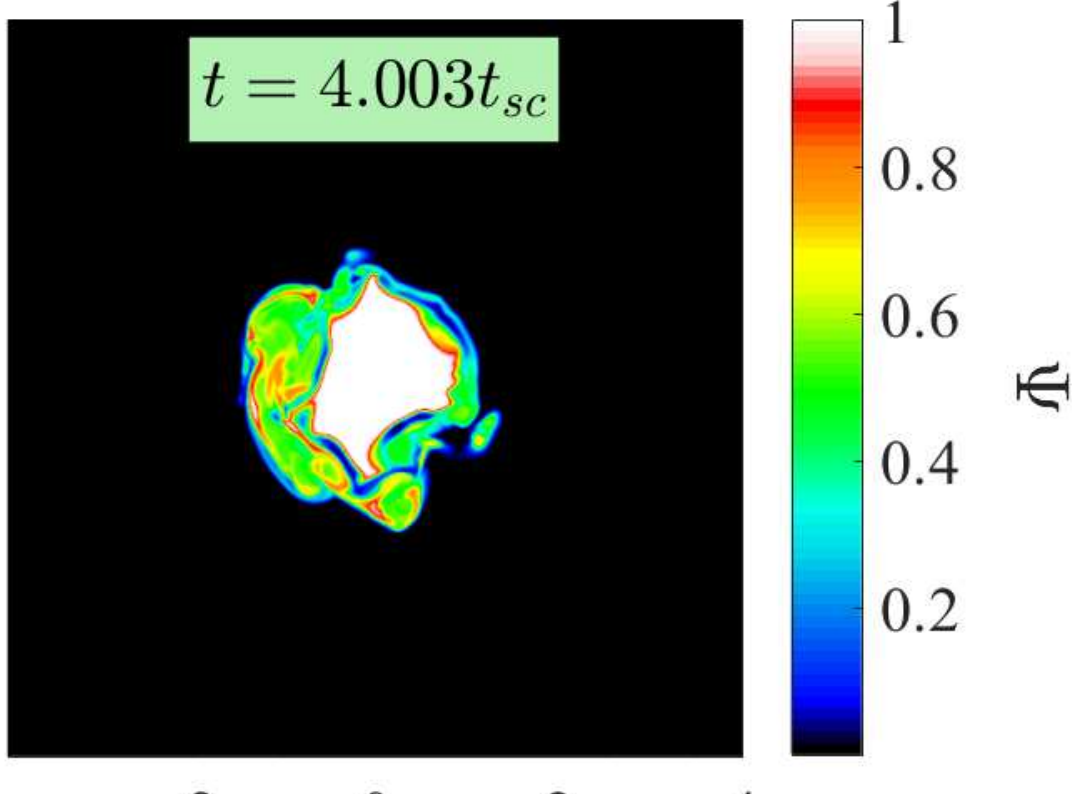}\\
\vspace{-0.09cm}
\includegraphics[trim={0.0cm 1.238cm 3.3cm 0.22cm}, clip, width =0.393 \textwidth]{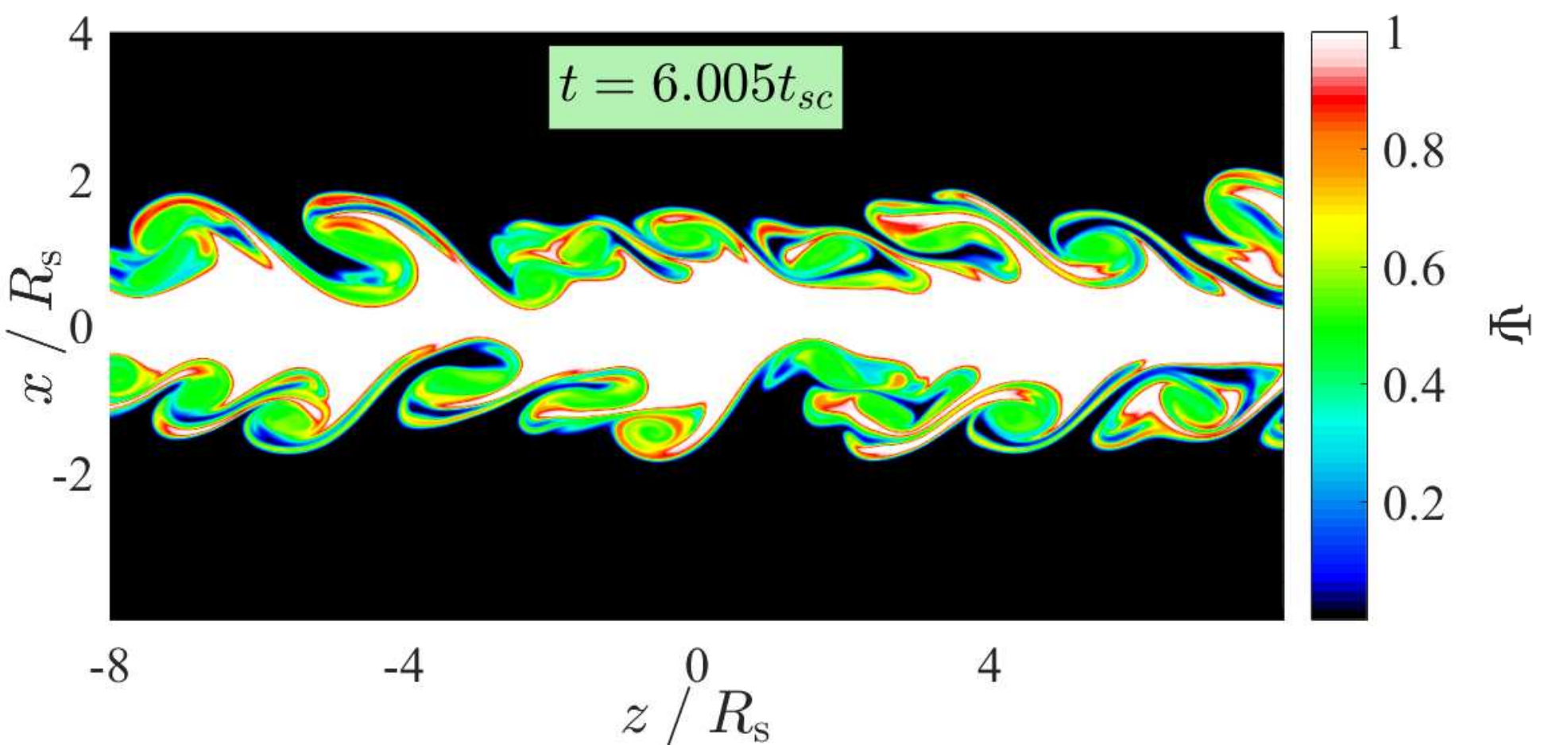}
\hspace{-0.3cm}
\includegraphics[trim={1.3cm 1.238cm 3.3cm 0.22cm}, clip, width =0.363 \textwidth]{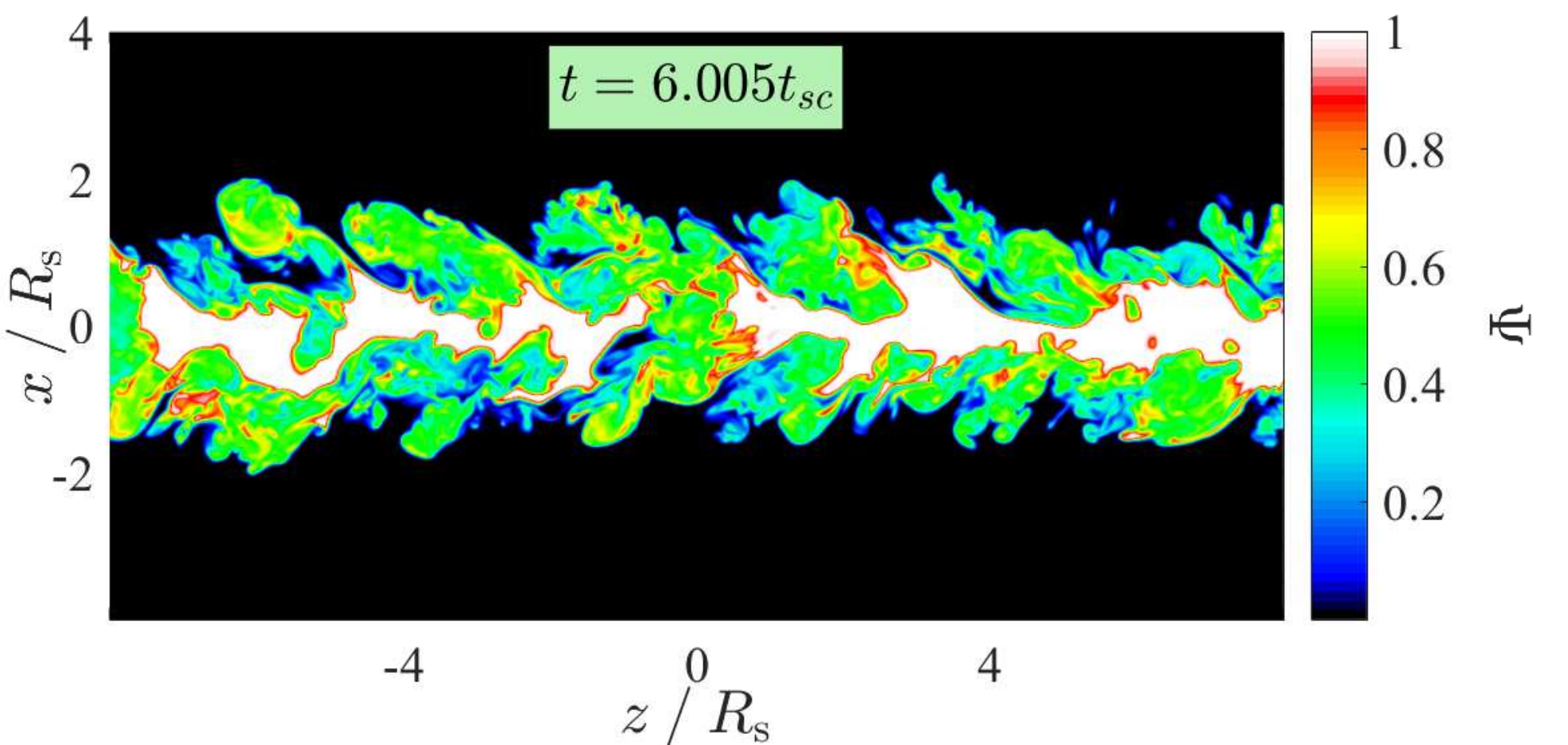}
\hspace{-0.29cm}
\includegraphics[trim={5.1cm 1.238cm 4.05cm 0.22cm}, clip, width =0.257 \textwidth]{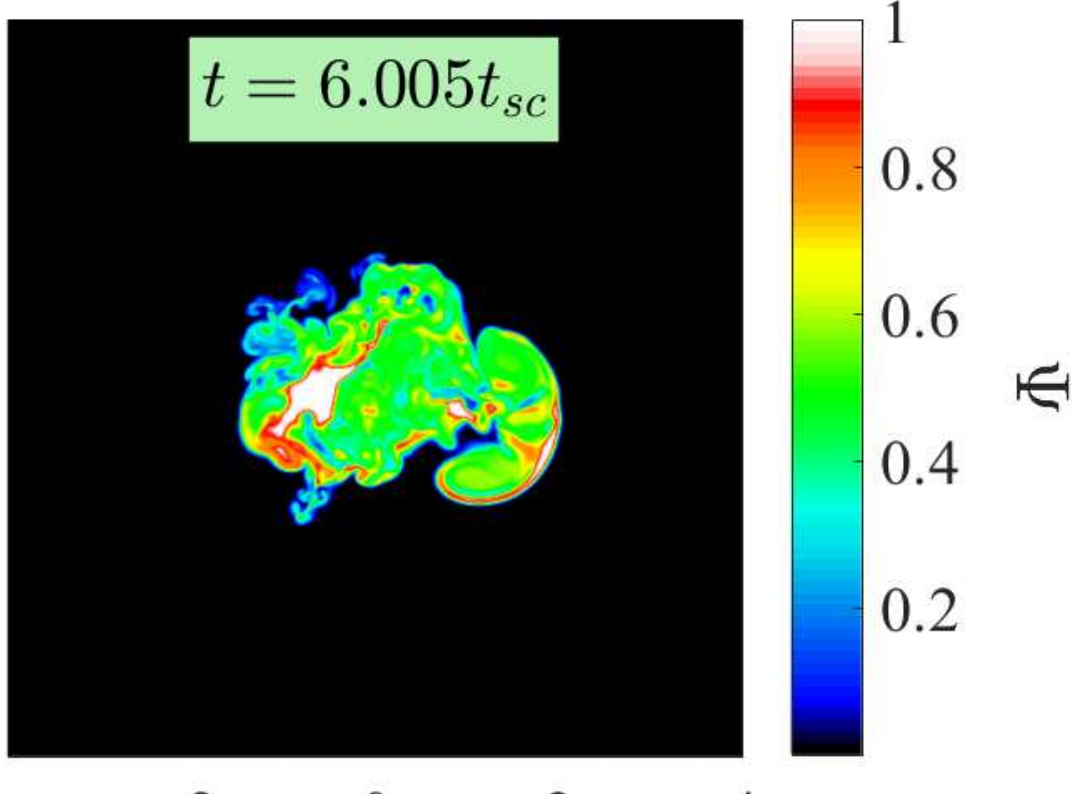}\\
\vspace{-0.09cm}
\includegraphics[trim={0.0cm 0.0cm 3.3cm 0.22cm}, clip, width =0.393 \textwidth]{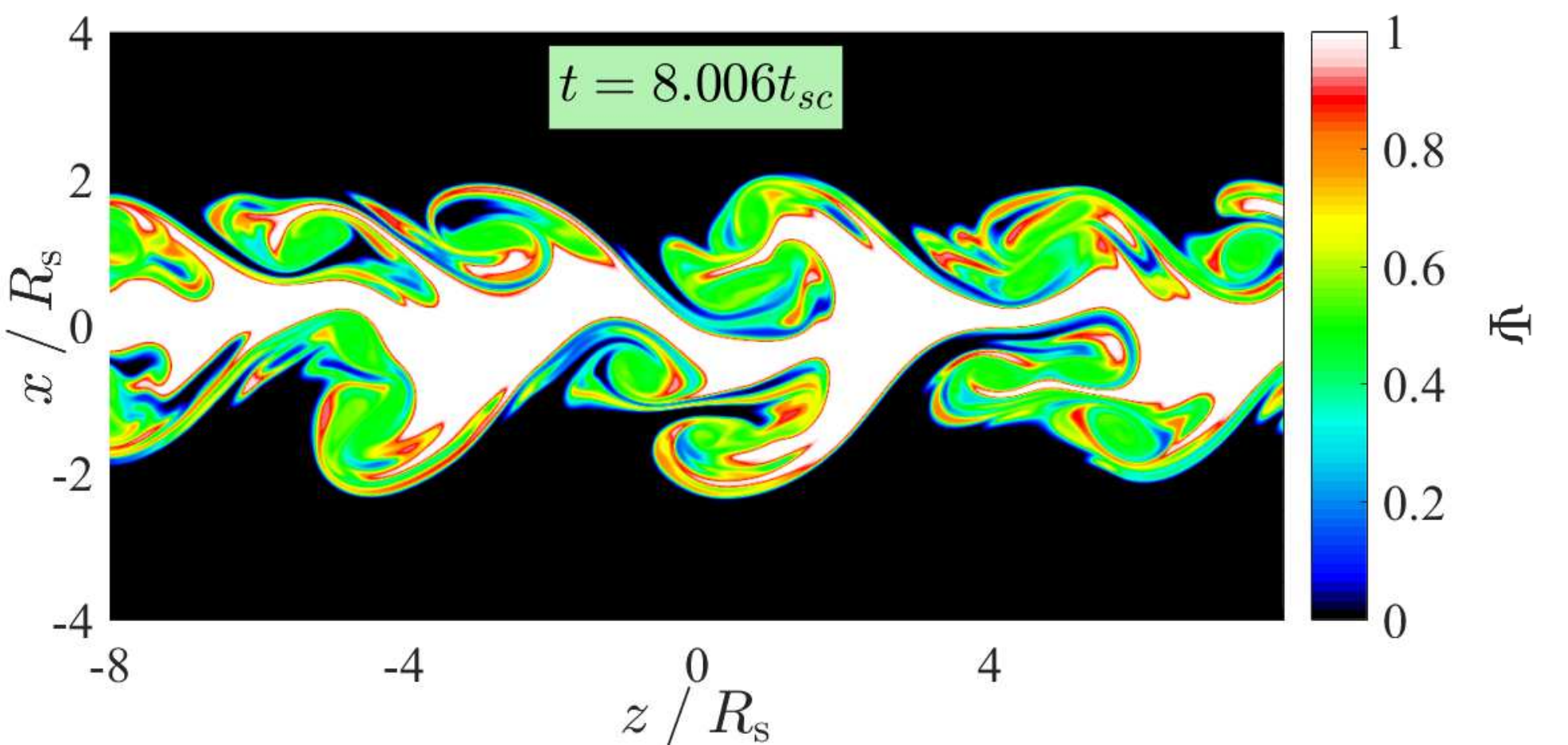}
\hspace{-0.3cm}
\includegraphics[trim={1.3cm 0.0cm 3.3cm 0.22cm}, clip, width =0.363 \textwidth]{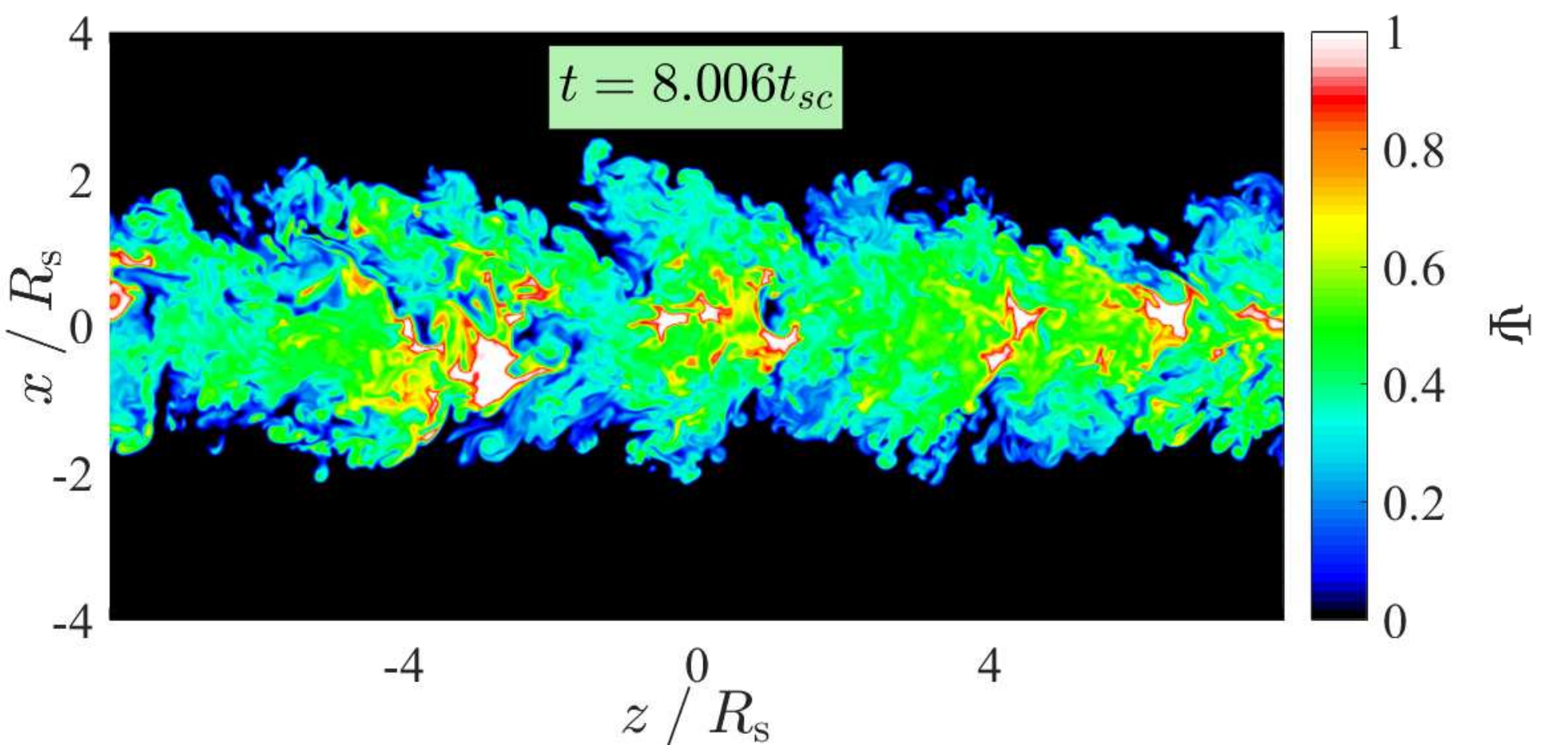}
\hspace{-0.29cm}
\includegraphics[trim={5.1cm 0.0cm 4.05cm 0.22cm}, clip, width =0.257 \textwidth]{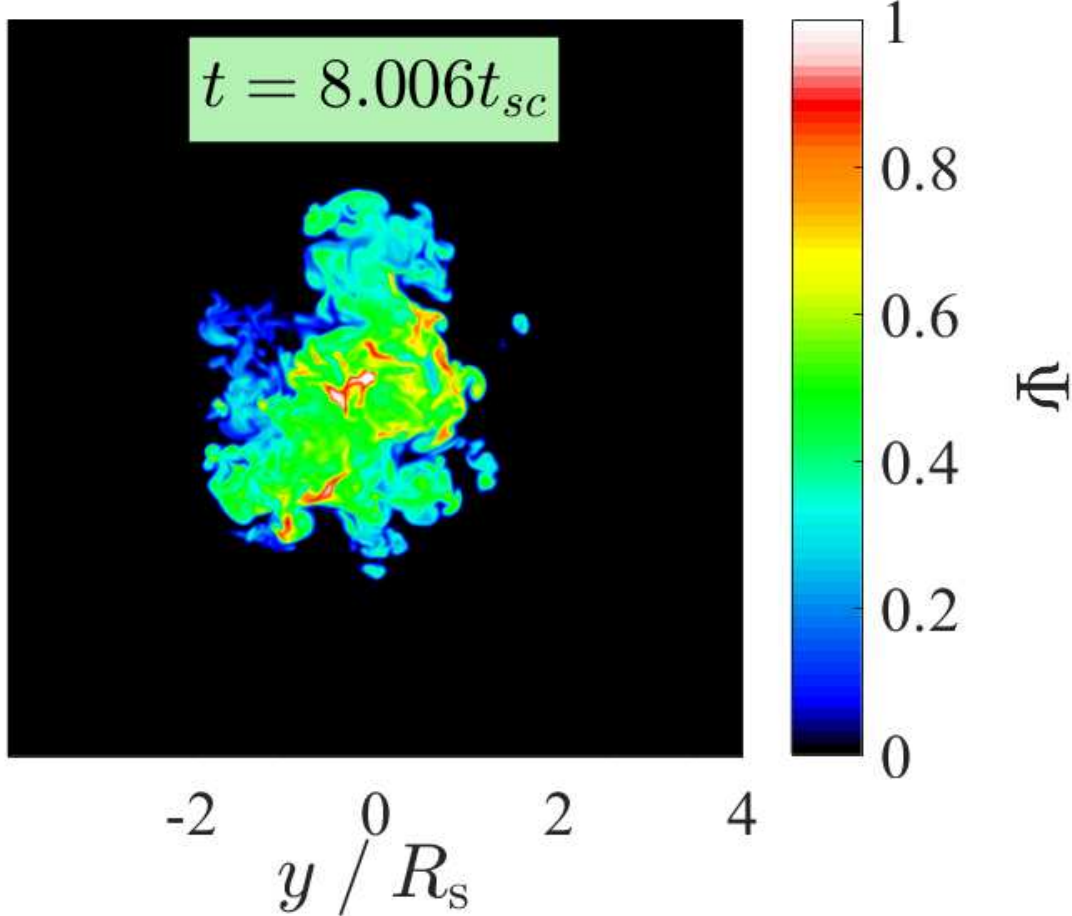}
\end{center}
\caption{Evolution of surface modes in a 2d slab (left) and a 3d cylinder (centre, and right). 
Shown are snapshots of $\psi$, a passive scalar field used as a Lagrangian tracer for the 
stream fluid. The values $\psi=0$ (background) and $\psi=1$ (stream) correspond to cells with 
a pure fluid composition, while cells with $0<\psi<1$ contain a mixture of both fluids. The 
unperturbed initial conditions were $(\Mb,\delta)=(1,1)$. The initial perturbations were identical 
in both simulations and are described in \se{methods}. The snapshot times in units of the stream 
sound crossing time, $\tsc=2\Rs/\cs$, are shown in each panel. The left-hand column shows the 
distribution of $\psi$ in a 2d slab simulation. The center column shows the distribution of $\psi$ 
in the $xz$ plane of a 3d simulation, an ``edge-on" view of the cylinder, while the right-hand 
column shows the distribution in the $xy$ plane, a ``face-on" view. At $t\sim 2\tsc$, the edge-on 
distribution of $\psi$ in the 3d cylinder appears nearly identical to its distribution in the 2d 
slab simulation, while the face-on distribution is dominated by symmetric, $m=0$, and antisymmetric, 
$m=1$, modes. At $t=4\tsc$, the face-on view of the cylinder reveals higher-order azimuthal modes, 
though the edge-on view remains similar to the slab. By $t=6\tsc$, the structure of the 3d 
simulation begins to deviate from that of the 2d slab. In 2d, the largest eddies remain coherent 
and continue to grow while the centre of the stream remains unmixed even at $t=8\tsc$. In 3d, 
by $6\tsc$ the largest eddies have broken up and generated small-scale turbulence, and by $8\tsc$ 
there is no unmixed fluid left in the stream. Similar figures showing stream evolution in 
simulations with different values of $(\Mb,\delta)$ can be found in \se{add_figures}.
}
\label{fig:colour_panel_M1D1} 
\end{figure*}

\smallskip
Initially, each interface is characterized by a sharp transition\footnote{Neglecting the smoothing 
introduced in \equ{ramp2}.} from $\overline{\psi}=1$ at $r<\Rs$ to $\overline{\psi}=0$ at $r>\Rs$. 
The nonlinear evolution of KHI mixes the fluids near the interface, and the initial discontinuity of 
$\overline{\psi}_{(r)}$ is smeared over a finite width around each interface (see \fig{colour_panel_M1D1}). 
The resulting profile is monotonic\footnote{Neglecting small fluctuations on the grid scale.} and can 
be used to define the edges of the perturbed region around an interface, $r(\overline{\psi}=\epsilon)$ 
on the background side and $r(\overline{\psi}=1-\epsilon)$ on the stream side, where $\epsilon$ is an 
arbitrary threshold. The background-side thickness of the perturbed region is then defined as 
\be 
\label{eq:hb_def}
\hb \equiv {\rm max_r}r(\overline{\psi}=\epsilon)-\Rs,
\ee
{\no}while the stream-side thickness is defined as 
\be 
\label{eq:hs_def}
\hs \equiv \Rs-{\rm min_r}r(\overline{\psi}=1-\epsilon).
\ee
{\no}The ${\rm max_r}$ and ${\rm min_r}$ in \equs{hb_def} and \equm{hs_def} are only necessary to avoid 
fluctuations in the $\overline{\psi}(r)$ profile, which in practice is smooth and monotonic, especially 
near the edges. While $\hb$ as defined in \equ{hb_def} is always well defined, at late times the perturbed 
region encompases the entire stream and $\overline{\psi}(r=0)<1-\epsilon$. In this case, we define $\hs=\Rs$. 
The total width of the perturbed region is given by $h\equiv \hb+\hs$.

\begin{figure*}
\begin{center}
\includegraphics[trim={0.0cm 1.238cm 3.3cm 0.2cm}, clip, width =0.45 \textwidth]{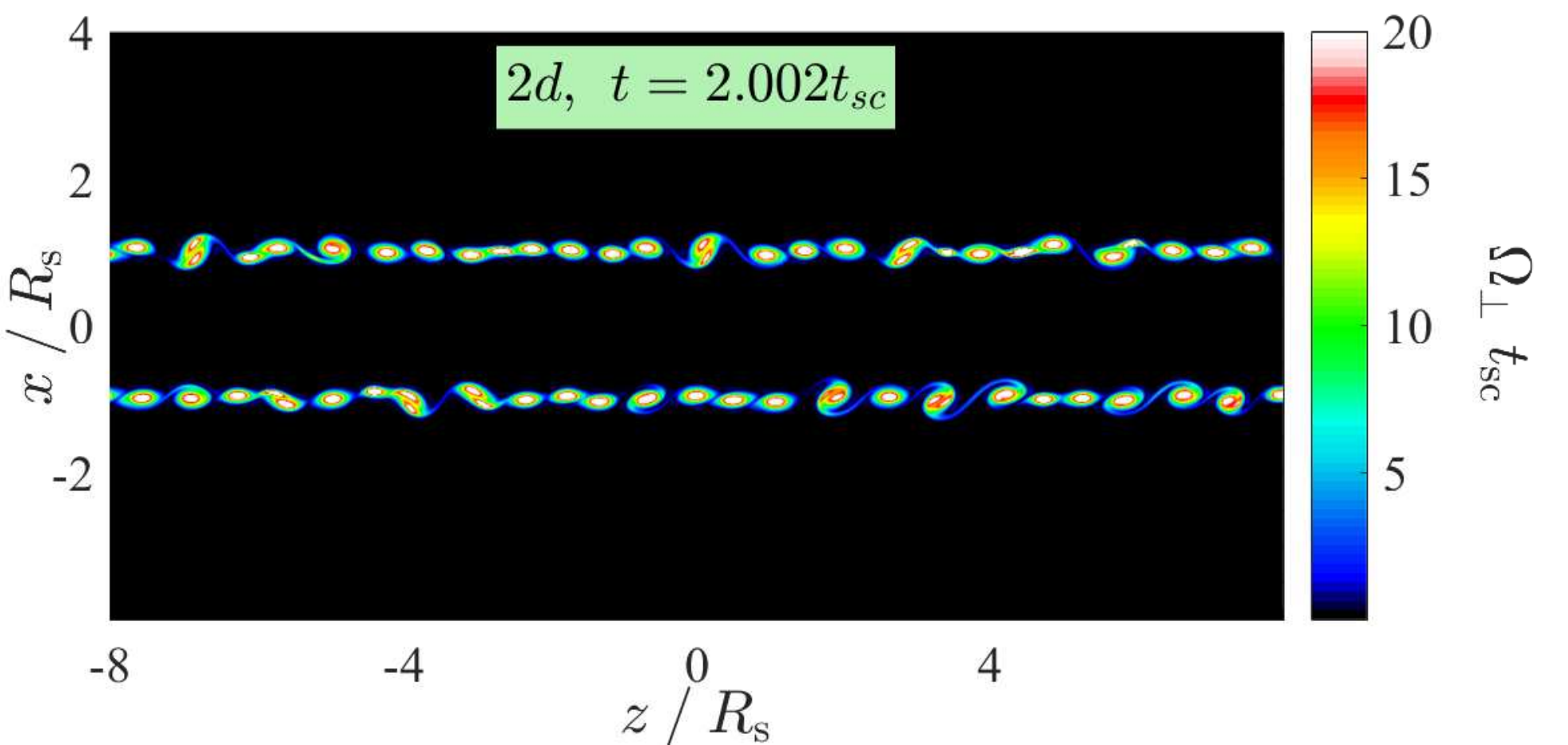}
\hspace{-0.3cm}
\includegraphics[trim={1.3cm 1.238cm 0.1cm 0.2cm}, clip, width =0.501 \textwidth]{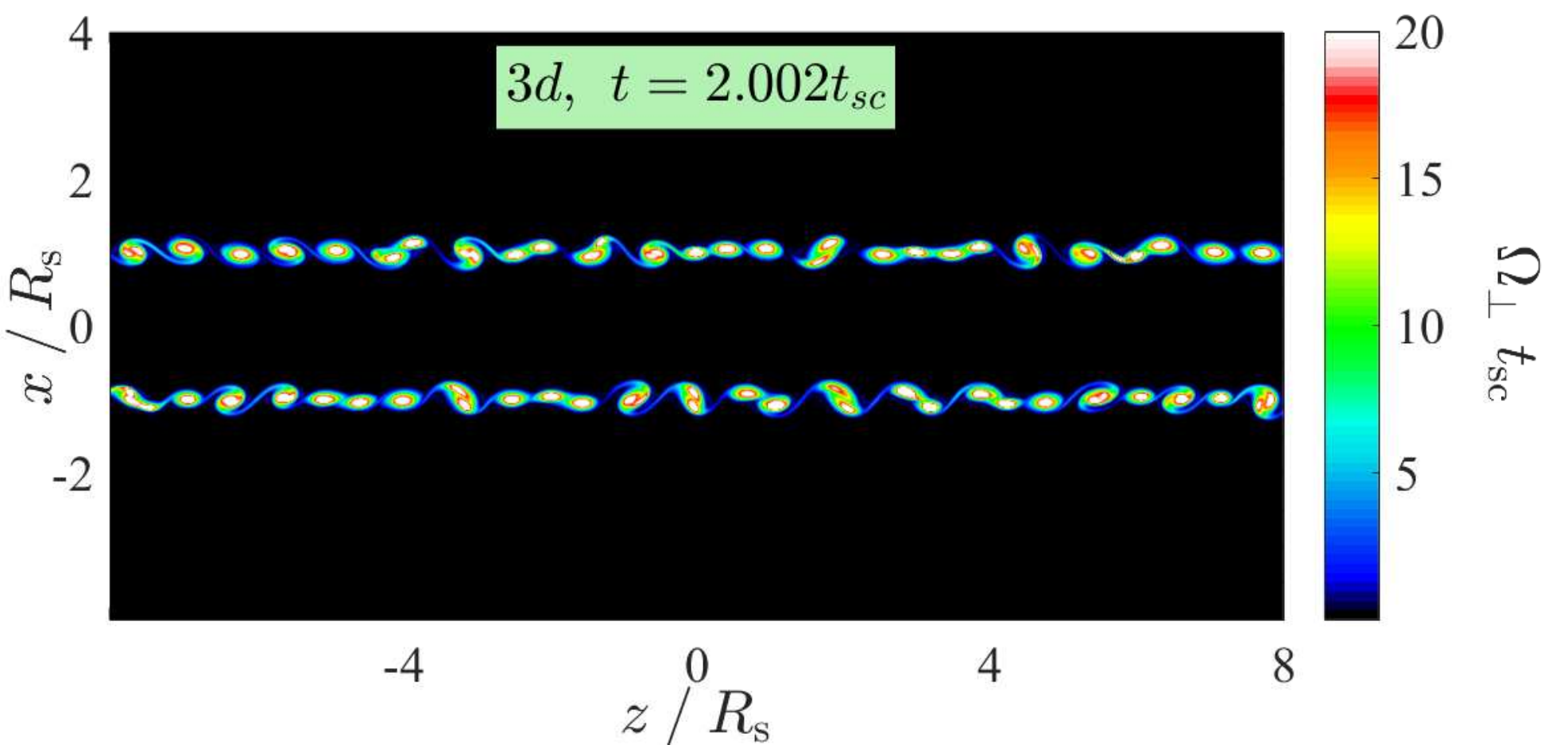}\\
\vspace{-0.06cm}
\includegraphics[trim={0.0cm 1.238cm 3.3cm 0.2cm}, clip, width =0.45 \textwidth]{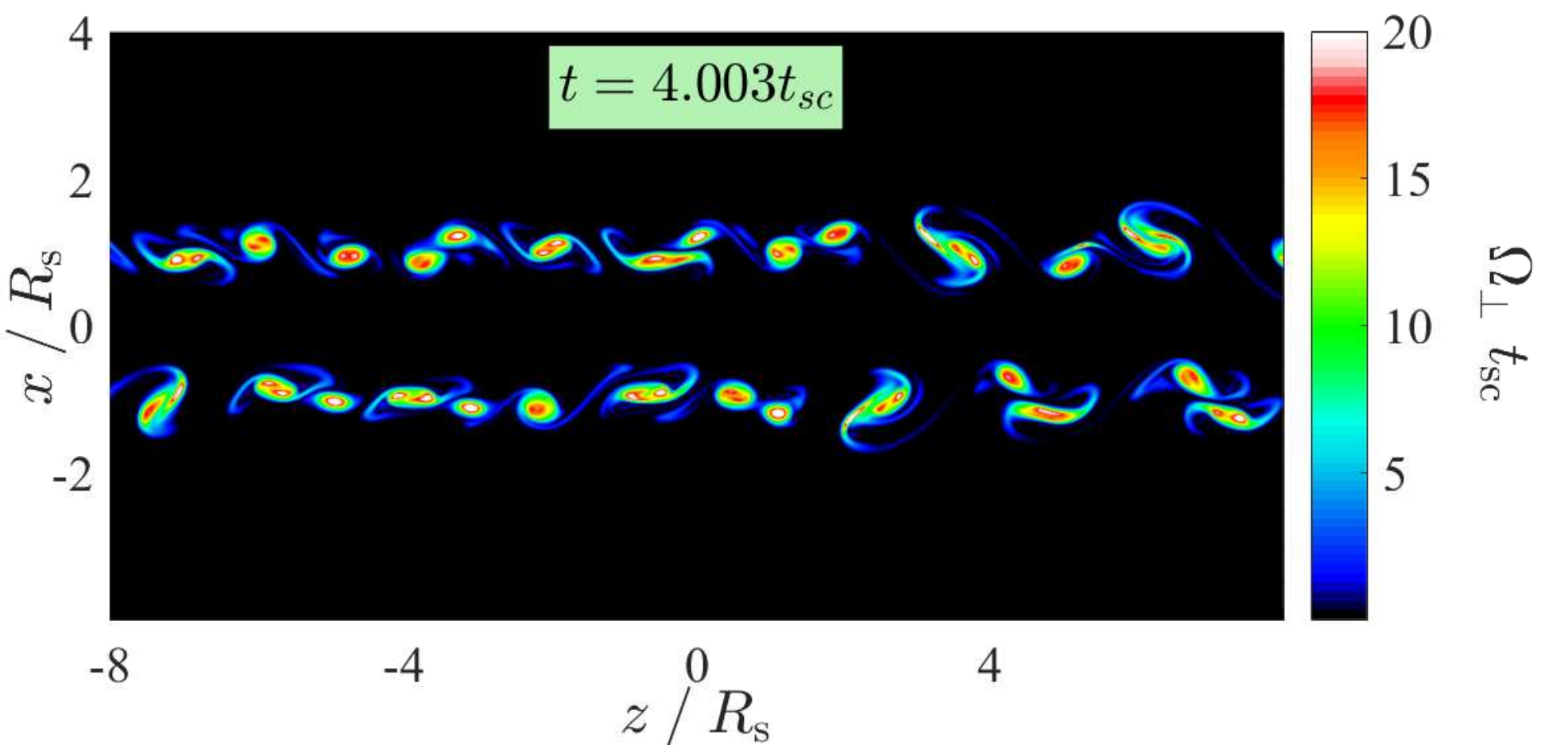}
\hspace{-0.3cm}
\includegraphics[trim={1.3cm 1.238cm 0.1cm 0.2cm}, clip, width =0.501 \textwidth]{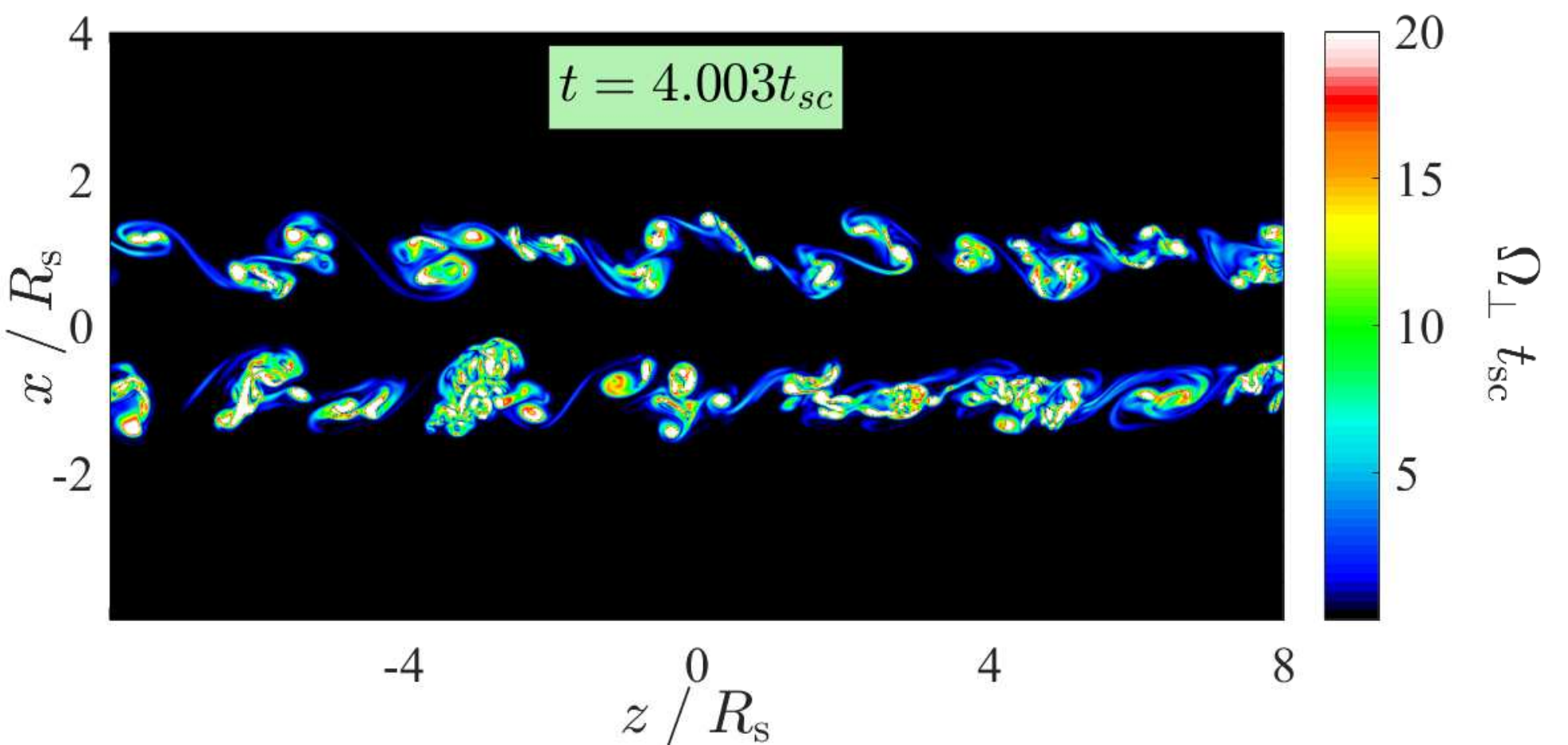}\\
\vspace{-0.06cm}
\includegraphics[trim={0.0cm 1.238cm 3.3cm 0.2cm}, clip, width =0.45 \textwidth]{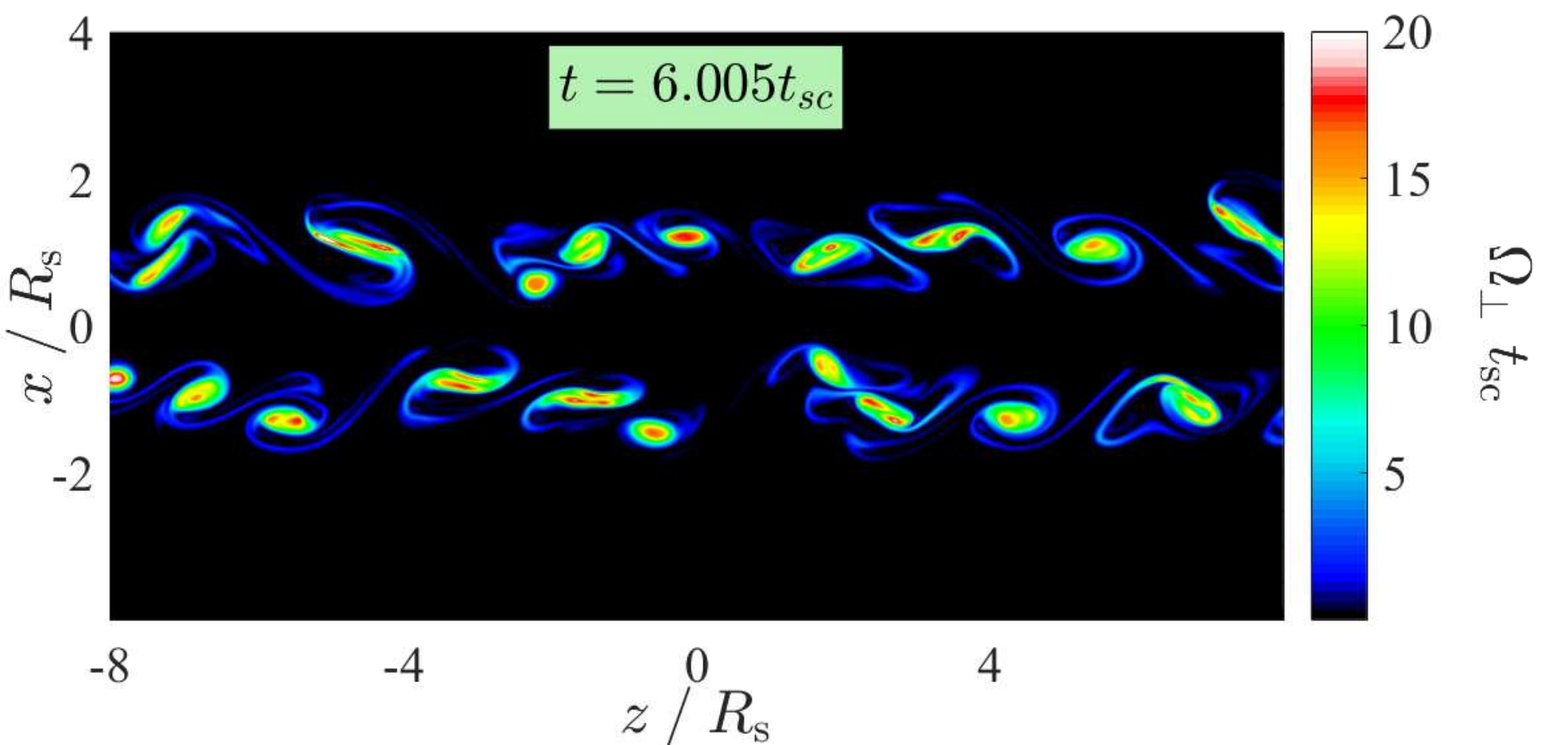}
\hspace{-0.3cm}
\includegraphics[trim={1.3cm 1.238cm 0.1cm 0.2cm}, clip, width =0.501 \textwidth]{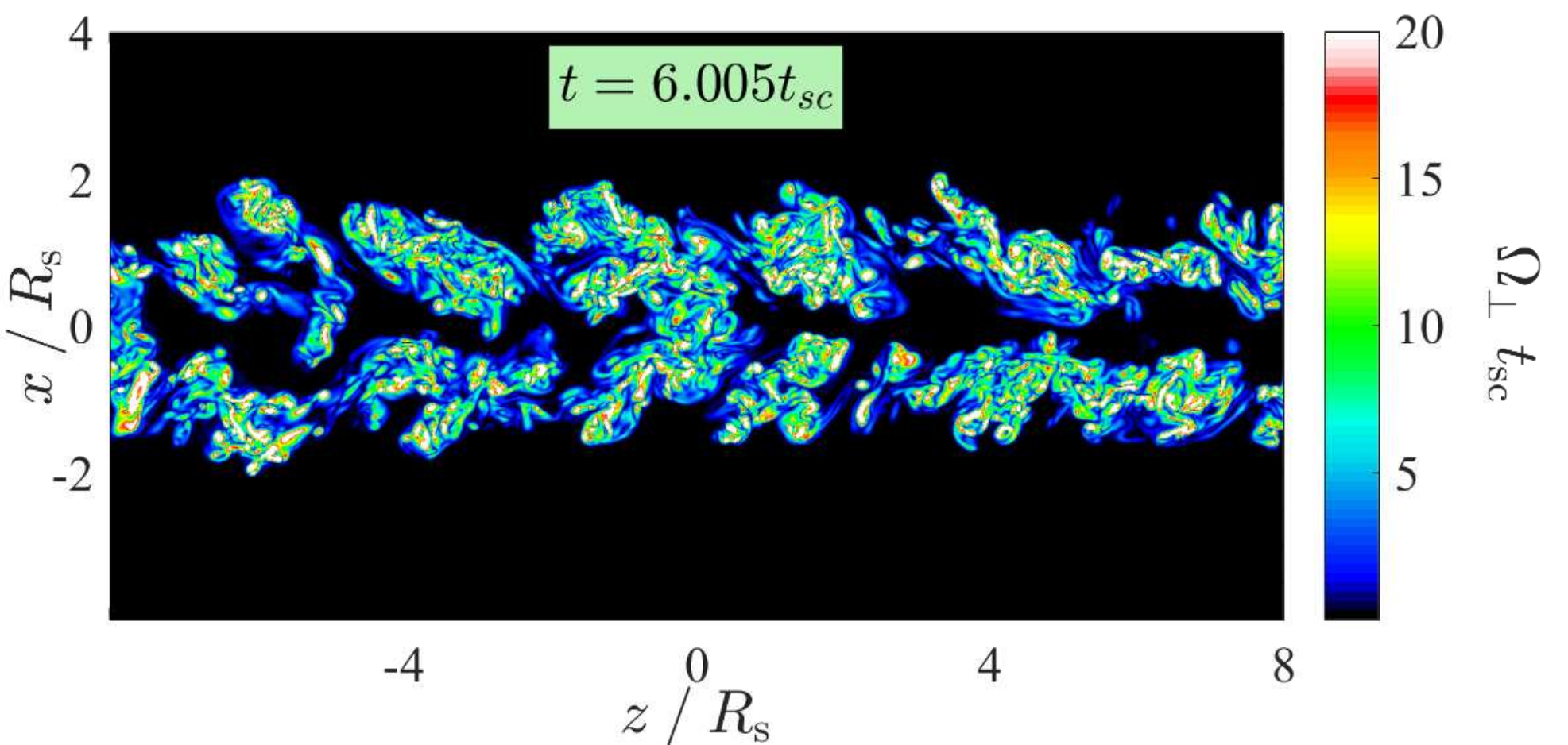}\\
\vspace{-0.06cm}
\includegraphics[trim={0.0cm 0.0cm 3.3cm 0.2cm}, clip, width =0.45 \textwidth]{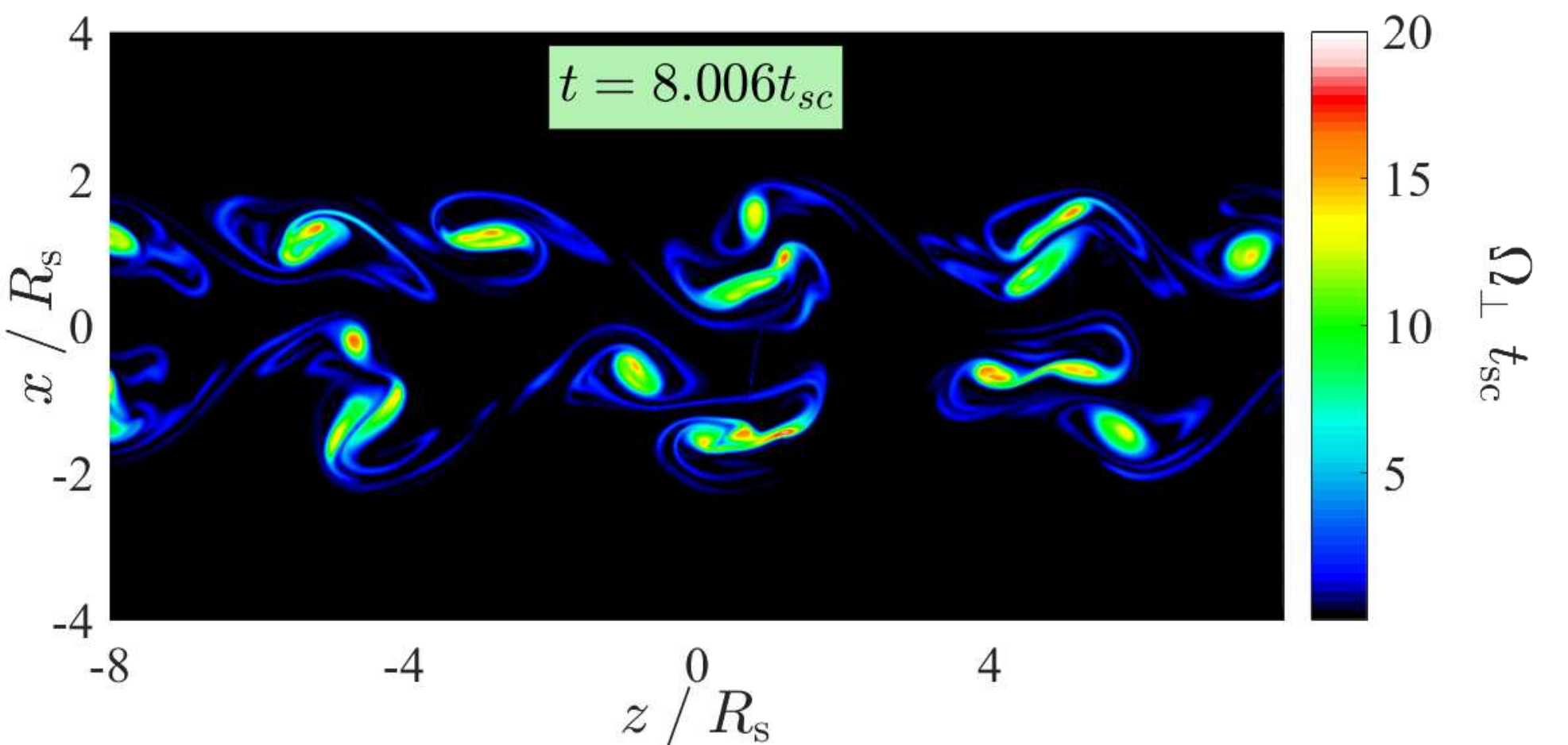}
\hspace{-0.3cm}
\includegraphics[trim={1.3cm 0.0cm 0.1cm 0.2cm}, clip, width =0.501 \textwidth]{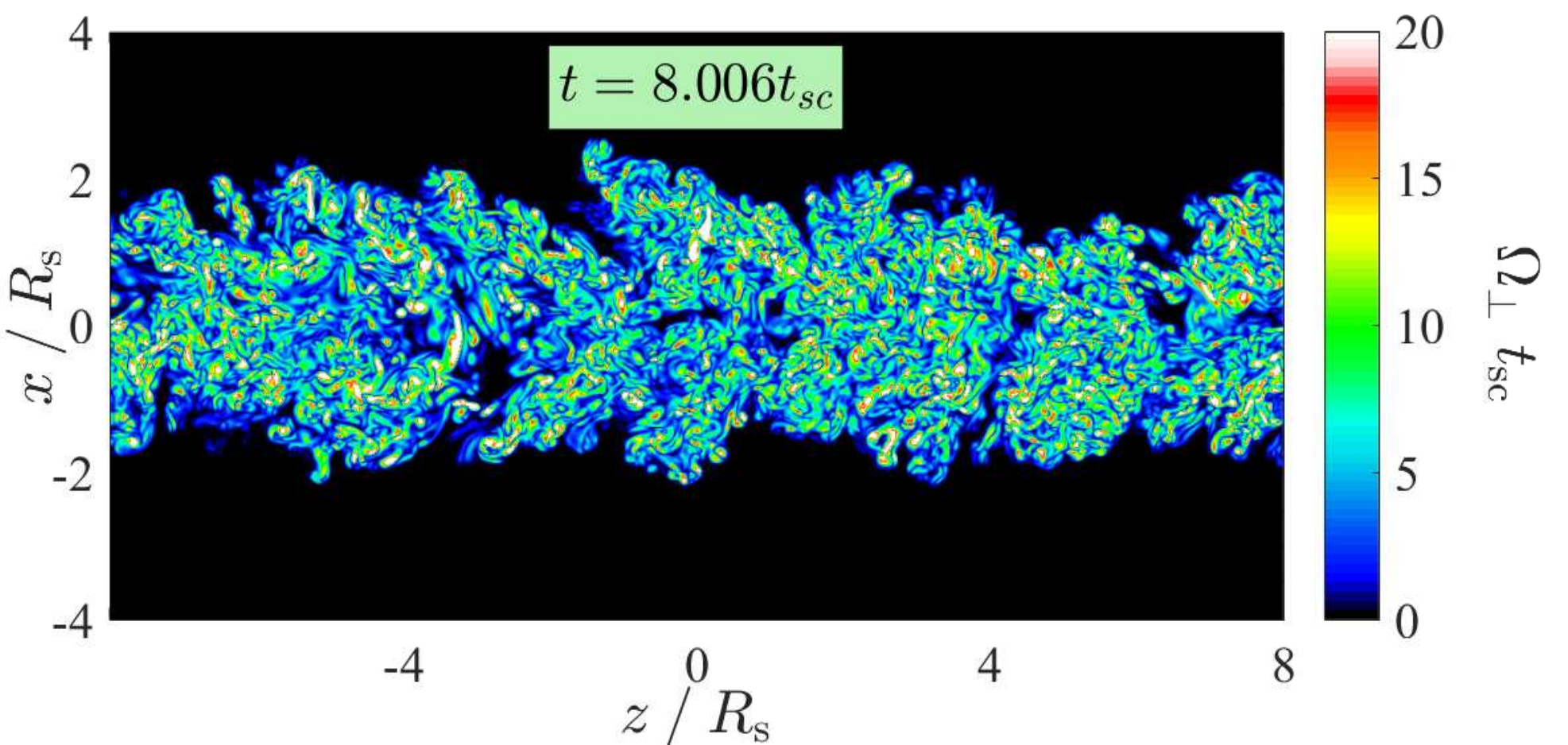}\\
\end{center}
\caption{Evolution of vorticity in surface mode simulations of a 2d slab (left) and a 
3d cylinder (right). Shown are snapshots of $\Omega_{\perp}$, the magnitude of the vorticity 
component perpendicular to the stream axis, normalized by the stream sound crossing time. In 
2d, $\Omega_{\perp}=|\Omega_{\rm y}|$ while in 3d $\Omega_{\perp}=(\Omega_{\rm r}^2+\Omega_{\varphi}^2)^{1/2}$. 
The simulation and the snapshots shown are the same as in \fig{colour_panel_M1D1}, and the 
slices shown are the same as the left-two columns in that figure, namely for the 3d simulation 
we show a single slice through the $xz$ plane. Similar to the distribution of $\psi$ shown in 
\fig{colour_panel_M1D1}, at $t\sim 2\tsc$ the vorticity in the 3d cylinder appears nearly identical 
to the 2d slab. At later times, the vortices in the 2d slab merge to form larger vortices which remain 
coherent until $8\tsc$ with no noticeable small scale strcture. On the other hand, in the 3d cylinder 
the vortices begin to break up and small scale structure is evident already at $4\tsc$. By $8\tsc$ the 
medium appears completely turbulent and the largest vortices are barely visible. This exemplifies the 
qualitative difference between vortex evolution and turbulence generation in two- and three-dimensions.
}
\label{fig:vorticity_slice_panel_M1D1} 
\end{figure*}

\begin{figure*}
\begin{center}
\includegraphics[trim={3.79cm 0.0cm 7.0cm 0}, clip, width =0.2616 \textwidth]{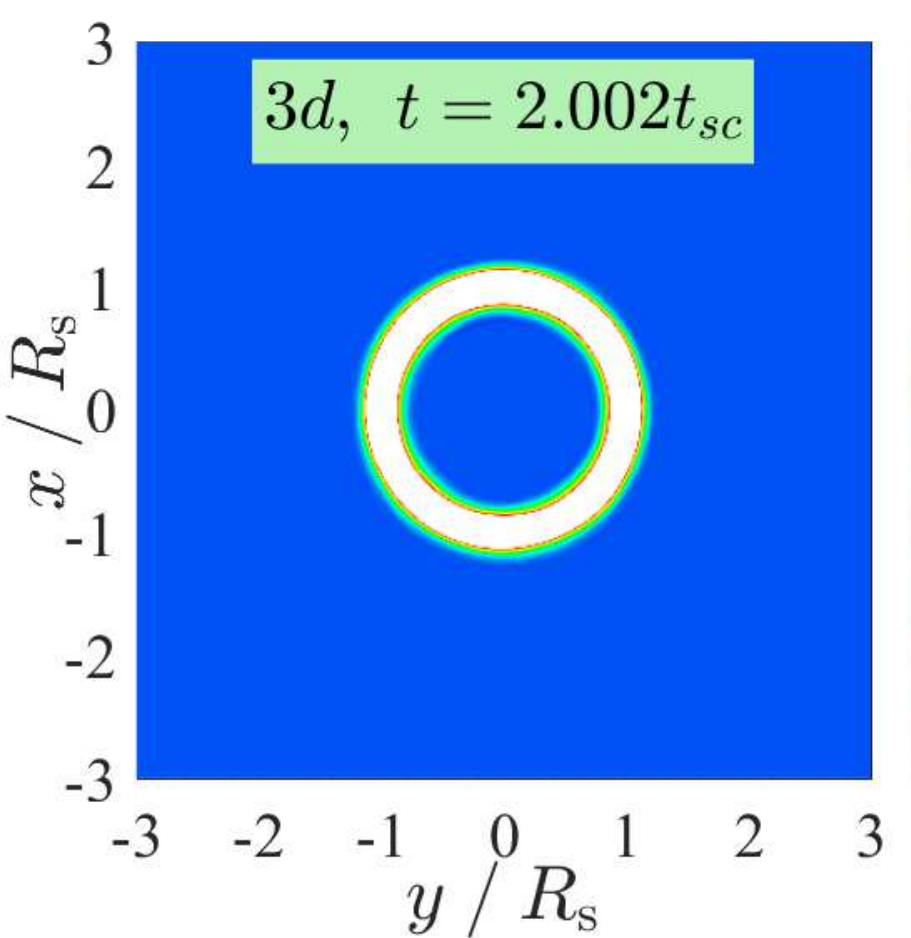}
\hspace{-0.3cm}
\includegraphics[trim={5.1cm 0.0cm 7.0cm 0}, clip, width =0.2247 \textwidth]{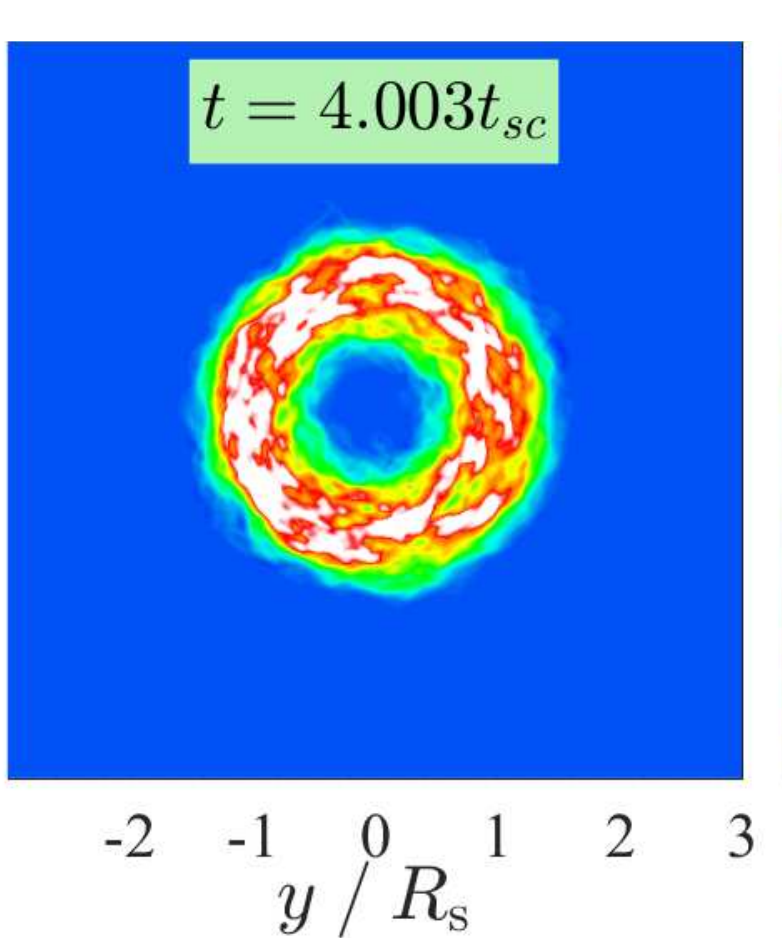}
\hspace{-0.3cm}
\includegraphics[trim={5.1cm 0.0cm 7.0cm 0}, clip, width =0.2247 \textwidth]{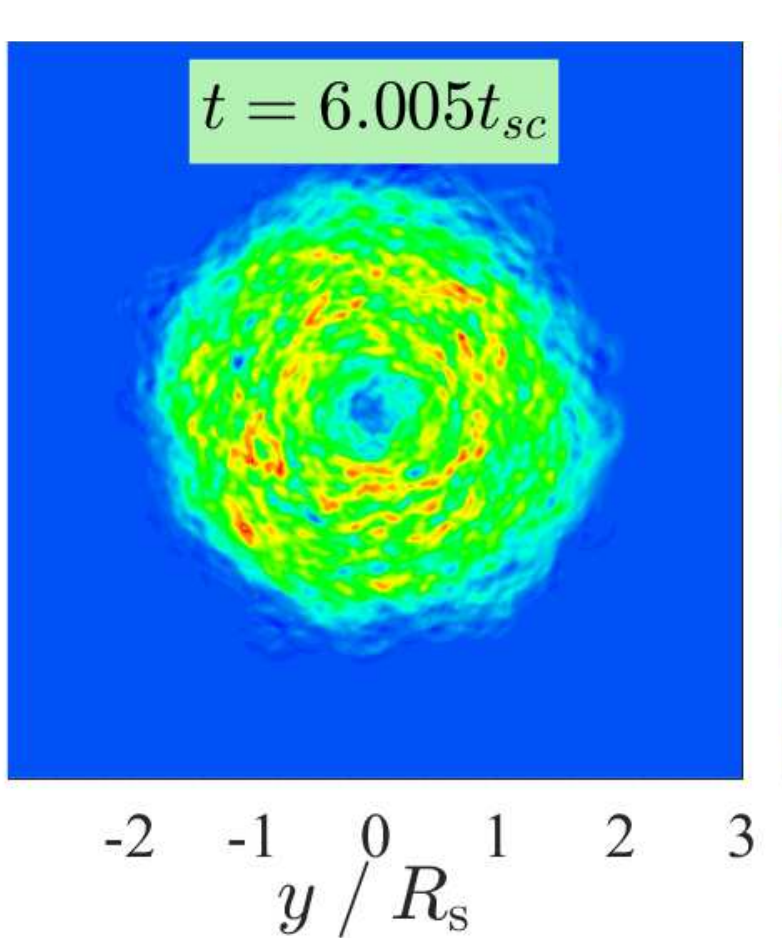}
\hspace{-0.3cm}
\includegraphics[trim={5.1cm 0.0cm 4.05cm 0}, clip, width =0.3081 \textwidth]{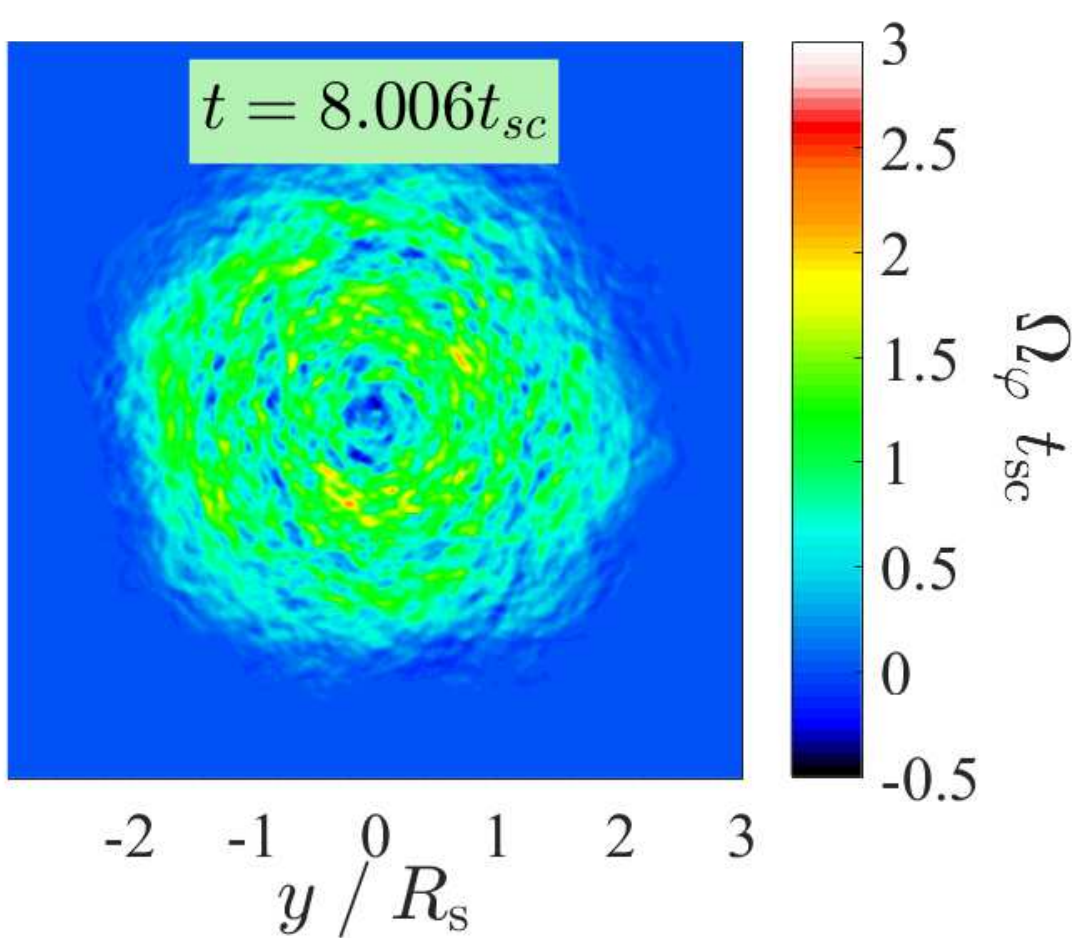}
\end{center}
\caption{Evolution of mean vorticity in 3d surface mode simulations. 
Shown are maps of the azimuthal components of the vorticity, 
$\Omega_{\varphi}$, 
averaged over the entire length of the box in the $z$ direction 
and projected onto the $xy$ plane. The simulation and the snapshots 
shown are the same as in \figs{colour_panel_M1D1} and \figss{vorticity_slice_panel_M1D1}. 
%We show the azimuthal component of the vorticity, $\Omega_{\rm \varphi}$ on the top, the radial component, $\Omega_{\rm r}$ in the middle, and the longitudinal component, $\Omega_{\rm z}$ on the bottom. 
At $t=2\tsc$, the vorticity is entirely in the azimuthal direction 
as in an ideal vortex ring, implying that the fluid motion is confined 
to the $zr$ plane at each azimuthal angle $\varphi$, and consistent with 
the similarity between the 2d slab and 3d cylinder at this time seen in 
\figs{colour_panel_M1D1} and \figss{vorticity_slice_panel_M1D1}. 
$\Omega_{\rm \varphi}$ remains the dominant component of vorticity through 
$t=8\tsc$, maintaining a ring-like structure within the shearing layer between 
the two fluids. At $t\ge 4\tsc$, small fluctuations in $\Omega_{\rm r}$ and 
$\Omega_{\rm z}$ are present as well (see text). 
}
\label{fig:vorticity_panel_M1D1} 
\end{figure*}

\smallskip
The above definitions for $h$, $\hs$, and $\hb$ depend on $\epsilon$. The exact dependence depends on 
the $\overline{\psi}(r)$ profile, which varies somewhat with $(\Mb,\delta)$. In general, larger values 
of $\epsilon$ correspond to smaller values of $h$. We adopt $\epsilon=0.04$ as our fiducial value, though 
we experimented as well with $\epsilon=0.03$ and $0.02$, which was the fiducial choice in P18. In almost 
all our simulations the difference between these values was very small, on the order of $\lsim 10\%$. The 
exceptions are simulations with $\delta=100$, where we find that $\hs$ estimated with $\epsilon=0.02$ 
fluctuates with time rather than monotonically growing. These fluctuations are damped when using $\epsilon=0.04$, 
and so we adopt this value for all cases.

%-------------------------------------------
\subsection{Comparison to P18} 
\label{sec:methods-comp}

\smallskip
There are several differences between the numerical setup used in the 2d slab simulations presented here 
and those presented in P18. Firstly, the fiducial stream radius in P18 was $\Rs=1/64$ and $1/128$ for surface 
and body modes respectively, while the cell size in the highest resolution region was $\Delta=2^{-13}$. This 
yields 256 and 128 cells per stream diameter for surface mode and body mode simulations respectively, while our 
simulations have 128 cells per diameter for both cases. Furthermore, the high resolution region was much larger 
in P18 than in our simulations, extending to $10-20\Rs$. These changes were all necessary in order to make the 
transition to 3d, and were adopted in our 2d simulations as well for consistency. In Appendix \se{convergence}, 
we show that our 3d results are converged with respect to the number of cells per stream diameter and the size 
of the high resolution region, as was shown in P18 for the 2d case. The only potential effect of making the 
stream wider with respect to the box is that we are more sensitive to boundary effects. However, as discussed 
in \se{methods-boundary}, we do not expect this to be an issue. 

\smallskip
An additional difference is the width of the initial smoothing layer between the stream and the background 
(\equnp{ramp2}). In P18, our fiducial value for $\sigma$ was $\Rs/128$ for surface modes and $\Rs/32$ for 
body modes, while we used $\Rs/32$ for surface modes and values in the range $\Rs/32-\Rs/8$ for body modes. 
We showed in P18 that the results of 2d simulations were not strongly dependent on the precise value of 
$\sigma$, so long as this was of order a few cells. However, as we show in \se{body}, when $\Mb>M_{\rm crit}$ 
the results of 3d simulations depend strongly on the choice of $\sigma$, and for consistency we also adjust 
the values for our 2d slab simulations. 

\smallskip
The most important differences between our setup and that of P18 concern the initial perturbations. 
We initiate perturbations in the radial component of the velocity dubbed \textit{velocity-only 
perturbations} in P18, where the fiducial method of perturbing the stream was \textit{interface-only 
perturbations}, periodic perturbations to the shape of the stream-background interface. The onset of 
shear layer growth or stream deformation requires a perturbation in the stream-background 
interface to grow to nonlinear amplitude. Before this can happen, the initial velocity perturbation 
must evolve into eigenmodes and trigger the growth of interface perturbations, which takes of order 
the perturbation sound crossing time for surface modes and of order the stream sound crossing time 
for body modes (M16). Furthermore, while we seeded perturbations with wavelengths in the range $(1/2-2)\Rs$, 
the wavelength range for surface mode simulations in P18 was $(1/16-1/2)\Rs$, while for body modes 
it was $(1/4-16)\Rs$. Recall that for body modes, the transition to nonlinearity is dominated by the 
critical perturbation with a wavelength of order $\sim 10\Rs$ (\se{theory_body}). Since our initial 
conditions contain no power on scales larger than $2\Rs$, we must wait for the inverse cascade to 
transfer energy to large scales before the critical perturbation can begin to grow. In P18, on the 
other hand, the initial conditions already contained power at these scales, so growth could begin 
immediately. As we will see in \se{results}, both of these effects lead to a delay in the onset of 
nonlinear growth in our simulations. However, once nonlinear growth begins the evolution is insensitive 
to the initial perturbations, as was demonstrated in P18. As discussed in \se{body}, we explicitly test 
this by performing one 3d simulation with interface only perturbations (\equnp{pertr}) in the wavelength 
range $(1/2-16)\Rs$. 

%\footnote{This is different from P18, where each individual perturbation mode was initiated on only one of the slab interfaces. We make this change here in order to perturb cylinders and slabs in a self-consistent way.}
%\footnote{In P18, velocity-only perturbations with a broadband white noise spectrum (see table 2 in P18) have a rms amplitude of $\sim 0.01V$, where V is the slab velocity. Since $V\sim \cb=\delta^{1/2}\cs$, for large density contrasts, this can be significanly larger than the amplitude assumed here.}

%%%%%%%%%%%%%%%%%%%%%%%%%%%%%%%%%%%%%%%%%%%%%%%%  
\section{Simulation Results}
\label{sec:results}
\smallskip
We now present the results of our numerical simulations. In \se{surface} we address the 
nonlinear evolution of surface modes. In \se{body} we discuss the nonlinear evolution of 
body modes and high-$m$ surface modes at $\Mb>M_{\rm crit}$.

%!!!!!!!!!!!!!!!!!!!!!!!!!!!!!!!
\subsection{The Nonlinear Evolution of Surface Modes}
\label{sec:surface}
\smallskip
The range of parameters studied in the numerical simulations presented in this section is listed in \tab{surface}. 
These span the range of density contrast and Mach number relevant to cosmic cold streams, $0.5 < \Mb < 1.5$ and 
$1 < \delta < 100$, and also include an effectively incompressible case with $\Mb = 0.1$. For each case, we 
simulated both a 3d cylinder and a 2d slab as described in \se{methods}. Additionally, we simulated several 
cases with different resolution or refinement schemes to check convergence. The results of these convergence 
studies are presented in Appendix \se{convergence}. All of the results presented in this section are converged 
with respect to the grid.

\subsubsection{Stream Morphology}
\smallskip
\Fig{colour_panel_M1D1} shows a time sequence of the evolution of the passive scalar field, $\psi$, for the 
case $(\Mb,\delta)=(1,1)$. We show snapshots at $t/\tsc\sim 2$, $4$, $6$, and $8$ for the 2d slab simulation 
(left), a slice through the $y=0$ plane of a 3d cylinder (an edge-on view, centre), and a slice through the 
$z=0$ plane of the cylinder (a face-on view, right). At $t\sim 2\tsc$, the distribution of $\psi$ in the $xz$ 
plane of the 3d cylinder appears nearly identical to its distribution in the 2d slab, as expected from 
\se{theory_surface}. The distribution in the $xy$ plane appears dominated by a combination of symmetric and 
antisymmetric modes, with $m=0$ and $1$, as initialized. At $t=4\tsc$, the edge-on view of the cylinder remains 
very similar to the slab, with large-scale coherent eddies surrounding a relatively unmixed core. On the other 
hand, in the face-on view the symmetry seems to have broken and azimuthal modes with $m$ of a few are present. 
By $t=6\tsc$, the structure of the 3d simulation begins to deviate from that of the 2d slab. In 2d, the large-scale 
eddies remain coherent and continue to grow while the inner $\sim 20\%$ of the slab remains unmixed. On the 
other hand, in 3d the largest eddies have begun to break-up and cascade towards smaller scales, while the high-$m$ 
modes continue to grow, generating a more turbulent and mixed structure with very little unmixed fluid in the stream, 
concentrated along its axis. By $t=8\tsc$, the 3d simulation is completely turbulent and the stream contains no unmixed 
fluid, while in the 2d simulation the largest eddies are still coherent and the inner $\sim 10\%$ of the stream is 
still relatively unmixed. This highlights that while the initial shear layer growth in 3d cylinders is very similar 
to 2d slabs, there is a qualitative difference between 2d and 3d, as discussed in \se{theory_surface}. In 2d, there 
exists only an inverse cascade to larger scales which is why the largest eddies remain coherent and only grow larger 
as they merge. In 3d, the inverse cascade coexists with a direct cascade to smaller scales which breaks up the largest 
eddies, generates turbulence, and enhances mixing. Nevertheless, the overall thickness of the shear layer is very similar 
in 2d and 3d, as predicted in \se{theory_surface}.

\subsubsection{Vorticity}
\smallskip
\Fig{vorticity_slice_panel_M1D1} compares the evolution of vorticity, $\vec{\Omega}=(\vec{\nabla}\times\vec{v})$, 
in 2d slab and 3d cylinders. We focus on the same case shown in \fig{colour_panel_M1D1}, $(\Mb,\delta)=(1,1)$, but 
these results apply to all simulations. We show slices through the $y=0$ plane of the magnitude of the vorticity 
component perpendicular to the stream axis, at the same times as in \fig{colour_panel_M1D1}. This is 
$\Omega_{\perp}=|\Omega_{\rm y}|$ and $(\Omega_{\rm r}^2+\Omega_{\varphi}^2)^{1/2}$ in 2d and 3d respectively, 
and these have been normalized by the stream sound crossing time. 
Similar to the distribution of $\psi$ seen in \fig{vorticity_slice_panel_M1D1}, at $t\sim 2\tsc$ the vorticity in 
the 2d and 3d simulations are nearly identical. In the 2d simulation the vorticity remains concentrated in well 
defined vortices which continue to grow by mergers until $t>8\tsc$. However, in the 3d simulation the vortices begin 
to break up and transfer power to smaller scales already at $t\sim 4\tsc$, and by $t=8\tsc$ the situation is completely 
turbulent, with a homogeneous and isotropic distribution of vorticity. This highlights the qualitative difference between 
vortex evolution and turbulence generation in 2d and 3d discussed above.

\smallskip
\Fig{vorticity_panel_M1D1} shows the evolution of the mean azimuthal component of the vorticity, 
$\Omega_{\varphi}=(\nabla\times \vec{v})\cdot {\hat{\varphi}}$, in the same 3d simulation and 
at the same times as shown in \figs{colour_panel_M1D1} and \figss{vorticity_slice_panel_M1D1}. 
%We show the three cylindrical components of the vorticity, $\Omega_{\rm r}$, $\Omega_{\rm varphi}$, and $\Omega_{\rm z}$. Each component has been averaged over the full length of the box in the $z$ direction and normalized by the stream sound crossing time. 
$\Omega_{\varphi}$ has been averaged over the full length of the box in the $z$ direction and 
normalized by the stream sound crossing time. This averaging highlights the vorticity of the 
largest eddies while removing most small scale random motions. At $t=2\tsc$, the vorticity is 
entirely in the $\varphi$ direction and concentrated in a well defined ring, as expected from 
the discussion in \se{theory_surface} and consistent with the similarity between the 2d slab 
and 3d cylinder at this time (\figs{colour_panel_M1D1} and \figss{vorticity_slice_panel_M1D1}). 
At later times, the vortex ring expands radially into both the stream and the background as the shear 
layer grows. Its structure remains relatively coherent, with nearly all of the large-scale vorticity 
in the $\varphi$ direction. 
%, despite the appearance of small scale vorticity in the $r$ and $z$ directions as well. 
At $t=8\tsc$, the volume-weighted mean values of $\Omega_{\rm \varphi}$, $\Omega_{\rm r}$ 
and $\Omega_{\rm z}$, averaged over the full length of the box and within $r<2\Rs$, are 
$\sim 0.77$, $\sim 0.0001$ and $0.0022\tsc^{-1}$ respectively, while their standard deviations 
are $\sim 0.36$, $0.29$ and $0.40\tsc^{-1}$. However, in any single slice along $z$, the 
vorticity is nearly isotropic at $t\gsim 4\tsc$, once velocity perturbations in all three 
directions have grown, as evident in \fig{vorticity_slice_panel_M1D1}. 

\begin{figure*}
\begin{center}
\includegraphics[width =0.47 \textwidth]{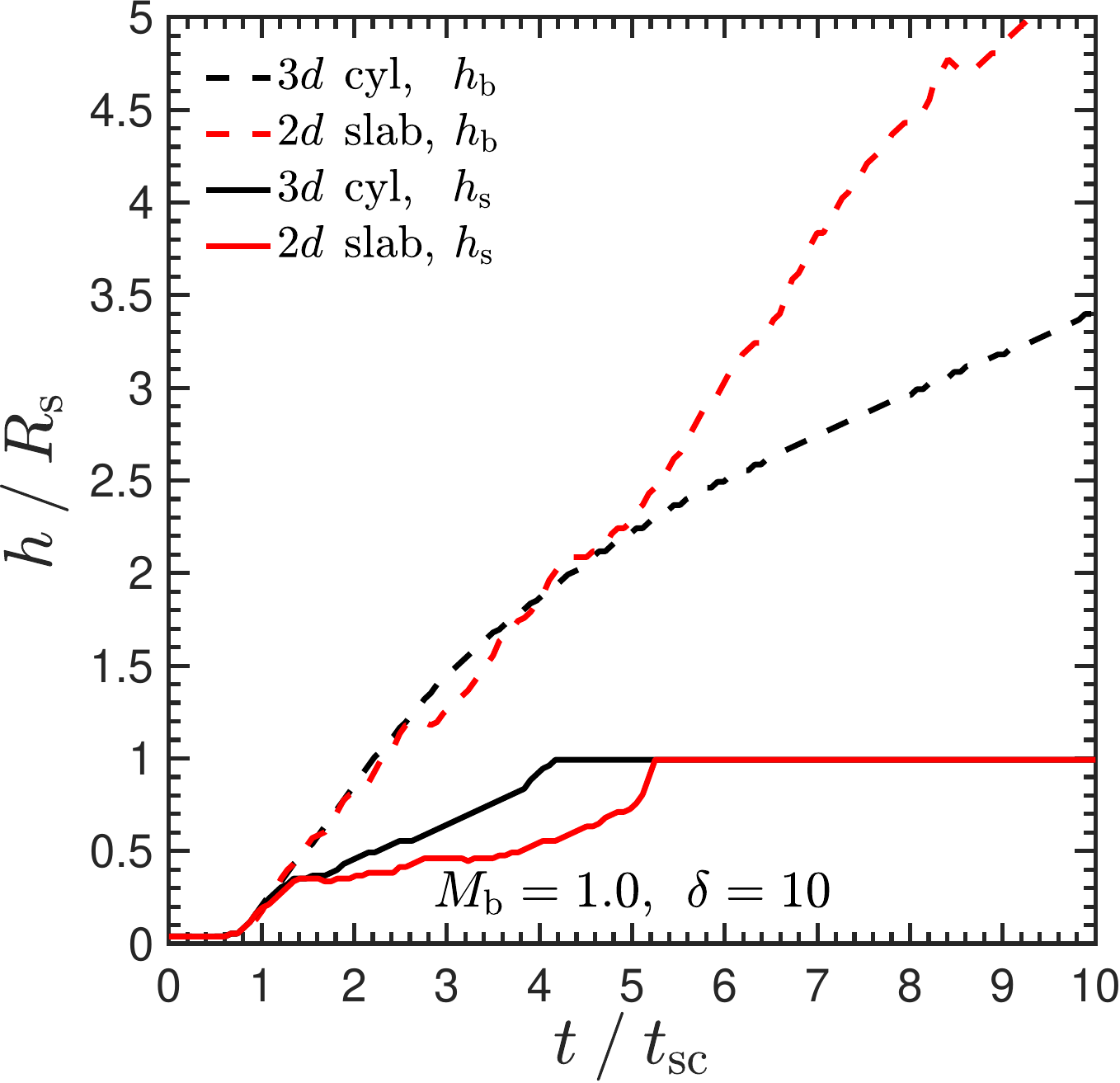}
\hspace{-0.3cm}
\includegraphics[trim={-0.5cm -0.2cm 0.0cm 0}, clip, width =0.495 \textwidth]{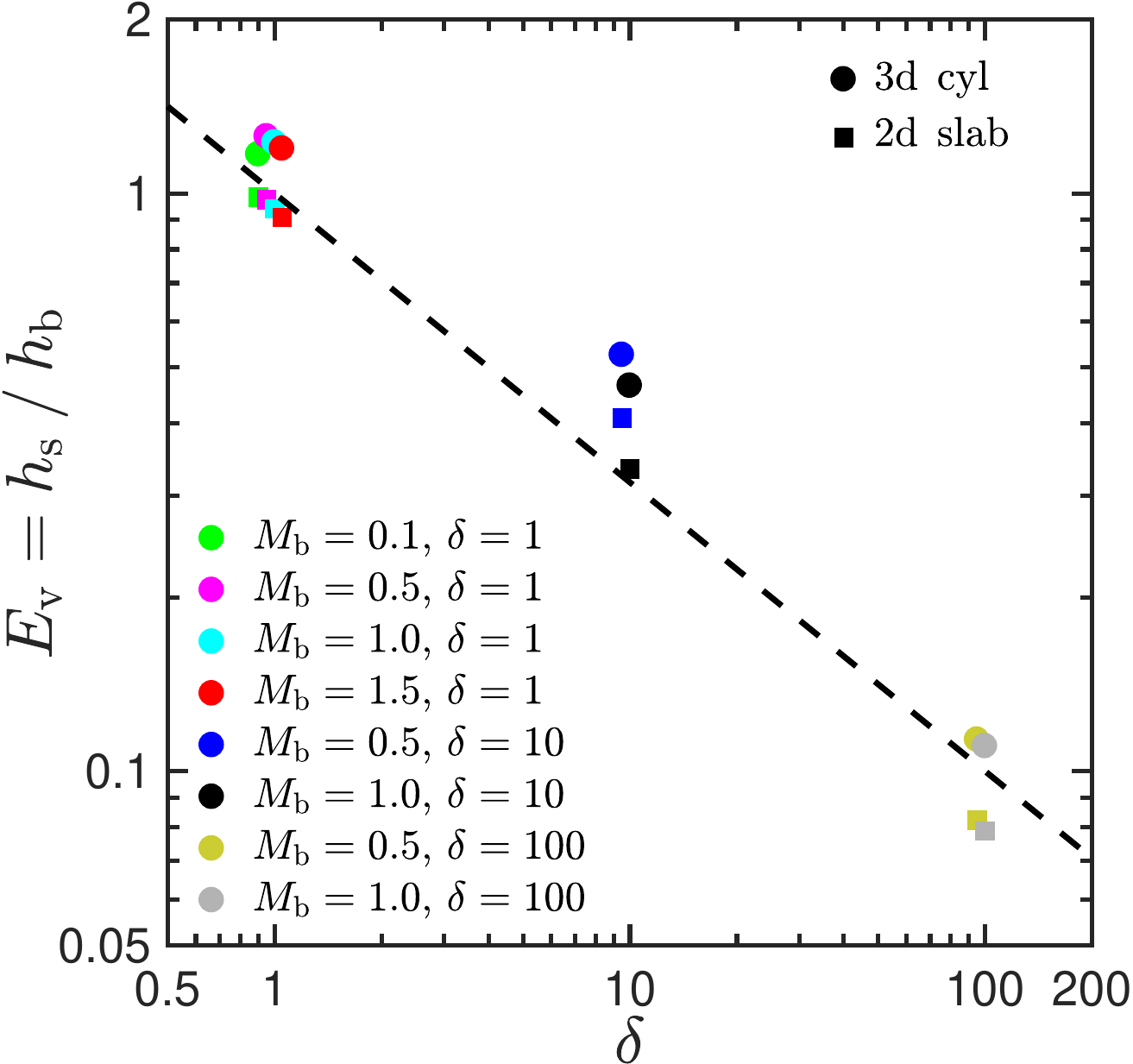}
\end{center}
\caption{Shear layer growth in 2d slabs versus 3d cylinders, for subsonic and transonic streams with 
respect to the sum of the sound speeds in both fluids, $M_{\rm tot}\le 1$. \textit{On the left} we show 
the penetration of the shearing layer into the stream and the background, $\hs/\Rs$ (solid) and $\hb/\Rs$ 
(dashed) respectively, as a function of time, $t/\tsc$, for a 2d slab (red) and a 3d cylinder (black) 
with $(\Mb,\delta)=(1.0,10)$. $\hs$ behaves similarly in 2d and 3d until it reaches $\sim 0.3\Rs$, 
after which it grows $\sim 30\%$ faster in 3d. $\hb$ remains extremely similar in 2d and 3d until it 
reaches $\gsim 1.5\Rs$, after which the growth rate in 3d is reduced by a factor of $\sim 2$, while the 
2d growth rate remains constant. Very similar behaviour is seen in all simulations (see text). \textit{The 
centre panel} shows, for all simulations, the average value of the entrainment ratio, $E_{\rm v}=\hs/\hb$, 
during the period when $t>2\tsc$ and $\hs<0.9\Rs$. Different colours mark different combinations of $(\Mb,\delta)$ 
while squares and circles mark 3d and 2d simulations respectively. The dashed line shows the analytic prediction, 
$\Ev=\delta^{-1/2}$ (\equnp{Ev}). While $\Ev$ is systematically $\sim 30\%$ higher in 3d due to differences 
in $\hs$, the simulation results agree very well with the analytic predictions in both 2d and 3d.}
\label{fig:surface_h} 
\end{figure*}

\begin{figure*}
\begin{center}
\includegraphics[width =0.252 \textwidth]{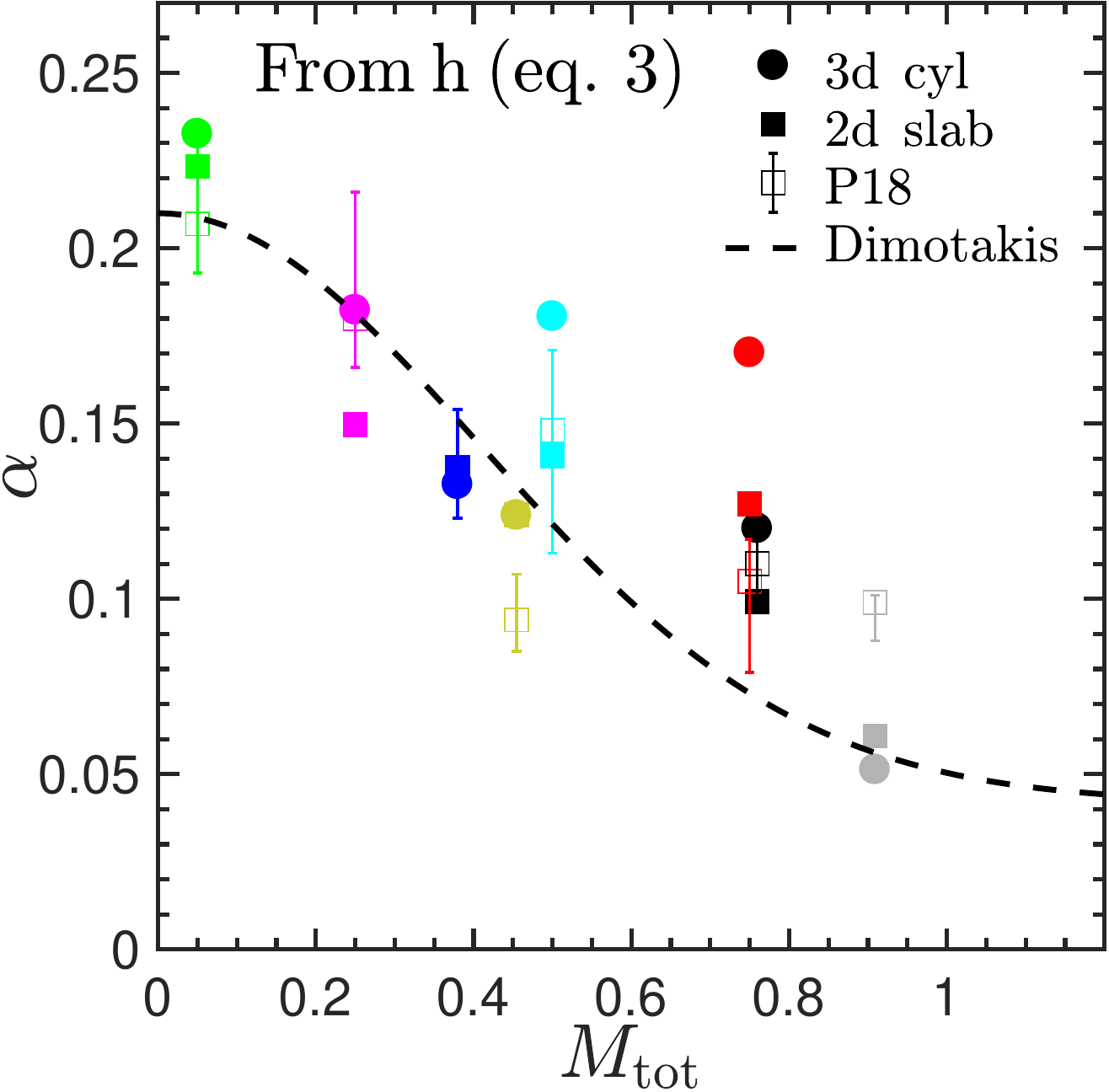}
\hspace{-0.2cm}
\includegraphics[trim={1.7cm 0.0cm 0.0cm 0}, clip, width =0.22 \textwidth]{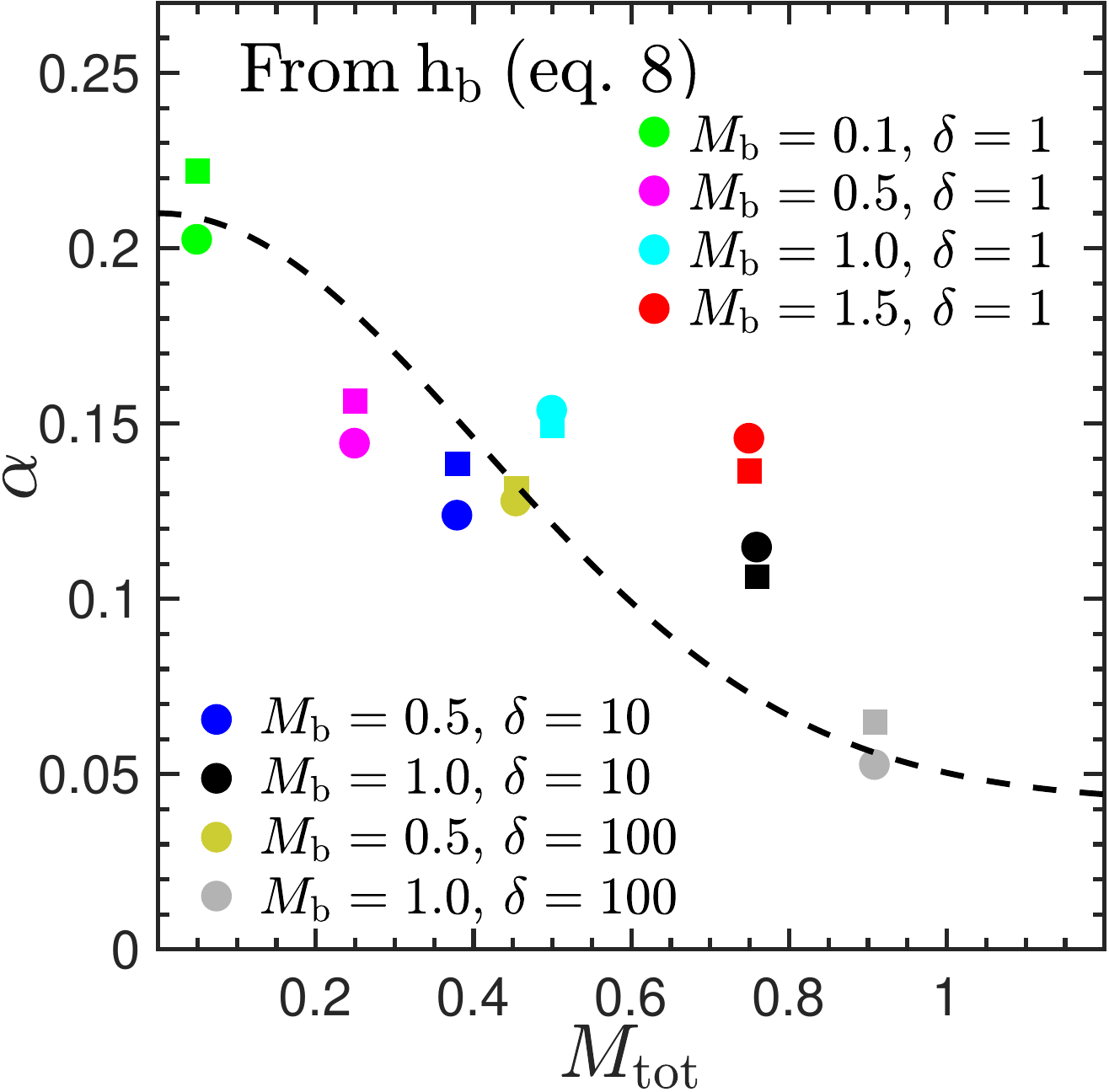}
\hspace{-0.2cm}
\includegraphics[trim={1.7cm 0.0cm 0.0cm 0}, clip, width =0.2265 \textwidth]{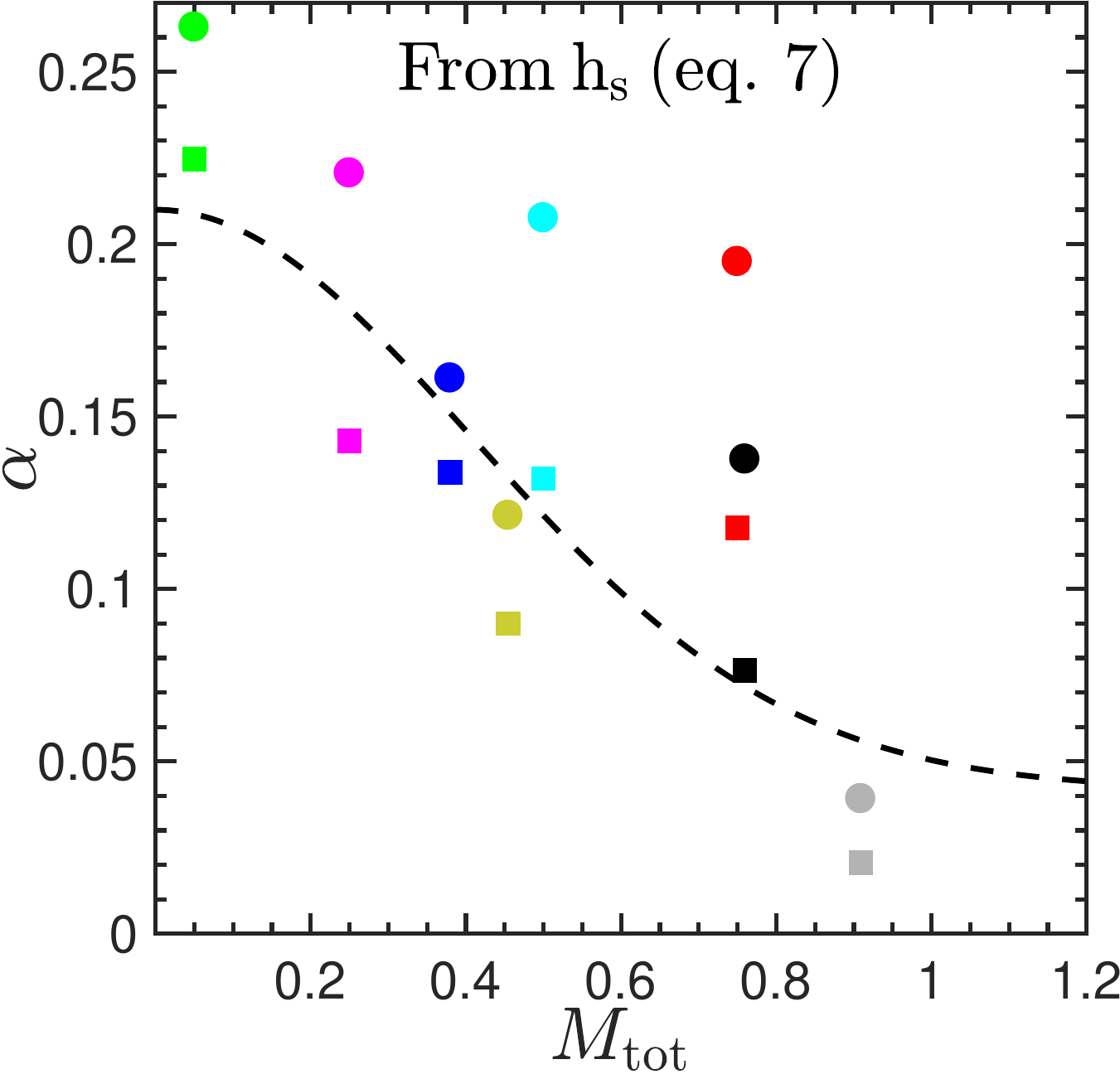}
\hspace{-0.25cm}
\includegraphics[trim={-0.2cm 0.0cm 0.0cm 0}, clip, width =0.267 \textwidth]{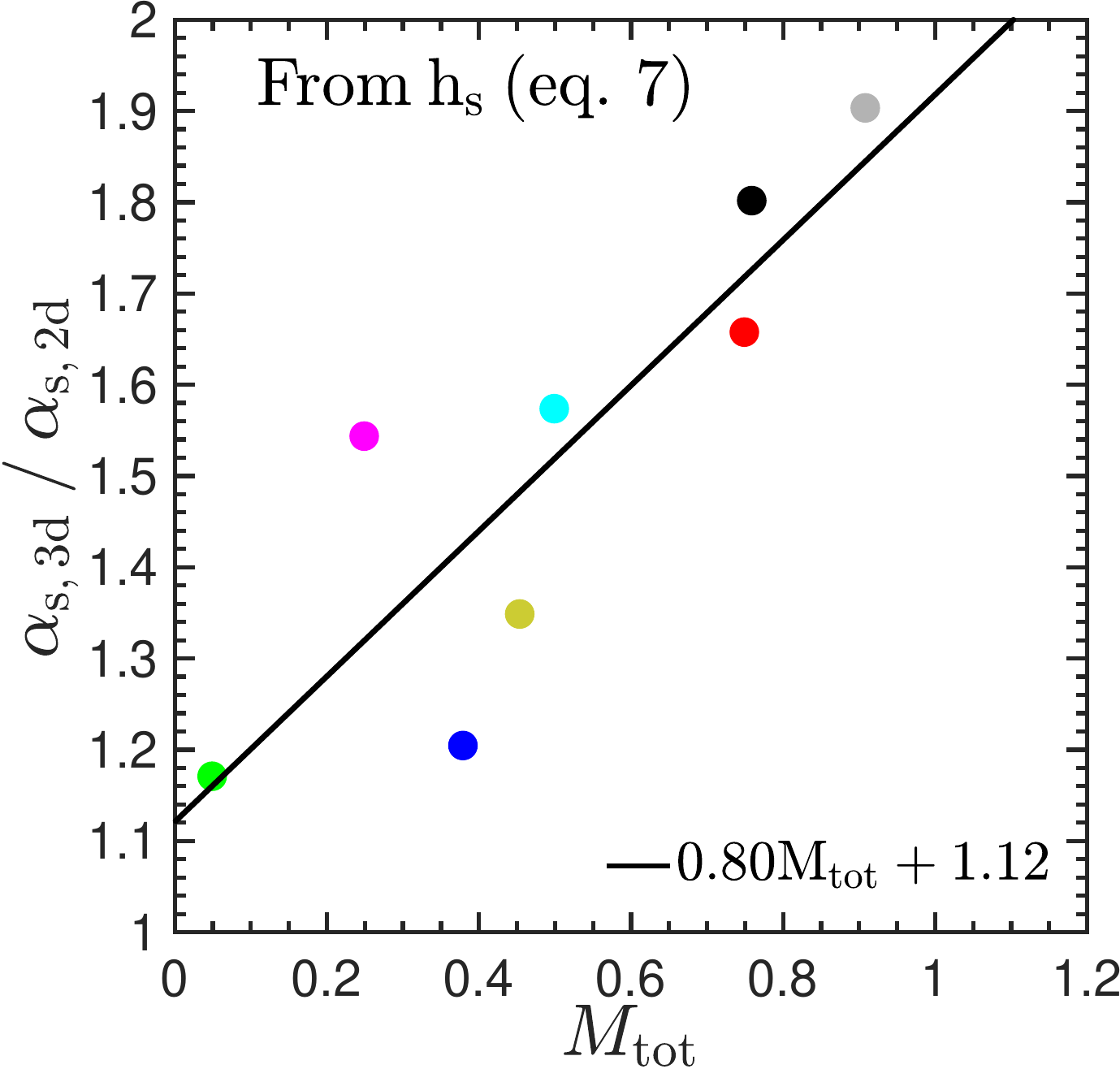}
\end{center}
\caption{The normalized shear layer growth rate, $\alpha$, as a function of compressibility parametrized 
by $M_{\rm tot}$. We show the values of $\alpha$ derived using \equ{shear_growth} (left), \equ{hb_growth} (centre-left), 
and \equ{hs_growth} (centre-right), after measuring $h$, $\hb$, and $\hs$ as a function of time when $t>2\tsc$ and $\hs<0.9\Rs$.
Different colours mark different combinations of $(\Mb,\delta)$ while filled squares and circles mark 3d and 2d simulations 
respectively. Empty squares in the left-hand panel show the results of P18 with error bars denoting the scatter among 
three random realizations of the initial perturbation spectrum. In our 2d simulations, all three methods for measuring 
$\alpha$ yield very similar results, which are consistent with both the results of P18 and the empirical fit by 
\citet{Dimotakis91} (\equnp{alpha_fit}), shown by the dashed line. Using $\hb$ to determine $\alpha$ yields very similar 
results in 2d and 3d simulations, though using $\hs$ yields larger values in 3d cylinders. In the right-hand panel 
we show the ratio of $\alpha_{\rm 3d}$ to $\alpha_{\rm 2d}$ as determined using $\hs$. This ratio is roughly proportional 
to $M_{\rm tot}$, ranging from $\sim (1.1-1.9)$ in the range $M_{\rm tot}\sim (0-1)$. We present the best-fit linear relation 
with a solid line.}
\label{fig:alpha} 
\end{figure*}

\subsubsection{Shear Layer Growth}
\smallskip
\Fig{surface_h} examines the growth of the shear layer in 2d slab and 3d cylinder simulations. 
On the left we show the one-sided thicknesses, $\hb/\Rs$ and $\hs/\Rs$, as a function of time for 
for 2d slab and 3d cylinder simulations with $(\Mb,\delta)=(1.0,10)$. Very similar behaviour is 
seen in all $(\Mb,\delta)$ combinations. While $\hs\lsim 0.3\Rs$, its behaviour in 2d and 
in 3d are very similar. At later times, the growth of $\hs$ in the 2d simulation undergoes a 
series of slight stalls where its amplitude is nearly constant for a brief time followed by 
continued growth at roughly the same rate. This is not seen in the 3d simulations, where $\hs$ 
grows at a roughly constant rate until $\hs\sim 0.8\Rs$ at which point the growth rate increases 
until the whole stream is engulfed by the shear layer. This may partly be due to the fact that 
there are many more eddies in a 3d cylinder than a 2d slab, so discrete eddy mergers (the main 
mechanism for shear layer growth) average out. Furthermore, as $\hs$ grows towards the stream 
axis, the distance between vortex planes along different azimuthal cuts of a cylinder decreases, 
and interactions between vortex planes may occur, generating turbulence. Overall, $\hs\sim (30-40)\%$ 
larger in 3d than in 2d for most of the evolution. On the other hand, the behaviour of $\hb$ is 
identical in 2d and in 3d until $\hb\sim (1.5-2)\Rs$, consistent with our expectations from 
\se{theory_surface}. At later times, $\hb$ continues to grow at the same rate in 2d, while the 
growth rate in 3d decreases by a factor of $\sim 2$. In all simulations, the decrease in the growth 
rate of $\hb$ occurs when $\hb\sim (1.5-2)\Rs$, and does not appear to be correlated with a fixed 
number of sound crossing times or with a fixed decrease in the stream velocity (discussed below). 
We therefore speculate that it is due to turbulence transfering energy from large to small scales, 
thereby decreasing the energy available to the largest eddies which drive the shear layer growth. 
This cascade exists only in 3d, and ``kicks in'' once the largest eddies have grown comparable to 
the stream width. 

\begin{figure*}
\begin{center}
\includegraphics[width =0.47 \textwidth]{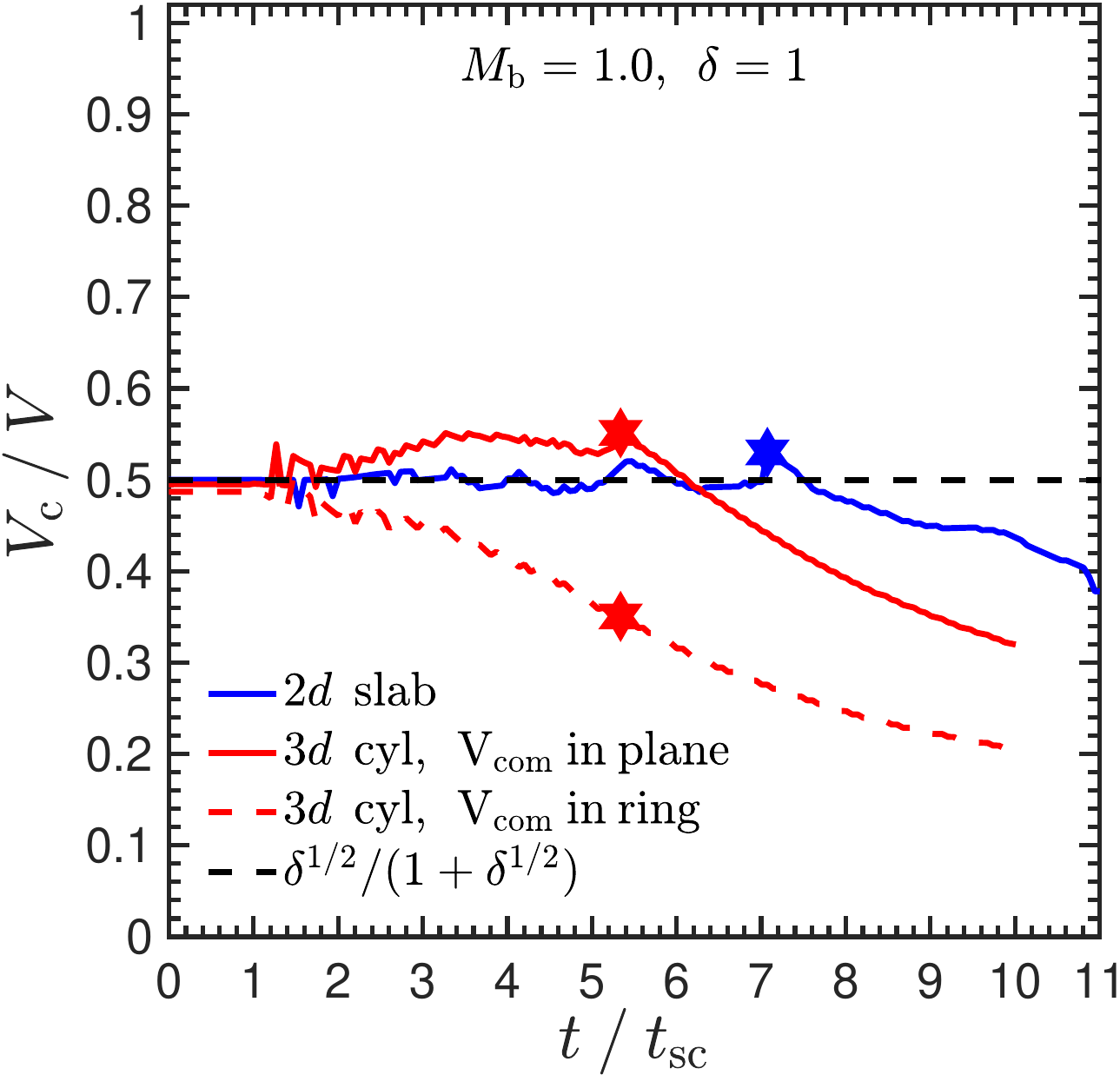}
\hspace{-0.3cm}
\includegraphics[trim={-0.5cm -0.2cm 0.0cm 0}, clip, width =0.4975 \textwidth]{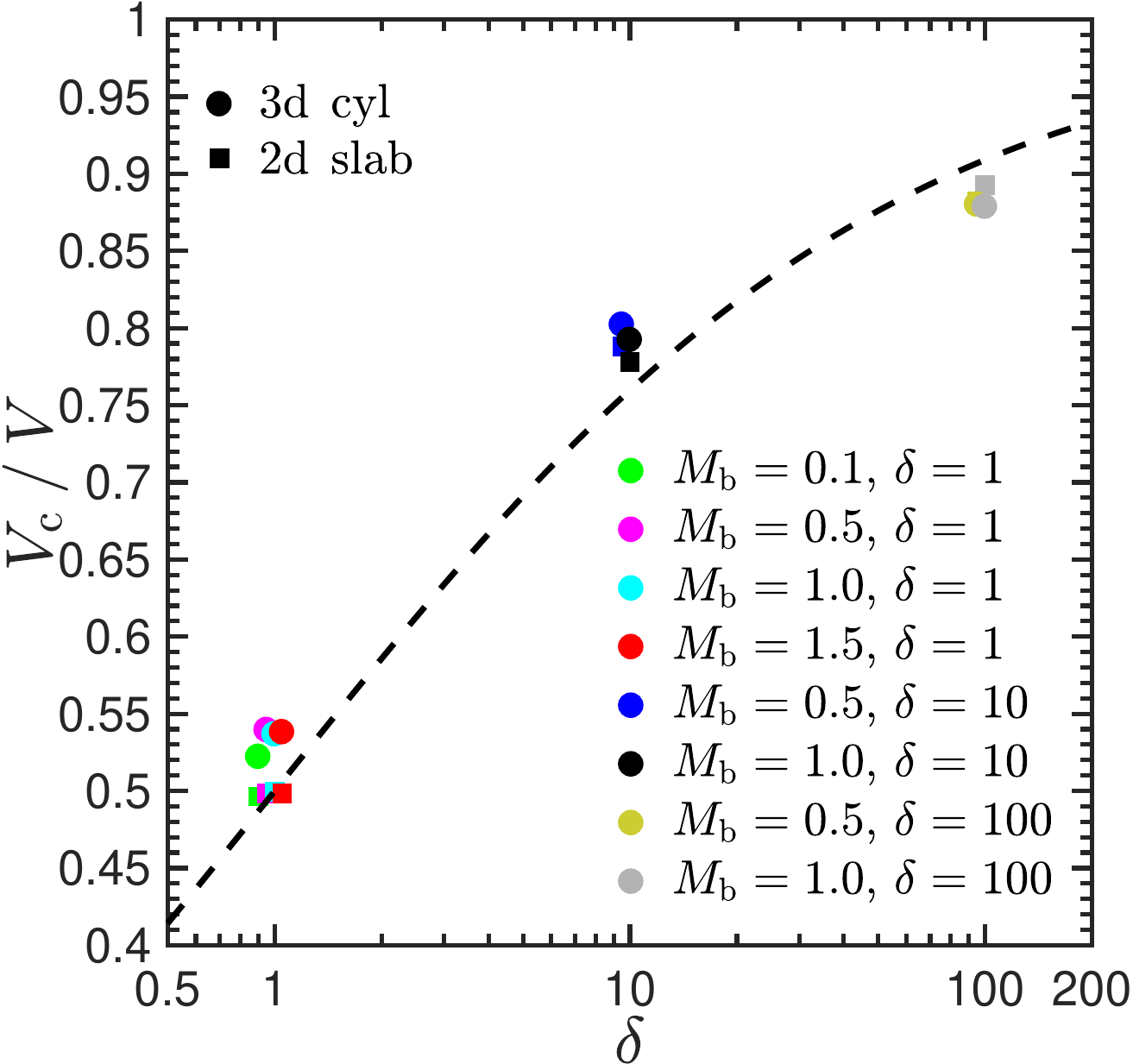}
\end{center}
\caption{Convection velocity, $V_{\rm c}$, of 2d slabs and 3d cylinders, for subsonic and transonic streams with 
$M_{\rm tot}\le 1$. $V_{\rm c}$ is the velocity of the largest eddies in the shear layer between the stream and 
the background, and is predicted to follow \equ{convective}, shown by the dashed lines in both panels. In the 
left-hand panel we show the evolution of $V_{\rm c}$ as a function of time for the simulations with $(\Mb,\delta)=(1,1)$. 
The blue line shows the 2d slab case, where $V_{\rm c}$ is evaluated by the centre of mass velocity in the shear layer, 
following P18. This matches the analytic prediction until $\hs=\Rs$, which is marked by the blue star. The 
red lines represent the 3d cylinder simulation. The solid line shows the centre of mass velocity in the shear 
layer within a planar slice through the stream midplane, following \equ{Vcom_plane}, while the dashed line shows 
the centre of mass velocity within the full cylindrical shear layer, following \equ{Vcom_cyl}. The former matches 
the analytic prediction until $\hs=\Rs$, marked by the red star, while the latter is systematically lower and 
monotonically decreasing from $t\gsim \tsc$. The right-hand panel shows the results for all simulations as a 
function of density contrast, averaged when $t>2\tsc$ and $\hs<0.9\Rs$. Symbols are as in \fig{surface_h}. 
The results closely follow the analytic prediction, in both 2d and 3d.
}
\label{fig:Vc} 
\end{figure*}

\begin{figure}
\includegraphics[width =0.47 \textwidth]{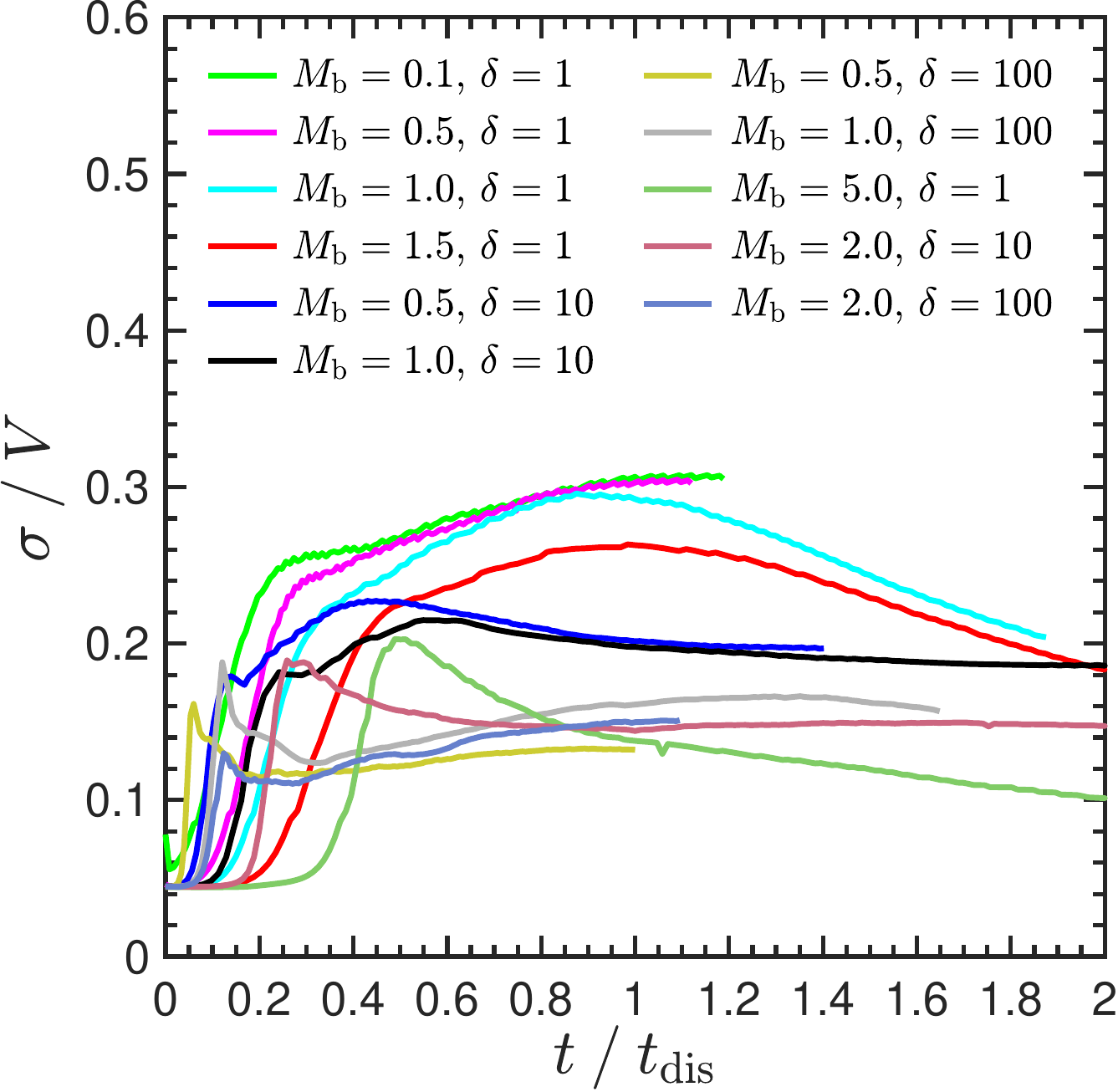}
\caption{Turbulence induced by shear layer growth. We show the turbulent velocities inside the shear layer 
as a function of time for all 3d simulations unstable to surface modes. These include three simulations with 
$\Mb>M_{\rm crit}$ that are unstable to high-$m$ surface modes, discussed in \se{body}. The turbulent velocities, 
$\sigma$, have been normalized to the initial stream velocity, $V$, while the time has been normalized to the 
disruption time, $t_{\rm dis}$, when the shear layer consumes the entire stream. All simulations, regardless 
of $\Mb$ and $\delta$, follow the same trend. The turbulence increases at early times, reaches a peak value of 
$\sigma/V\sim 0.2-0.3$ at early times, and then decreases to an assymptotic value of $\sigma/V\lsim 0.2$. 
2d slab simulations do not exhibit such universal behaviour.}
\label{fig:turbulence} 
\end{figure}

\smallskip
In the right panel of \fig{surface_h} we show the entrainment ratio (\equnp{Ev}) for all simulations, both 2d and 3d. 
These were evaluated at $t>2\tsc$ and while $\hs<0.9\Rs$, following an initial transient phase and before the behaviour 
of $\hs$ becomes too stochastic. We show for comparison the analytic prediction $E_{\rm v}=\delta^{-1/2}$ (\equnp{Ev}). Our 
simulations match the analytic prediction very well, and in all cases the 3d results are very similar to the 2d results. 
There is a systematic trend for $E_{\rm v}$ to be $\sim (30-40)\%$ larger in 3d than in 2d. This is driven primarily by 
$\hs$ being larger in 3d, while $\hb$ is very similar in 2d and 3d, as seen in the left-hand panel and described above. 

\begin{figure*}
\begin{center}
\includegraphics[trim={-0.7cm -0.0cm 0.0cm 0}, clip, width =0.345 \textwidth]{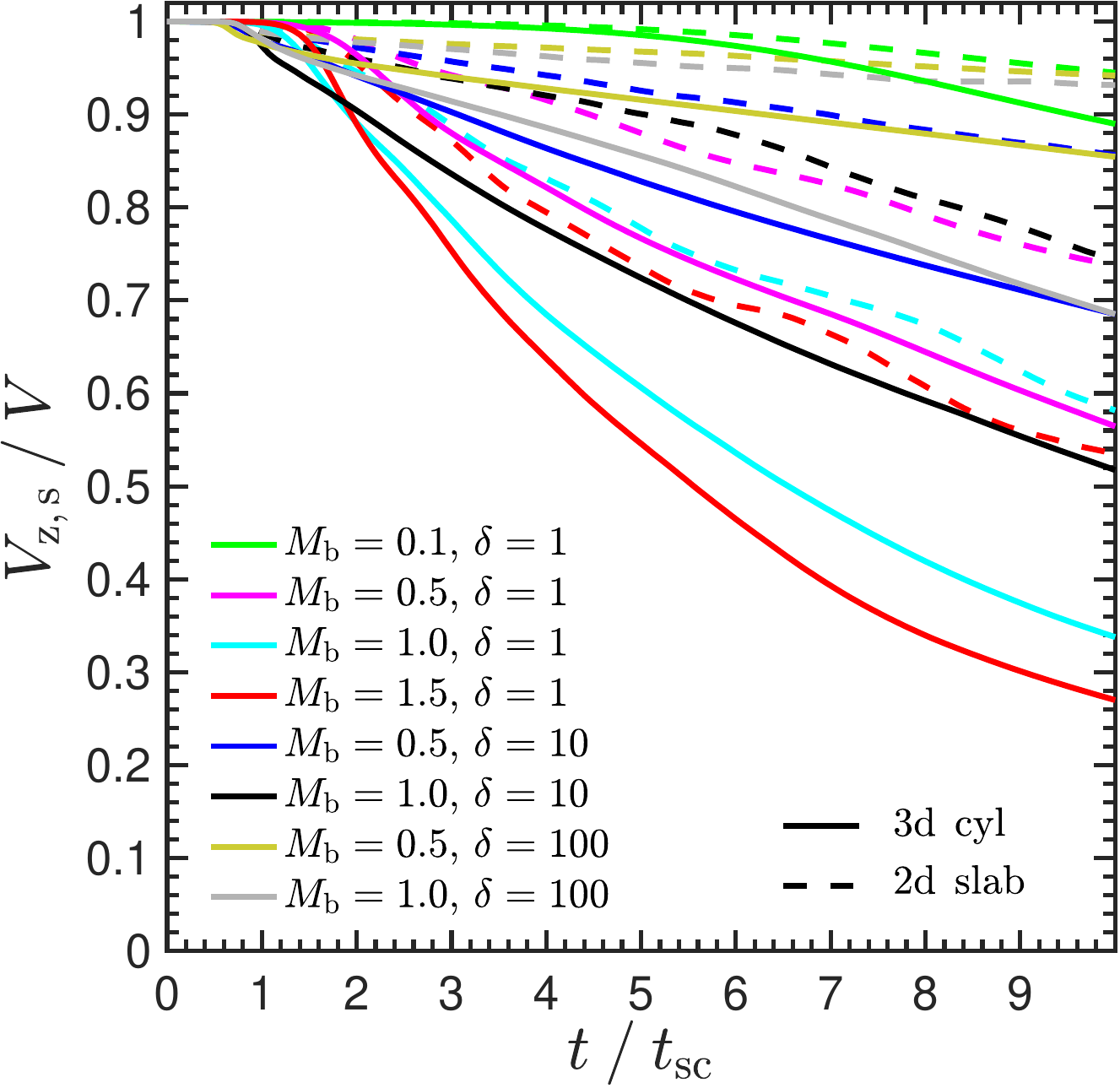}
\hspace{-0.175cm}
\includegraphics[trim={1.9cm -0.0cm 0.0cm 0}, clip, width =0.29 \textwidth]{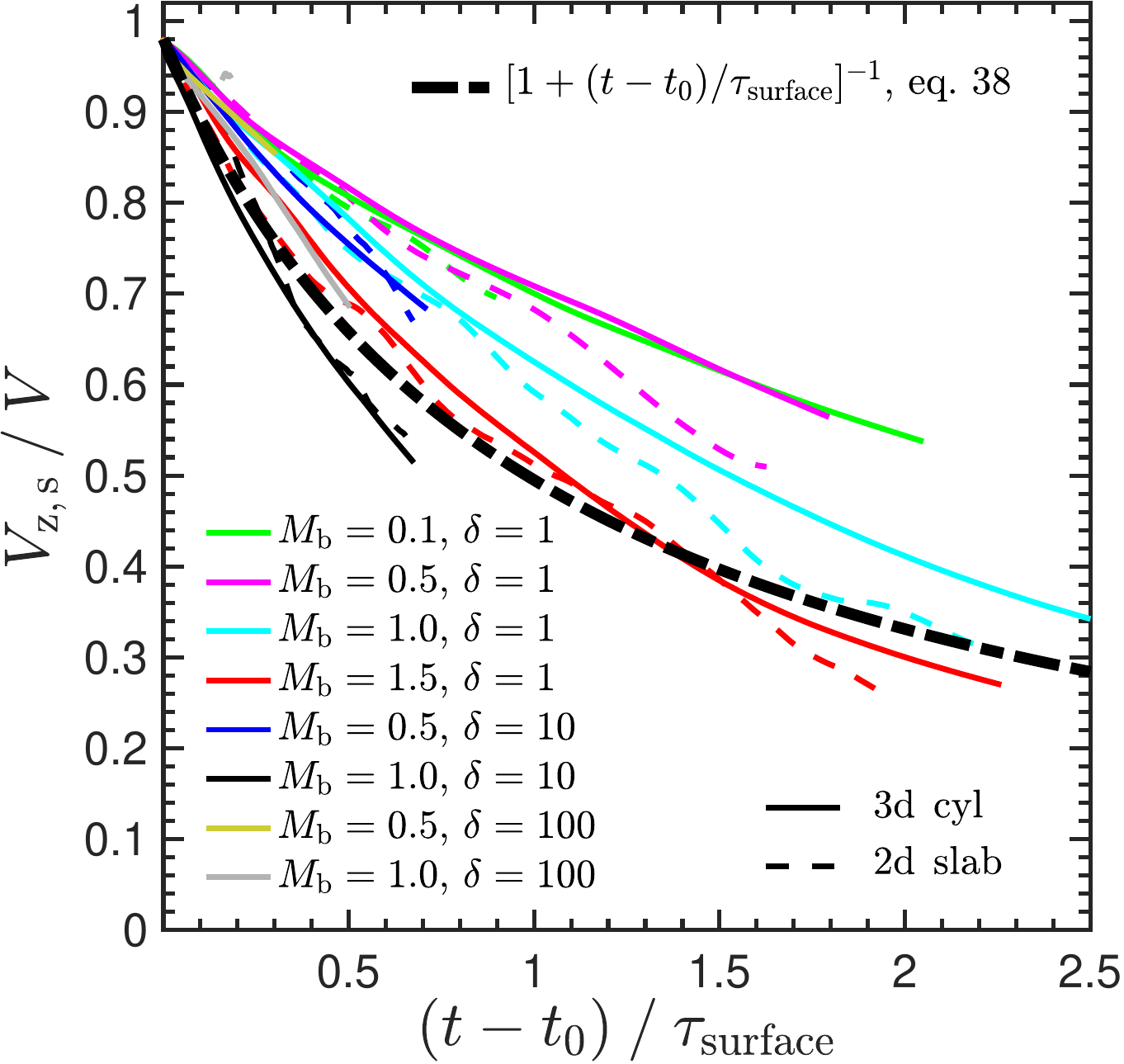}
\hspace{-0.4cm}
\includegraphics[trim={-0.7cm -0.0cm 0.0cm 0}, clip, width =0.3525 \textwidth]{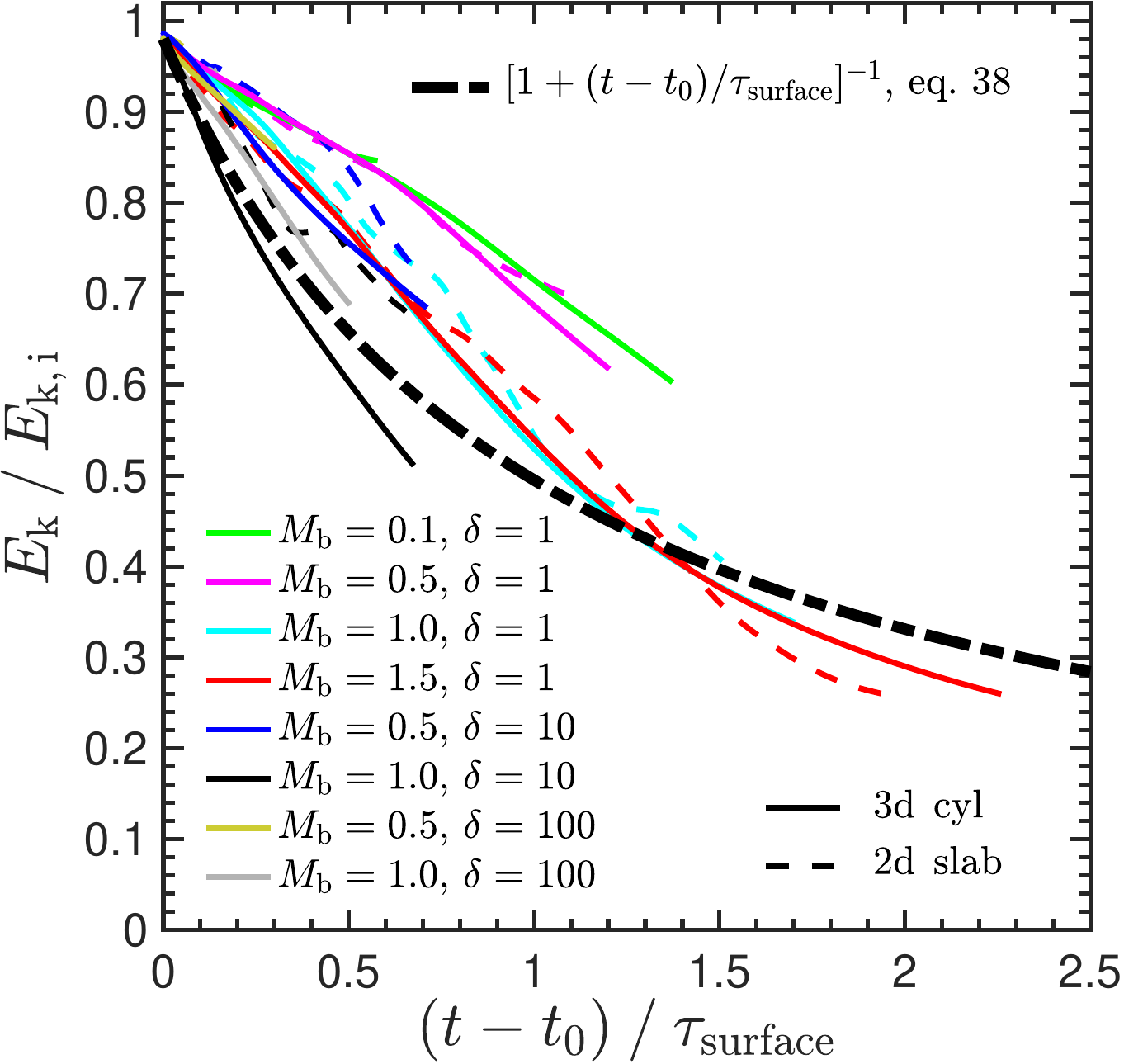}
\end{center}
\caption{Deceleration and dissipation of bulk kinetic energy in streams due to 
surface modes, for subsonic and transonic streams with $M_{\rm tot}\le 1$. \textit{In 
the left and centre panels}, we show the centre of mass velocity of the stream fluid (weighted by 
the passive scalar $\psi$) normalized by its initial value, as a function of time for all 
simulations. Different colours mark the different combinations of $(\Mb,\delta)$. Solid 
and dashed lines show the results of 3d and 2d simulations respectively. In the left-hand 
panel the time axis is normalized by the sound crossing time, which is the same in 2d and 
3d. Clearly, 3d cylinders decelerate much faster than 2d slabs in all cases, though the 
time at which deceleration begins, $t_0$, is of order the sound crossing time in all cases. 
In the right-hand panel we renormalize the time axis by the predicted deceleration timescales, 
$\tau_{\rm surface}$, given by \equ{tau_surface_2d} and \equ{tau_surface_3d} for 2d and 3d 
respectively. For $\alpha$, we use \equ{alpha_fit} for 2d slabs and 3d cylinders with $\delta<8$, 
while for 3d cylinders with $\delta>8$ we use half that value (see text for details). Both the 
2d and 3d simulations reach half their initial velocity at $t\sim t_0 + \tau_{\rm surface}$, 
as predicted. The thick dot-dashed line in the right-hand panel shows the predicted velocity 
profile from \equ{dec_model}. Except for the cases with $\delta=1$ and $\Mb\le 1$, which are 
the least relevant for cold streams, the model is a good match to the simulated results. 
\textit{In the right-hand panel}, we show the total kinetic energy associated 
with laminar flow of both stream and background fluid normalized by its initial value, as a 
function of time normalized by $\tau_{\rm surface}$. The rate of dissipation of bulk kinetic 
energy in 3d cylinders is extremely similar to the stream deceleration rate, as predicted.
}
\label{fig:deceleration_surface} 
\end{figure*}

\smallskip
\Fig{alpha} shows the value of $\alpha$ (\equnp{shear_growth}) measured in our simulations. Similar to $E_{\rm v}$, 
these were evaluated at $t>2\tsc$ and while $\hs<0.9\Rs$. As discussed in P18, $\alpha$ is expected to be mainly a 
function of the compressibility, and is empirically found to correlate with $M_{\rm tot}$, shown on the $x$ axis. We show 
the values of $\alpha$ measured from the simulations in three different ways, using the total width of the shear layer, $h$, 
following \equnp{shear_growth} (left), using the one-sided thickness in the background, $\hb$, following \equnp{hb_growth} 
(centre-left), and using the one-sided thickness in the stream, $\hs$, following \equnp{hs_growth} (centre-right). 
We note that the growth of $\hs$ is related to stream disruption (\equnp{tau_diss_2d}), while the growth of $\hb$ is related 
to stream deceleration (\equsnp{tau_surface_2d} and \equmnp{tau_surface_3d}). In the left hand panel, we also show the results 
from P18 (figure 7), where the error bars represent the scatter obtained in three random realizations of the initial perturbation 
spectrum.
%In practice, to avoid the systematic $\sim 20-30\%$ differences in $\hs$ between 2d and 3d, we measured the growth rate of $\hb$ and evaluated $\alpha$ from \equ{hb_growth}. This is justified since we have seen above that the ratio of $\hs/\hb$ follows \equ{Ev} in both 2d and 3d. 
For comparison, in each panel we show the empirical fit proposed by \citet{Dimotakis91} (\equnp{alpha_fit}). 
When measuring the full width of the shear layer (left), the results of our 2d simulations are roughly consistent 
with \equ{alpha_fit}, at least to the same degree as the results of P18. In fact, our results at $\delta=100$ are 
actually more consistent with \equ{alpha_fit} than those of P18. However, for $\delta=1$ the values of $\alpha$ 
measured in this way are systematically higher in 3d simulations. This is due to the more rapid growth of the shear 
layer into the stream, which is a larger fraction of the total shear layer growth for smaller $\delta$. Therefore, 
when using $\hb$ to determine $\alpha$, as in the centre panel, the results of 3d and 2d simulations are in excellent 
agreement, though we stress that $\alpha$ here is measured primarily when $\hb < 2\Rs$, \textit{before} the growth 
rate in 3d simulations decreases by a factor of 2 relative to 2d simulations, as discussed above. On the other hand, 
when using $\hs$ to determine $\alpha$ in the centre-right panel, the discrepancy between 2d and 3d is much larger, 
with the ratio $\alpha_{\rm 3d}/\alpha_{\rm 2d}$ scaling roughly linearly with $M_{\rm tot}$, as shown in the right-most 
panel. We find the best-fit linear relation to be 
\be 
\label{eq:alpha_ratio}
\frac{\alpha_{\rm s,\,3d}}{\alpha_{\rm s,\,2d}} \simeq 0.80M_{\rm tot}+1.12,
\ee
{\no}which is shown by the solid line in the right-most panel of \fig{alpha}.

\smallskip
We conclude that shear layer growth in 3d cylinders is overall very similar to 2d slabs, as expected, though the 
penetration into the stream is systematically more rapid following \equ{alpha_ratio}. Furthermore, similar 
to P18, we conclude that the shear layer growth rate is reasonably well fit by \equ{alpha_fit}, which is an excellent 
match to our simulations at high and low values of $M_{\rm tot}$, and shows a $\sim 50\%$ non-systematic discrepancy 
at intermediate values. As discussed in Appendix B of P18, \equ{alpha_fit} does appear to fit experimental data better 
than numerical simulations. The reasons for this are not entirely clear, and left for future study. However, since 
the shear layer growth rate is linear with $\alpha$, such a \textit{non-systematic} $50\%$ discrepancy will be 
negligible compared to other uncertanties in our model when applied to astrophysical scenarios in \se{application}, 
and also when compared to the \textit{systematic} differences between 2d slabs and 3d cylinders.

\subsubsection{Convection Velocity}
\smallskip
\Fig{Vc} examines the convection velocity, $V_{\rm c}$ (\equnp{convective}), the typical velocity 
of the largest eddies in the shear layer. In P18 we showed that in 2d slab simulations this could 
be evaluated from the centre of mass velocity of the fluid inside the shear layer, shown as the blue 
line in the left-hand panel for the case $(\Mb,\delta)=(1,1)$. All other simulations show similar 
behaviour. The 2d result is consistent with the analytic prediction, shown by the dashed line, until 
the shear layer encompasses the entire stream, $\hs=\Rs$, marked by the star. Based on the discussion 
in \se{theory_surface}, we expect that for 3d cylinders the same will be true for the centre of mass 
velocity of the fluid in the shear layer in a 2d slice through the stream axis at constant $\varphi$. 
The scatter between different $\varphi$ values is very small, as expected, of the order of a few percent. 
Therefore, we average over all angles to obtain better statistics. In practice, we compute 
\be 
\label{eq:Vcom_plane}
V_{\rm com,\,in\,plane} = \dfrac{\int_{\Rs-\hs}^{\Rs+\hb} \overline{\rho}_{(r)} \widetilde{v_{\rm z}}_{(r)} ~{\rm dr}}{\int_{\Rs-\hs}^{\Rs+\hb} \overline{\rho}_{(r)} ~{\rm dr}},
\ee
{\no}where the density profile, $\overline{\rho}(r)$, is obtained analogously to 
\equ{volume-averaged-colour-3d}, and the velocity profile is density-weighted, i.e. 
$\widetilde{v_{\rm z}}\equiv \overline{\rho v_{\rm z}}/\overline{\rho}$. This is shown 
by the red solid line in the left-hand panel of \fig{Vc}, and closely matches the analytic 
prediction and the 2d result, until $\hs=\Rs$ which is marked by the red star. 
%We have verified that the results obtained in this way are extremely similar to the results obtained by computing the center of mass velocity in the shear layer in a slice through the $xz$ plane or the $yz$ plane. 
Clearly, this is different than computing the centre of mass velocity within the cylindrical shear 
layer, given by 
\be 
\label{eq:Vcom_cyl}
V_{\rm com,\,in\,ring} = \dfrac{\int_{\Rs-\hs}^{\Rs+\hb} \overline{\rho}_{(r)} \widetilde{v_{\rm z}}_{(r)} r ~{\rm dr}}{\int_{\Rs-\hs}^{\Rs+\hb} \overline{\rho}_{(r)} r ~{\rm dr}},
\ee
{\no}and shown by the red dashed line in the same panel. While this begins at similar values to the prediction 
for $V_{\rm c}$ at $t\lsim \tsc$, it decreases monotonically in time. Given the similarity between 2d slabs 
and slices through 3d cylinders discussed above, we speculate that $V_{\rm c}$ may still be associated 
with the drift velocity of the largest eddies in the shear layer of 3d cylinders, as it is for 2d slabs (\se{theory_surface}). 
We thus conclude that the largest eddies in a cylindrical shear layer move \textit{faster} than the centre of mass 
velocity within the shearing ring.

\smallskip
In the right-hand panel of \fig{Vc} we show the convection velocity measured in all our simulations as 
a function of the density contrast. These were evaluated at $t>2\tsc$ and while $\hs<0.9\Rs$, similar 
to \fig{surface_h}. The results in 2d and 3d are extremely similar, and both follow the analytic prediction 
of \equ{convective}, shown by the dashed line. This strengthens the picture described in \se{theory_surface}, 
further highlighting the similarity of shear layer growth in 2d slabs and 3d cylinders.

\subsubsection{Turbulence}
\smallskip
The growth of the shear layer drives turbulence, fascilitating the mixing of the two fluids. This can be 
seen visually in \fig{colour_panel_M1D1} (and in \fig{colour_panel_M5D1} discussed in \se{body}). We evaluate 
the magnitude of turbulence within the shear layers in our 3d cylinder simulations as 
\be 
\label{eq:turbulence}
\sigma^2 = \dfrac{\int_{\Rs-\hs}^{\Rs+\hb} \overline{\rho}_{(r)} \left[(v_{\rm z}-\widetilde{v_{\rm z}}_{(r)} )^2 + v_{\rm x}^2+v_{\rm y}^2\right] r~ {\rm dr}}{\int_{\Rs-\hs}^{\Rs+\hb} \overline{\rho}_{(r)}\, r ~{\rm dr}}.
\ee
{\no}We subtract the longitudinal velocity at each radius in order to remove any ordered shear inherited 
from the initial conditions and focus on small scale turbulence. 

\smallskip
In \fig{turbulence} we show the evolution of 
$\sigma/V$, where $V=\Mb\cb$ is the initial stream velocity, for all 3d cylinder simulations studied here, as 
well as three simulations with $\Mb>M_{\rm crit}$ that are unstable to high-$m$ surface modes, discussed in 
\se{body} below. The latter have $(\Mb,\delta)=(5,1),\,(2,10),\,(2,100)$. The time axis in \fig{turbulence} has 
been normalized by $t_{\rm dis}$, the time when the shear layer consumes the entire stream, i.e. when $\hs=\Rs$. 
All simulations follow the same trend. $\sigma/V$ increases to a maximum of $\sim 0.2-0.3$ at early times, 
and then decays to an assymptotic value of $\sigma/V\lsim 0.2$. Simulations with larger $\delta$ peak at smaller 
values and at earlier times, but the variance is small considering the range of $\delta$ values simulated. 
Weighting the density in \equ{turbulence} (and in the evaluation of the shearing profile $\widetilde{v_{\rm z}}_{(r)}$) 
by the passive scalar $\psi$ in order to focus on turbulent motions of the stream fluid yields very similar 
results at late times, $t>t_{\rm dis}$, showing that at these late times both fluids are well mixed within the 
shear layer. 
Note that this means that one \textit{cannot} directly infer the total turbulent energy relative to the initial 
kinetic energy of the stream from \fig{turbulence}, because the mass contained within the shear layer grows with 
time, such that at $t>t_{\rm dis}$ the mass contained within the shear layer is larger than the initial stream mass. 
In practice, we find in all simulations that the total turbulent energy assymptotes at $\sim 10\%$ of the initial 
kinetic energy of the stream.

\smallskip
Fourier analysis reveals that on scales $\lsim \Rs$, the turbulence is very close to isotropic for all cases. 
We defer a more detailed analysis of the turbulence to future work which will include radiative cooling, as well 
as the gravity of the halo into which the stream is flowing which may be an additional source of turbulence (see 
\se{phys}). We note that 2d slabs do not exhibit such universal evolution. At $t\sim t_{\rm dis}$, $\sigma/V$ 
varies by more than a factor of 3 and does not have a well defined maximum for $\delta>1$. However, since 
it is well known that turbulence manifests itself in a qualitatively different way in 2d than in 3d, we do not 
dwell on this comparison here.

\subsubsection{Deceleration and Kinetic Energy Dissipation}
\smallskip
In \fig{deceleration_surface} we show the deceleration of streams in 2d slab and 3d cylinder simulations. 
We evaluate this by computing the centre of mass velocity of stream fluid, 
\be 
\label{Vstream}
V_{\rm z,s} = \dfrac{\int_{0}^{L/2} \overline{\psi}_{(r)}\, \overline{\rho}_{(r)}\, \widetilde{v_{\rm z}}_{(r)}\, r~{\rm dr}}{\int_{0}^{L/2} \overline{\psi}_{(r)}\, \overline{\rho}_{(r)}\, r~{\rm dr}}.
\ee
{\no}In the left-hand panel we show the stream velocity normalized to its initial value, $V$, as a function 
of time normalized by the stream sound crossing time. It is evident that 3d cylinders decelerate faster than 
2d slabs, as predicted in \se{theory_surface}. However, the time at which deceleration begins is very similar 
in 2d and 3d for all cases. We denote this time as $t_0$, and define it as the time when the stream velocity 
has dropped to $98\%$ its initial value\footnote{This delay in the onset of deceleration was not seen in P18, 
and is due to our seeding perturbations in the radial velocity rather than the stream-background interface, as 
explained in \se{methods-comp}.}. For all values of $(\Mb,\delta)$, we find $t_0 \sim 0.5-1.5\tsc$.

\begin{table} 
	\centering
	\caption{
	Parameters of simulations with $M_{\rm tot}>1$ (\equnp{Mtot}), studying the nonlinear 
	evolution of body modes and of high-$m$ surface modes. The columns show, from left 
	to right, the Mach number of the stream with respect to the background sound speed, 
	$\Mb$, the density contrast between the stream and the background, $\delta$, the 
	Mach number of the stream with respect to the sum of the two sound speeds, $M_{\rm tot}$, 
	the number of cells per stream radius, $\Rs/\Delta$ where $\Delta$ is the cell size 
	in the 	highest resolution region, the ratio of the stream radius to the smoothing 
	parameter $\sigma$ 	(\equnp{ramp2}), and the refinement scheme (see \tab{surface} 
	for details). The first three rows in the table represent simulations with the same 
	smoothing as in those presented in \se{surface}. The following six represent simulations 
	with more aggressive smoothing of the initial conditions. The next five rows represent 
	tests to check convergence with resolution and refinement strategy, and are presented in 
	Appendix \se{convergence}. The final two rows represent simulations where the initial 
	perturbation spectrum was changed, as described in the text. For these two models, 
	only a 3d cylinder was simulated.}
	\label{tab:body}
	\begin{tabular}{ccccccc}
		\hline
		$\Mb$&$\delta$&$M_{\rm tot}$&$\Rs/\Delta$&$\Rs/\sigma$&$\Delta_{\rm ref}/\Rs$&Note \\
		\hline
		5.0&1&2.50&64&32&3.0& \\
		2.0&10&1.52&64&32&3.0& \\
		2.0&100&1.82&64&32&3.0& \\
		\hline
		5.0&1&2.50&64&8&1.5& \\
		2.5&5&1.72&64&8&1.5& \\
		2.0&10&1.52&64&16&1.5& \\
		2.0&10&1.52&64&8&1.5& \\
		2.5&20&2.04&64&8&1.5& \\
		2.0&100&1.82&64&8& 1.5& \\
		\hline
		2.0&10&1.52&32&32&1.5& \\
		2.0&10&1.52&64&32&1.5& \\
		2.0&10&1.52&128&32&1.5& \\
		2.0&10&1.52&32&8&1.5& \\
		2.0&10&1.52&128&8&1.5& \\
		\hline
		5.0&1&2.50&64&32&3.0&$m=0-4$ \\
		2.5&5&1.72&64&8&1.5&\equ{pertr}, $16\Rs$ \\
		\hline
	\end{tabular}
\end{table}

\smallskip
In the centre panel of \fig{deceleration_surface}, we rescale the time axis by the predicted deceleration 
timescale, $\tau_{\rm surface}$, given by \equ{tau_surface_2d} and \equ{tau_surface_3d} for 2d and 3d respectively. 
Before rescaling the time axis, we move its origin to $t_0$ in order to focus on the deceleration itself rather 
than the ``incubation" period before it begins. For the value of $\alpha$ in the expressions for $\tau_{\rm surface}$, 
we use \equ{alpha_fit} for 2d slabs and 3d cylinders with $\delta<8$, while for 3d cylinders with $\delta>8$ we 
use half this value to account for the fact that the growth rate of $\hb$ in 3d cylinders decreases by a factor 
of $\sim 2$ once $\hb\sim 2\Rs$. This occurs at $\sim \tau_{\rm surface}$ for $\delta\sim 8$, while for denser 
streams it happens earlier. Since the deceleration rate is proportional to the background mass flux into the shear 
layer, which is proportional to $(\Rs+\hb)^2$, we approximate that all the deceleration occurs with the reduced value 
of $\alpha$. This approximation is valid for cold streams with $\delta>10$. As can be seen from the centre panel of  \fig{deceleration_surface}, when scaled to the predicted deceleration timescales, the results of 2d and 3d simulations 
are in agreement, and the stream velocity decreases to half its initial value at $t\sim \tau_{\rm surface}$, as predicted. 

\smallskip
The curvature in the deceleration profiles can be approximated by the following simple toy model. We assume that the deceleration 
can be modeled as 
\be 
\label{eq:akhi}
\frac{{\rm d}V_{\rm s}}{{\rm d}t}=-\frac{V_{\rm s}}{\tau_{\rm surface}} = -\frac{\alpha\sqrt{\delta}}{(1+\sqrt{\delta})(\sqrt{1+\delta}-1)}\frac{V_{\rm s}^2}{\Rs},
\ee
{\no}where $V_{\rm s}(t)$ is the time-dependent stream velocity, and the final equality is valid for 3d cylinders using 
\equ{tau_surface_3d} for $\tau_{\rm surface}$, which we now assume to be time dependent through $V_{\rm s}$ rather than 
constant. An analogous expression can be derived for 2d slabs using \equ{tau_surface_2d}. The solution to \equ{akhi} with 
the initial condition $V_{\rm s}(t=0)=V_0$ is 
\be 
\label{eq:dec_model}
V_{\rm s}(t) = V_0(1+t/\tau_{\rm surface,\,0})^{-1},
\ee
{\no}where $\tau_{\rm surface,\,0}$ is the inital value of $\tau_{\rm surface}$, i.e. with the initial stream velocity $V_0$. 
\Equ{dec_model} is valid for both 3d cylinders and 2d slabs. We show this model as the thick dash-dotted line in the centre 
panel of \fig{deceleration_surface}. Except for the cases with $\delta=1$ and $\Mb\le 1$, where the flow is subsonic with respect 
to the sound speed in the stream and which are the least relevant for actual cold streams, this simple model is a good fit to 
the simulations. 
The subsonic cases can be brought into good agreement with the model if we lower the value of $\alpha$ used 
in $\tau_{\rm surface}$ by a factor $\sim 2/3$. This seems to imply that for subsonic flow, the transfer of momentum from the stream 
to the background in the shearing layer is not instantaneous, which was implicit in the derivation of $\tau_{\rm surface}$. For 
higher Mach number flows, shocks fascilitate the momentum transfer.

\smallskip
In addition to stream deceleration, we wish to evaluate the decrease of kinetic energy associated with laminar streaming 
motions, of both stream and background fluid. This is not obviously inferred from the left and centre panels of 
\fig{deceleration_surface}, since the stream transfers some of its momentum to the background gas as it decelerates. As noted 
above, the stream decelerates by distributing its initial momentum over more and more mass as background fluid is continuously 
added to the shear layer. If the material within the outer envelope of the shear layer moves at a characteristic velocity $v$, 
and the total mass within this region is $m$, we thus have 
\be 
\label{eq:m_acc}
\dot{v}/v \sim -\dot{m}/m.
\ee
{\no}Since the kinetic energy associated with this streaming motion is $E_{\rm K} \propto mv^2$, we have 
\be 
\label{eq:EK}
\frac{\dot{E_{\rm K}}}{E_{\rm K}} \sim \frac{\dot{v}}{v}.
\ee
{\no}Making the simplifying assumption that the characteristic velocity in the outer shear layer, $v$, is 
equal to the stream velocity, $V_{\rm s}$, we thus predict that the kinetic energy associated with the laminar 
flow of stream and background fluid decreases at the same rate as the stream velocity. While this assumption 
is clearly simplified, since there is a velocity profile within the shear layer, it actually reproduces the kinetic 
energy loss seen in simulations rather well. This is shown in the right-hand panel of \fig{deceleration_surface}, 
where we plot the kinetic energy normalized by its initial value as a function of time normalized by $\tau_{\rm surface}$. 
The kinetic energy is evaluated from the simulations as 
\be 
\label{eq:EK_sim}
E_{\rm K}=\pi~\int_{0}^{L/2} \overline{\rho}_{(r)}\, \widetilde{v_{\rm z}}_{(r)}^2\, r~{\rm dr},
\ee
{\no}where by using $\widetilde{v_{\rm z}}_{(r)}$ we are focusing only on the laminar flow, without turbulent motions. 
We note that the results are identical whether we limit the integration in \equ{EK_sim} to $r=\Rs+\hb$ rather than $L/2$, 
showing that there is no bulk laminar flow in the background outside the shear layer. By comparing the centre and 
right-hand panels of \fig{deceleration_surface} we see that for 3d cylinders the total bulk kinetic energy decreases 
at a very similar rate to the stream velocity, as predicted by our simple toy model. We also note that in 2d slabs, 
the kinetic energy seems to decrease at a slower rate. This implies that the background fluid accreted onto the shear 
layer does not mix as efficiently in 2d, and moves at an overall slower velocity than the bulk of the stream fluid.

\smallskip
As the bulk kinetic energy decreases, it is transferred into several channels. Some of it goes into turbulent motions, 
though as discussed in relation to \fig{turbulence}, this only accounts for $\sim 10\%$ of the initial kinetic energy in 
our simulations. Additional channels for the lost kinetic energy are thermal energy, of both the stream and background fluids, 
and sound waves which propagate away from the stream and may eventually leave the simulation domain. We defer a detailed 
analysis of the relative importance of these various channels to future work which will include radiative cooling (Mandelker 
et al., in prep.). In particular, we will address there what fraction of the lost kinetic energy is eventually dissipated 
into radiation. For the moment, all we can say is that \equ{EK} and \fig{deceleration_surface} represent an upper limit to the 
total energy that may be dissipated and subsequently radiated due to KHI.

%!!!!!!!!!!!!!!!!!!!!!!!!!!!!!!!
\subsection{The Nonlinear Evolution of Body Modes}
\label{sec:body}
\smallskip
We now turn to study streams with $M_{\rm tot}>1$, i.e. streams which are supersonic with respect to the 
sum of the sound speeds in both fluids. As described in \se{theory}, in 2d such streams are dominated by 
body modes as surface modes stabilize, while in 3d there exist unstable surface modes with azimuthal modes 
$m>1$. The numerical simulations presented in this section is listed in \tab{body}. These span the range of 
density contrast and Mach number relevant to cosmic cold streams, $2.0 < \Mb < 2.5$ and $10 < \delta < 100$, 
with additional simulations at lower density contrast and higher Mach number, $\delta=1,\,5$ and $\Mb=5$, in 
order to test the general validity of our results. For each case, we simulated both a 3d cylinder and a 2d 
slab as described in \se{methods}. Furthermore, we examine the effect of varying the width of the initial 
smoothing layer, parametrized by $\sigma$ in \equ{ramp2}, as well as several cases with different resolution 
or refinement scheme to check convergence, and present these results in Appendix \se{convergence}. Finally, 
we performed two simulations where we varied the initial perturbation spectrum. In the first of these we 
included azimuthal modes with $m=(0-4)$ instead of $m=(0-1)$. In the second we initiated perturbations to the 
stream-background interface following \equ{pertr} rather than to the radial velocity, and extended the wavelength 
range to $(0.5-16)\Rs$ rather than $(0.5-2)\Rs$ (see \se{methods-pert}). Unlike all other cases, these final 
two models include only 3d cylinder simulations.

\smallskip
We stress that varying the width of initial transition layer, $\sigma$, is not merely a numerical test, but 
rather has physical meaning, since increasing the width of the transition layer damps short wavelength 
perturbations. This has been described in \se{methods-unperturbed} and is demonstrated below. In reality, 
the width will be determined by additional physics, such as thermal conduction, diffusion, or the radial 
accretion of gas onto the stream. As none of these processes are included in our current simulations, we 
vary $\sigma$ as a crude way to test this effect, delaying a more detailed consideration of these physical 
processes to future work (see \se{phys}).

\subsubsection{Appearance of high-$m$ modes}
\begin{figure*}
\begin{center}
\includegraphics[trim={3.1cm 0.0cm 6.85cm 0}, clip, width =0.351 \textwidth]{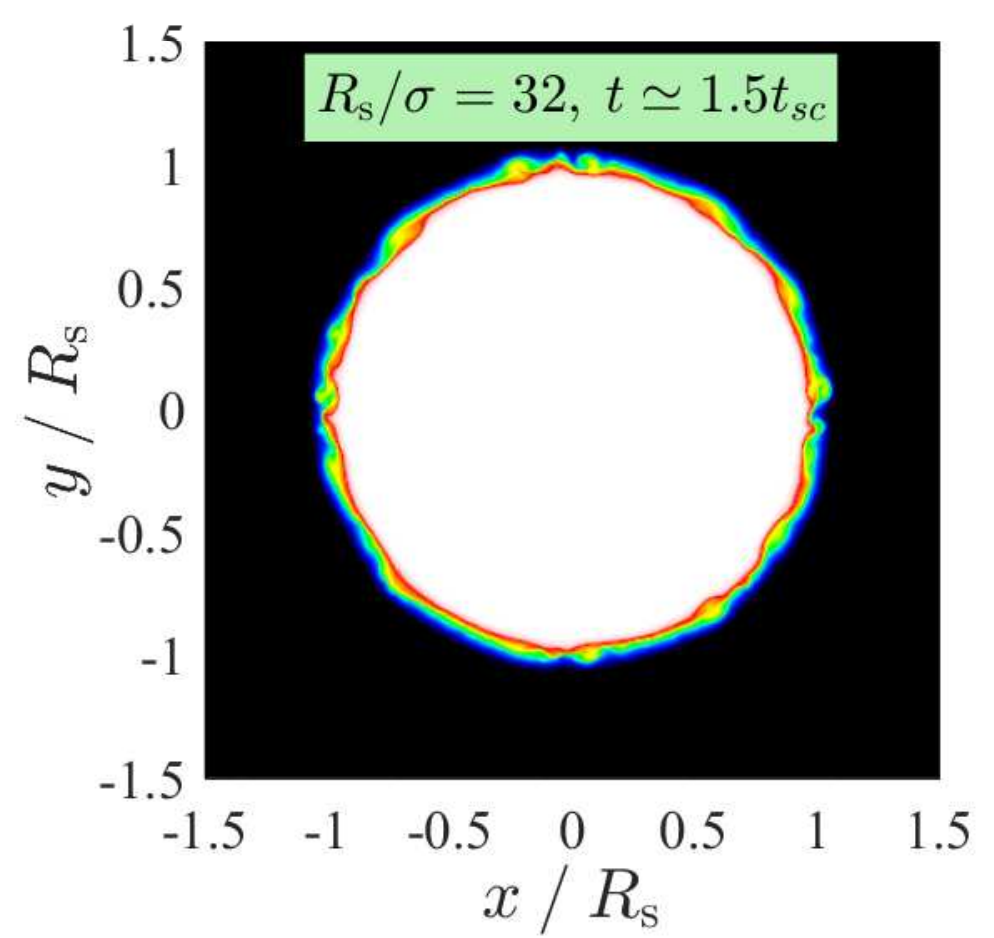}
\hspace{-0.35cm}
\includegraphics[trim={5.1cm 0.0cm 6.85cm 0}, clip, width =0.281 \textwidth]{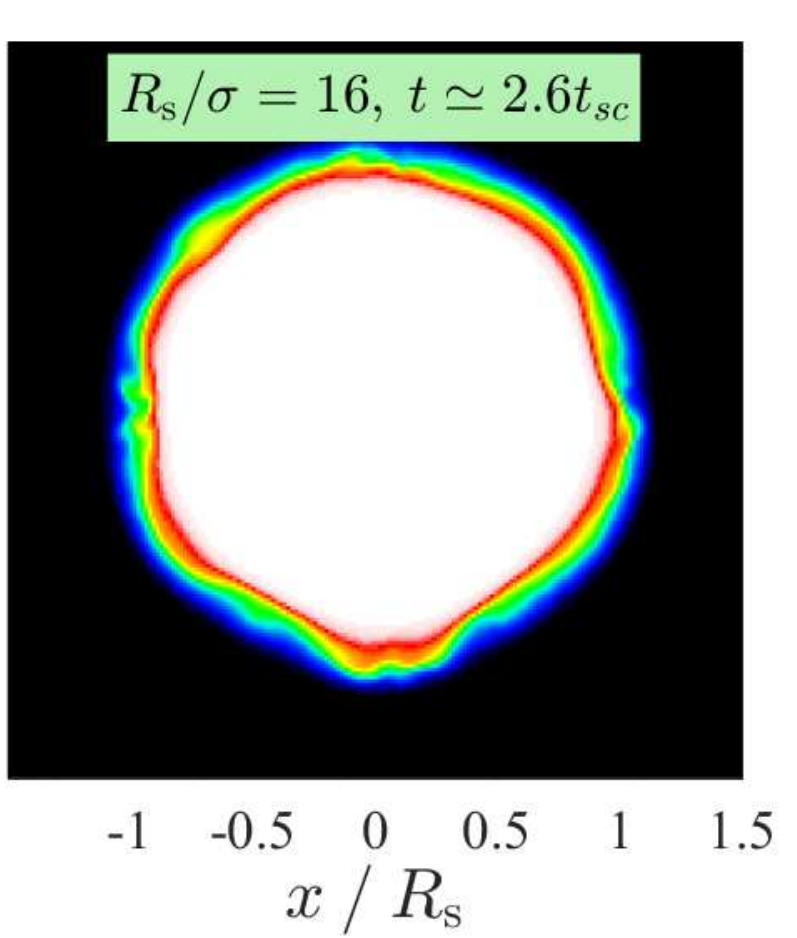}
\hspace{-0.35cm}
\includegraphics[trim={5.1cm 0.0cm 3.85cm 0}, clip, width =0.386 \textwidth]{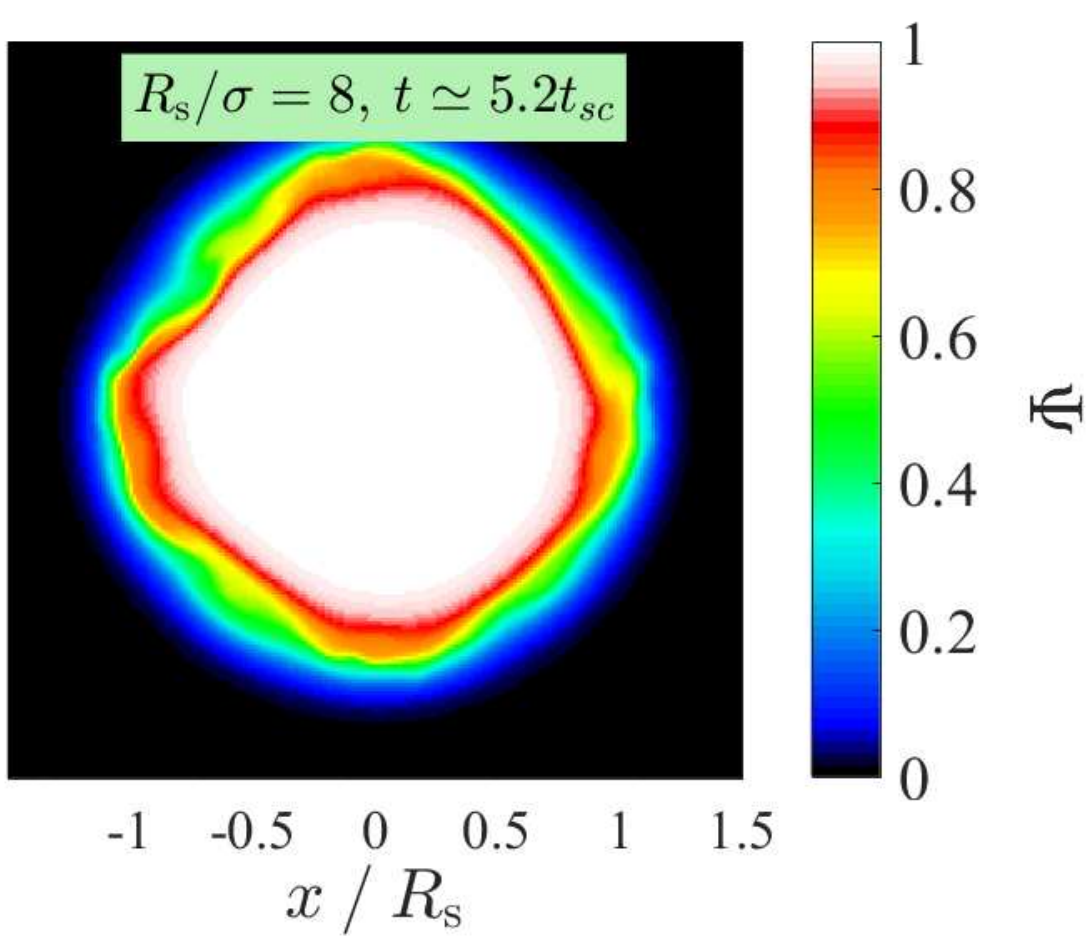}
\end{center}
\caption{Excitation of high-$m$ surface modes in cylinders with $\Mb>M_{\rm crit}$. Shown are slices through the 
$z=0$ plane of a simulation with $(\Mb,\delta)=(2,10)$, with different values of the smoothing parameter used in 
\equ{ramp2}, $\sigma/\Rs=1/32$ (left), $1/16$ (centre), and $1/8$ (right). colour shows the value of the passive 
scalar $\psi$, as in \fig{colour_panel_M1D1}. For each case, we show a snapshot shortly after $\hb$ has begun to 
grow (see \fig{h_body}), at $t\sim 1.5$, $2.6$ and $5.2\tsc$ respectively. The dominant modes are $m\sim 12$, 
$m=6$ and $m=4$. Based on \equ{mcrit}, at a longitudinal wavelength of $\lambda=2\Rs$, the largest wavelength 
seeded in the initial perturbation, surface modes are unstable for $m>2$. However, the smoothing layer with 
$\sigma=\Rs/32$ suppresses the growth of surface modes with $\lambda\lsim 0.5\Rs$. The azimuthal wavelength 
associated with mode number $m$ is $\lambda_{\rm \varphi,\,m}=2\pi\Rs/m$, which is $\sim 0.5\Rs$ for $m=12$, 
explaining why this is the dominant mode in the left-hand panel. Increasing $\sigma$ by a factor of $2$ increases 
the shortest unstable wavelength by a factor of $2$, resulting in a decrease in the dominant $m$ by a factor of 
$\sim 2$, as seen in the other two panels. Since $t_{\rm KH}\propto \lambda$ for surface modes, this also leads 
to a factor of $\sim 2$ reduction in growth rate, consistent with the simulation results. 
}
\label{fig:smoothing_panel} 
\end{figure*}

\smallskip
In \fig{smoothing_panel} we show the effects of varying the width of the initial smoothing layer. We show slices 
through the $z=0$ plane of a 3d simulation with $(\Mb,\delta)=(2,10)$, using three different values for $\sigma$: 
$\Rs/32$, which was the value used in the simulations presented in \se{surface}, $\Rs/16$ and $\Rs/8$. For each 
case, we show a snapshot shortly after the shearing layer between the stream and the background begins to expand, 
marked by the growth of $\hb$ in \fig{h_body} below. This happens at $t/\tsc\sim 1.5$, $2.6$, and $5.2$ for 
$\Rs/\sigma=32$, $16$, and $8$ respectively. The dominant modes at these times appear to be $m\sim 12$, $m=6$ and 
$m=4$, corresponding to azimuthal wavelengths $\lambda_{\rm \varphi}=2\pi\Rs/m\sim 0.5\Rs$, $1.0\Rs$, and $1.6\Rs$. 
For $\delta=10$, $M_{\rm crit}\sim 1.77$ (\equnp{Mcrit}). Therefore, according to \equ{mcrit}, for $\Mb=2$ and longitudinal 
wavelengths $\lambda= 2\Rs$, the longest wavelength in the initial perturbation spectrum, surface modes are unstable 
for $m>2$. However, the initial smoothing layer suppresses the growth of surface modes where either $\lambda$ or 
$\lambda_{\varphi}$ is of order the smoothing layer thickness, $\sim (3-4)\sigma$, or smaller. For $\sigma=\Rs/32$, 
the total width of the smoothing layer is $\gsim 0.1\Rs$, which substantially reduces the growth rates of surface 
modes with $\lambda_{\rm \varphi}\lsim 0.5\Rs$, i.e. $m\gsim 12$ making this the fastest growing mode. Increasing 
$\sigma$ by a factor of 2 increases the shortest unstable wavelength by a factor of $\sim 2$, which in turn decreases 
the growth rate of the fastest growing mode by a factor of $\sim 2$ because $\omega\propto k$ for surface modes. This 
is consistent with both the dominant mode and with the time when it dominates the stream morphology for $\sigma=\Rs/16$ 
and $\Rs/8$. We note that in the latter case, while the $m=4$ surface mode does eventually dominate the stream morphology, 
this only occurs after the critical body mode perturbation has become nonlinear, so the transition to nonlinearity is 
dominated by body modes even though the nonlinear phase contains a mix of both surface and body modes.

\begin{figure*}
\begin{center}
\includegraphics[width =0.265 \textwidth]{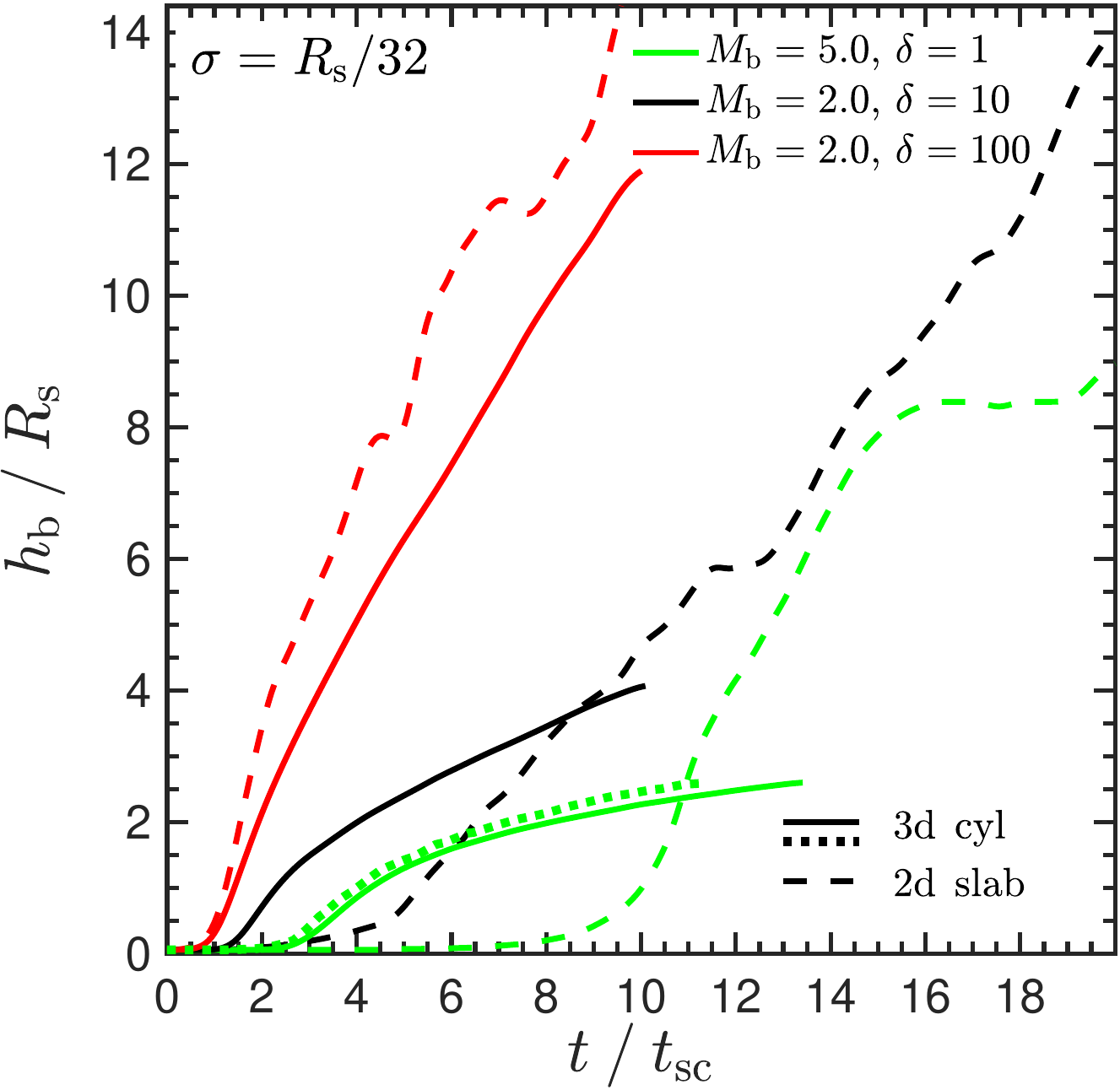}
\hspace{-0.15cm}
\includegraphics[trim={1.9cm 0.0cm 0.0cm 0}, clip, width =0.2323 \textwidth]{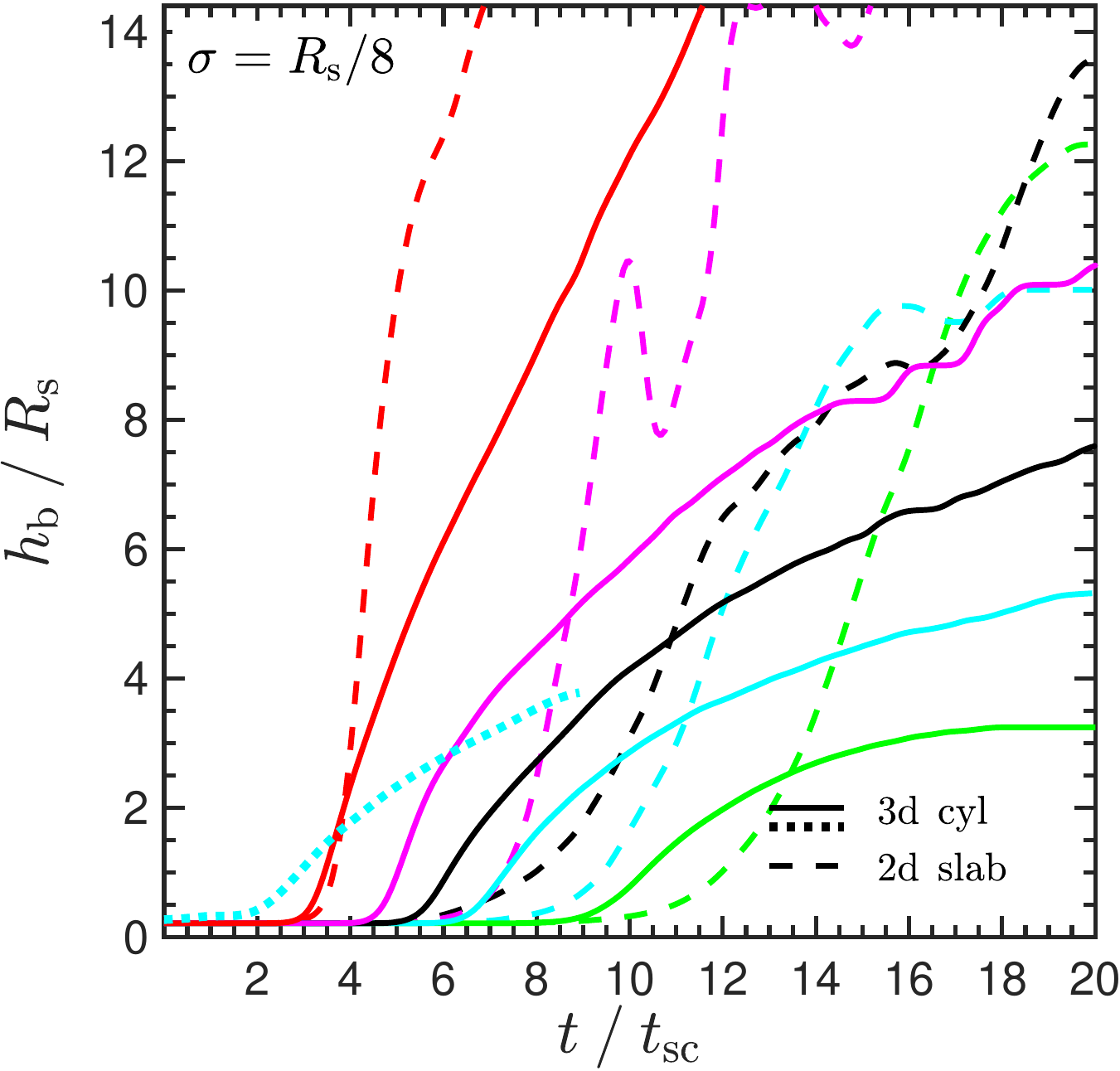}
\hspace{-0.19cm}
\includegraphics[width =0.265 \textwidth]{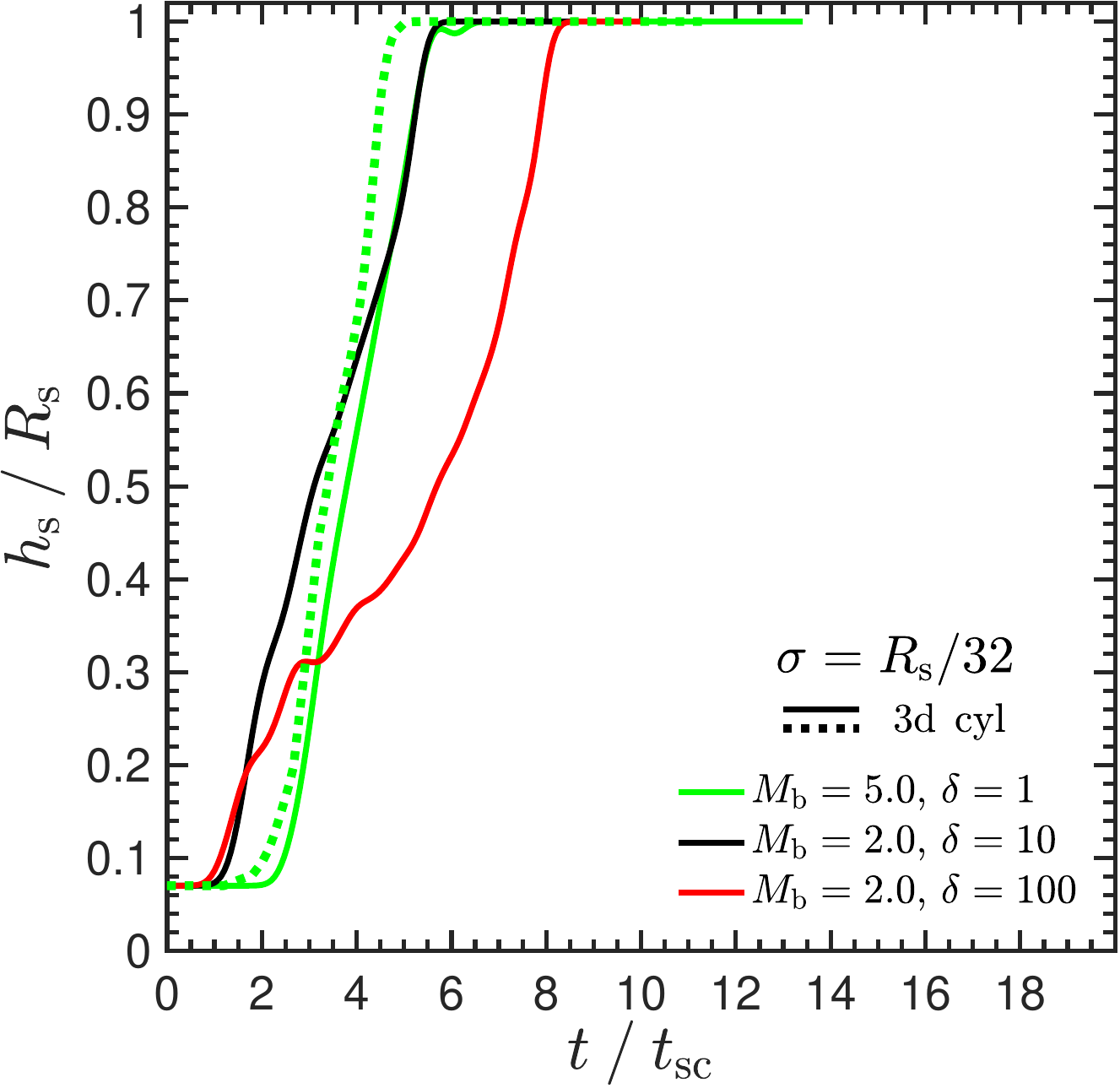}
\hspace{-0.15cm}
\includegraphics[trim={1.9cm 0.0cm 0.0cm 0}, clip, width =0.2323 \textwidth]{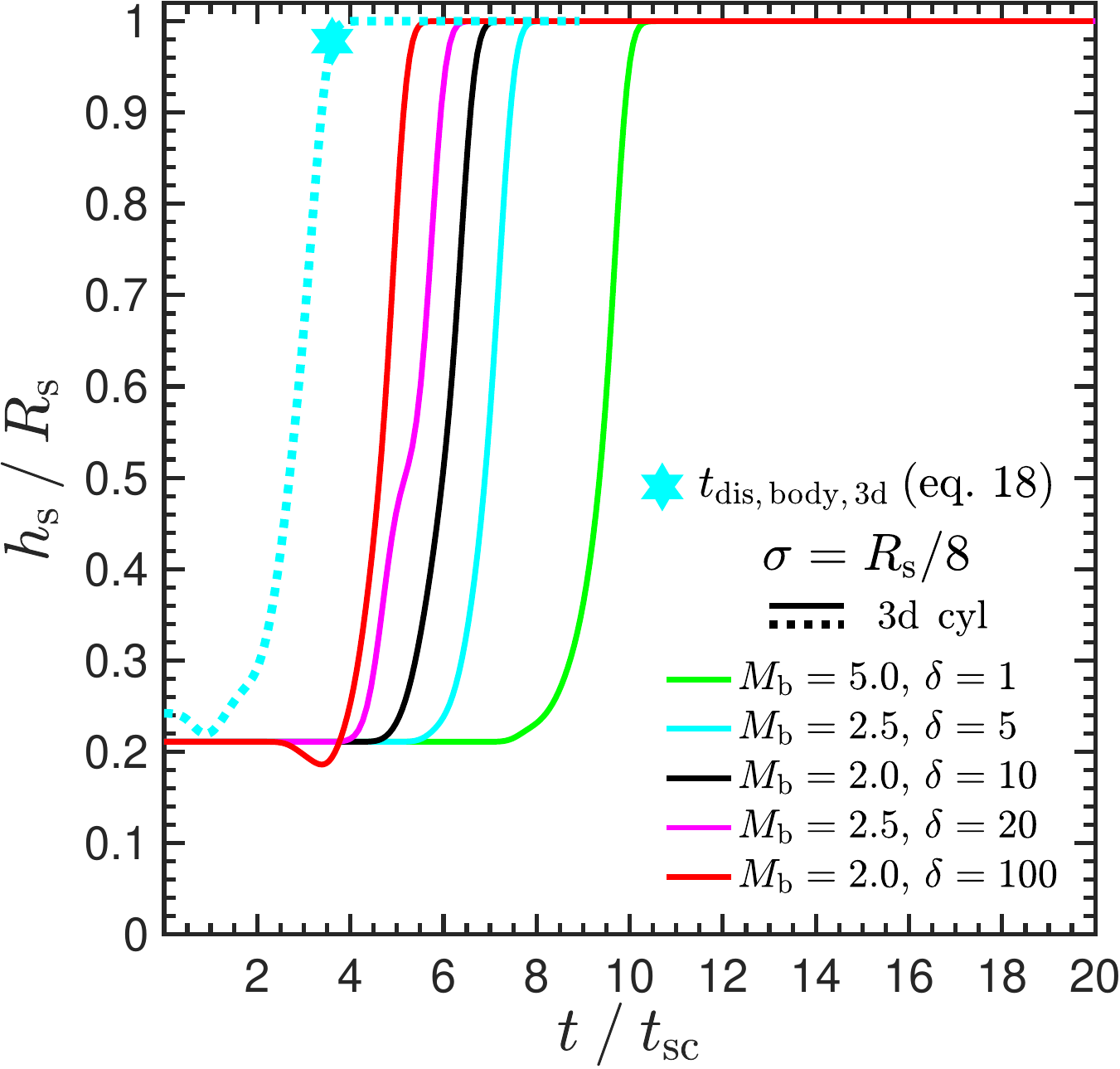}
\end{center}
\caption{Expansion and shear layer growth in streams where $\Mb>M_{\rm crit}$ (\equnp{Mcrit}). 
We show $\hb/\Rs$ in the left two panels and $\hs/Rs$ in the right two panels, each as a function 
of time normalized by the stream sound crossing time, $\tsc=2\Rs/\cs$. The first and third panels 
from the left show simulations with a narrow initial smoothing layer, $\sigma=\Rs/32$, while the 
second and fourth panels show simulations with a wide initial smoothing layer, $\sigma=\Rs/8$. 
The large initial values of $\hs$ and $\hb$ in the simulations with $\sigma=\Rs/8$ are due 
to the initial smoothing of the contact discontinuity, which extends to roughly $1.5\sigma$ in 
each medium. 
Solid lines in all panels represent 3d simulations while dashed lines in the two $\hb$ panels represent 
2d simulations (not shown in the two $\hs$ panels). Different colours represent different combinations of 
$(\Mb,\delta)$ (the colour legend in the second and fourth panels is the same). The dotted 
lines represent simulations with a different initial perturbation spectrum (\tab{body}). The 2d simulations 
behave similarly for both values of $\sigma$, though the onset of expansion is slighly delayed in the case 
with large $\sigma$. In the case with $\sigma=\Rs/8$, the 2d expansion rates are roughly constant until 
$\hb\sim (8-10)\Rs$. The 3d simulations with $\sigma=\Rs/8$ behave similarly to their 2d counterparts at 
early times, though they begin expanding slightly earlier. The simulation with interface perturbations at 
long wavelengths (dotted cyan line) begins expanding at $t_{\rm NL}$ predicted from \equ{tNL} (see text), 
and is destroyed, in the sense that $\hs=\Rs$, at $t_{\rm dis,\,body,\,3d}$ predicted from \equ{tau_break_body_3d}, 
which is marked with a star in the rightmost panel. This is unlike its counterpart with velocity perturbations 
at shorter wavelengths (solid cyan line). Unlike in 2d, the 3d expansion rates decrease once $\hb\gsim 2\Rs$, 
similar to the simulations described in \se{surface}, while a shear layer rapidly penetrates into the stream, 
consuming it within $1-2\tsc$. When $\sigma=\Rs/32$, the 3d simulations become dominated by high-$m$ surface 
modes before body modes develop. As a result, $\hs$ and $\hb$ begin growing at $t\sim 1-2\tsc$ and are initially 
well fit by \equs{hs_growth}-\equm{hb_growth} with $\alpha=0.05$. The shear layer consumes the stream within 
$4-5\tsc$. The simulation which was initiated with $m=0-4$ modes (dotted green line) is extremely similar to 
its counterpart with $m=0-1$ (solid green line).}
\label{fig:h_body} 
\end{figure*}

\subsubsection{Stream Expansion and Shear Layer Growth}
\smallskip
In the left two panels of \fig{h_body} we examine the evolution of the stream thickness in 2d slabs and 3d 
cylinders with $\sigma=\Rs/32$ and $\Rs/8$. For body modes, the onset of rapid expansion of the stream can in 
principle be predicted from \equ{tNL}. However, this assumes an initial perturbation in the stream-background 
interface at the critical wavelength ($\gsim 10\Rs$ in all cases), with an initial amplitude $H_0$. This was 
the case in the body mode simulations presented in P18, where perturbations in the stream-background interface 
were seeded with a spectrum extending out to $16\Rs$ and constant initial amplitude for each seeded wavelength. 
As a result, $t_{\rm NL}$ could be directly evaluated from the initial conditions, and coincided with the onset 
of rapid stream expansion seen in the simulations (P18, figure 17). However, as highlighted in \se{methods-comp}, 
this is not the case in most of the simulations presented here, where we seeded perturbations in the radial velocity 
with wavelengths not exceeding $2\Rs$. Before the critical perturbation can be excited, interface perturbations 
at the critical wavelength must be excited. This is discussed further below.

\smallskip
However, we did perform one simulation with interface perturbations at wavelengths up to $16\Rs$ (see \tab{body}). 
This is shown as the dotted cyan line in the second and fourth panels from the left of \fig{h_body}. 
This simulation consisted of 126 modes with an rms amplitude of $0.1\Rs$ (\se{methods-pert}), yielding an 
amplitude per wavelength of $\sim 0.009\Rs$. The critical wavelength in this case is $\sim 9\Rs$ with a KH time 
of $\sim 0.55\tsc$ (see \tab{body2}). According to \equ{tNL}, this corresponds to $t_{\rm NL}\sim 2.6\tsc$, 
in excellent agreement with the onset of rapid stream expansion. Furthermore, the predicted time for stream 
disruption based on \equ{tau_break_body_3d} is $\sim 3.6\tsc$, which is marked by the star in the right-most panel 
of \fig{h_body} and is in excellent agreement with the time at which $\hs=\Rs$. Once the stream begins expanding, 
its expansion rate is extremely similar to that of a simulation with identical parameters but fiducial initial 
perturbations, shown by the solid cyan line. We therefore conclude that while our fiducial initial perturbations 
lead to a delayed onset of stream expansion, the subsequent behaviour is well captured. 

\smallskip
While we cannot predict the precise disruption times in our fiducial simulations with velocity 
perturbations, we can qualitatively understand the order in which they disrupt and the approximate 
ratios between their disruption times. These stem primarily from the fact that the initial 
radial velocity perturbation amplitudes correspond to different interface dispacement amplitudes for different 
values of $\Mb$ and $\delta$. In Appendix \se{ur_to_h} we derive the displacement amplitude 
corresponding to a given radial velocity perturbation amplitude and wavelength. As shown there, if 
the amplitude of the radial velocity perturbation at the critical wavelength is the same for all simulations, 
and defining $t_{\rm NL}$ as unity for $(\Mb,\delta)=(5.0,1)$, then for $(\Mb,\delta)=(2.5,5),\:(2.0,10),\:(2.5,20)$, 
and $(2.0,100)$ we have $t_{\rm NL}\sim 0.73,\:0.68,\:0.67,\:0.62$. In the simulations, stream expansion begins 
at $t\sim 0.72,\:0.61,\:0.50,\:0.33$ relative to the $(\Mb,\delta)=(5.0,1)$ simulation. Our estimate is thus 
in good quantitative agreement with the simulation results for $(\Mb,\delta)=(2.5,5)$, and in reasonable agreement 
for $(\Mb,\delta)=(2.0,10)$. As $\Ms=\delta^{1/2}\Mb$ is increased, our estimate overshoots the ratio in simulations 
by a larger amount. This is because our estimate is based on the incorrect assumption that all simulations begin 
with the same radial velocity amplitude at the critical wavelength. While the amplitude is the same in the seeded 
wavelength range of $(0.5-2.0)\Rs$, the critical wavelength at $\sim 10\Rs$ is excited after one sound crossing time 
has elapsed due to shocks in the stream. These shocks are stronger for larger Mach number flows (with respect to the 
stream sound speed), leading to larger initial amplitudes of radial velocity which correspond to larger initial amplitudes 
in the interface displacement, which result in shorter $t_{\rm NL}$. The details of how exactly the 
critical wavelength is triggered are beyond the scope of this paper. However, the qualitative agreement of this analysis 
with the simulation results together with the quantitative agreement in the case where we seeded the simulation with long 
wavelength interface perturbations, as well as the analogous analysis presented in P18 for the 2d case, lead us to conclude 
that our predictions for stream disruption due to body modes from \se{theory_body} are supported by numerical simulations.

\smallskip
The 2d simulations behave qualitatively similar with both values of $\sigma$, though the onset of growth in $\hb$ 
is slightly delayed in the simulations with larger $\sigma$, by $\sim (1-2)\tsc$ in all cases. We note that in 
the case with $\sigma=\Rs/8$, all 2d simulations exhibit very similar growth rates when time is normalized by 
$t_{\rm s}=\Rs/V$\footnote{Note that this does not imply that all simulations have the same value of $\alpha$ 
in \equ{hb_growth}.} (not shown), roughly $\hb/\Rs\sim (0.15-0.2)\times t/t_{\rm s}$. This is sensible, 
since once the critical mode has taken over and the stream has begun to expand, the stream sound crossing time 
is no longer meaningful. Rather, the time scale dominating stream evolution is naturally $t_{\rm s}$. 

\begin{figure*}
\begin{center}
\includegraphics[trim={0.0cm 1.238cmcm 3.3cm 0}, clip, width =0.45 \textwidth]{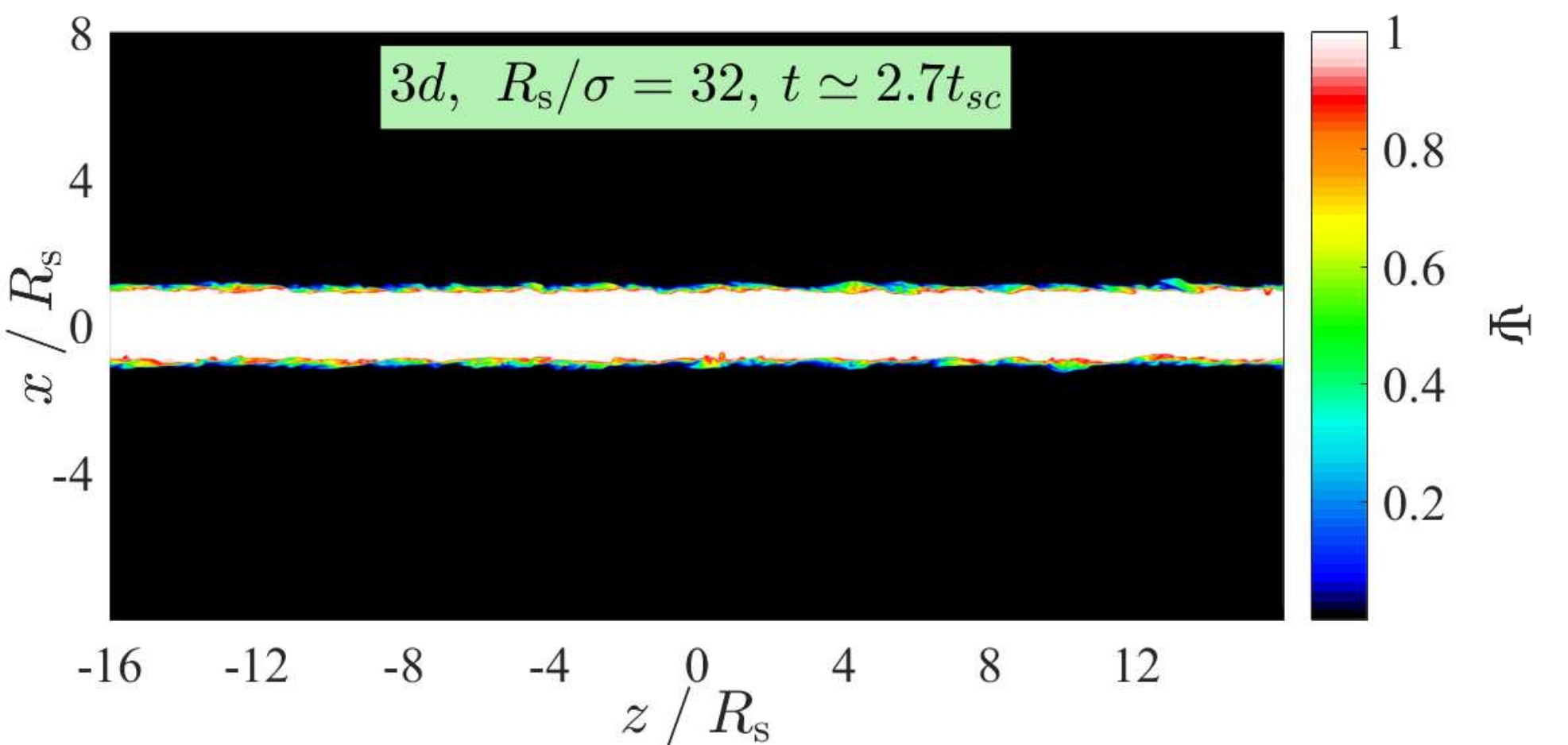}
\hspace{-0.3cm}
\includegraphics[trim={1.3cm 1.238cmm 0.1cm 0}, clip, width =0.501 \textwidth]{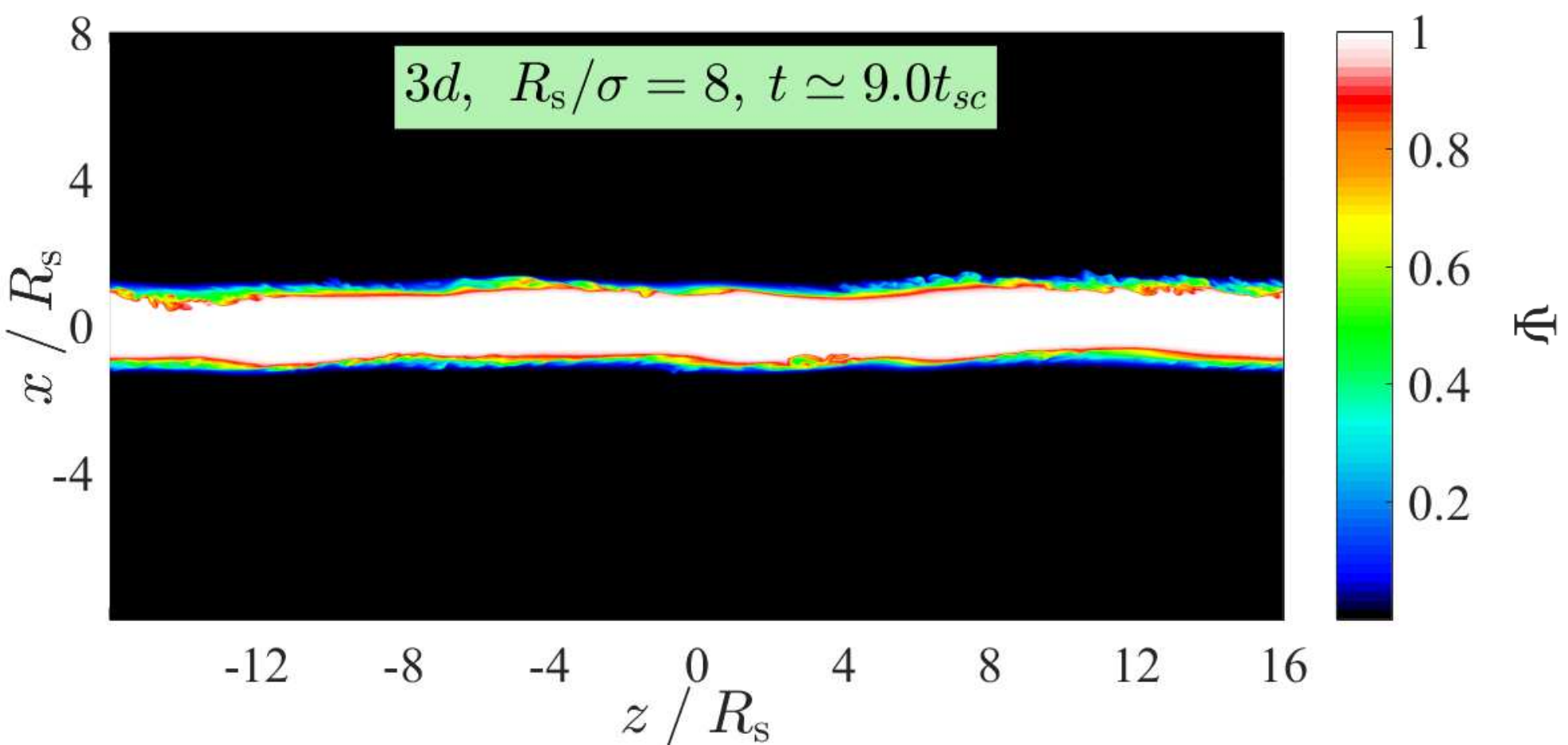}\\
\vspace{-0.09cm}
\includegraphics[trim={0.0cm 1.238cmcm 3.3cm 0}, clip, width =0.45 \textwidth]{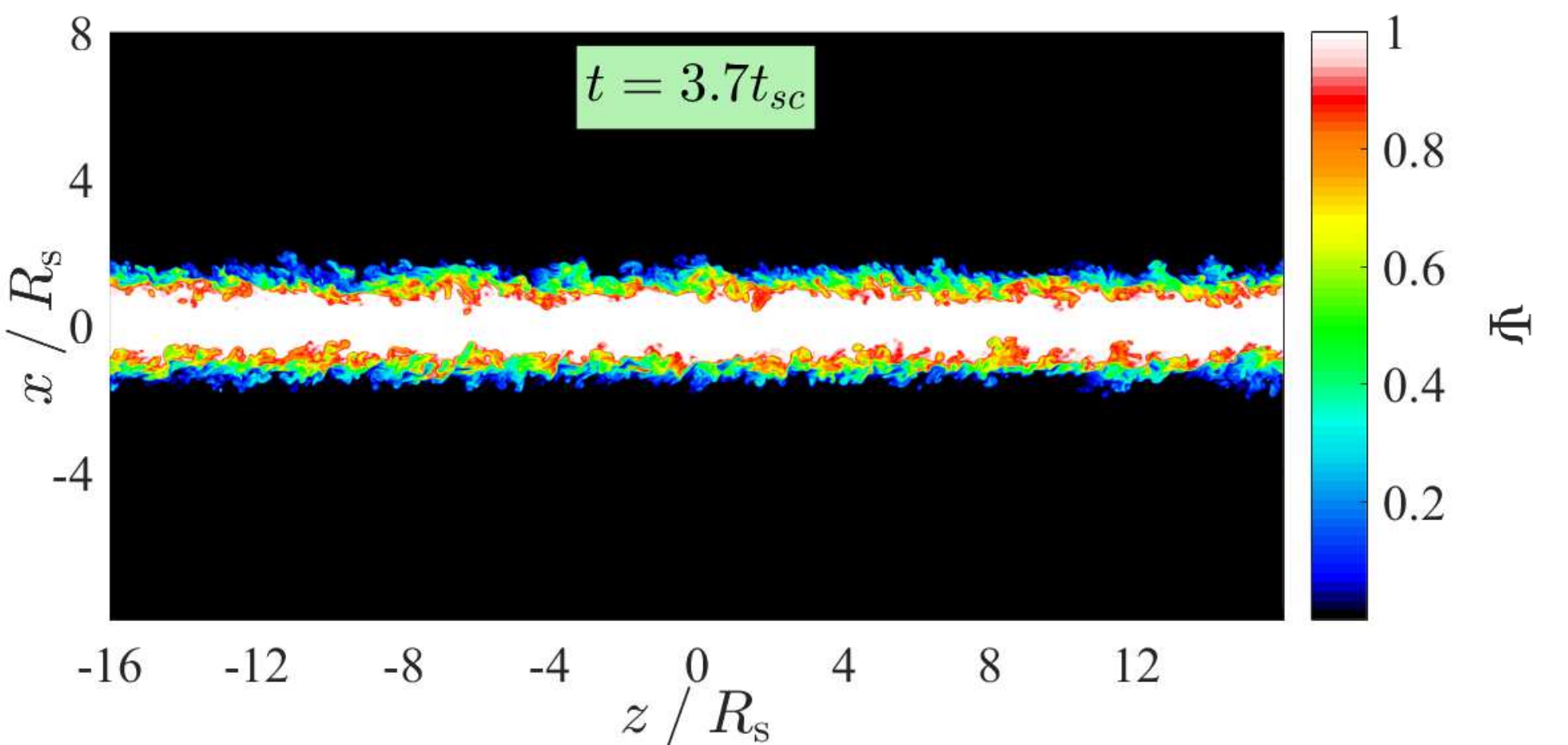}
\hspace{-0.3cm}
\includegraphics[trim={1.3cm 1.238cmcm 0.1cm 0}, clip, width =0.501 \textwidth]{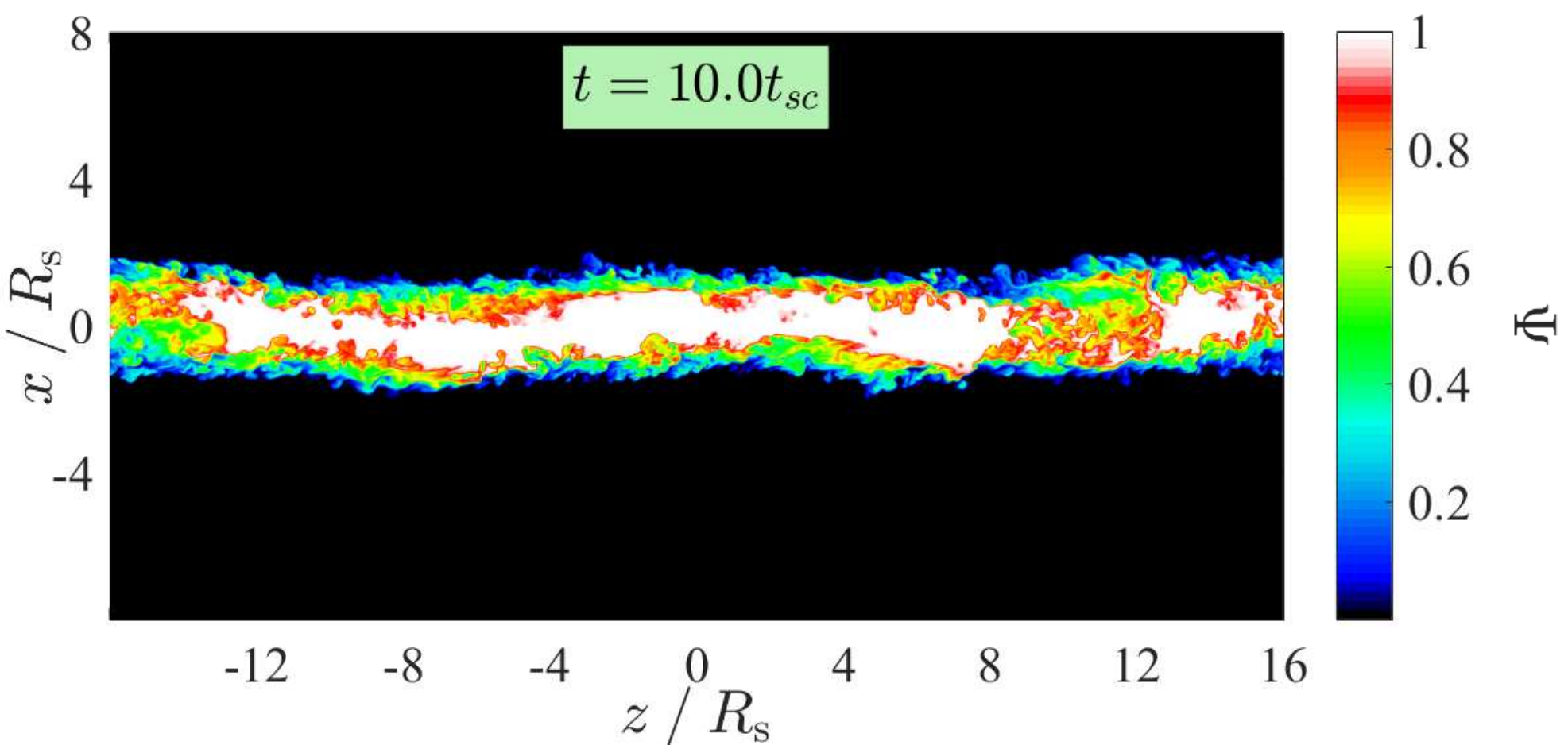}\\
\vspace{-0.09cm}
\includegraphics[trim={0.0cm 0.0cm 3.3cm 0}, clip, width =0.45 \textwidth]{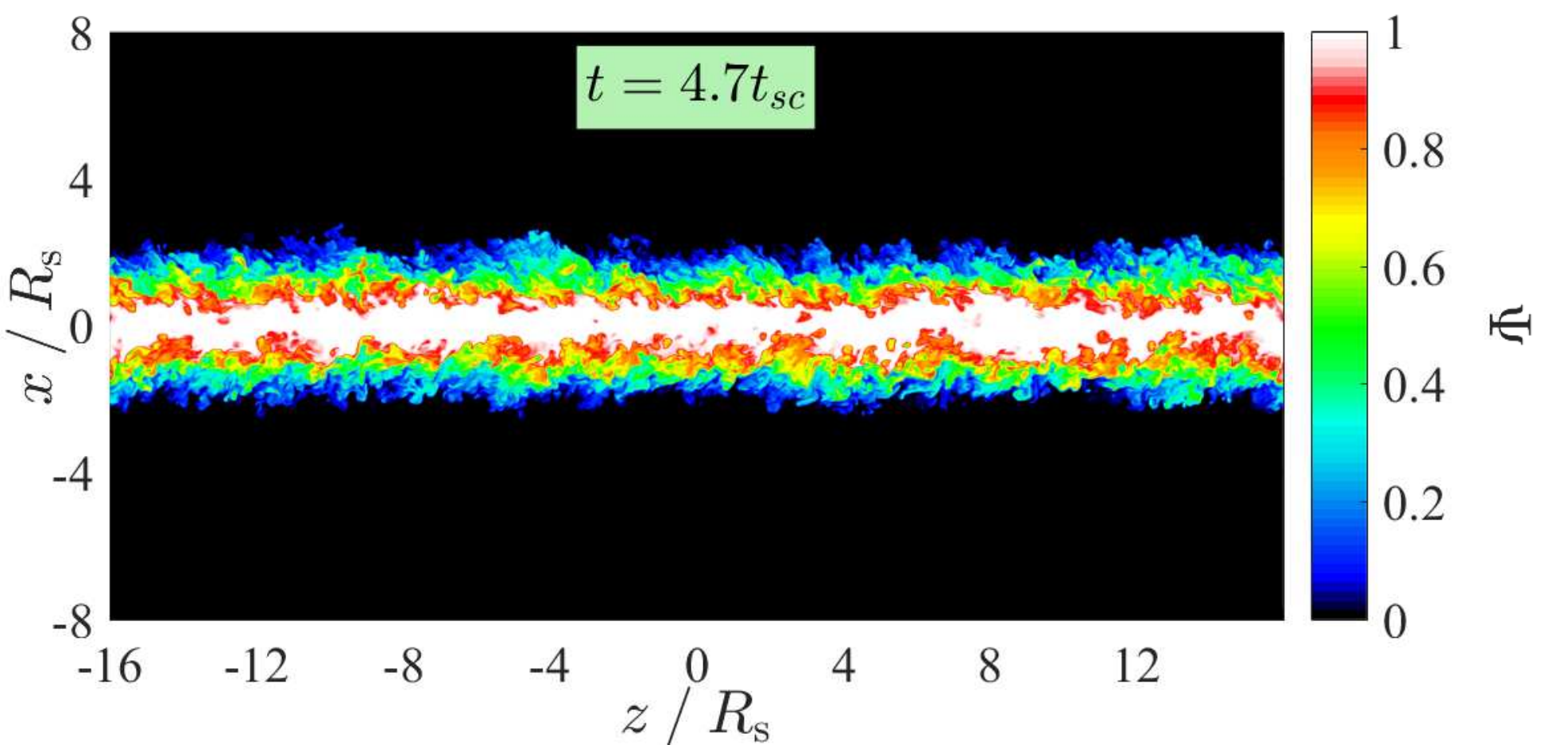}
\hspace{-0.3cm}
\includegraphics[trim={1.3cm 0.0cm 0.1cm 0}, clip, width =0.501 \textwidth]{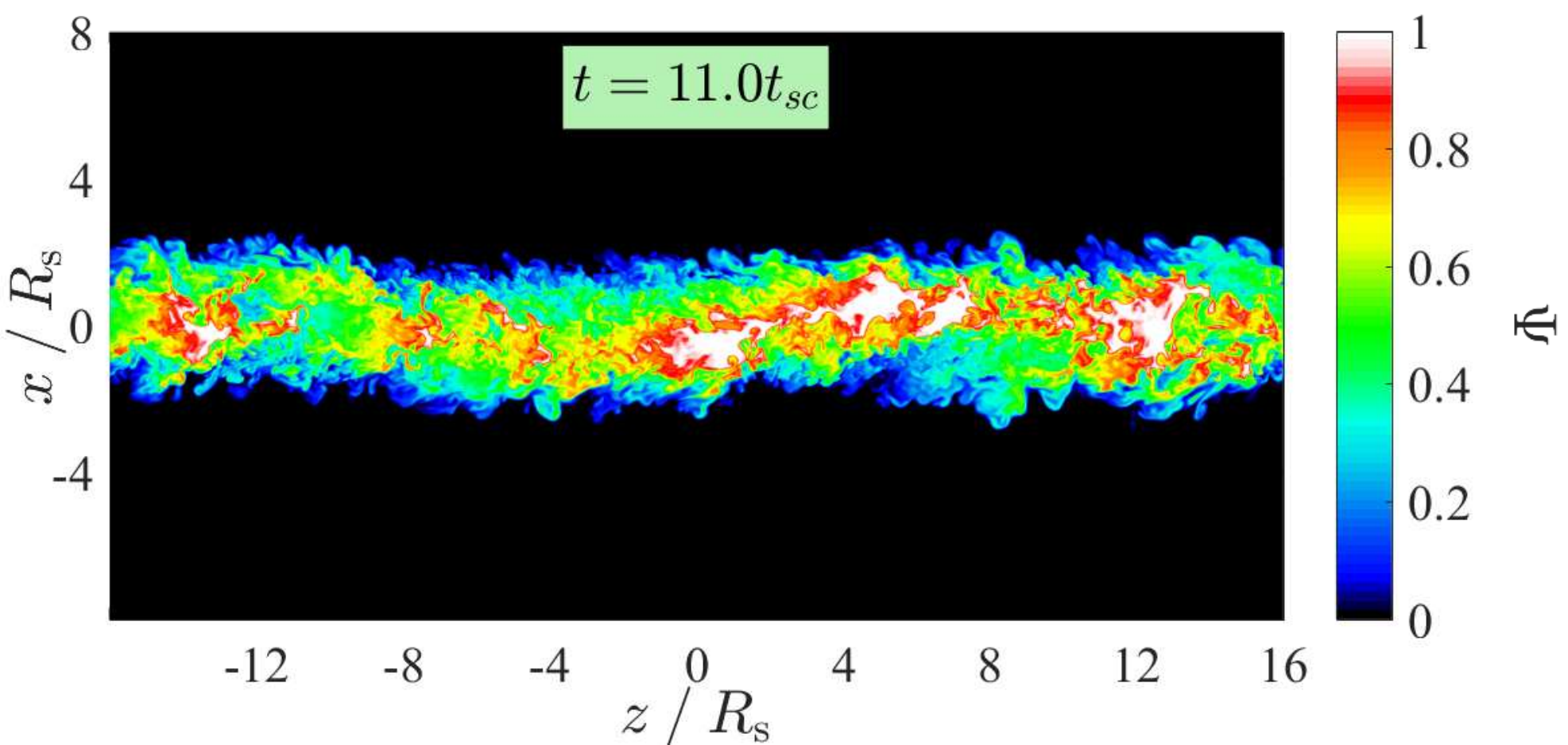}\\
\end{center}
\caption{Nonlinear stream evolution and disruption by high-$m$ surface modes vs. body modes. We show slices 
through the $y=0$ plane of a simulation with $(\Mb,\delta)=(5,1)$, with two different values of the smoothing 
parameter (\equnp{ramp2}), $\sigma/\Rs=1/32$ (left), and $1/8$ (right). Colour shows the value of the passive 
scalar $\psi$, as in \fig{colour_panel_M1D1}. For each case, the top row shows a snapshot just as $\hb$ begins 
to grow (\fig{h_body}), while the middle and bottom rows show snapshots 1 and 2 sound crossing times later, 
respectively. When $\sigma=\Rs/32$, the stream-background interface is dominated by small-scale structure when 
$\hb$ begins growing, at $t\sim 2.7\tsc$. The ensuing growth of $\hb$ is due to an expanding shear layer, as 
in the surface mode simulations described in \se{surface}. At $\sim 4.7\tsc$ there is still an unmixed core at 
the centre of the stream, which does not fully mix until $t\sim 6.5\tsc$. On the other hand, when $\sigma=\Rs/8$, 
the stream-background interface is dominated by a large-scale sinusoidal perturbation with relatively little small 
scale structure when $\hb$ begins growing, at $t\sim 9\tsc$. By $t\sim 10\tsc$ a sinusoidal structure with 
$\lambda\sim 16\Rs$ clearly dominates, in agreement with the predicted critical perturbation (\se{theory_body}). 
However, at the same time small scale structure has begun to develop, due to unstable surface modes with high-$m$. 
Within one additional sound crossing time these small-scale perturbations have efficiently mixed the stream and the 
background fluids, though the large scale sinusoidal mode is still visible. By $t\sim 11.5\tsc$ there is no unmixed 
fluid left in the stream. Similar figures showing stream evolution in simulations with different values of 
$(\Mb,\delta)$ can be found in \se{add_figures}.
}
\label{fig:colour_panel_M5D1} 
\end{figure*}

\smallskip
Unsurprisingly, the dependence of the 3d simulations on $\sigma$ is more significant. When $\sigma=\Rs/32$, stream 
expansion in all three simulations begins at $(1-2)\tsc$, similar to the surface mode simulations. The initial growth 
rate of $\hb$, until it reaches $\sim 2\Rs$, is consistent with \equ{hb_growth} with $\alpha\sim 0.05$, as expected 
from \equ{alpha_fit}. We note that a simulation initiated with $m=(0-4)$ modes (dotted green line) is extremely similar 
to the corresponding simulation with $m=0-1$ only (solid green line). However, when $\sigma=\Rs/8$ and high-$m$ surface 
modes are suppressed, stream expansion in 3d begins only shortly before the corresponding 2d simulation. The ratio of 
onset times in 3d versus 2d simulations is $\lsim 0.7$, roughly consistent with the shorter $t_{\rm KH}$ for the critical 
perturbation in cylindrical versus slab geometry. Unlike the 2d simulations, which maintain a roughly constant growth 
rate until $\hb\sim (8-10)\Rs$, the growth rate of $\hb$ in 3d simulations declines smoothly once $\hb\gsim 2\Rs$. In 
the case of $\delta=1$, $\hb$ seems to have saturated at a maximal value of $\sim 3.5\Rs$. As we shall see below, similarly 
to what was seen in \se{surface}, the decline in the stream expansion rate coincides with the onset of turbulence, which 
transfers energy from the large scales driving the stream expansion, to the small scales efficiently mixing the stream 
and background fluids. It is not entirely clear why the decrease in the growth rate of $\hb$ in 3d cylinders compared 
to 2d slabs seems so much larger here than it did for surface modes (see \fig{surface_h}). However, we recall that for body 
modes in 2d slabs, $\hb$ does not expand due to shear layer growth but rather due to the global deformation of the stream 
into a long wavelength sinusoidal, so the expansion mechanism is completely different. We defer a more detailed study of 
this to future work.

\smallskip
In the right two panels of \fig{h_body} we examined the penetration of the shear layer into the stream, $\hs/\Rs$, 
for simulations with $\sigma=\Rs/32$ and $\Rs/8$. We only show here results for 3d cylinders, since $\hs$ is not a particularly 
meaningful quantity for body modes in slabs. As detailed in P18 and described in \se{theory_body}, shear layer growth is 
suppressed for body modes, and stream disruption appears through global deformation of the stream into a large sinusoidal 
shape. However, this is not the case for 3d cylinders due to the appearance of high-$m$ surface modes which create a shear 
layer that penetrates into the stream (see also \fig{colour_panel_M5D1} below). The onset of shear layer penetration into 
the stream is coincident with the onset of rapid expansion. As mentioned above, in simulations with $\sigma=\Rs/32$ (second 
panel from the right) the shear layer is well described by \equs{hb_growth} and \equm{hs_growth} with $\alpha=0.05$, and in 
practice for the stream parameters studied here the shear layer consumes the entire stream $\sim (4-5)\tsc$ after the onset of 
growth. On the other hand, simulations with $\sigma=\Rs/8$ show much more rapid shear layer growth after the onset, with the 
stream fully consumed within $\sim (1-2)\tsc$, consistent with \equ{tau_break_body_3d} (see also \fig{colour_panel_M5D1} below). 
As noted above, the simulation with interface perturbations at wavelengths up to $16\Rs$, shown by the dashed cyan line, 
reaches $\hs=\Rs$ at the predicted stream disruption time from \equ{tau_break_body_3d}, which is marked by the cyan star.

\subsubsection{Stream Morphology}
\smallskip
In \fig{colour_panel_M5D1} we compare the stream morphology in simulations with $\sigma=\Rs/32$ 
and $\Rs/8$, at the onset of stream expansion. We focus on the case $(\Mb,\delta)=(5,1)$, but the 
results presented apply to all other cases. We show a slice edge-on through the midplane of the 
stream, in the $y=0$ plane, just as $\hb$ begins growing, $t\sim 2.7\tsc$ and $9\tsc$ for $\sigma=\Rs/32$ 
and $\Rs/8$ respectively. For each simulation we also show two additional snapshots, one and two 
sound crossing times after the first. The simulation with narrow smoothing is dominated by small 
scale structure concentrated at the stream-background interface. This is visible in the first snapshot, 
and expands with time into both the stream and the background, forming a stratified structure with an 
unmixed core surrounded by a mixed shear layer. This shear layer is formed by high-$m$ surface modes, 
and obeys the same physics as described in \se{surface}. By $t\sim 6.5\tsc$ the shear layer has consumed 
the entire stream. This implies a disruption time of $\sim 3.8\tsc$ from the onset of shear layer growth, 
consistent with \equ{tau_diss_2d} with $\alpha\sim 0.05$. 

\begin{figure*}
\begin{center}
\includegraphics[width =0.2585 \textwidth]{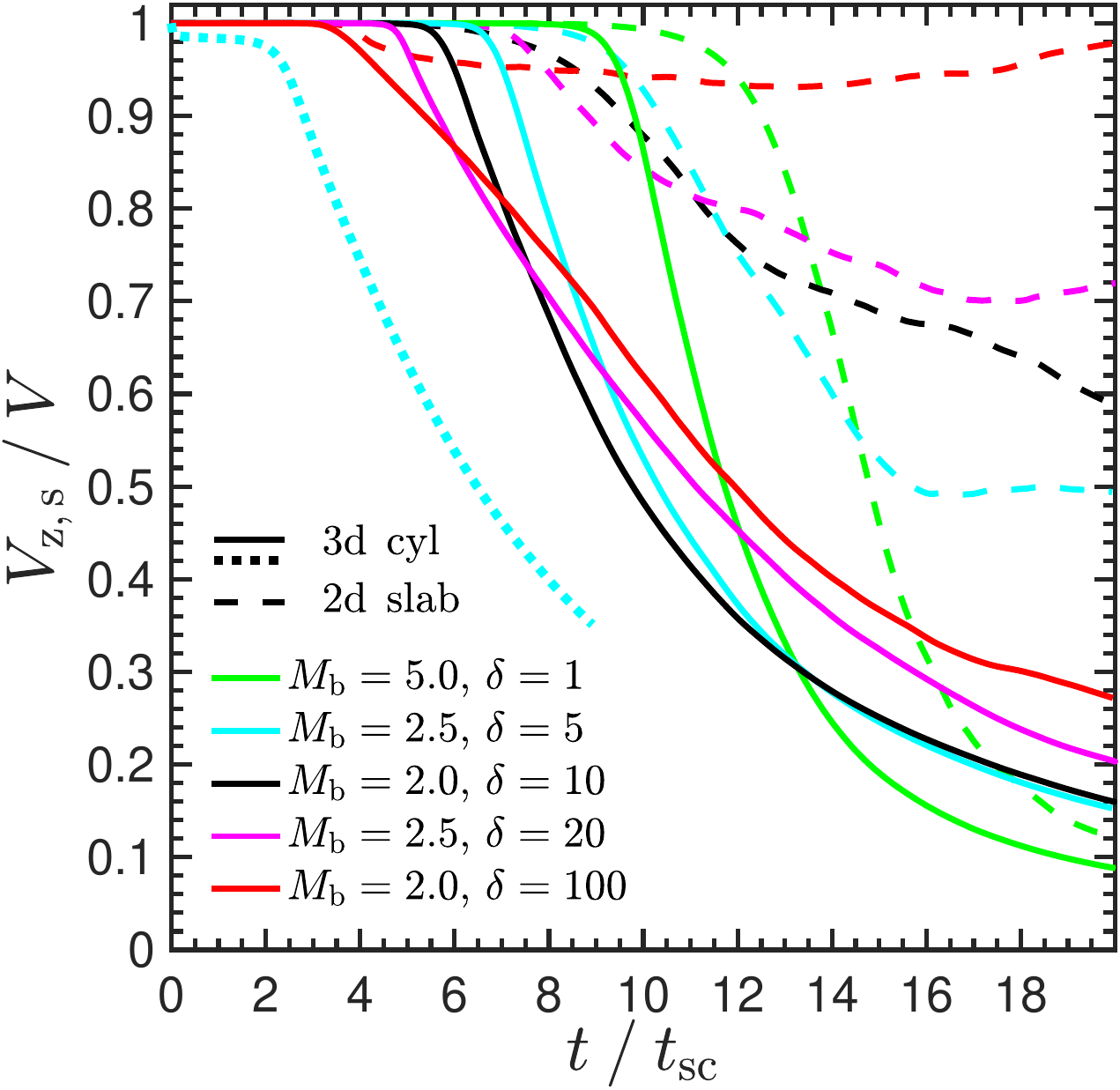}
\hspace{-0.19cm}
\includegraphics[trim={1.9cm 0.01cm 0.0cm 0}, clip, width =0.2205 \textwidth]{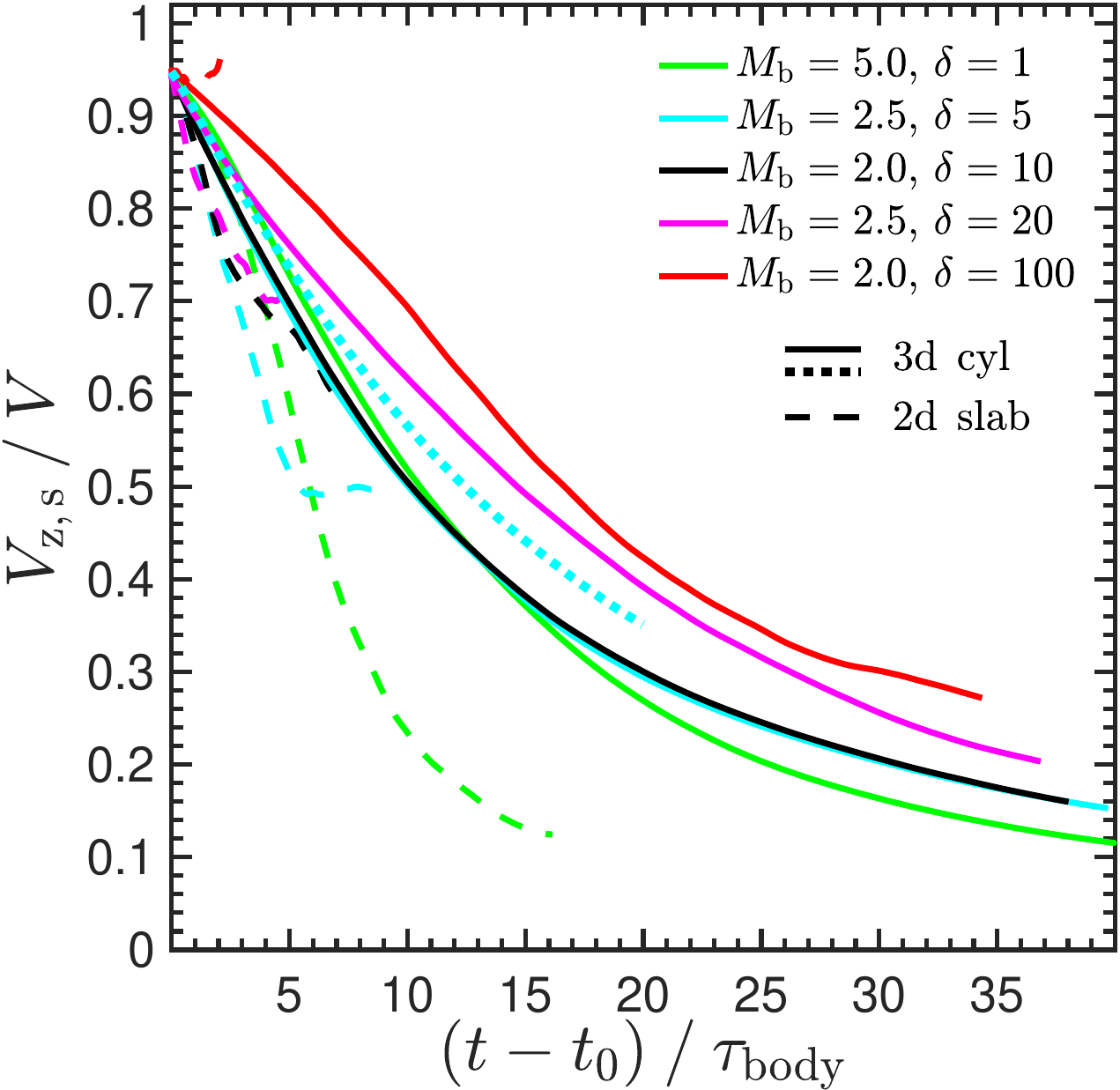}
\hspace{-0.19cm}
\includegraphics[trim={1.9cm -0.12cm 0.0cm 0}, clip, width =0.222 \textwidth]{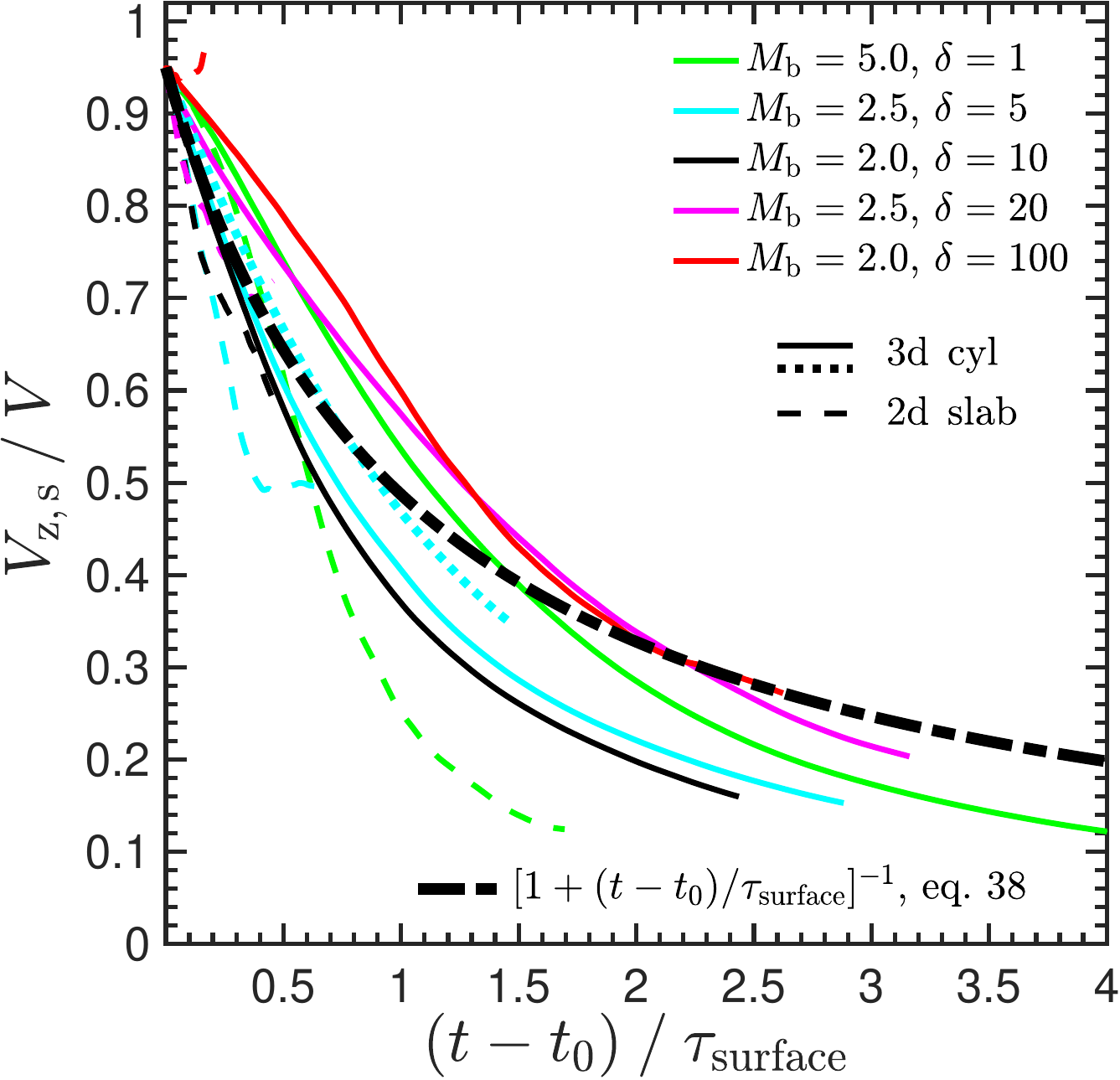}
\hspace{-0.32cm}
\includegraphics[trim={-0.65cm -0.10cm 0.0cm 0}, clip, width =0.2715 \textwidth]{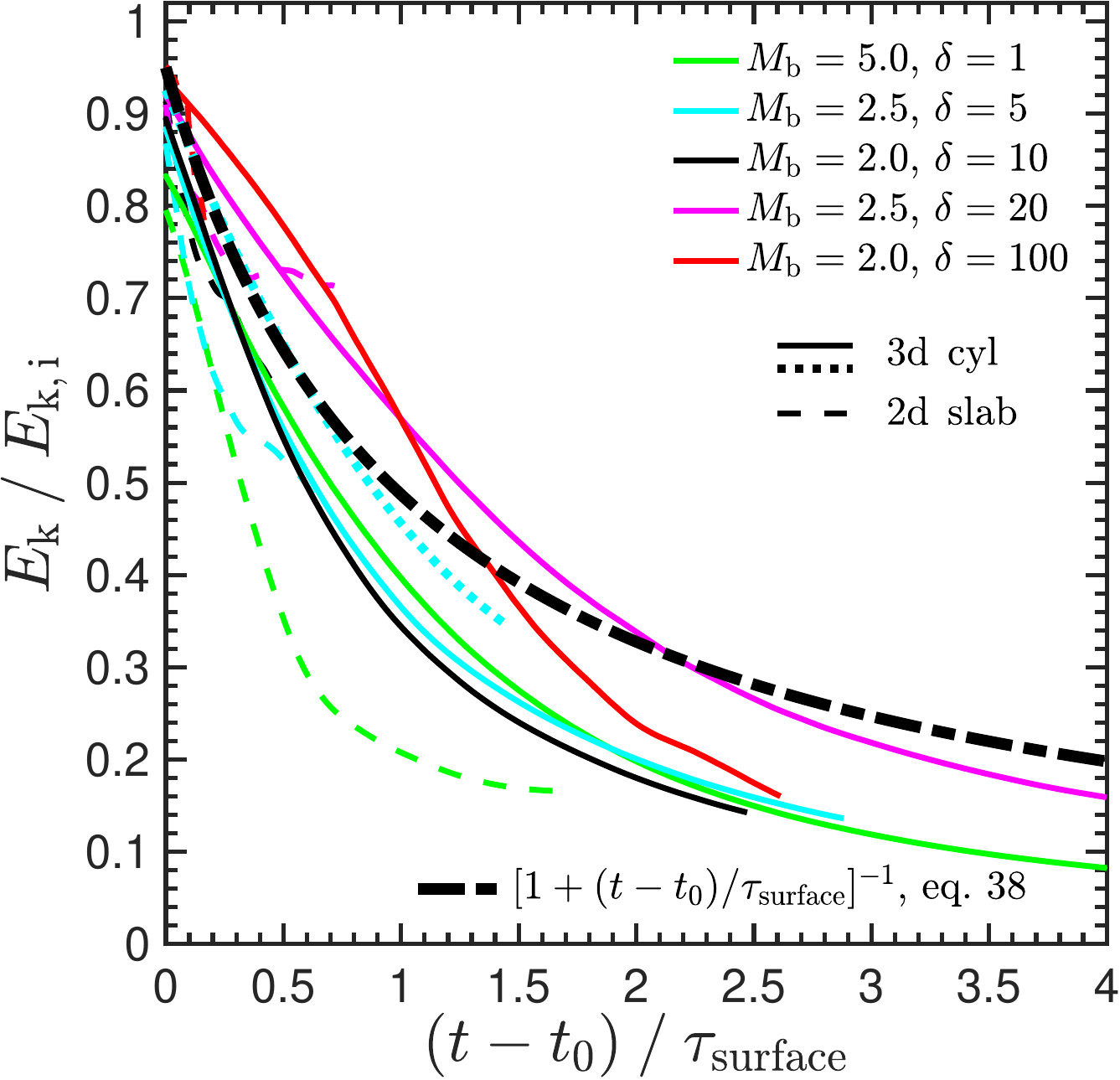}
\end{center}
\caption{Deceleration and dissipation of bulk kinetic energy in streams due to body modes, 
for supersonic streams with $\Mb>M_{\rm crit}$. \textit{In the left three panels}, we show the 
centre of mass velocity of the stream fluid (weighted by the passive scalar $\psi$) normalized 
by its initial value, as a function of time for all body mode simulations with $\sigma=\Rs/8$. 
Line styles and colours are the same as in \fig{h_body}. 
In the leftmost panel time is normalized by the sound crossing time. The deceleration begins 
together with the growth of $\hb$ (\fig{h_body}). With the exception of $\delta=1$, the 3d cylinders 
have decelerated significantly more than the 2d slabs by the end of the simulation at $20\tsc$. In the 
second panel from the left, we renormalize the time axis by the predicted body mode deceleration 
timescale, $\tau_{\rm body}$ (\equnp{tau_body_2d} and \equnp{tau_body_3d} for 2d and 3d respectively), 
and show all curves beginning at $t_0$, the time when the velocity has reached $0.95$ its initial value. 
The 2d simulations all follow the same slope at first, roughly $V_{\rm z,\,s}\sim -0.1 Vt/\tau_{\rm body}$. 
The deceleration rate sharply declines once $\hb\sim (8-10)\Rs$, which is almost always before the 
slab reaches half its initial velocity. \textit{The increase in slab velocity at $t>15\tsc$ for the simulation with 
$\delta=100$ is due to stream fluid leaving the simulation domain and is not to be trusted.} Similarly, 
the 3d simulations all have very similar slopes, roughly $V_{\rm z,\,s}\sim -0.05Vt/\tau_{\rm body}$. 
While the simulation with $\delta=100$ has a shallower slope at first, it coincides with the other 
simulations once $V_{\rm z,\,s}/V\lsim 0.75$. The transition of 3d simulations to a shallower slope 
always occurs after the stream reaches half its initial velocity. In the third panel, we renormalize 
the time axis by the predicted deceleration timescale for surface modes, $\tau_{\rm surface}$ 
(\equnp{tau_surface_2d} and \equnp{tau_surface_3d} for 2d and 3d respectively), and again show all 
curves beginning at $t_0$. The 3d simulations all roughly follow the predicted velocity profile for 
surface modes from \equ{dec_model}, shown by the thick dot-dashed line, though this is not as good a 
match for 2d slabs. This is consistent with the late time evolution of 3d cylinders, after the critical 
body mode begins to grow, being dominated by high-$m$ surface modes and efficient mixing (\fig{colour_panel_M5D1}). 
\textit{In the rightmost panel}, we show the total kinetic energy associated with laminar flow of 
both stream and background fluid normalized by its initial value, as a function of time normalized by 
$\tau_{\rm surface}$. The rate of dissipation of bulk kinetic energy in 3d cylinders is extremely 
similar to the stream deceleration rate, as predicted.
}
\label{fig:momentum_body} 
\end{figure*}

\smallskip
On the other hand, in the simulation with wide smoothing, there is very little small-scale structure 
present at the onset of stream expansion, at $t=9\tsc$. Rather, the stream is dominated by a sinusoidal 
wave with a wavelength of $\lambda\sim 16\Rs$, consistent with the critical perturbation for this case 
which is the fundamental helical mode [$(m,n)=(1,0)$] with $\lambda_{\rm crit}\sim 13\Rs$. After a sound 
crossing time, the critical perturbation has clearly taken over and the stream is dominated by a large 
sinusoidal wave with an amplitude of $\gsim \Rs$. At the same time, small-scale turbulence has appeared 
throughout the stream, triggerred by unstable high-$m$, long wavelength surface modes. Within less than 
one additional sound crossing time, this turbulence has consumed the entire stream. While the long wavelength 
sinusoidal is still visible, the stream has effectively dissintegrated and mixed with the background. The 
total disruption time from the onset of shear layer growth is thus $<2\tsc$, shorter than the case with narrow 
smoothing, which was dominated by surface modes from the beginning. The accelerated shear layer growth in 
this case is likely due to the increased surface area of the stream-background interface caused by the 
large scale sinus wave. Of course, we recall that the onset of shear layer growth was significantly delayed 
in this case, until after the critical perturbation had grown. Finally, we note that the turbulent velocities 
in the shear layer exhibit the same behaviour as in the surface mode simulations shown in \fig{turbulence}, 
increasing to a maximum of $\sim 0.2-0.3$ at early times, and then decaying to an assymptotic value of 
$\sigma/V\lsim 0.2$.
%However, even in the case of a narrow initial smoothing layer, if the criticl perturbation is seeded with a large enough initial amplitude it can reach its critical amplitude before the shear layer consumes the entire stream, thus accelerating the stream disruption.

\subsubsection{Deceleration and Kinetic Energy Dissipation}
\smallskip
In \fig{momentum_body} we show the stream deceleration in 2d slab and 3d cylinder simulations with 
$\sigma=\Rs/8$. In the leftmost panel we show the stream velocity as a function of time normalized by the 
stream sound crossing time. For all cases, deceleration begins together with stream expansion and the growth 
of $\hb$ in \fig{h_body}. Clearly, stream deceleration is more rapid in 3d. The only exception is the case with 
$\delta=1$, where the final stream velocity in 2d and 3d are comparable. For $\delta\ge 10$, the stream velocity 
at the end of the 3d simulations is a factor $\sim 3-4$ lower than in the 2d simulations, where the deceleration 
rate greatly decreases once $\hb\sim (8-10)\Rs$ (see \fig{h_body}), consistent with the results of P18. The apparent 
increase in stream velocity in the 2d simulation with $\delta=100$ at $t>15\tsc$ is due to stream material leaving 
the box at this time, \textit{and is not to be trusted}. All other simulations end before any stream material leaves 
the box. We note that the simulation initiated with interface perturbations at long wavelengths (dotted cyan line) 
has a very similar deceleration rate to the corresponding simulation with the same parameters but a fiducial perturbation 
spectrum (solid cyan line), though it begins decelerating earlier, in agreement with the discussion following \fig{h_body}.

\smallskip
In the second panel of \fig{momentum_body}, we normalize the time axis to $\tau_{\rm body}$ from \equ{tau_body_2d} 
or \equ{tau_body_3d} for 2d and 3d respectively. Before rescaling the time axis, we move its origin to $t_0$, 
defined as the time when the velocity reaches $0.95$ of its original value, in order to focus on the deceleration 
itself rather than the ``incubation" period before it begins. When scaled in this way, all the 2d simulations exhibit 
the same behaviour. The initial deceleration rate, while $\hb<8\Rs$, is approximately 
$V_{\rm z,\,s}\sim -0.1 Vt/\tau_{\rm body,\,2d}$, consistent with the results of P18. The 3d simulations with 
$\delta\le 10$ all exhibit the same behaviour as well, though different from their 2d counterparts. The stream 
velocity approximately follows $V_{\rm z,\,s}\sim -0.05 Vt/\tau_{\rm body,\,3d}$, meaning that deceleration is 
half as efficient per $\tau_{\rm body}$ in 3d cylinders than in 2d slabs, though this timescale is objectively 
much shorter in 3d than in 2d. The cases with $\delta\ge 20$ exhibit slightly shallower slopes, though we note 
that the simulation with $\delta=100$ exhibits the same slope as those with $\delta\le 10$ once the velocity has 
decreased to $\sim 0.75$ its initial value. Similar to the 2d simulations, the deceleration rate in 3d eventually 
decreases, but this only occurs once the velocity has decreased to less than half its initial value. Unlike the 
2d simulations, the decrease in the deceleration rate in 3d is not associated with a critical $\hb$, but rather 
with a decrease in the growth rate of $\hb$, coincident with the onset of small scale turbulence and mixing.

\smallskip
Based on the discussion above, a 3d cylinder reaches half its initial velocity in $\sim 10\tau_{\rm body}$. 
Based on \equ{tau_ratio}, the ratio between this and $\tau_{\rm surface}$ is $\sim 2(1+\delta^{-1/2})/Mb$, 
where we have assumed $\alpha=0.05$. For $\delta\gsim 10$ and $\Mb\sim 2$ this ratio is $\sim 1$. Indeed, 
when plotted against $t/\tau_{\rm surface}$ with $\alpha=0.05$, the different 3d simulations align just as 
well as when plotted against $t/\tau_{\rm body}$, as seen in the third panel of \fig{momentum_body}. 
The streams reach half their initial velocity at $t\sim \tau_{\rm surface}$, and are reasonably well fit by 
our toy model for stream deceleration due to surface modes, \equ{dec_model}. We conclude that it is equally 
valid to describe 3d cylindrical streams as decelerating due to shear layer mixing with the appropriate 
effective growth rate, consistent with the efficient mixing following the growth of the critical perturbation 
seen in \fig{colour_panel_M5D1}. Based on \fig{momentum_body}, this desctription is less satisfactory for 2d 
slab simulations, as expected.

\smallskip
In the rightmost panel of \fig{momentum_body}, we show the total kinetic energy associated with laminar 
flow, evaluated according to \equ{EK_sim}, and normalized to its initial value. The bulk kinetic energy 
decreases at the same rate as the stream velocity, as was the case for surface modes (\fig{deceleration_surface}), 
and as predicted by our simple toy model of shear layer growth and accretion of background material 
which begins flowing at roughly the same velocity as the stream (\equnp{EK}). 
This further supports the description of 3d cylindrical streams as decelerating due to shear layer mixing with 
the appropriate effective growth rate, as described above.

%%%%%%%%%%%%%%%%%%%%%%%%%%%%%%%%%%%%%%%%%%%%%%%%  
\section{Application to Cold Streams in Massive SFGs at High-$z$}
\label{sec:application} 

\smallskip
In this section we use the analytic theory of \se{theory}, which is broadly supported by our simulations 
presented in \se{results}, to predict the effect of KHI on cold streams 
feeding massive galaxies at high redshift. We begin in \se{app_param} by providing estimates for the relevant 
stream parameters. In \se{app_diss} we estimate the potential for stream disruption, and in \se{app_decel} 
we address the effect on their inflow rate.

%!!!!!!!!!!!!!!!!!!!!!!!!!!!!!!!
\subsection{Stream Parameters}
\label{sec:app_param}
Our predictions depend primarily on three dimensionless parameters: The Mach number, $\Mb$, the density 
contrast, $\delta$, and the ratio of stream radius to the halo virial radius, $\Rs/\Rv$. These were 
crudely estimated in M16 and refined in P18, based on simple analytic arguments and cosmological 
simulations. We repeat the main arguments here for completeness, and refer the reader to P18 (section 
5.1) for further details. 

\smallskip
The stream velocity is proportional to the halo virial velocity, which can be related to the sound speed 
of gas at the virial temperature, yielding $\Mb\sim 1$. The density contrast is obtained by assuming pressure 
equilibrium between hot gas at $\Tb\sim \Tv$, the virial temperature of an NFW halo, and cold gas at 
$\Ts \sim 10^4 {\rm K}$, set by the steep drop in the cooling rate below that temperature \citep{SD93}. 
If both the halo and the stream are roughly isothermal, then this ratio is constant throughout the halo. 
In practice, the stream temperature can be as high as $\sim 3\times 10^4{\rm K}$ due to photo-heating 
from the UV background \citep[e.g.][]{Goerdt10}, while the post-shock temperature in the hot halo near $\Rv$ 
may only be $\sim 0.5\Tv$ \citep[][P18]{db06}. Finally, the stream radius is related to its velocity and 
density through the mass accretion rate along the stream, ${\dot {M}}_{\rm s}\simeq \pi \Rs^2 \rhos V_{\rm s}$. 
Cosmological simulations suggest that this is typically a fixed fraction of the total accretion rate onto 
the halo virial radius \citep{Danovich12}, which is constrained by cosmology \citep[e.g.][]{Dekel13}. The 
final expressions, including uncertainties in model parameters, are 
\be 
\label{eq:Mb_stream}
\Mb \simeq 0.75-2.25,
\ee
\be 
\label{eq:del_stream}
\delta \simeq (10-100)\times M_{12}^{2/3} (1+z)_3^{-1},
\ee
\be 
\label{eq:Rs_stream}
\frac{\Rs}{\Rv} \simeq (0.01-0.1)\times \left(\delta_{75} M_{\rm b,1.5}\right)^{-1/2},
\ee
{\no}where $M_{12}=\Mv/10^{12}\Msun$, $(1+z)_3=(1+z)/3$, $\delta_{75}=\delta/75$ and $M_{\rm b,1.5}=\Mb/1.5$. 
We note that $\delta\sim 10$ implies that the backgrond gas has a temperature of $\sim 3\times 10^5\K$, 
which would make it thermally unstable unless it is illuminated by a very hard ionizing background (e.g. 
\citealp{Efstathiou92}, though see also \citealp{Binney09} who argue that buoyancy effects may stabilize 
the halo against thermal instability even without a photoionizing source). This implies that $\delta\gsim 30$ 
is likely the most physically plausible regime.

%!!!!!!!!!!!!!!!!!!!!!!!!!!!!!!!
\subsection{Stream disruption}
\label{sec:app_diss}

\smallskip
For 2d slabs, stream disruption manifests itself in different ways for surface and body modes. 
For the former, disruption occurs when the shear layer consumes the entire slab, while for 
the latter the stream expads to several times its initial width and breaks up before significant 
mixing occurs. However, for 3d cylinders there is no such distinction. Even if body modes dominate 
the transition to nonlinearity, small scale surface modes quickly develop and efficiently mix the 
stream and background fluids after the critical perturbation takes over (\fig{colour_panel_M5D1}).

\begin{figure*}
\begin{center}
\includegraphics[trim={0.0cm 0.9cm 3.0cm 0}, clip, width =0.4285 \textwidth]{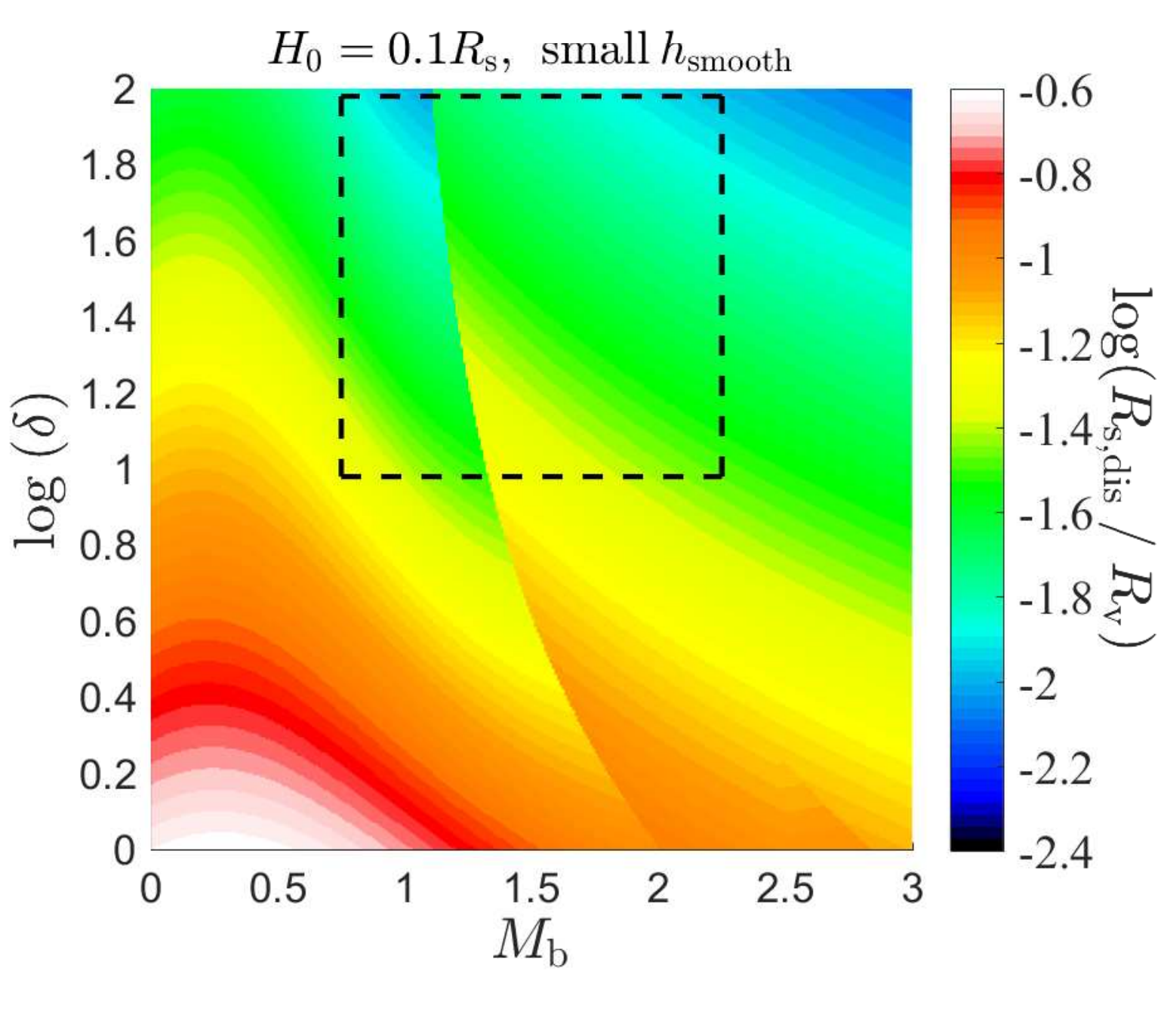}
\hspace{-0.30cm}
\includegraphics[trim={1.85cm 0.9cm 0.0cm 0}, clip, width =0.4675 \textwidth]{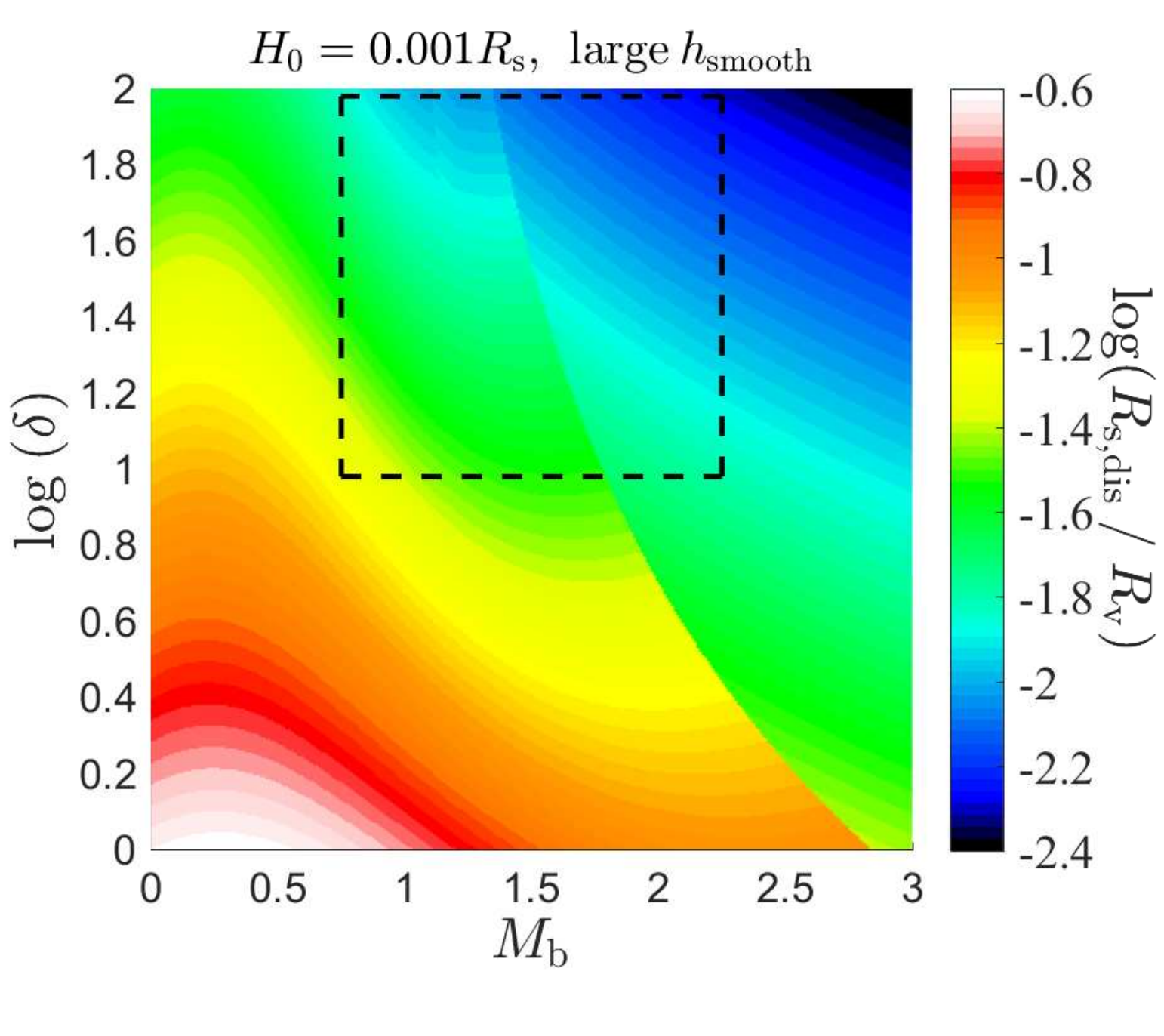}
\end{center}
\caption{The critical stream radius for disruption due to KHI, $R_{\rm s,dis}$, as a function of $\Mb$ 
and $\delta$. Streams with $\Rs<R_{\rm s,dis}$ are expected to dissintegrate in the CGM before reaching 
the central galaxy. The critical radius is given by either \equ{stream_dis_surface} or \equ{stream_dis_body}, 
depending on whether surface or body modes dominate the instability, as described in the text. The two panels 
represent two extreme limits for the regime where body modes are unstable, $\Mb>(1+\delta^{-1/2})$. In the 
left-hand panel, the initial amplitude of the critical body mode is assumed to be large, $H_0=0.1\Rs$, and 
the initial smoothing layer between the stream and the background is small enough to allow high-$m$ surface 
modes to grow, so both surface and body modes are unstable. In the right-hand panel, the initial amplitude 
of the critical perturbation is assumed to be small, $H_0=0.001\Rs$, and the initial smoothing layer is large 
enough to stabilize all surface modes when $\Mb>M_{\rm crit}$ (\equnp{Mcrit}), so only body modes are unstable 
in this regime. When $\Mb<(1+\delta^{-1/2})$, these two limits are identical by definition. At larger Mach 
numbers, the critical radii in the left-hand panel are a factor of $\sim 2$ larger than those in the right-hand 
panel throughout the range of $(\Mb,\delta)$ values relevant for cold streams, marked by dashed squares. 
$R_{\rm s,dis}$ ranges from $\sim (0.005-0.05)\Rv$, decreasing towards higher $\delta$ and higher $\Mb$, 
save for a jump to larger values when body modes destabilize. This is $\sim 70\%$ larger than the corresponding 
range of critical radii for disruption of 2d slabs.
}
\label{fig:stream_disruption} 
\end{figure*}

\smallskip
In 2d slabs, surface modes lead to stream disruption after a time $\sim t_{\rm dis,\,surface,\,2d}$ 
(\equnp{tau_diss_2d}), with $\alpha_{\rm s,\,2d}$ given by \equ{alpha_fit}, where the subscript 
$s$ denotes that this value of $\alpha$ corresponds to the penetration of the shear layer into the stream 
rather than the background. However, we find in our simulations that the penetration of the shear layer 
into 3d cylinders is more rapid than in 2d slabs (\fig{alpha}, right-two panels), leading to disruption 
by $t_{\rm dis,\,surface,\,3d}\sim (\alpha_{\rm s,\,2d}/\alpha_{\rm s,\,3d})~t_{\rm dis,\,surface,\,2d}$,  
where the ratio of $\alpha$ values is given by \equ{alpha_ratio}. Since the largest 
eddies in the shear layer move at the convection velocity, $V_{\rm c}$ (\equnp{convective}), the shear layer will 
consume the stream before reaching the halo centre if $t_{\rm dis,\,surface,\,3d}\le \Rv/V_{\rm c}$. 
Following P18, we write this as a condition on the stream radius, $\Rs\le R_{\rm s,dis}^{\rm surface}$, where 
\be 
\label{eq:stream_dis_surface}
\frac{R_{\rm s,dis}^{\rm surface}}{\Rv} = \frac{\alpha_{\rm s,\,3d}(\Mb,\delta)}{\sqrt{\delta}}.
\ee

\smallskip
If the instability is initially dominated by body modes, stream disruption occurs after 
$t_{\rm dis,\,body,\,3d}$ (\equnp{tau_break_body_3d}), roughly one sound crossing time 
after the critical perturbation becomes nonlinear (see \fig{colour_panel_M5D1}). Since 
body modes cause negligible deceleration prior to this, the stream disintegrates prior 
to reaching the halo centre if $t_{\rm dis,\,body,\,3d}\le \Rv/V$, or $\Rs\le R_{\rm s,dis}^{\rm body}$ 
where 
\be 
\label{eq:stream_dis_body}
\frac{R_{\rm s,dis}^{\rm body}}{\Rv} = \left[2\sqrt{\delta}\Mb\left(0.5~{\rm ln}\left(\dfrac{\Rs}{H_0}\right)+1\right)\right]^{-1}.
\ee
{\no}Note that unlike \equ{stream_dis_surface}, this depends on the initial 
conditions through the initial amplitude of the critical perturbation, $H_0$.

\smallskip
If $\Mb<(1+\delta^{-1/2})$, then $M_{\rm tot}<1$ (\equnp{Mtot}) and body modes 
are stable, so the critical radius for stream disruption is $R_{\rm s,dis}^{\rm surface}$. 
If $(1+\delta^{-1/2})<\Mb<M_{\rm crit}=(1+\delta^{-1/3})^{3/2}$ (\equnp{Mcrit}), 
surface and body modes are both unstable, and the critical stream radius for disruption 
is given by ${\rm max}(R_{\rm s,dis}^{\rm surface},\,R_{\rm s,dis}^{\rm body})$. 
If $M_{\rm crit}<\Mb$, the behaviour depends on the width of the initial smoothing 
layer between the stream and background fluids, $h_{\rm smooth}$. This width is set 
by various physical processes, such as thermal conduction or the dynamics of cylidrical 
accretion onto the stream outside the halo. Simple estimates using \citet{Spitzer56} 
conductivity show that it may vary from $h_{\rm smooth}\sim \Rs$ to $h_{\rm smooth}<<\Rs$, 
depending on whether one uses densities and temperatures characteristic of the cold streams 
or the hot background (M16). A more accurate determination of $h_{\rm smooth}$ is beyond the 
scope of this paper, and we instead examine two limits. If $h_{\rm smooth}>>2\pi\Rs/m_{\rm crit}$, 
with $m_{\rm crit}$ given by \equ{mcrit}, then surface modes are stable and the critical radius 
for stream disruption is $R_{\rm s,dis}^{\rm body}$. However, if $h_{\rm smooth}\lsim 2\pi\Rs/m_{\rm crit}$, 
then both surface and body modes are unstable, and the critical stream radius for disruption 
is given by ${\rm max}(R_{\rm s,dis}^{\rm surface},\,R_{\rm s,dis}^{\rm body})$.

\smallskip
\Fig{stream_disruption} shows the resulting critical stream radius for disruption, $R_{\rm s,dis}/\Rv$, 
as a function of $\Mb$ and $\delta$. The two panels show two extreme limits for the body mode regime. 
In the left-hand panel, the initial amplitude of the critical perturbation is assumed to be large, $H_0=0.1\Rs$, 
and the initial smoothing layer between the stream and the background is assumed to be small so that both 
surface and body modes are unstable. On the other hand, in the right-hand panel, the initial amplitude of 
the critical perturbation is very small, $H_0=0.001\Rs$, and the initial smoothing layer is large so 
surface modes are stable. The behaviour at $M_{\rm tot}<1$ is identical in these two limits by definition. 
At larger Mach numbers, the critical radius in the former limit is a factor of $\sim 2$ larger than in the 
latter limit throughout the parameter range relevant for cold streams, marked by a dashed square in each panel. 
We find that streams with radius up to $\sim 0.05\Rv$ may disentegrate prior to reaching the galaxy. The critical 
radius ranges from $R_{\rm s,dis}\sim (0.005-0.05)\Rv$, decreasing with $\delta$, and tending to decrease with 
$\Mb$ as well save for a jump to larger values when body modes destabilize. For 2d slabs, the critical radius 
for disruption was found to range from $\sim (0.003-0.03)\Rv$ (P18), so moving to 3d has increased the potential 
for stream disruption by $\sim 70\%$. 

\begin{figure*}
\begin{center}
\includegraphics[trim={0.0cm 0.9cm 3.0cm 0}, clip, width =0.423 \textwidth]{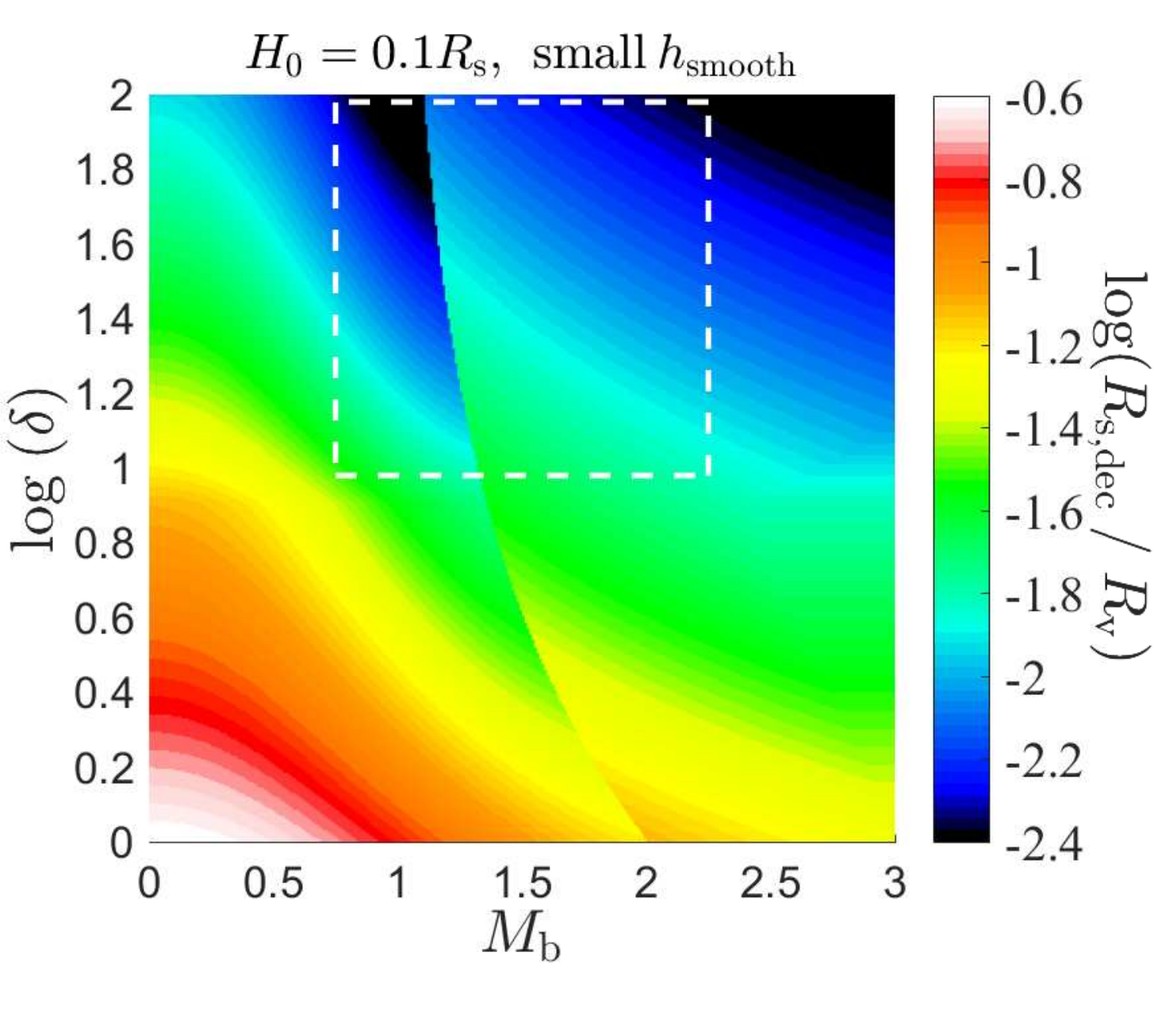}
\hspace{-0.30cm}
\includegraphics[trim={1.75cm 0.9cm 0.0cm 0}, clip, width =0.4675 \textwidth]{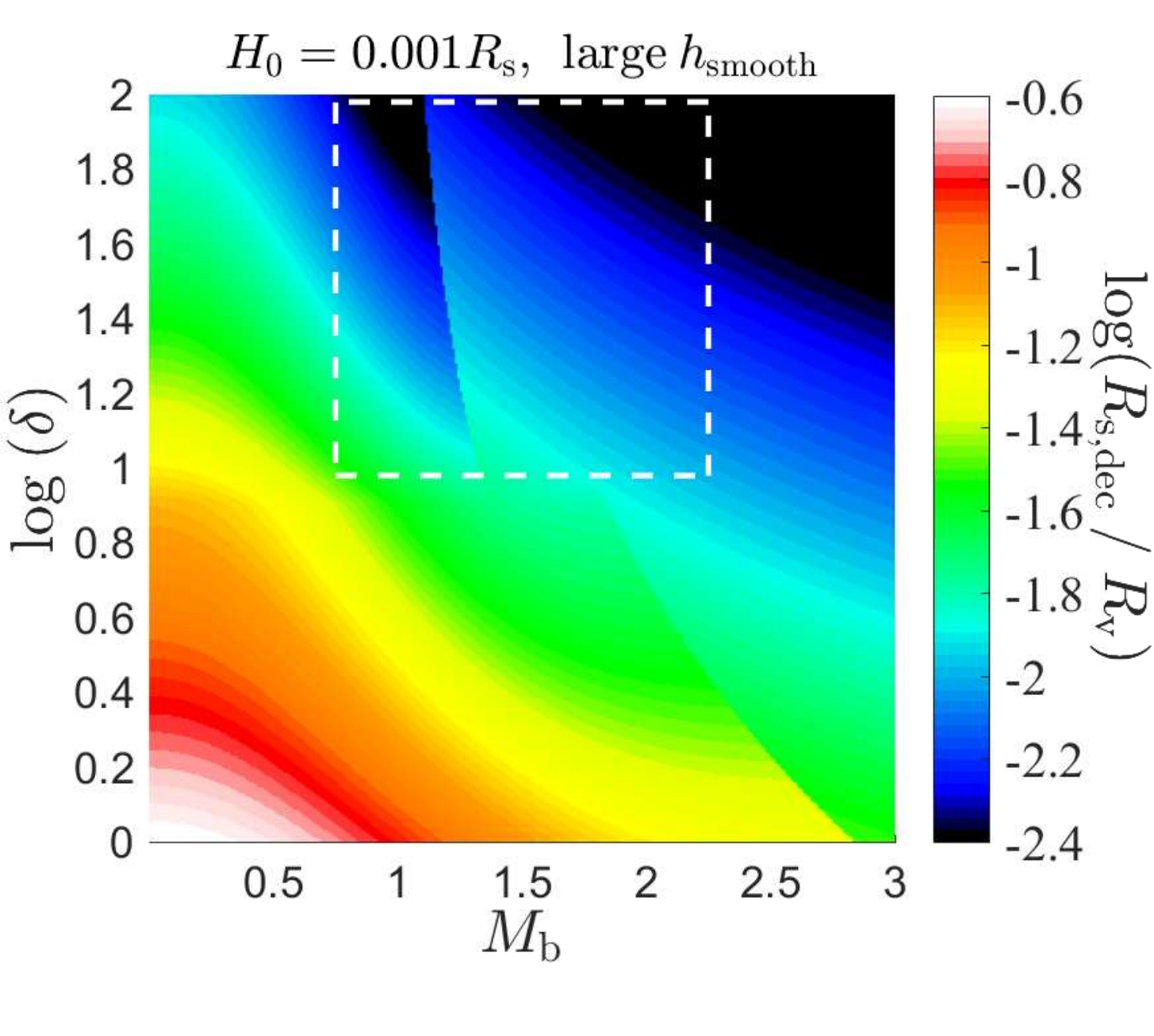}
\end{center}
\caption{The critical stream radius for a factor 2 reduction in the inflow velocity of cold gas onto the 
central galaxy due to KHI, $R_{\rm s,dec}$, as a function of $\Mb$ and $\delta$. In the absence of the 
gravitational acceleration of the dark matter halo, streams with $\Rs<R_{\rm s,dec}$ are expected to 
decelerate to less than half their initial velocity in a virial crossing time. $R_{\rm s,dec}$ is given 
by either \equ{stream_dec_surface} or \equ{stream_dec_body}, depending on whether surface or body modes 
dominate the instability, as described in the text. The two panels represent the same two extreme limits 
as in \fig{stream_disruption} for the regime where body modes are unstable. By definition, these two limits 
have no effect at $\Mb<(1+\delta^{-1/2})$. At larger Mach numbers, the critical radii in the left-hand 
panel are only $\sim 40\%$ larger than those in the right-hand panel throughout the range of $(\Mb,\delta)$ 
values relevant for cold streams, marked by dashed squares. $R_{\rm s,dec}$ ranges from $\sim (0.004-0.03)\Rv$, 
decreasing towards higher $\delta$ and higher $\Mb$, save for a jump to larger values when body modes destabilize. 
This is a factor of $3-4$ larger than the corresponding range of critical radii for disruption of 2d slabs.
}
\label{fig:stream_deceleration} 
\end{figure*}

%!!!!!!!!!!!!!!!!!!!!!!!!!!!!!!!
\subsection{Deceleration and Kinetic Energy Dissipation}
\label{sec:app_decel}

\smallskip
The deceleration of stream fluid due to KHI is interpreted in the cosmological case as a reduction in the 
inflow rate of cold gas onto the central galaxy. As a crude estimate of the magnitude of this effect, 
following P18 we estimate the critical stream radius for a factor of 2 reduction in inflow rate to 
occur prior to reaching the central galaxy, $R_{\rm s,dec}$. 

\smallskip
Surface modes reduce the stream velocity to half its initial value after a time $\sim \tau_{\rm surface,\,3d}$, 
which is proportional to $\Rs$ (\equnp{tau_surface_3d}). By requiring that $\tau_{\rm surface,\,3d}=\tv=\Rv/\Vv$, 
we obtain the critical stream radius 
\be 
\label{eq:stream_dec_surface}
\frac{R_{\rm s,dec}^{\rm surface}}{\Rv} = \dfrac{\alpha_{\rm b} \sqrt{\delta}}{\left(1+\sqrt{\delta}\right)\left(\sqrt{1+\delta}-1\right)},
\ee
{\no}where $\alpha_{\rm b}$ represents shear layer penetration into the background rather than the stream. 
For $\delta<8$ we approximate $\alpha_{\rm b}$ with \equ{alpha_fit} while for $\delta>8$ we use 
half of this value\footnote{In order for \equ{stream_dec_surface} to be continuous at $\delta=8$ 
when we decrease $\alpha_{\rm b}$ by a factor 2, the second term in the denominator becomes $\sqrt{1+\delta}-2$ 
for $\delta>8$.}, in order to account for the decrease in the expansion rate of the shear layer into 
the background once $\hb\sim (1.5-2)\Rs$, as explained in \se{surface} and shown in \fig{surface_h}. 
Our simulations show that this decreased value of $\alpha_{\rm b}$ reproduces stream deceleration rates 
at high $\delta$ (\fig{deceleration_surface}). 

\smallskip
If the instability is initially dominated by body modes, the stream only begins to 
decelerate after the critical perturbation becomes nonlinear, at $t_{\rm NL}$ 
(\equnp{tNL}). From this point, the stream decelerates at a roughly constant rate 
until it reaches half its initial velocity. In \se{body} we showed that the timescale 
for the stream to reach half its initial velocity can either be expressed as 
$\sim 10\tau_{\rm body,\,3d}$ (\equnp{tau_body_3d}) or as $\tau_{\rm surface,\,3d}$ 
with $\alpha_{\rm b}=0.05$. These two timescales are very similar for values of $(\Mb,\delta)$ 
typical of cold streams, while the former becomes smaller as either $\Mb$ or $\delta$ 
are increased. For the remainder of this analysis we adopt $\sim 10\tau_{\rm body,\,3d}$ 
as the relevant timescale, but stress that this choice has no significant effect on 
our conclusions. The resulting critical radius for deceleration due to body modes is 
\be 
\label{eq:stream_dec_body}
\frac{R_{\rm s,dec}^{\rm body}}{\Rv} = \left[10\Mb\left(\sqrt{1+\delta}-1\right) + \Mb\sqrt{\delta}~{\rm ln}\left(\dfrac{\Rs}{H_0}\right)\right]^{-1}.
\ee
{\no}Note that unlike \equ{stream_dec_surface}, this condition depends on the initial 
conditions through the initial amplitude of the critical perturbation, $H_0$.

%\begin{figure}
%\includegraphics[trim={0.3cm 0.2cm 1.3cm 0.0cm}, clip, width =0.47 \textwidth]{./deceleration_toy.pdf}
\begin{figure*}
\begin{center}
\includegraphics[trim={0.5cm 1.93cm 1.7cm 0}, clip, width =0.32 \textwidth]{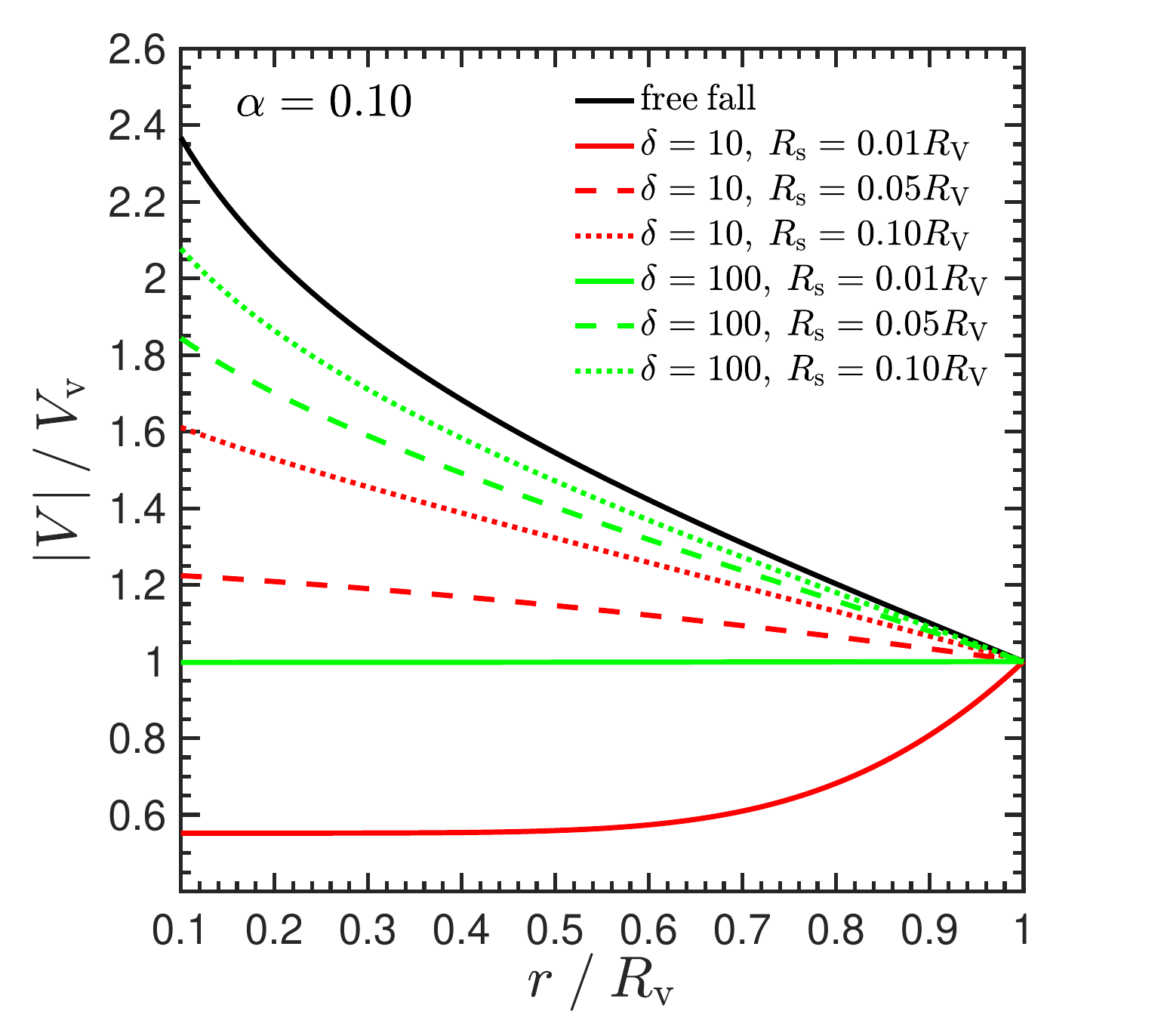}
\hspace{-0.23cm}
\includegraphics[trim={0.5cm 1.93cm 1.7cm 0}, clip, width =0.32 \textwidth]{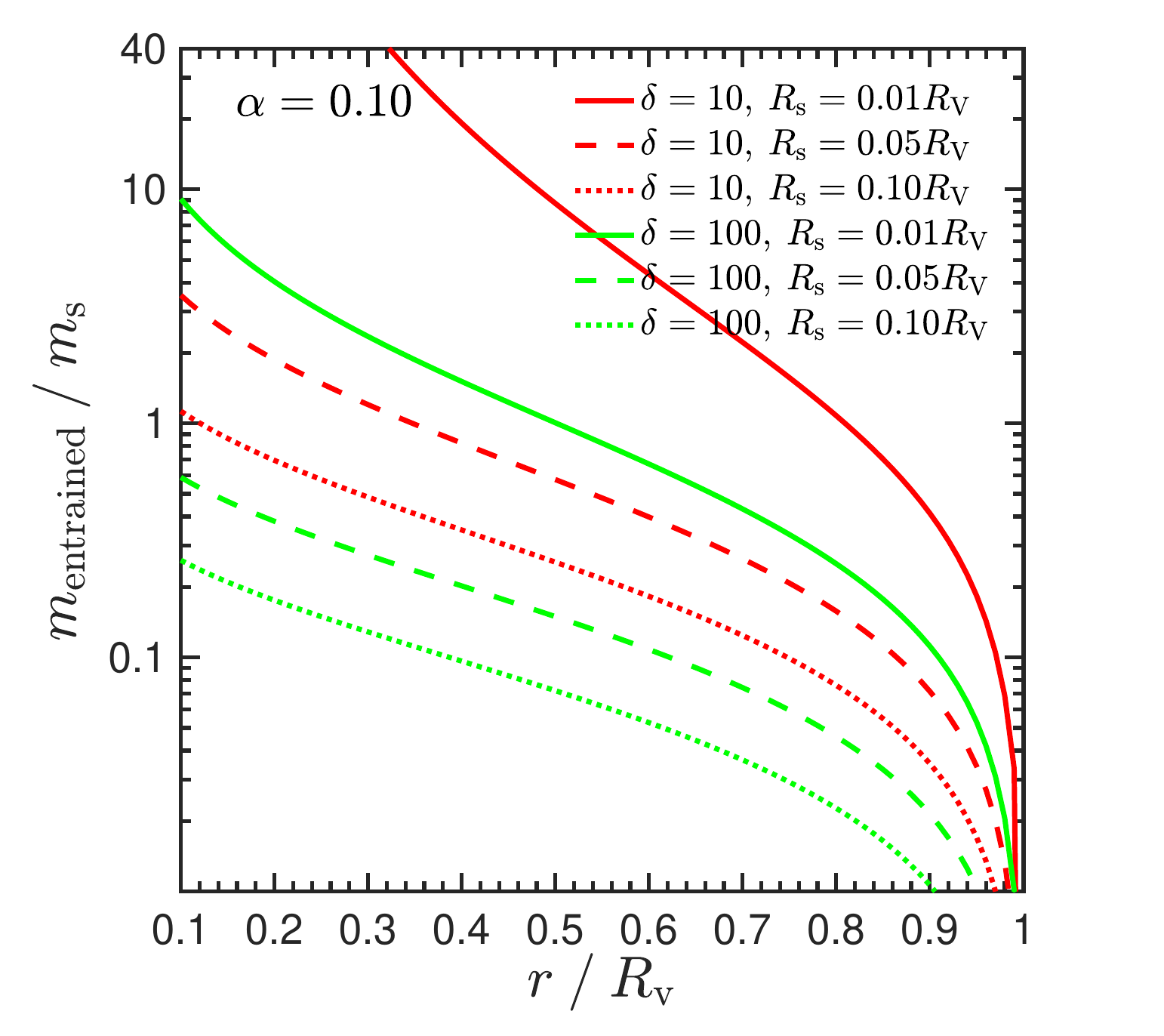}
\hspace{-0.18cm}
\includegraphics[trim={0.5cm 1.93cm 1.7cm 0}, clip, width =0.32 \textwidth]{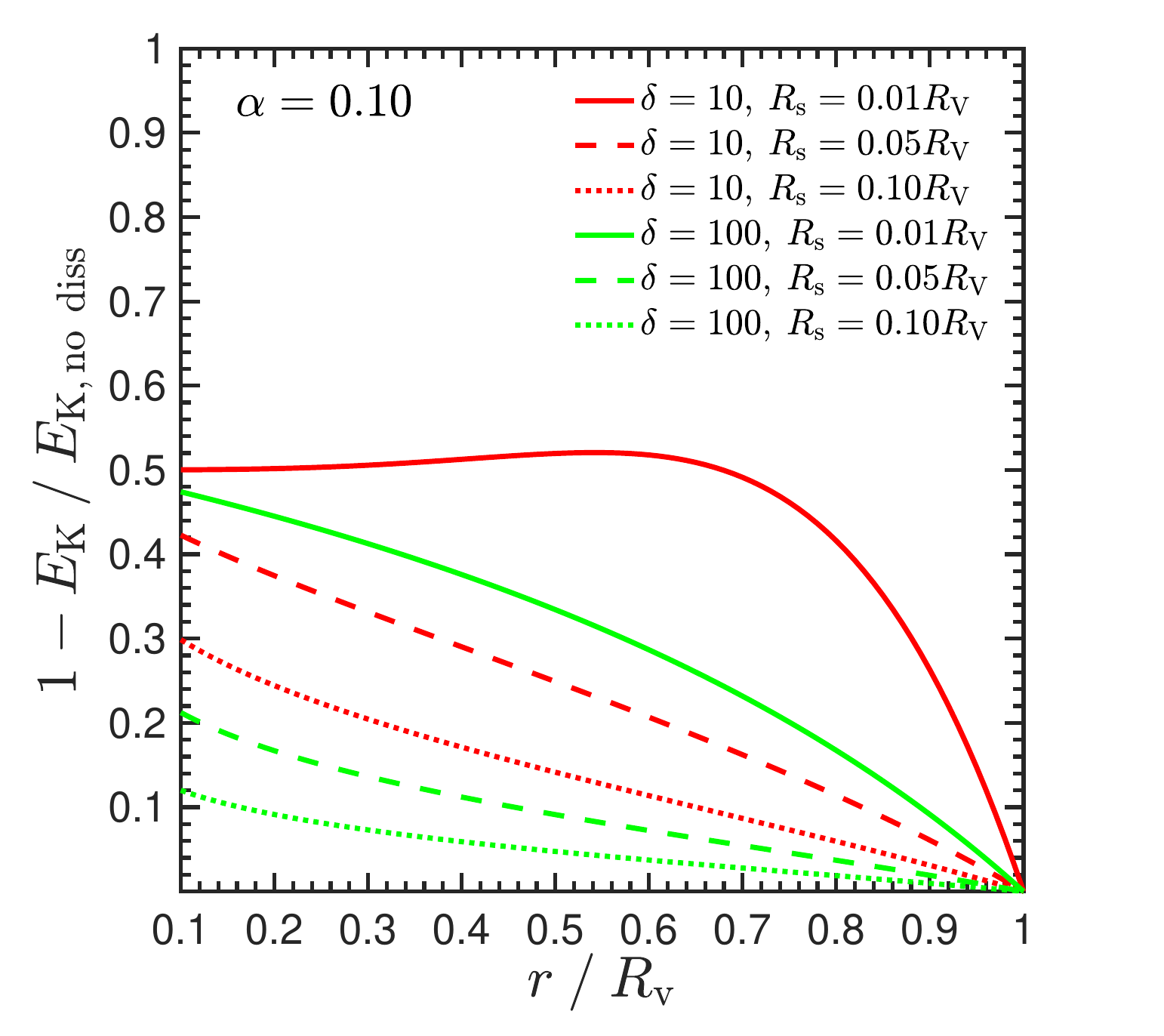}
\vspace{-0.50cm}
\includegraphics[trim={0.5cm 0.0cm 1.7cm 0.6cm}, clip, width =0.32 \textwidth]{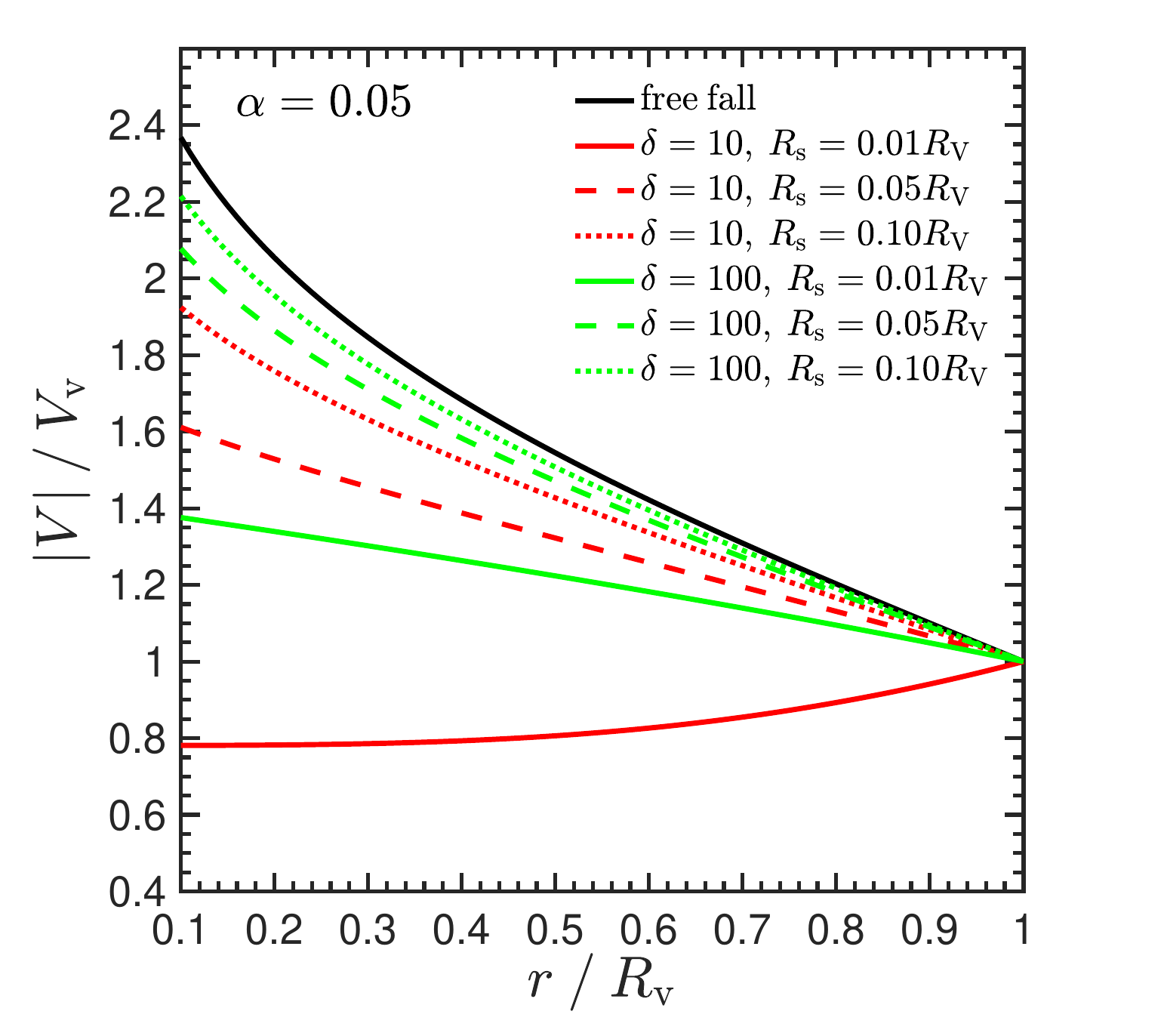}
\hspace{-0.23cm}
\includegraphics[trim={0.5cm 0.0cm 1.7cm 0.6cm}, clip, width =0.32 \textwidth]{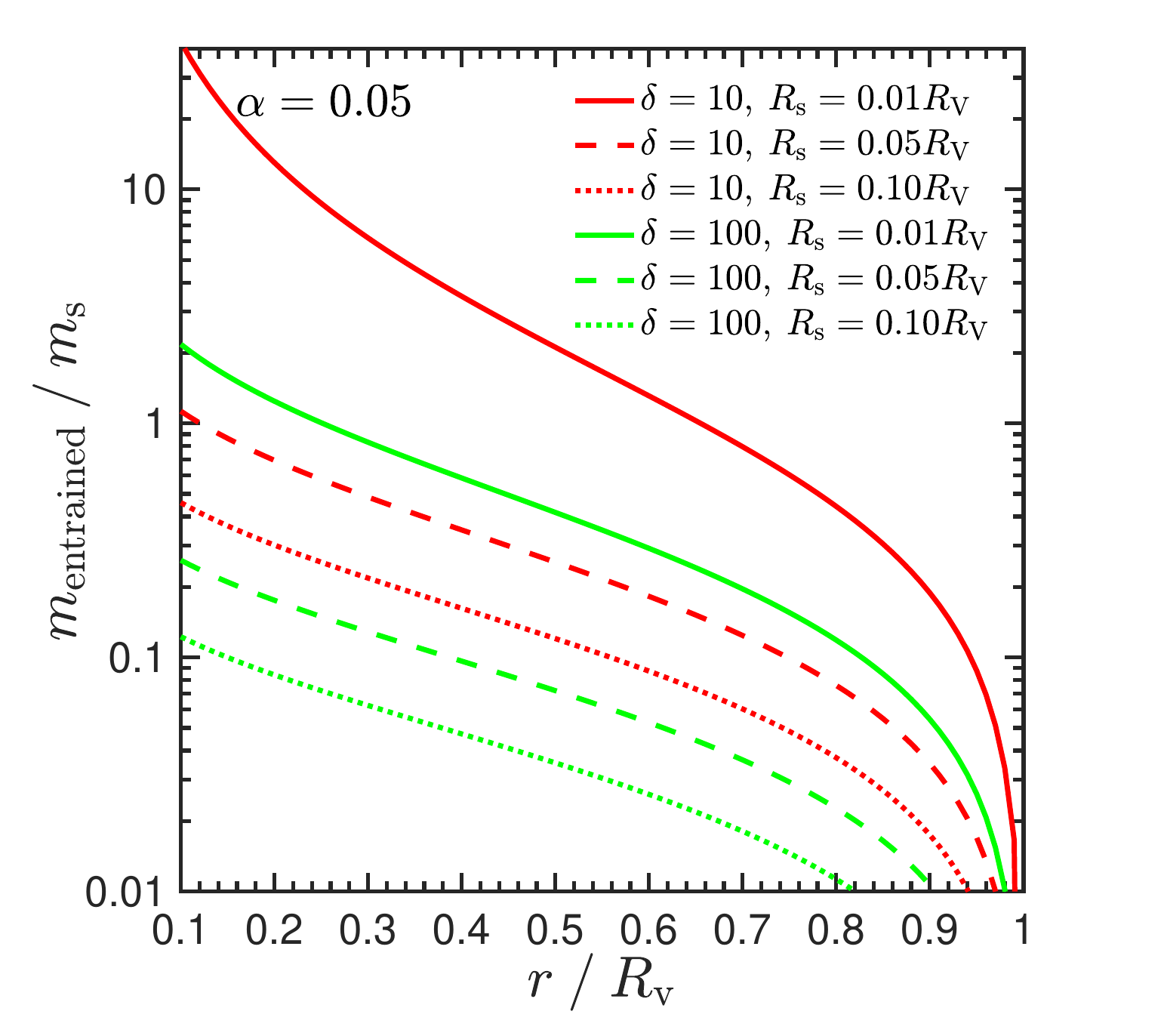}
\hspace{-0.18cm}
\includegraphics[trim={0.5cm 0.0cm 1.7cm 0.6cm}, clip, width =0.32 \textwidth]{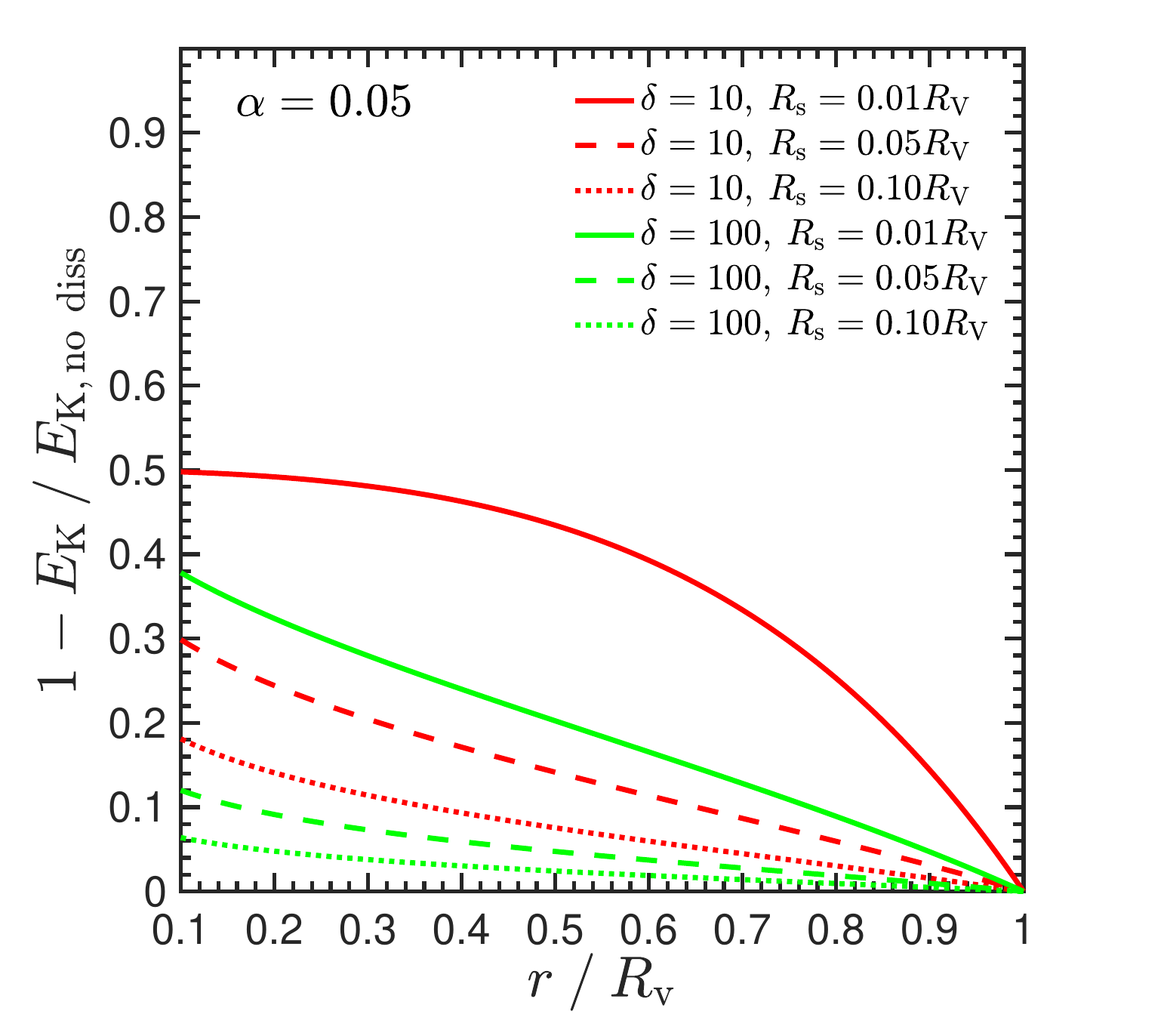}
\end{center}
\caption{Evolution of a stream in the CGM of a massive dark matter halo, according to solutions 
of a toy model including gravitational acceleration and a friction-like deceleration due to KHI.  
\equs{Vr_sol}-\equm{Bdef}. The horizontal axes show the halocentric radius normalized by the halo 
virial radius, and the vertical axes show the magnitude of the inflow velocity normalized to 
the halo virial velocity (left, with the black solid lines showing the free fall solution), the 
mass of backgroud gas entrained in the shear layer normalized by the stream mass (middle), and the 
fraction of bulk kinetic energy associated with laminar flow which is lost (right). In each panel, the 
red and green lines show solutions for $\delta=10$ and $100$ respectively, while the solid, dashed and 
dotted lines show solutions for $\Rs/\Rv=0.01$, $0.05$, and $0.10$ respectively. The top (bottom) row 
shows results assuming a constant value of $\alpha=0.1$ ($0.05$). Realistic stream values are likely 
bracketed by the solid and the dashed green lines and by these two values of $\alpha$ (see text). 
For these values, we see that KHI can significantly contribute to the deceleration of streams as they 
fall down the potential well of massive dark matter halos, while the mass entrained within the shear layer 
is a significant fraction of the stream mass, and can exceed it in some cases. Overall, $\sim 10-50\%$ of the 
bulk kinetic energy is lost to a combination of turbulence, thermal energy, and sound waves, and may subsequently 
be dissipated into radiation.
}
\label{fig:deceleration_toy} 
\end{figure*}

\smallskip
\Fig{stream_deceleration} shows the resulting critical stream radius for a factor of 2 reduction in inflow 
velocity, $R_{\rm s,dec}/\Rv$, as a function of $\Mb$ and $\delta$. The two panels show the same two extreme 
limits for the body mode regime as in \fig{stream_disruption}. As before, the behaviour at $M_{\rm tot}<1$ is 
identical in these two limits by definition. At larger Mach numbers, the critical radius in the former limit 
is only $\sim 40\%$ larger than in the latter limit throughout the parameter range relevant for cold streams, 
marked by a dashed square in each panels. We find that streams with radii up to $\sim 0.03\Rv$ can decelerate 
to half their initial velocity prior to reaching the galaxy due to KHI. The critical radius ranges from 
$R_{\rm s,dec}\sim (0.004-0.03)\Rv$, decreasing with $\delta$, and tending to decrease with $\Mb$ as well 
save for a jump to larger values when body modes destabilize. For 2d slabs, deceleration was found to be 
negligible, with the corresponding critical radius ranging from $R_{\rm s,dec}\sim (0.001-0.01)\Rv$ (P18), 
below the expected range of radii for cold streams (\equnp{Rs_stream}). The 3d geometry thus increases the 
potential for stream deceleration by a factor of $\sim 3-4$. 

\smallskip
The above discussion of stream deceleration ignores the gravitational acceleration due to the dark matter 
halo through which the stream is flowing. This can compensate for some or all of the deceleration caused 
by KHI, and may be related to the roughly constant inflow velocities of cold streams seen in cosmological 
simulations at much lower resolution \citep{Dekel09,Goerdt15a}, where KHI is at best marginally resolved. 
In P18 we introduced a simple toy model to estimate the radial velocity profile of a stream in the CGM of 
a dark matter halo, finding that unless $\delta\sim 10$ and $\Rs/\Rv\sim 0.01$, which is an unlikely 
combination according to \equ{Rs_stream}, the stream velocity profile was very close to the free-fall 
velocity. We revisit this model below and apply it to 3d cylindrical streams rather than 2d slabs. 

\smallskip
The stream fluid is modeled as a rigid body falling radially into the halo under the influence of gravity 
and a drag-like force associated with KHI. The dark matter halo is modeled as a singular isothermal 
sphere, with a mass profile $M(r)=\Vv^2 r /G$, where $r$ is the halocentric radius, $G$ is Newton's constant, 
and $\Vv={\rm const}$ is the halo virial velocity. The gravitational acceleration at radius $r$ is thus 
\be 
\label{eq:agrav}
a_{\rm g}=-\frac{\Vv^2}{r}.
\ee

\smallskip
We assume that within the halo the stream has a conical shape as it is focused towards the halo centre, so 
that its radius at halocentric distance $r$ is 
\be
\label{eq:rs_r}
r_{\rm s}(r)=\frac{\Rs}{\Rv}r.
\ee
{\no}This is consistent with cosmological simulations \citep{Dekel09,vdv12b}. We model the deceleration 
due to KHI, $\dot{V}_{\rm KHI}$, using \equ{akhi}, which we found to be a good fit for both surface modes 
(\fig{deceleration_surface}) and body modes (\fig{momentum_body}). We replace the constant stream radius, $\Rs$, 
in \equ{akhi} with by $r_{\rm s}(r)$, and assume that the density contrast, $\delta$, is independent of radius. 

\smallskip
Combining \equs{akhi}, \equm{agrav} and \equm{rs_r}, one obtains an equation of motion for the stream under 
the influence of gravity and KHI-induced drag. Since ${\rm d}V/{\rm d}t=0.5({\rm d}V^2/{\rm d}r)$, this 
can be converted into a differential equation for the radial velocity profile, $V(r)$, with the boundary condition 
$V(r=\Rv)=\Vv$. The solution is 
\be 
\label{eq:Vr_sol}
\frac{V(r)}{\Vv} = \left\{\begin{array}{c c}
-\sqrt{\dfrac{(B-2)x^B+2}{B}} & B\ne 0\\
\\
-\sqrt{1-2{\rm ln}(x)} & B=0
\end{array}\right. ,
\ee
{\no}where $x=r/\Rv$, and 
\be 
\label{eq:Bdef}
B = \frac{2\alpha\sqrt{\delta}\Rv}{(1+\sqrt{\delta})(\sqrt{1+\delta}-1)\Rs}.
\ee
{\no}The case $B=0$ represents gravitational free-fall, corresponding to $\alpha=0$. 

\smallskip
In the left-hand panels of \fig{deceleration_toy}, we show the resulting radial profiles for 
$\delta=10,100$ and $\Rs/\Rv=0.01,0.05,0.10$, as well as for free-fall. The top (bottom) row 
shows the results for $\alpha=0.1~(0.05)$, which bracket the range of $\alpha$ values relevant for 
cold streams (\fig{alpha}). For narrow streams, with $\Rs/\Rv=0.01$, the deceleration 
due to KHI is able to overcome the gravitational acceleration. With $\alpha=0.1$, dense streams 
with $\delta=100$ maintain a constant inflow velocity throughout the halo, reaching $0.1\Rv$ with a 
velocity $\sim 0.42$ times the free-fall velocity, hereafter $V_{\rm ff}$. Dilute streams with $\delta=10$ 
actually net \textit{decelerate} in the CGM, reaching $0.1\Rv$ at $\sim 0.55$ times their initial velocity, 
or $\sim 0.23V_{\rm ff}$. However, based on \equ{Rs_stream} such a narrow and dilute stream is an unlikely 
combination, and furthermore would be expected to disrupt while still in the CGM (\fig{stream_disruption}). 
Streams with $\Rs=0.05\Rv$, which are not likely to disrupt before reaching the central galaxy (\fig{stream_disruption}), 
also show non-negligible deceleration, reaching $0.1\Rv$ with velocities $\sim 0.52V_{\rm ff}$ and $0.78V_{\rm ff}$ 
for $\delta=10$ and $100$ respectively. Even very wide streams, with $\Rs/\Rv=0.10$, reach $0.1\Rv$ with 
$\sim 0.68V_{\rm ff}$ for $\delta=10$. For $\delta=100$, the velocity is $\sim 0.88V_{\rm ff}$, though such 
wide and dense streams are unlikely based on \equ{Rs_stream}. Using $\alpha=0.05$ in \equ{Bdef} increases 
the velocities at $0.1\Rv$ by $\sim (10-40)\%$ compared to $\alpha=0.1$. Overall, these results show that KHI 
can be significant in preventing the acceleration of streams as they fall down the potential well of dark matter 
halos, contrary to the results of the 2d analysis in P18, highlighting the importance of the 3d analysis presented 
here. 

\smallskip
While our results suggest that KHI is unlikely to maintain a constant inflow velocity unless the streams 
are very narrow, $\Rs\sim 0.01\Rv$, in Appendix \se{convergence} we found that in systems dominated by high-$m$ 
surface modes, the deceleration rates are overestimated in low resolution simulations. This may explain why 
cosmological simulations indicate constant velocities for streams which can be much wider than $0.01\Rv$.

\smallskip
We now extend the toy model beyond its original use in P18. In \se{results} we showed that stream deceleration 
can be described by a toy model where background mass is entrained in a shear layer, mixes with the original stream 
fluid and shares its initial momentum. As indicated by the total bulk kietic energy loss rates shown in \figs{deceleration_surface} 
and \figss{momentum_body}, the simplifying assumption that the velocity within the shear layer is constant and equal 
to the stream velocity seems consistent with simulation results. We thus evaluate the accretion rate of 
background fluid into the shear layer using \equ{m_acc} with $\dot{v}$ representing only KHI induced deceleration, 
$\dot{V}_{\rm KHI}$, without $a_{\rm g}$ from \equ{agrav}. Using $\dot{m}=V{\rm d}m/{\rm d}r$, and the boundary condition 
$m_{\rm entrained}(r=\Rv)=0$, we obtain
\be 
\label{eq:m_toy}
\frac{m_{\rm entrained}}{m_{\rm s}}=x^{-B/2}-1,
\ee
{\no}where $m_{\rm entrained}$ is the total background mass entrained in the shear layer and $m_{\rm s}$ is the 
initial stream mass. As in \equ{Vr_sol}, $x=r/\Rv$ and $B$ is given by \equ{Bdef}. This is shown in the centre panels 
of \fig{deceleration_toy}. In most cases the entrained mass at $0.1\Rv$ is a significant fraction of the 
stream mass, and for narrow streams, $\Rs\lsim 0.03\Rv$, the entrained mass can significantly outweigh the initial stream 
even for density contrasts of $\delta=100$. For $\delta=10$, the entrained mass equals the initial stream mass for stream 
radii $\Rs\sim (0.05-0.1)\Rv$ for $\alpha\sim (0.05-0.1)$. For $\delta=100$, this occurs at $\Rs\sim (0.02-0.03)\Rv$. 
Only for very dense and wide streams, which are unlikely based on \equ{Rs_stream}, does the entrained mass amount to 
only a small fraction, $\sim 10-20\%$, of the stream mass.

\smallskip
Combining \equs{Vr_sol} and \equm{m_toy}, we can estimate the kinetic energy of the inflowing material as 
$E_{\rm K}\sim 0.5(m_{\rm entrained}+m_{\rm s})V^2$, where as in \equs{m_acc} and \equm{EK} we have assumed 
that the entrained material and the stream material are fully mixed and moving at the same velocity, which 
seems consistent with simulations as shown in \se{results}. We wish to 
compare this to the total kinetic energy the system would have were there no deceleration. However, this is not 
simply $0.5m_{\rm s}V_{\rm ff}^2$, since by virtue of adding additional mass to the system and transporting 
it down the potential well, gravity is doing more work than it would have on the stream alone, and this must be 
accounted for when estimating the energy dissipation. In fact, since the entrained mass is such a large fraction 
of the stream mass and can even exceed it, the kinetic energy of the system can actually be larger than 
$0.5m_{\rm s}V_{\rm ff}^2$ even after energy losses. We will comment further on the physical implications of 
this below. The total kinetic energy of the system without any deceleration is given by 
\be 
\label{eq:E_no_dis}
\begin{array}{c}
E_{\rm K,\,no\: diss}(r)=\dfrac{m_{\rm s}V_{\rm ff}^2(r)}{2} \\
+ \displaystyle\int_{r}^{\Rv} \dfrac{{\rm d}m_{\rm entrained}}{{\rm d}\tilde{r}}[\Phi (r)-\Phi (\tilde{r})] ~{\rm d \tilde{r}},
\end{array}
\ee
{\no}where $\Phi (r)=-\Vv^2[{\rm ln}(r/\Rv)-1]$ is the potential of a singular isothermal sphere truncated at $\Rv$. The result is
\be 
\label{eq:E_no_dis2}
E_{\rm K,\,no\: diss}(x)= \frac{m_{\rm s}V_{\rm ff}^2(x)}{2}\left[1+\frac{4\left(x^{-B/2}-1\right)}{B}\right].
\ee

\smallskip
In the right-hand panels of \fig{deceleration_toy} we show the dissipation of bulk kinetic energy due to KHI, 
$1-E_{\rm K}/E_{\rm K,\,no\:diss}$. With the exception of the case $\delta=10,\Rs/\Rv=0.01$, which entrains many 
times its own mass, all cases very nearly have $E_{\rm K}/E_{\rm K,\,no\:diss}\sim V/V_{\rm ff}$, in agreement with 
\equ{EK}, showing the consistancy of our model. Ignoring the cases $\delta=10,\Rs/\Rv=0.01$ and $\delta=100,\Rs/\Rv=0.10$ 
which are unlikely based on \equ{Rs_stream}, we see that when the stream reaches $0.1\Rv$ between $10-40\%$ of the kinetic 
energy is lost for $\alpha=0.05$ and between $20-50\%$ for $\alpha=0.1$. As described in \se{surface}, the lost kinetic energy 
is split between turbulent energy, thermal energy of either the stream or the background, and sound waves propagating away 
from the stream, and as such represents an upper limit to the amount of energy which will eventually be dissipated into 
radiation.

\smallskip
It has been suggested that the observed Lyman-$\alpha$ blobs (LABs), with luminosities of 
$10^{42}-10^{44}~{\rm erg\,s^{-1}}$ around halos with masses of $\Mv\sim 10^{11}-10^{13}$ at $z=3$, can be powered by 
the dissipation of gravitational energy in cold streams feeding these galaxies, provided the fraction of energy radiated 
away exceeds $\sim 20\%$ \citep{Dijkstra09}. Similar conclusions were reached by \citet{Goerdt10} based on cosmological 
simulations (though see also \citealp{FG10} for a less optimistic view). Based on this, we conclude that \textit{KHI can 
contribute significantly to the observed Lyman-$\alpha$ luminosity of LABs.} We stress that this is contrary to the 
conclusions reached following the 2d analysis in P18, highlighting the importance of our current analysis. 
A potential caveat to this conclusion is that while the bulk of the stream may be self-shielded to the UVB 
\citep{Goerdt10,FG10}, this may not be true for the shear layer where the background fluid is entrained, which in turn 
may inhibit the production of Ly-$\alpha$ radiation. This will be studied in detail in an upcoming paper (Mandelker et al., 
in prep.), where we will include explicit radiative cooling in our calculations.

\smallskip
In our model, stream deceleration and energy dissipation are caused by background material, i.e. hot gas in the CGM, 
being entrained in and mixing with the stream, and eventually flowing toward the centre at the same velocity as the cold 
gas. Implicit in this is the assumption that the entrained gas is no longer held in hydrostatic equilibrium within the CGM. 
One likely expanation for this is the ram pressure of the cold gas. Cleary, in the unmixed stream core the ram pressure 
is significantly larger than the thermal pressure of the hot gas, since $\Mb\sim 1$ and $\delta>>1$, and even in the shear 
layer the ram pressure can still be larger than the thermal pressure. This will be studied in more detail in future work which 
explicitly accounts for the external halo potential. An additional intriguing possibility is the presence of thermal instabilities 
in the shear layer which may cause the hot gas to cool and condense onto the cold stream. This issue will be further studied 
in detail in and upcoming paper (Mandelker et al., in prep.), where we will examine the nonlinear evolution of KHI with radiative 
cooling. If this indeed occurs, the cooling radiation of the entrained gas as it mixes with the stream could be another powerful 
source of radiation not considered in our current work.

%%%%%%%%%%%%%%%%%%%%%%%%%%%%%%%%%%%%%%%%%%%%%%%%  
\section{Additional Physics and Future Work}
\label{sec:phys} 

\smallskip
While the analysis presented here has been thorough and exhaustive in terms of adiabatic hydrodynamics, 
it is worth noting potential caveats to our conclusions due to additional physics not included in our 
current analysis. The detailed effects of these processes on the stability of cold streams feeding massive 
galaxies at high redshift will be the subject of future work, where we will add them one by one to our 
current framework. 

\smallskip
We begin by considering radiative cooling, which is clearly very important for the evolution of cold streams. 
Previous studies have found that radiative cooling can either enhance or suppress the growth rates of KHI in 
the linear regime, depending on the slope of the cooling function and on the ratio of the cooling time in each 
fluid to the sound crossing time \citep{Massaglia92,Bodo93,Vietri97,Hardee97,Xu00}. In our case, the cooling 
time in the hot halo is typically much longer than the stream sound crossing time, while the cooling time in 
the streams is typically much shorter. Under these conditions, the linear growth rates at wavelengths $\lambda\gsim \Rs$ 
are very similar to the adiabatic case considered here (Mandelker et al., in prep.). However, radiative cooling 
can substantially alter the nonlinear evolution of KHI in dense streams \citep{Vietri97,Stone97,Xu00,Micono00}, 
though the net effect again depends on the details of the cooling function and the stream parameters. Some authors 
find that radiative cooling leads to more violent disruption of the stream \citep{Stone97,Xu00}, while others 
find that cooling prevents stream disruption by limiting the penetration of the shear layer into the stream 
\citep{Vietri97,Micono00}, though the deceleration rates remain similar to the adiabatic case \citep{Micono00}. 
It is also found that KHI in a cooling medium leads to much larger density fluctuations, and to the formation of 
dense knots and filaments inside the stream. In an upcoming paper (Mandelker et al., in prep.) we will examine 
the nonlinear evolution of KHI with radiative cooling for parameters typical of cold streams. Preliminary results 
suggest that the stream remains more colimated and denser, and is less prone to dissintegarion, though the 
deceleration rates are slightly larger than in the adiabatic case.

\smallskip
The self-gravity of the gas streams may also affect their evolution. It has been suggested that for a wide range 
of typical parameters cold streams are gravitationally unstable, as their mass per unit length, $M_{\rm L}\simeq \pi\Rs^2\rhos$, 
is larger than the critical mass for filament stability \citep{M18}. This can lead to stream fragmentation in the 
halo irrespective of KHI, which would completely alter our analysis. Density fluctuations induced by KHI may also 
trigger local collapse before the entire stream has fragmented. On the other hand, if the streams are marginally 
gravitationally stable, the self-gravity can stabilize the KHI and prevent stream disruption. This has been demonstrated 
in the context of cold, dense, spherical clouds moving through a hot medium, meant to represent giant molecular clouds 
moving through the interstellar medium \citep{Murray93}. In an upcoming paper (Aung et al., in prep) we will examine 
the evolution KHI in cold streams with self-gravity. Preliminary results suggest that so long as the stream is 
globally gravitationally stable, self-gravity suppresses stream disruption though has only a small effect on stream 
deceleration. Furthermore, the gravitational potential of the host dark matter filament can affect the gravitational 
stability of the stream. It may stabilize the stream by making it more buoyant, or it may destabilize the stream by 
increasing the inward radial gravitational force, thus requiring non-thermal turbulent motions to support the stream 
against radial collapse. These issues will be explored in future work.

\smallskip
The central gravity of the dark matter halo into which the streams are flowing may also affect their 
evolution. In \se{app_decel} we attempted to estimate the net deceleration of streams falling down 
the potential well of a dark matter halo while undergoing KHI. While we showed that non-negligible 
net deceleration may still occur, this model was very simplistic and a more detailed study is warranted. 
Furthermore, as the stream is focused towards the centre of the potential it assumes a roughly conical 
shape, with its radius decreasing towards the halo centre \citep{Dekel09,vdv12b}. As narrower streams 
disrupt faster (\fig{stream_disruption}), this can greatly enhance the potential for stream disruption 
in the CGM. These issues will be the focus of future work.

\smallskip
As noted in \se{app_diss}, thermal conduction will set the scale of the transition region between the 
stream and the background, which can greatly influence the stability of short wavelength perturbations 
and high-$m$ surface modes. While this has been qualitatively accounted for in our current analysis, it 
is unclear whether thermal conduction may further alter the evolution of the instability, especially when 
cooling is included. This is further complicated by the fact that the conduction length in the hot halo is 
likely orders of magnitude larger than in the cold stream (M16). 
Finally, magnetic fields parallel to the flow have been found to stabilize high-$m$ modes due to an effective 
surface-tension which suppresses the presence of small scale structure on the stream boundary \citep{Ferrari81,
Birkinshaw90}. However, in the high redshift streams we are considering, the magnetic field is expected to 
be very weak, with the plasma beta parameter\footnote{the ratio of thermal to magnetic pressure} possibly as 
large $\beta\gsim 1000$ (M16). We therefore do not expect magnetic fields to fundamentally alter our conclusions. 
However, even a dynamically unimportant magnetic field can significantly weaken thermal conductivity and viscosity, 
so it will be important to account for these effects simultaneously in future work.

\smallskip
Finally, we speculate that the spectrum and amplitude of initial perturbations in the stream will be determined 
primarily by the cosmological context, such as a non-trivial accretion geometry and non-zero impact parameter 
with respect to the central galaxy, deviations from pressure equilibrium as the stream enters the virial shock, 
galactic or AGN feedback, turbulence in the halo, etc. To study this, efforts are underway to perform fully cosmological 
simulations with tailored refinement along streams in order to properly resolve their structure and stability (Cornuault 
et al., in prep). In this context, cosmological simulations without star-formation and feedback may also be employed 
to study the effects of accretion geometry and entry into the shock region.

%%%%%%%%%%%%%%%%%%%%%%%%%%%%%%%%%%%%%%%%%%%%%%%%  
\section{Summary and Conclusions}
\label{sec:conc} 

\smallskip
Massive star-forming galaxies (SFGs) at redshifts $z\gsim 1$ are thought to be fed by cold streams ($T\sim 10^4\K$) 
of dense gas that flow along cosmic web filaments, through the hot circumgalactic medium (CGM) containing dilute 
gas at the virial temperature ($T\sim 10^6\K$). These streams have Mach numbers of $\Mb\sim 0.75-2.25$ with respect 
to the sound speed of the hot halo, density contrasts of $\delta \sim 10-100$ compared to the hot halo, and radii of 
$\Rs\sim 0.01-0.10$ times the halo virial radius. While this is yet to be confirmed observationally, it is a robust 
theoretical concept, and is seen in numerous different cosmological simulations \citep{Keres05,db06,Ocvirk08,Dekel09,
CDB,FG11,vdv11}. In order to study the evolution of these streams as they travel through the CGM towards the central 
galaxy, we presented here a detailed study of the nonlinear stage of purely-hydrodynamic, adiabatic Kelvin-Helmholtz 
Instability (KHI) in 3d cylinders, using both analytic models and RAMSES numerical simulations. This extended our 
previous work on the linear phases of the instability \citep{M16}, and the nonlinear instability in 2d planar slabs 
\citep{P18}. Our main results from this paper can be summarized as follows.

\begin{enumerate}
\smallskip
\item 
For streams which are subsonic with respect to the sum of the two sound speeds, i.e. $M_{\rm tot}<1$ 
(\equnp{Mtot}), the instability is dominated by surface modes, and the nonlinear evolution of 3d cylinders 
is qualitatively similar to 2d slabs. It is driven by vortex mergers that lead to self-similar shear layer 
growth in a ring-like structure around the stream. The shear layer growth rate, the convection (drift) velocity 
of the largest eddies along the stream, and the ratio of shear layer thickness in the stream/background are all 
very similar in 2d slabs and 3d cylinders, and follow our analytic predictions. However, the timescale for the 
shear layer to encompases the entire stream width is roughly $\sim 0.5$ as long in 3d as it is in 2d. Furthermore, 
3d cylinders decelerate significantly faster than 2d slabs, by up to a factor of $\sim 10$ for density contrasts 
of order $100$. This is because the shear layer expanding into the background fluid sweeps up mass at a higher rate. 
Overall, as either the density contrast between the stream and background, $\delta$, or the Mach number of the stream 
with respect to the background sound speed, $\Mb$, are increased, the shear layer growth rates decreases, the ratio of 
shear layer thickness in the two fluids diverges from unity, and the deceleration rates decrease. 

\smallskip
\item 
For 3d cylindrical streams with $M_{\rm tot}<1$, once the shear layer has expanded to roughly $2\Rs$ the largest eddies 
begin to break up and cascade towards smaller scales. This drives turbulence and enhances mixing of the stream and background 
fluids, while at the same time decreasing both the expansion rate of the shear layer and the deceleration rate of the stream.
This drives turbulence, enhances mixing of the stream and background fluids, and slows the growth of the shear layer. The 
turbulence in the shear layer reaches a maximal value of $\sigma/V\sim (0.2-0.3$) and then decays to an assymptotic value 
of $\sim(0.1-0.2)$, independent of $\Mb$ or $\delta$, where $V$ is the initial velocity of the stream. At late times, 
$\sim 10\%$ of the initial bulk kinetic energy is in turbulent motions of both stream and background fluid, and on scales 
$\lsim \Rs$ the turbulence is roughly isotropic. This is unlike the situation in 2d where the largest eddies do 
not break up, the shear layer continues to expand by mergers of larger and larger eddies, and the magnitude of turbulence is not 
universal. 

\smallskip
\item 
For streams which are supersonic with respect to the sum of the two sound speeds, i.e. $M_{\rm tot}>1$, 
there is a qualitative difference between 2d slabs and 3d cylinders. In 2d geometry, 
surface modes have completely stabilized, while 3d cylinders can still be unstable to azimuthal surface 
modes, characterized by a mode number, $m$, and associated azimuthal wavelength, $2\pi\Rs/m$. Such modes 
are unstable above a critical mode number, $m>m_{\rm crit}$ (\equnp{mcrit}), which depends on $\Mb$, $\delta$, 
and the longitudinal wavelength of the perturbation, with shorter wavelengths having larger $m_{\rm crit}$. 
However, the $m$-th mode is only unstable provided the initial transition region between the pure stream and 
background fluids is narrower than $2\pi\Rs/m$. These modes behave very similarly to the ``standard" surface 
modes described above, and their nonliner growth is dominated by self-similar shear layer growth. However, the 
largest eddies are smaller, $\propto m^{-1}$, rendering the turbulent cascade difficult to resolve. Low 
resolution simulations in this regime thus overestimate both shear layer growth and stream deceleration (appendix 
\se{convergence}).

\smallskip
\item 
If the width of the initial transition region between the stream and background is large enough, high-$m$ surface 
modes are stable. For streams with $M_{\rm tot}>1$, the instability will thus be dominated by body modes, as in the 
2d case. The nonlinear evolution of body modes is driven by long-wavelength sinusoidal perturbations. We find good 
agreement between the analytical predictions for the critical perturbation mode that is expected to break the stream, 
and the dominant mode observed at late times in numerical simulations. While these are similar for 2d slabs and 3d 
cylinders, the timescale for this perturbation to become nonlinear is $\sim 0.5$ as long in 3d as it is in 2d (\tab{body2}). 
Unlike for surface modes, the timescales for nonlinear evolution of body modes explicitly depend on the initial conditions 
through the initial amplitude of the long-wavelength critical perturbation. Once this perturbation becomes non-linear, 
a higly turbulent structure rapidly develops, similar to subsonic streams, encompasing the entire stream and mixing it 
with the background within roughly one sound crossing time. This is contrary to the evolution in 2d slabs, where the slab 
expands to $\sim 8-10$ times its initial width and breaks up into discrete blobs which remain unmixed for several sound 
crossing times. Similar to surface modes in subsonic streams, body modes in supersonic streams also lead to much faster 
deceleration in 3d compared to 2d, due to similar geometrical effects.

\smallskip
\item
We predict that for a significant region of the allowed parameter range, 
KHI can cause cold streams feeding massive galaxies at high redshfit to 
disintegrate prior to reaching the central galaxy. The upper limit for 
the stream radius that results in disintegration depends on the Mach number, 
the density contrast, the initial amplitude of the critical body-mode 
perturbation, and the initial width of the transition region between 
the stream and the background. For typical parameters, this critical 
radius ranges from $\sim (0.005-0.05)\Rv$, roughly $70\%$ larger than 
the critical radius deduced for 2d slabs.

\smallskip
\item 
Unlike 2d slabs, where KHI was found to have little to no impact on the 
stream velocities in the CGM, for 3d cylinders KHI can cause significant 
stream deceleration before they reach the central galaxy. The upper limit 
for the stream radius that results in a factor of $\ge 2$ decrease in the 
inflow velocity ranges from $\sim (0.004-0.03)\Rv$, a factor of $\sim 4$ 
larger than the critical radius for 2d slabs. Furthermore, using a simple 
toy model to estimate the balance between the gravitational acceleration 
towards the halo centre and the deceleration induced by KHI in the CGM, 
we estimate that the inflow velocities of streams can be significantly 
less than the free-fall velocity. Since the deceleration is overestimated in 
low resolution simulations where high-$m$ surface modes dominate, (appendix 
\se{convergence}), this may explain the roughly constant inflow velocities 
of streams seen in cosmological simulations.

\smallskip
\item 
Using a simple toy model to estimate the loss of bulk kinetic energy assocaited with 
laminar flow of both the stream and background fluids induced by KHI in the CGM, we estimate 
that typical streams can loose up to $\sim (10-50)\%$ of the gravitational energy gained between 
$\Rv$ and $0.1\Rv$. This suggests that KHI can be a viable power source for the observed emission 
of Lyman-$\alpha$ blobs. However, this is only an upper limit on the total amount of energy which 
will actually be radiated away, as some of it can go into sound waves which escape the system, or 
into thermal energy of the hot medium. This will be explored in detail in future work which will 
incorporate radiative cooling (Mandelker et al., in prep.).

\smallskip
\item 
Potential caveats to our results stem from the lack of important physical 
processes beyond adiabatic hydrodynamics, such as radiative cooling, the 
self-gravity of the stream, the external gravity of the halo, magnetic fields 
and thermal conduction. It is very likely that these will have a significant 
impact on the stream morphology and eventual disruption, though we speculate 
that our results regarding stream deceleration may not change as significantly. 
In ongoing and future work, we will explicitly account for these processes 
and more accurately address the evolution and fate of cold streams in massive, 
hot halos.

\end{enumerate}

%%%%%%%%%%%%%%%%%%%%%%%%%%%%%%%%% 
\section*{Acknowledgments} 
We sincerely thank the anonymous referee for his or her insightful 
comments which greatly improved the quality of this manuscript. 
We are grateful to Romain Teyssier, both for making \texttt{RAMSES} 
publicly available, and for his many helpful suggestions and advice 
when running the simulations. We thank Frank van den Bosch for his 
careful reading and helpful comments on an earlier draft of this manuscript. 
We thank Frederic Bournaud, Frank van den Bosch, Andi Burkert, Nicolas 
Cornuault, Mark Krumholz, Santi Roca-Fabrega, and Elad Steinberg for 
helpful discussions. The simulations were performed on the Omega cluster 
at Yale. NM acknowledges support from the Klauss Tschira Foundation through 
the HITS Yale Program in Astropysics (HYPA). This work is supported in part 
by NSF AST-1412768 and the facilities and staff of the Yale Center for Research 
Computing. This work was partly supported by ISF grant 1059/14, by BSF grant 
2014-273, by GIF grant I-1341-303.7/2016, by DIP grant STE1869/2-1 GE625/17-1, 
by PICS grant 3-12391 by the I-CORE Program of the PBC, and by NSF grant AST-1405962. 
YB and DP acknowledge support from ISF grant 1059/14.

%%%%%%%%%%%%%%%%%%%%%%%%%%%%%%%%%%%% 
\bibliographystyle{mn2e} 
\bibliography{biblio}

\begin{thebibliography}{76}
\expandafter\ifx\csname natexlab\endcsname\relax\def\natexlab#1{#1}\fi

\bibitem[{{Arrigoni Battaia} {et~al}\mbox{.}(2018){Arrigoni Battaia},
  {Prochaska}, {Hennawi}, {Obreja}, {Buck}, {Cantalupo}, {Dutton}, \&
  {Macci{\`o}}}]{Arrigoni18}
{Arrigoni Battaia} F., {Prochaska} J.~X., {Hennawi} J.~F., {Obreja} A., {Buck}
  T., {Cantalupo} S., {Dutton} A.~A., {Macci{\`o}} A.~V., 2018, \mnras, 473,
  3907

\bibitem[{{Bassett} \& {Woodward}(1995)}]{Bassett95}
{Bassett} G.~M., {Woodward} P.~R., 1995, \apj, 441, 582

\bibitem[{{Binney}, {Nipoti} \& {Fraternali}(2009){Binney}, {Nipoti}, \&
  {Fraternali}}]{Binney09}
{Binney} J., {Nipoti} C., {Fraternali} F., 2009, \mnras, 397, 1804

\bibitem[{{Birkinshaw}(1990)}]{Birkinshaw90}
{Birkinshaw} M., 1990, {The Stability of Jets}

\bibitem[{{Birnboim} \& {Dekel}(2003)}]{bd03}
{Birnboim} Y., {Dekel} A., 2003, \mnras, 345, 349

\bibitem[{{Bodo} {et~al}\mbox{.}(1993){Bodo}, {Massaglia}, {Rossi}, {Trussoni},
  \& {Ferrari}}]{Bodo93}
{Bodo} G., {Massaglia} S., {Rossi} P., {Trussoni} E., {Ferrari} A., 1993,
  Physics of Fluids, 5, 405

\bibitem[{{Bodo} {et~al}\mbox{.}(1998){Bodo}, {Rossi}, {Massaglia}, {Ferrari},
  {Malagoli}, \& {Rosner}}]{Bodo98}
{Bodo} G., {Rossi} P., {Massaglia} S., {Ferrari} A., {Malagoli} A., {Rosner}
  R., 1998, \aap, 333, 1117

\bibitem[{{Bogey}, {Marsden} \& {Bailly}(2011){Bogey}, {Marsden}, \&
  {Bailly}}]{Bogey11}
{Bogey} C., {Marsden} O., {Bailly} C., 2011, Physics of Fluids, 23, 035104

\bibitem[{{Bond}, {Kofman} \& {Pogosyan}(1996){Bond}, {Kofman}, \&
  {Pogosyan}}]{Bond96}
{Bond} J.~R., {Kofman} L., {Pogosyan} D., 1996, \nat, 380, 603

\bibitem[{{Borisova} {et~al}\mbox{.}(2016){Borisova}, {Cantalupo}, {Lilly},
  {Marino}, {Gallego}, {Bacon}, {Blaizot}, {Bouch{\'e}}, {Brinchmann},
  {Carollo}, {Caruana}, {Finley}, {Herenz}, {Richard}, {Schaye}, {Straka},
  {Turner}, {Urrutia}, {Verhamme}, \& {Wisotzki}}]{Borisova16}
{Borisova} E. {et~al.}, 2016, \apj, 831, 39

\bibitem[{{Bouch{\'e}} {et~al}\mbox{.}(2016){Bouch{\'e}}, {Finley},
  {Schroetter}, {Murphy}, {Richter}, {Bacon}, {Contini}, {Richard}, {Wendt},
  {Kamann}, {Epinat}, {Cantalupo}, {Straka}, {Schaye}, {Martin}, {P{\'e}roux},
  {Wisotzki}, {Soto}, {Lilly}, {Carollo}, {Brinchmann}, \&
  {Kollatschny}}]{Bouche16}
{Bouch{\'e}} N. {et~al.}, 2016, \apj, 820, 121

\bibitem[{{Bouch{\'e}} {et~al}\mbox{.}(2013){Bouch{\'e}}, {Murphy}, {Kacprzak},
  {P{\'e}roux}, {Contini}, {Martin}, \& {Dessauges-Zavadsky}}]{Bouche13}
{Bouch{\'e}} N., {Murphy} M.~T., {Kacprzak} G.~G., {P{\'e}roux} C., {Contini}
  T., {Martin} C.~L., {Dessauges-Zavadsky} M., 2013, Science, 341, 50

\bibitem[{{Cantalupo} {et~al}\mbox{.}(2014){Cantalupo}, {Arrigoni-Battaia},
  {Prochaska}, {Hennawi}, \& {Madau}}]{Cantalupo14}
{Cantalupo} S., {Arrigoni-Battaia} F., {Prochaska} J.~X., {Hennawi} J.~F.,
  {Madau} P., 2014, \nat, 506, 63

\bibitem[{{Ceverino}, {Dekel} \& {Bournaud}(2010){Ceverino}, {Dekel}, \&
  {Bournaud}}]{CDB}
{Ceverino} D., {Dekel} A., {Bournaud} F., 2010, \mnras, 404, 2151

\bibitem[{{Coles}(1985)}]{Coles85}
{Coles} D., ed., 1985, {The uses of coherent structure (Dryden Lecture)}

\bibitem[{{Danovich} {et~al}\mbox{.}(2015){Danovich}, {Dekel}, {Hahn},
  {Ceverino}, \& {Primack}}]{Danovich15}
{Danovich} M., {Dekel} A., {Hahn} O., {Ceverino} D., {Primack} J., 2015,
  \mnras, 449, 2087

\bibitem[{{Danovich} {et~al}\mbox{.}(2012){Danovich}, {Dekel}, {Hahn}, \&
  {Teyssier}}]{Danovich12}
{Danovich} M., {Dekel} A., {Hahn} O., {Teyssier} R., 2012, \mnras, 422, 1732

\bibitem[{{Dekel} \& {Birnboim}(2006)}]{db06}
{Dekel} A., {Birnboim} Y., 2006, \mnras, 368, 2

\bibitem[{{Dekel} {et~al}\mbox{.}(2009){Dekel}, {Birnboim}, {Engel},
  {Freundlich}, {Goerdt}, {Mumcuoglu}, {Neistein}, {Pichon}, {Teyssier}, \&
  {Zinger}}]{Dekel09}
{Dekel} A. {et~al.}, 2009, \nat, 457, 451

\bibitem[{{Dekel} \& {Mandelker}(2014)}]{DM14}
{Dekel} A., {Mandelker} N., 2014, \mnras, 444, 2071

\bibitem[{{Dekel} {et~al}\mbox{.}(2013){Dekel}, {Zolotov}, {Tweed}, {Cacciato},
  {Ceverino}, \& {Primack}}]{Dekel13}
{Dekel} A., {Zolotov} A., {Tweed} D., {Cacciato} M., {Ceverino} D., {Primack}
  J.~R., 2013, \mnras, 435, 999

\bibitem[{{Dijkstra} \& {Loeb}(2009)}]{Dijkstra09}
{Dijkstra} M., {Loeb} A., 2009, \mnras, 400, 1109

\bibitem[{{Dimotakis}(1986)}]{Dimotakis86}
{Dimotakis} P.~E., 1986, AIAA Journal, 24, 1791

\bibitem[{{Dimotakis}(1991)}]{Dimotakis91}
{Dimotakis} P.~E., 1991, {Turbulent free shear layer mixing and combustion}.
  Tech. rep.

\bibitem[{{Efstathiou}(1992)}]{Efstathiou92}
{Efstathiou} G., 1992, \mnras, 256, 43P

\bibitem[{{Elmegreen} {et~al}\mbox{.}(2007){Elmegreen}, {Elmegreen},
  {Ravindranath}, \& {Coe}}]{Elmegreen07}
{Elmegreen} D.~M., {Elmegreen} B.~G., {Ravindranath} S., {Coe} D.~A., 2007,
  \apj, 658, 763

\bibitem[{{Faucher-Gigu{\`e}re} {et~al}\mbox{.}(2010){Faucher-Gigu{\`e}re},
  {Kere{\v s}}, {Dijkstra}, {Hernquist}, \& {Zaldarriaga}}]{FG10}
{Faucher-Gigu{\`e}re} C.-A., {Kere{\v s}} D., {Dijkstra} M., {Hernquist} L.,
  {Zaldarriaga} M., 2010, \apj, 725, 633

\bibitem[{{Faucher-Gigu{\`e}re}, {Kere{\v s}} \&
  {Ma}(2011){Faucher-Gigu{\`e}re}, {Kere{\v s}}, \& {Ma}}]{FG11}
{Faucher-Gigu{\`e}re} C.-A., {Kere{\v s}} D., {Ma} C.-P., 2011, \mnras, 417,
  2982

\bibitem[{{Ferrari}, {Trussoni} \& {Zaninetti}(1981){Ferrari}, {Trussoni}, \&
  {Zaninetti}}]{Ferrari81}
{Ferrari} A., {Trussoni} E., {Zaninetti} L., 1981, \mnras, 196, 1051

\bibitem[{{Fielding} {et~al}\mbox{.}(2017){Fielding}, {Quataert}, {McCourt}, \&
  {Thompson}}]{Fielding17}
{Fielding} D., {Quataert} E., {McCourt} M., {Thompson} T.~A., 2017, \mnras,
  466, 3810

\bibitem[{{F{\"o}rster Schreiber} {et~al}\mbox{.}(2006){F{\"o}rster Schreiber},
  {Genzel}, {Lehnert}, {Bouch{\'e}}, {Verma}, {Erb}, {Shapley}, \& {et
  al.,}}]{Forster06}
{F{\"o}rster Schreiber} N.~M., {Genzel} R., {Lehnert} M.~D., {Bouch{\'e}} N.,
  {Verma} A., {Erb} D.~K., {Shapley} A.~E., {et al.,}, 2006, \apj, 645, 1062

\bibitem[{{Freund}, {Lele} \& {Moin}(2000){Freund}, {Lele}, \&
  {Moin}}]{Freund00}
{Freund} J.~B., {Lele} S.~K., {Moin} P., 2000, Journal of Fluid Mechanics, 421,
  229

\bibitem[{{Fumagalli} {et~al}\mbox{.}(2017){Fumagalli}, {Mackenzie},
  {Trayford}, {Theuns}, {Cantalupo}, {Christensen}, {Fynbo}, {M{\o}ller},
  {O'Meara}, {Prochaska}, {Rafelski}, \& {Shanks}}]{Fumagalli17}
{Fumagalli} M. {et~al.}, 2017, \mnras, 471, 3686

\bibitem[{{Fumagalli} {et~al}\mbox{.}(2011){Fumagalli}, {Prochaska}, {Kasen},
  {Dekel}, {Ceverino}, \& {Primack}}]{Fumagalli11}
{Fumagalli} M., {Prochaska} J.~X., {Kasen} D., {Dekel} A., {Ceverino} D.,
  {Primack} J.~R., 2011, \mnras, 418, 1796

\bibitem[{{Genzel} {et~al}\mbox{.}(2008){Genzel}, {Burkert}, {Bouch{\'e}},
  {Cresci}, {F{\"o}rster Schreiber}, {Shapley}, {Shapiro}, \& {et
  al.,}}]{Genzel08}
{Genzel} R., {Burkert} A., {Bouch{\'e}} N., {Cresci} G., {F{\"o}rster
  Schreiber} N.~M., {Shapley} A., {Shapiro} K., {et al.,}, 2008, \apj, 687, 59

\bibitem[{{Genzel} {et~al}\mbox{.}(2006){Genzel}, {Tacconi}, {Eisenhauer},
  {F{\"o}rster Schreiber}, {Cimatti}, {Daddi}, {Bouch{\'e}}, \& {et
  al.,}}]{Genzel06}
{Genzel} R., {Tacconi} L.~J., {Eisenhauer} F., {F{\"o}rster Schreiber} N.~M.,
  {Cimatti} A., {Daddi} E., {Bouch{\'e}} N., {et al.,}, 2006, \nat, 442, 786

\bibitem[{{Goerdt} \& {Ceverino}(2015)}]{Goerdt15a}
{Goerdt} T., {Ceverino} D., 2015, \mnras, 450, 3359

\bibitem[{{Goerdt} {et~al}\mbox{.}(2010){Goerdt}, {Dekel}, {Sternberg},
  {Ceverino}, {Teyssier}, \& {Primack}}]{Goerdt10}
{Goerdt} T., {Dekel} A., {Sternberg} A., {Ceverino} D., {Teyssier} R.,
  {Primack} J.~R., 2010, \mnras, 407, 613

\bibitem[{{Goerdt} {et~al}\mbox{.}(2012){Goerdt}, {Dekel}, {Sternberg}, {Gnat},
  \& {Ceverino}}]{Goerdt12}
{Goerdt} T., {Dekel} A., {Sternberg} A., {Gnat} O., {Ceverino} D., 2012,
  \mnras, 424, 2292

\bibitem[{{Hardee}, {Clarke} \& {Howell}(1995){Hardee}, {Clarke}, \&
  {Howell}}]{Hardee95}
{Hardee} P.~E., {Clarke} D.~A., {Howell} D.~A., 1995, \apj, 441, 644

\bibitem[{{Hardee} \& {Stone}(1997)}]{Hardee97}
{Hardee} P.~E., {Stone} J.~M., 1997, \apj, 483, 121

\bibitem[{{Kere{\v s}} {et~al}\mbox{.}(2005){Kere{\v s}}, {Katz}, {Weinberg},
  \& {Dav{\'e}}}]{Keres05}
{Kere{\v s}} D., {Katz} N., {Weinberg} D.~H., {Dav{\'e}} R., 2005, \mnras, 363,
  2

\bibitem[{{Leclercq} {et~al}\mbox{.}(2017){Leclercq}, {Bacon}, {Wisotzki},
  {Mitchell}, {Garel}, {Verhamme}, {Blaizot}, {Hashimoto}, {Herenz}, {Conseil},
  {Cantalupo}, {Inami}, {Contini}, {Richard}, {Maseda}, {Schaye}, {Marino},
  {Akhlaghi}, {Brinchmann}, \& {Carollo}}]{Leclercq17}
{Leclercq} F. {et~al.}, 2017, \aap, 608, A8

\bibitem[{{Mandelker} {et~al}\mbox{.}(2016){Mandelker}, {Padnos}, {Dekel},
  {Birnboim}, {Burkert}, {Krumholz}, \& {Steinberg}}]{M16}
{Mandelker} N., {Padnos} D., {Dekel} A., {Birnboim} Y., {Burkert} A.,
  {Krumholz} M.~R., {Steinberg} E., 2016, \mnras, 463, 3921

\bibitem[{{Mandelker} {et~al}\mbox{.}(2018){Mandelker}, {van Dokkum}, {Brodie},
  {van den Bosch}, \& {Ceverino}}]{M18}
{Mandelker} N., {van Dokkum} P.~G., {Brodie} J.~P., {van den Bosch} F.~C.,
  {Ceverino} D., 2018, \apj, 861, 148

\bibitem[{{Martin} {et~al}\mbox{.}(2014{\natexlab{a}}){Martin}, {Chang},
  {Matuszewski}, {Morrissey}, {Rahman}, {Moore}, \& {Steidel}}]{Martin14a}
{Martin} D.~C., {Chang} D., {Matuszewski} M., {Morrissey} P., {Rahman} S.,
  {Moore} A., {Steidel} C.~C., 2014{\natexlab{a}}, \apj, 786, 106

\bibitem[{{Martin} {et~al}\mbox{.}(2014{\natexlab{b}}){Martin}, {Chang},
  {Matuszewski}, {Morrissey}, {Rahman}, {Moore}, {Steidel}, \&
  {Matsuda}}]{Martin14b}
{Martin} D.~C., {Chang} D., {Matuszewski} M., {Morrissey} P., {Rahman} S.,
  {Moore} A., {Steidel} C.~C., {Matsuda} Y., 2014{\natexlab{b}}, \apj, 786, 107

\bibitem[{{Massaglia} {et~al}\mbox{.}(1992){Massaglia}, {Trussoni}, {Bodo},
  {Rossi}, \& {Ferrari}}]{Massaglia92}
{Massaglia} S., {Trussoni} E., {Bodo} G., {Rossi} P., {Ferrari} A., 1992, \aap,
  260, 243

\bibitem[{{Matsuda} {et~al}\mbox{.}(2006){Matsuda}, {Yamada}, {Hayashino},
  {Yamauchi}, \& {Nakamura}}]{Matsuda06}
{Matsuda} Y., {Yamada} T., {Hayashino} T., {Yamauchi} R., {Nakamura} Y., 2006,
  \apjl, 640, L123

\bibitem[{{Matsuda} {et~al}\mbox{.}(2011){Matsuda}, {Yamada}, {Hayashino},
  {Yamauchi}, {Nakamura}, {Morimoto}, {Ouchi}, {Ono}, {Kousai}, {Nakamura},
  {Horie}, {Fujii}, {Umemura}, \& {Mori}}]{Matsuda11}
{Matsuda} Y. {et~al.}, 2011, \mnras, 410, L13

\bibitem[{{Micono} {et~al}\mbox{.}(2000){Micono}, {Bodo}, {Massaglia}, {Rossi},
  {Ferrari}, \& {Rosner}}]{Micono00}
{Micono} M., {Bodo} G., {Massaglia} S., {Rossi} P., {Ferrari} A., {Rosner} R.,
  2000, \aap, 360, 795

\bibitem[{{Murray} {et~al}\mbox{.}(1993){Murray}, {White}, {Blondin}, \&
  {Lin}}]{Murray93}
{Murray} S.~D., {White} S.~D.~M., {Blondin} J.~M., {Lin} D.~N.~C., 1993, \apj,
  407, 588

\bibitem[{{Nelson} {et~al}\mbox{.}(2016){Nelson}, {Genel}, {Pillepich},
  {Vogelsberger}, {Springel}, \& {Hernquist}}]{Nelson16}
{Nelson} D., {Genel} S., {Pillepich} A., {Vogelsberger} M., {Springel} V.,
  {Hernquist} L., 2016, \mnras, 460, 2881

\bibitem[{{Nelson} {et~al}\mbox{.}(2013){Nelson}, {Vogelsberger}, {Genel},
  {Sijacki}, {Kere{\v s}}, {Springel}, \& {Hernquist}}]{Nelson13}
{Nelson} D., {Vogelsberger} M., {Genel} S., {Sijacki} D., {Kere{\v s}} D.,
  {Springel} V., {Hernquist} L., 2013, \mnras, 429, 3353

\bibitem[{{Ocvirk}, {Pichon} \& {Teyssier}(2008){Ocvirk}, {Pichon}, \&
  {Teyssier}}]{Ocvirk08}
{Ocvirk} P., {Pichon} C., {Teyssier} R., 2008, \mnras, 390, 1326

\bibitem[{{Padnos} {et~al}\mbox{.}(2018){Padnos}, {Mandelker}, {Birnboim},
  {Dekel}, {Krumholz}, \& {Steinberg}}]{P18}
{Padnos} D., {Mandelker} N., {Birnboim} Y., {Dekel} A., {Krumholz} M.~R.,
  {Steinberg} E., 2018, ArXiv e-prints

\bibitem[{{Prochaska}, {Lau} \& {Hennawi}(2014){Prochaska}, {Lau}, \&
  {Hennawi}}]{Prochaska14}
{Prochaska} J.~X., {Lau} M.~W., {Hennawi} J.~F., 2014, \apj, 796, 140

\bibitem[{{Rees} \& {Ostriker}(1977)}]{Rees77}
{Rees} M.~J., {Ostriker} J.~P., 1977, \mnras, 179, 541

\bibitem[{{Robertson} {et~al}\mbox{.}(2010){Robertson}, {Kravtsov}, {Gnedin},
  {Abel}, \& {Rudd}}]{Robertson10}
{Robertson} B.~E., {Kravtsov} A.~V., {Gnedin} N.~Y., {Abel} T., {Rudd} D.~H.,
  2010, \mnras, 401, 2463

\bibitem[{{Spitzer}(1956)}]{Spitzer56}
{Spitzer} L., 1956, {Physics of Fully Ionized Gases}

\bibitem[{{Springel}(2010)}]{Springel10}
{Springel} V., 2010, \mnras, 401, 791

\bibitem[{{Springel} {et~al}\mbox{.}(2005){Springel}, {White}, {Jenkins},
  {Frenk}, {Yoshida}, {Gao}, {Navarro}, {Thacker}, {Croton}, {Helly},
  {Peacock}, {Cole}, {Thomas}, {Couchman}, {Evrard}, {Colberg}, \&
  {Pearce}}]{Springel05}
{Springel} V. {et~al.}, 2005, \nat, 435, 629

\bibitem[{{Stark} {et~al}\mbox{.}(2008){Stark}, {Swinbank}, {Ellis}, {Dye},
  {Smail}, \& {Richard}}]{Stark08}
{Stark} D.~P., {Swinbank} A.~M., {Ellis} R.~S., {Dye} S., {Smail} I.~R.,
  {Richard} J., 2008, \nat, 455, 775

\bibitem[{{Steidel} {et~al}\mbox{.}(2000){Steidel}, {Adelberger}, {Shapley},
  {Pettini}, {Dickinson}, \& {Giavalisco}}]{Steidel00}
{Steidel} C.~C., {Adelberger} K.~L., {Shapley} A.~E., {Pettini} M., {Dickinson}
  M., {Giavalisco} M., 2000, \apj, 532, 170

\bibitem[{{Stone}, {Xu} \& {Hardee}(1997){Stone}, {Xu}, \& {Hardee}}]{Stone97}
{Stone} J.~M., {Xu} J., {Hardee} P., 1997, \apj, 483, 136

\bibitem[{{Sutherland} \& {Dopita}(1993)}]{SD93}
{Sutherland} R.~S., {Dopita} M.~A., 1993, \apjs, 88, 253

\bibitem[{{Teyssier}(2002)}]{Teyssier02}
{Teyssier} R., 2002, \aap, 385, 337

\bibitem[{{Toro}, {Spruce} \& {Speares}(1994){Toro}, {Spruce}, \&
  {Speares}}]{Toro94}
{Toro} E.~F., {Spruce} M., {Speares} W., 1994, Shock Waves, 4, 25

\bibitem[{{van de Voort} \& {Schaye}(2012)}]{vdv12b}
{van de Voort} F., {Schaye} J., 2012, \mnras, 423, 2991

\bibitem[{{van de Voort} {et~al}\mbox{.}(2012){van de Voort}, {Schaye},
  {Altay}, \& {Theuns}}]{vdv12}
{van de Voort} F., {Schaye} J., {Altay} G., {Theuns} T., 2012, \mnras, 421,
  2809

\bibitem[{{van de Voort} {et~al}\mbox{.}(2011){van de Voort}, {Schaye},
  {Booth}, {Haas}, \& {Dalla Vecchia}}]{vdv11}
{van de Voort} F., {Schaye} J., {Booth} C.~M., {Haas} M.~R., {Dalla Vecchia}
  C., 2011, \mnras, 414, 2458

\bibitem[{{van Leer}(1977)}]{vanLeer77}
{van Leer} B., 1977, Journal of Computational Physics, 23, 263

\bibitem[{{Vietri}, {Ferrara} \& {Miniati}(1997){Vietri}, {Ferrara}, \&
  {Miniati}}]{Vietri97}
{Vietri} M., {Ferrara} A., {Miniati} F., 1997, \apj, 483, 262

\bibitem[{{Vogelsberger} {et~al}\mbox{.}(2012){Vogelsberger}, {Sijacki},
  {Kere{\v s}}, {Springel}, \& {Hernquist}}]{Vogelsberger12}
{Vogelsberger} M., {Sijacki} D., {Kere{\v s}} D., {Springel} V., {Hernquist}
  L., 2012, \mnras, 425, 3024

\bibitem[{{White} \& {Rees}(1978)}]{White78}
{White} S.~D.~M., {Rees} M.~J., 1978, \mnras, 183, 341

\bibitem[{{Xu}, {Hardee} \& {Stone}(2000){Xu}, {Hardee}, \& {Stone}}]{Xu00}
{Xu} J., {Hardee} P.~E., {Stone} J.~M., 2000, \apj, 543, 161

\end{thebibliography}

\appendix 
%%%%%%%%%%%%%%%%%%%%%%%%%%%%%%%%%%%%%%%%%%%%%%%%  
\section{Simulating the linear regime}
\label{sec:linear_sims} 

\begin{figure}
\includegraphics[trim={0.4cm 0.0cm 1.3cm 0.0cm}, clip, width =0.47 \textwidth]{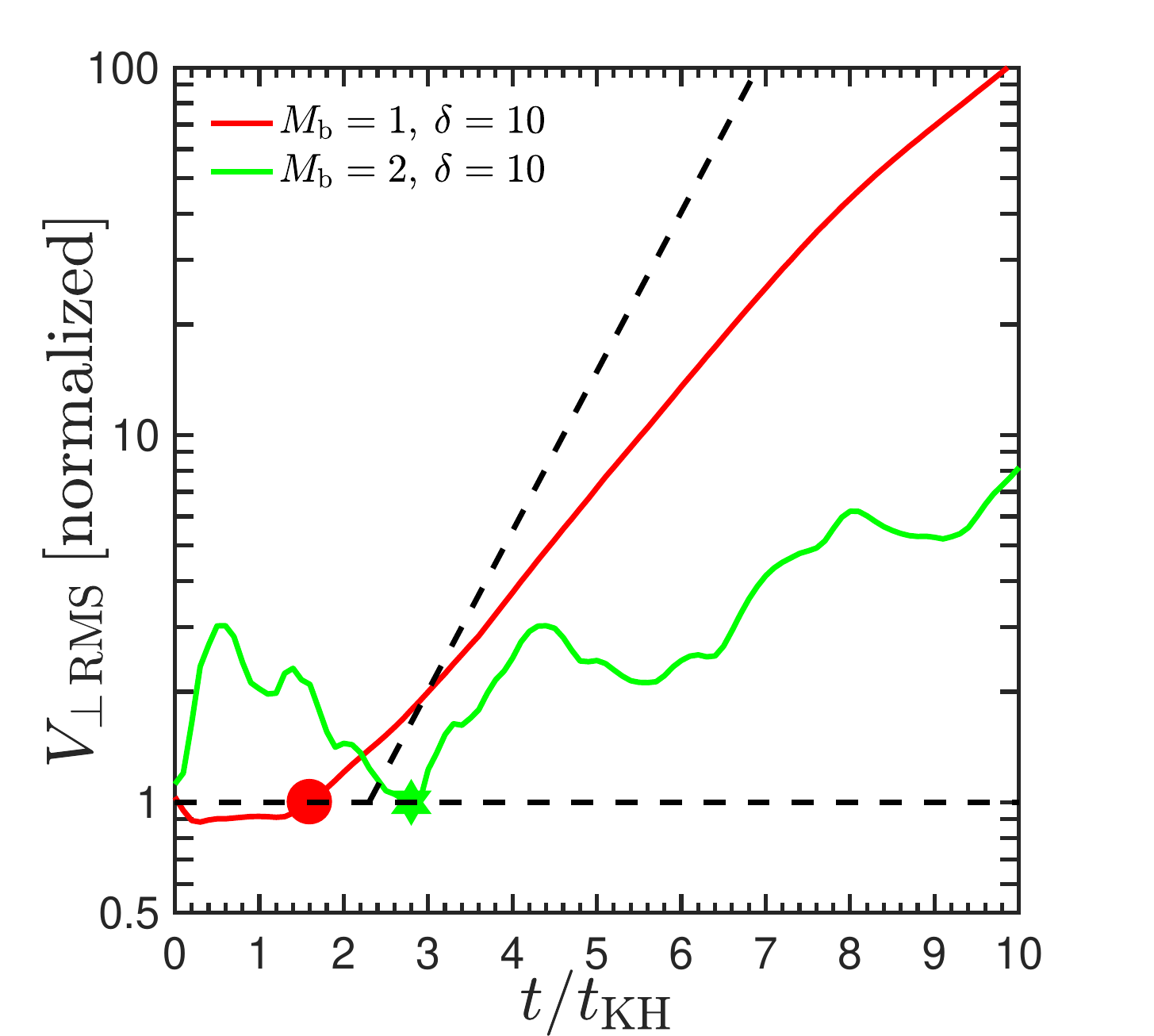}
\caption{Perturbaion amplitude as a function of time in the linear regime. Shown are results 
for two simulations, $(\Mb,\delta)=(1,10)$ in red and $(2,10)$ in green. Each simulation 
was initiated with a single wavelength ($\lambda=\Rs$), helical ($m=1$), perturbation in 
the radial velocity, as described in the text. The time in each panel has been normalized 
by the KH time of the fastest growing eigenmode, and the perturbation amplitude is estimated 
by the RMS of the transverse velocity. Perturbations must evolve into eigenmodes before they 
can grow. For surface modes, as in the $\Mb=1$ case, the timescale for this is the perturbation 
sound crossing time, $t_{\rm \lambda}=\lambda/\cb$, marked with a red circle. For body modes, as 
in the $\Mb=2$ case, it is the stream sound crossing time, $\tsc$, marked with a green star. The 
perturbation amplitude have been normalized to unity at these times. The slope of the dashed line 
marks the expected exponential growth at later times (the zero-point has been shifted for clarity). 
The $\Mb=1$ simulation shows a good match to the analytic prediction, underestimating the growth rate 
by less than $20\%$, due to the smoothing in the initial conditions. The growth in the $\Mb=2$ case 
is oscillatory, due to the large smoothing in the unperturbed solution and to mode interactions, 
though during periods of growth the slope is a good match to the analytic prediction.}
\label{fig:lin_growth} 
\end{figure}
%\end{figure*}

\begin{figure*}
\begin{center}
\includegraphics[trim={0.0cm 0.5cm 3.0cm 0.3cm}, clip, width =0.261 \textwidth]{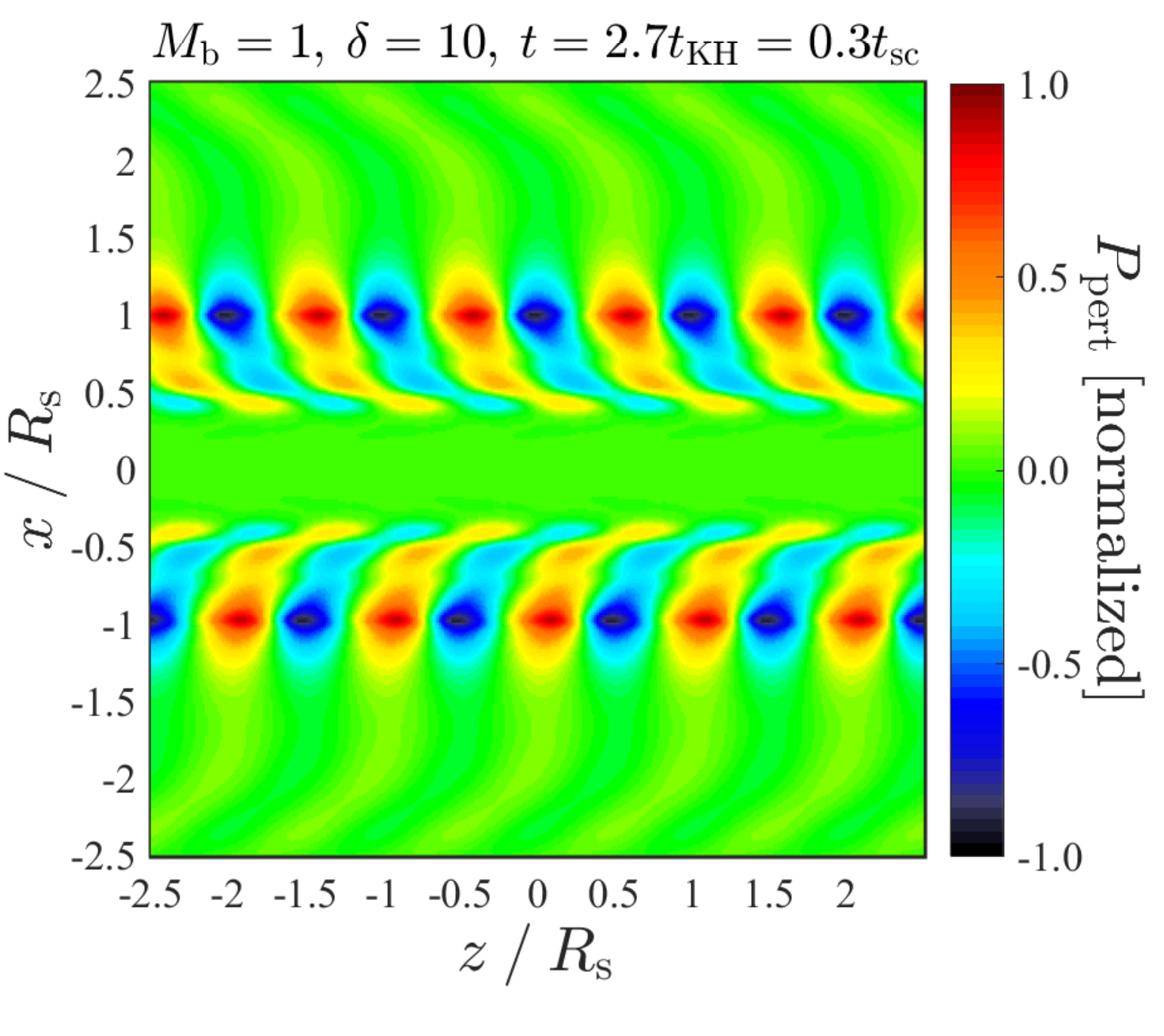}
\hspace{-0.25cm}
\includegraphics[trim={1.6cm 0.5cm 3.0cm 0.3cm}, clip, width =0.2275 \textwidth]{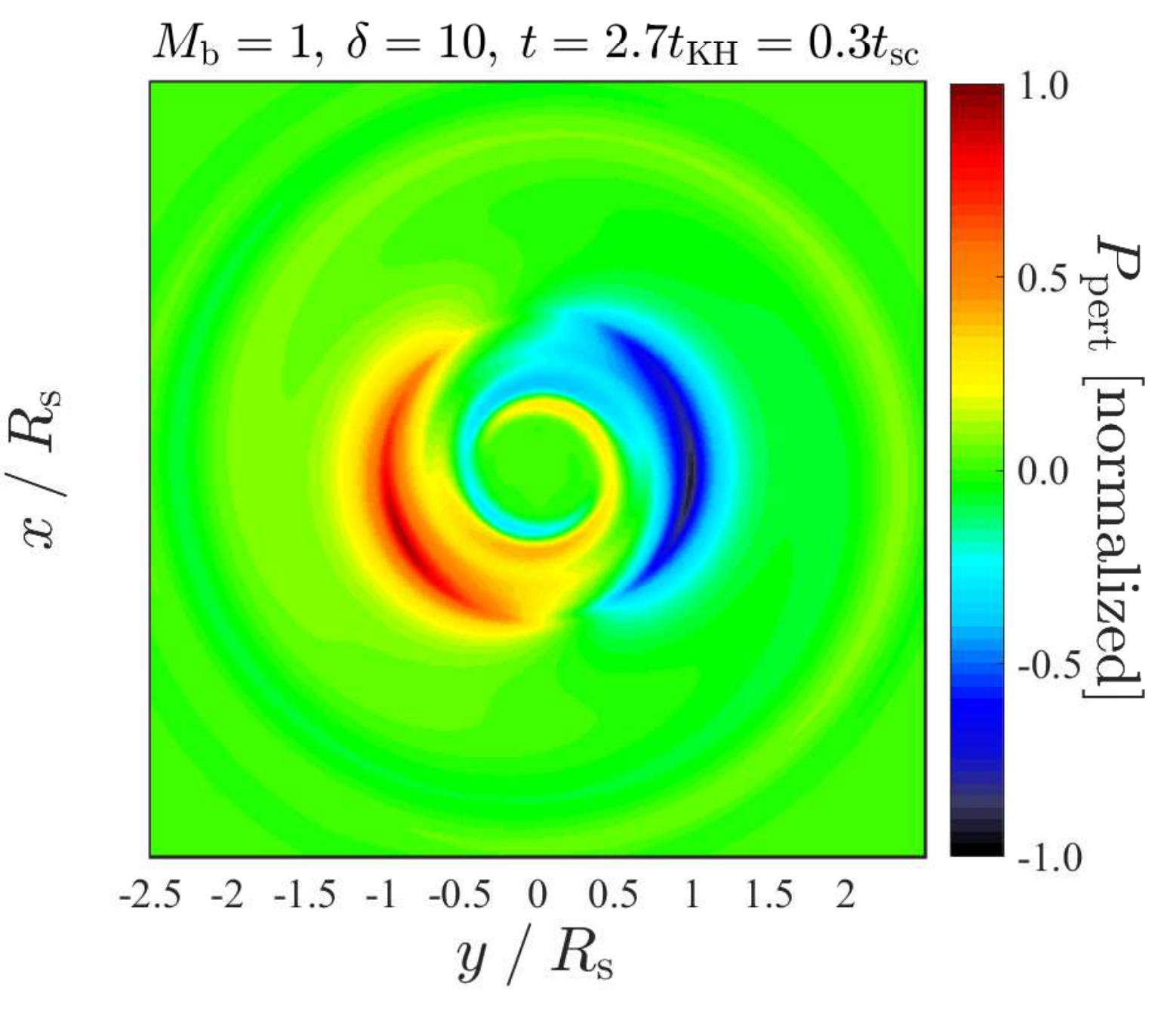}
\hspace{-0.15cm}
\includegraphics[trim={1.6cm 0.5cm 3.0cm 0.3cm}, clip, width =0.2275 \textwidth]{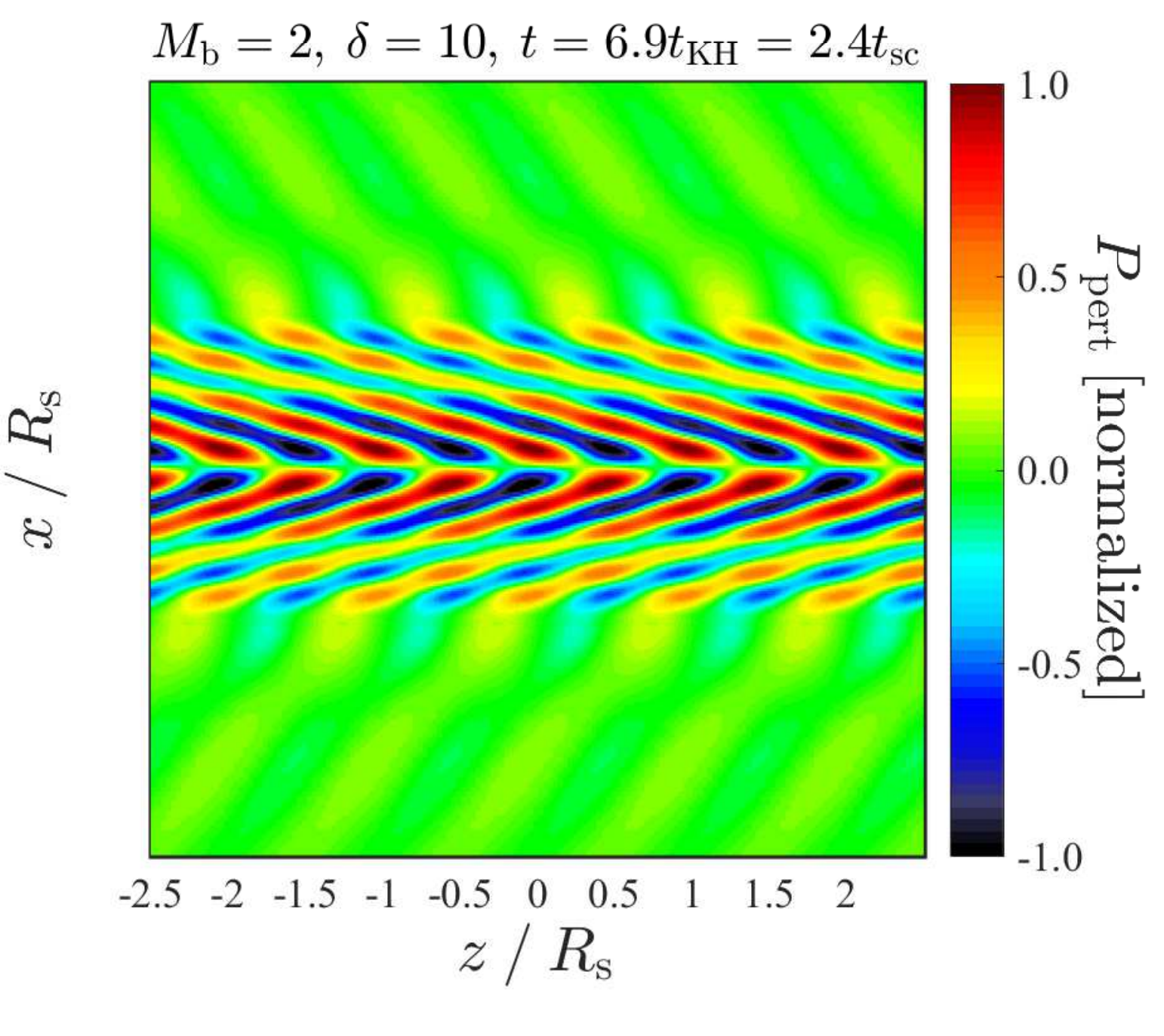}
\hspace{-0.25cm}
\includegraphics[trim={1.6cm 0.5cm 0.2cm 0.3cm}, clip, width =0.285 \textwidth]{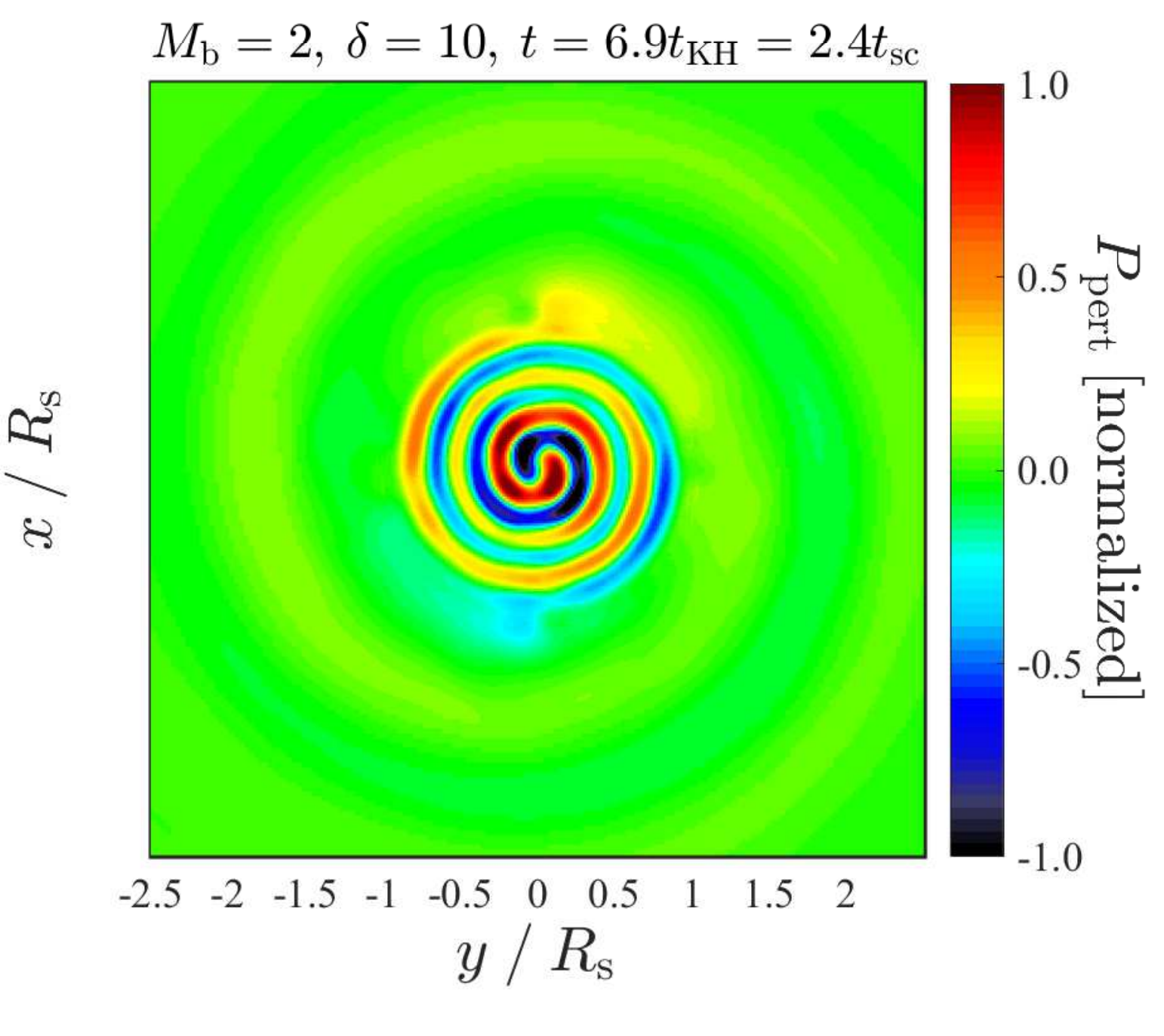}\\
\vspace{+0.1cm}
\includegraphics[trim={0.0cm 0.5cm 3.0cm 0.3cm}, clip, width =0.261 \textwidth]{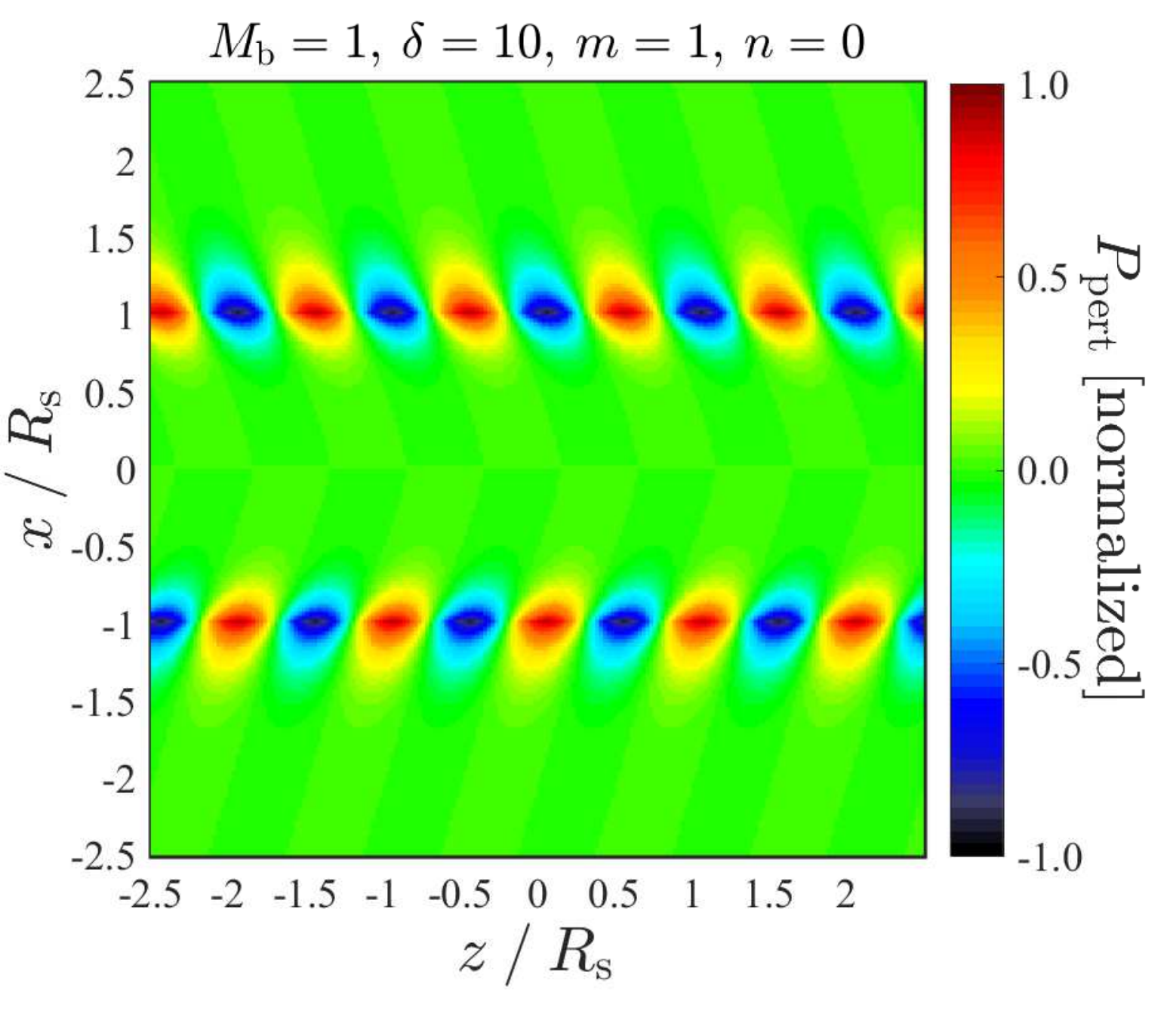}
\hspace{-0.25cm}
\includegraphics[trim={1.6cm 0.5cm 3.0cm 0.3cm}, clip, width =0.2275 \textwidth]{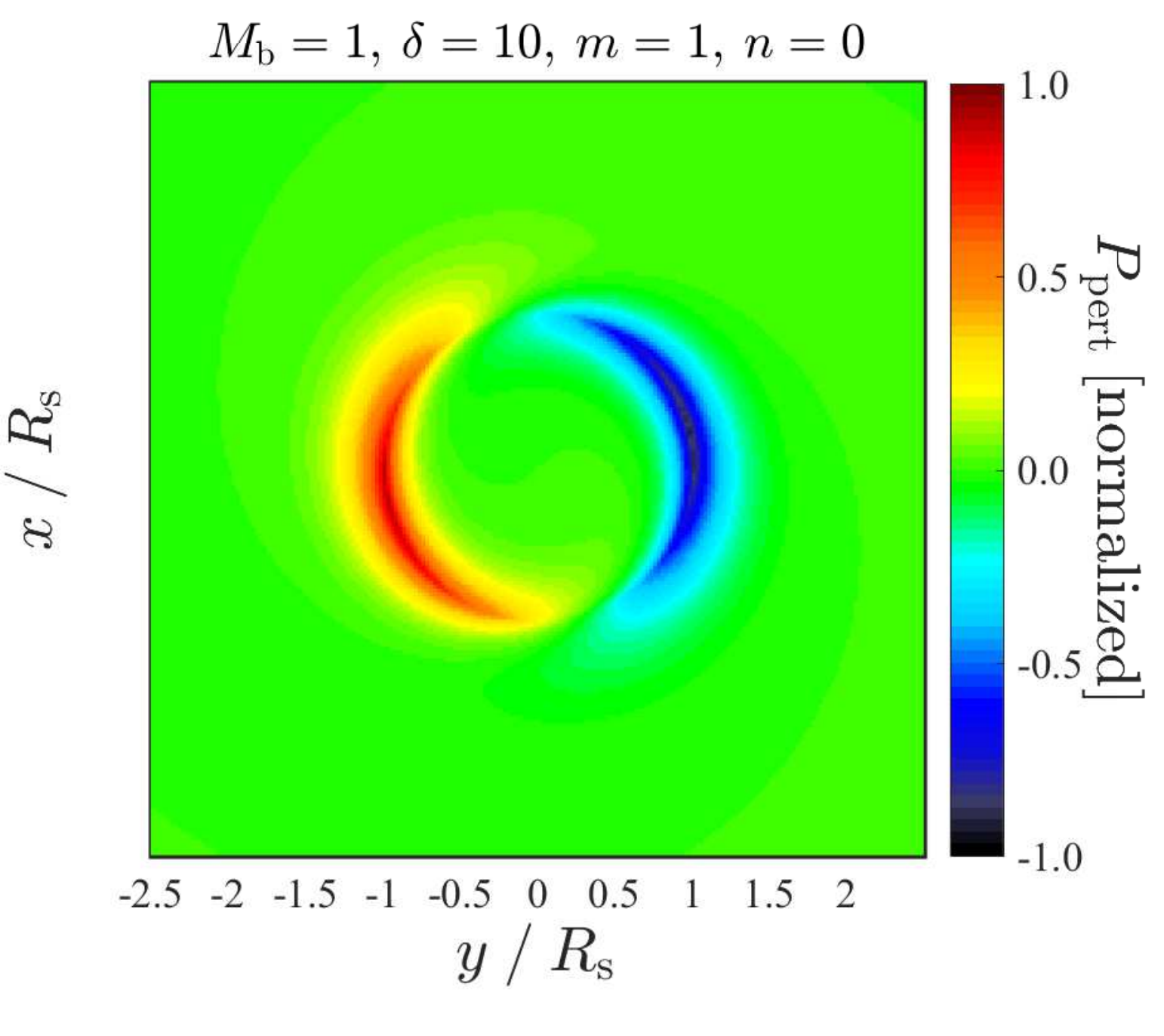}
\hspace{-0.15cm}
\includegraphics[trim={1.6cm 0.5cm 3.0cm 0.3cm}, clip, width =0.2275 \textwidth]{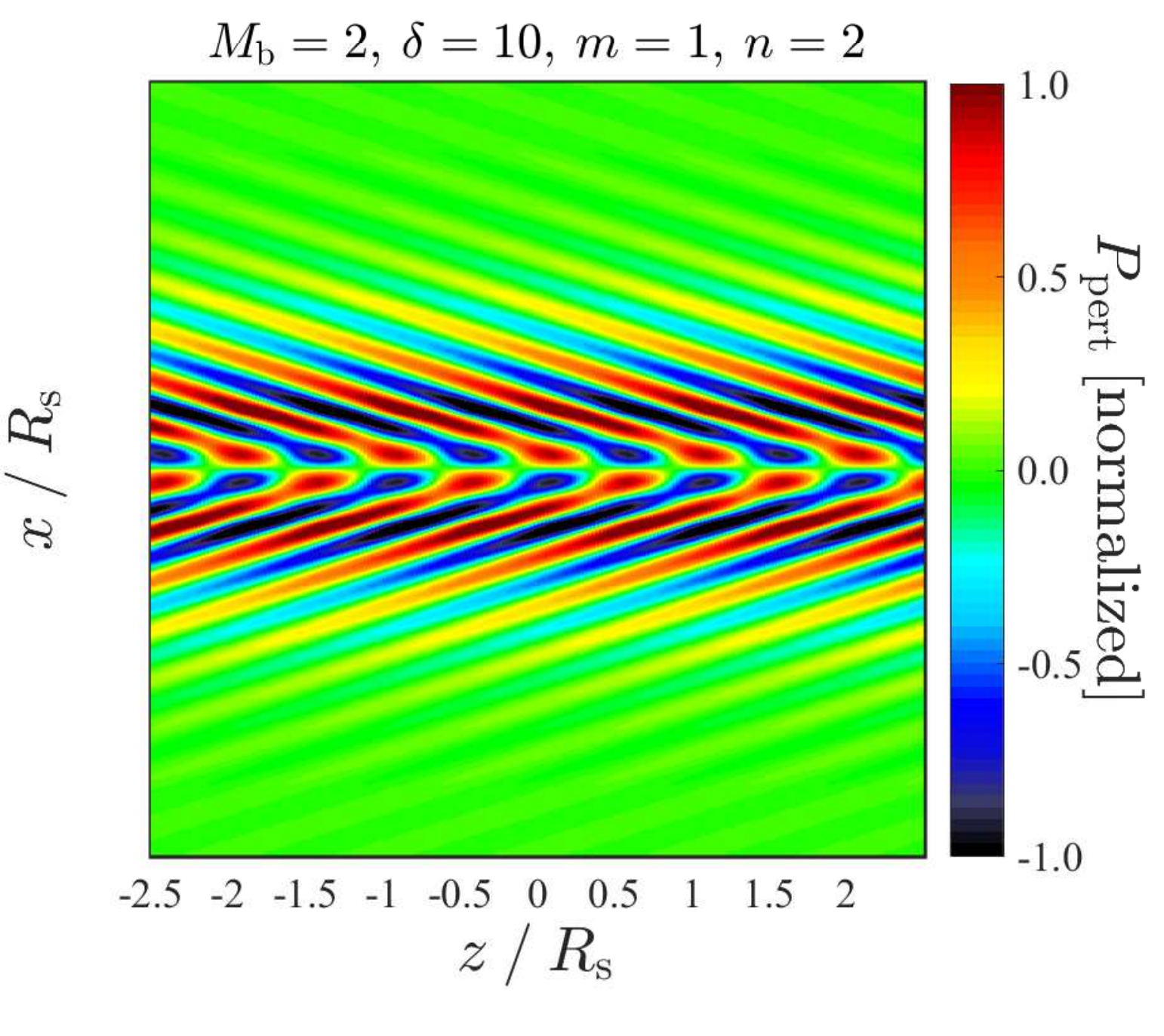}
\hspace{-0.25cm}
\includegraphics[trim={1.6cm 0.5cm 0.2cm 0.3cm}, clip, width =0.285 \textwidth]{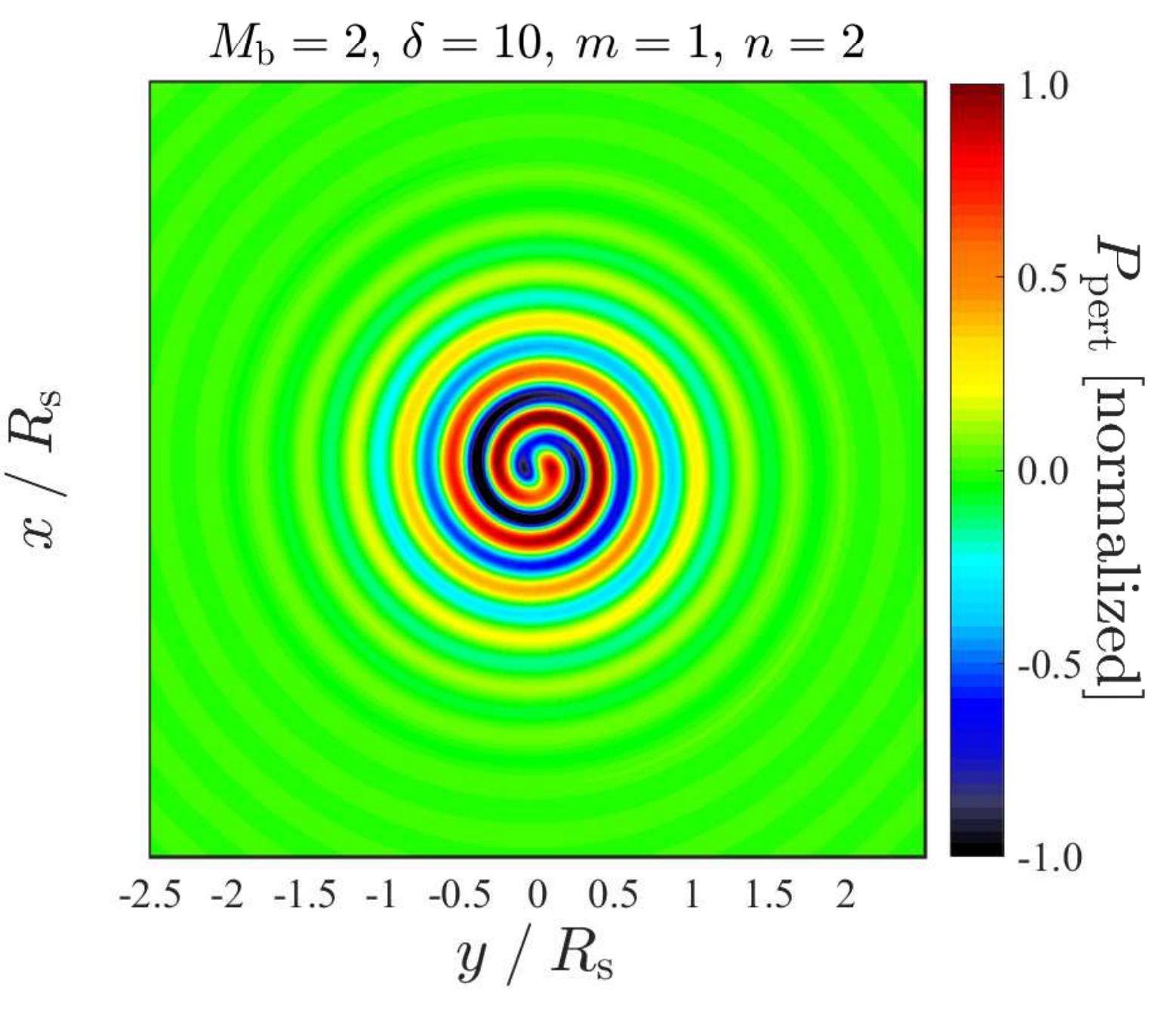}\\
\end{center}
\caption{Development of eigenmodes in simulations with a single wavelength perturbation. 
The top row shows the pressure perturbation for the same two simulations as in \fig{lin_growth}. 
The left two panels are for the $\Mb=1$ case, while the right two panels are for the $\Mb=2$ case. 
For each simulation we show both a slice through the $zy$ plane (1st and 3rd panels), and a slice 
through the $xy$ plane (2nd and 4th panels). The simulations are shown shortly after the fastest 
growing eigenmode has developed and begun to grow, $t\sim 0.3\tsc$ for $\Mb=1$ and $t\sim 2.4\tsc$ 
for $\Mb=2$. The bottom row shows the predicted pressure distribution for the eigenmodes whose growth 
rates were used in \fig{lin_growth}. The resemblance of the simulated perturbation to the analytic mode 
is striking, showing that our simulations properly capture the linear evolution.}
\label{fig:lin_panel} 
\end{figure*}

\smallskip 
In this section we present simulations of the linear evolution of single wavelength 
perturbations in 3d cylinders. This serves both to test the numerical scheme we use 
throughout the paper, as well as to test the analytic predictions for the growth rate 
and eigenmode structure of linear perturbations in 3d cylinders presented in M16 
(section 2.4).

\smallskip
We perform two simulations, with $(\Mb,\delta)=(1,10),$ and $(2,10)$, meant to represent surface 
and body modes, respectively. The unperturbed initial conditions were set up as described in 
\se{methods-unperturbed}. For the simulation with $\Mb=1$ we used $\sigma=\Rs/32$ in \equ{ramp2}, 
as in the surface-mode simulations presented in \se{surface}. For the $\Mb=2$ case we used 
$\sigma=\Rs/8$, as in the body-mode simulations presented in \se{body}. In both simulations, 
we initiate a perturbation in the radial component of the velocity, following \equs{pertx}-\equm{perty}, 
with a single wavelength, $\lambda=\Rs$ so $k=2\pi/\Rs$, and helical symmetry, $m=1$. The perturbation 
amplitude is $0.01\cs$ as in our main analysis. For the $\Mb=1$ case we set $\sigma_{\rm pert}=\Rs/16$, 
while for $\Mb=2$ we use $\sigma_{\rm pert}=\Rs/8$, as for the simulations in \se{surface} and \se{body} 
respectively. 
%For these value of $k$ and $m$, the fastest growing mode in the $\Mb=1$ case is a surface mode with a Kelvin-Helmholtz time $\tkh\simeq 0.10\tsc$, while in the $\Mb=2$ case it is a body mode with $\tkh\simeq 0.35\tsc$.

\smallskip
\Fig{lin_growth} shows the perturbation amplitude as a function of time, normalized by the respective 
value of $\tkh$ for the fastest growing mode in each simulation. For $\Mb=1$ this is the only available 
mode, the $n=0$ surface mode. For $\Mb=2$ this is the $n=2$ body mode (see M16 for details). 
%\footnote{For $\Mb=2$, the fastest growing mode for $\lambda=\Rs$ and $m=1$ is the $n=2$ body mode. However, the simulation seems to have converged to the $n=5$ body mode, both in terms of the growth rate (\fig{lin_growth}) and the distribution of pressure (\fig{lin_panel}). The reason for this discrepancy is unclear, but is likely related to the relatively large smoothing used in the unperturbed initial conditions, causing the system to deviate from the step-function solution. It should be noted that the growth rate of the $n=5$ mode is $\sim 0.62$ times that of the $n=2$ mode, a relatively small difference.}. 
The perturbation amplitude is estimated by the root-mean-squared (RMS) value of the transverse velocity, 
$v_{\perp}=(v_{\rm x}^2+v_{\rm y}^2)^{1/2}$. Before the perturbation can grow, it must evolve into an 
eigenmode of the system. For surface modes, the timescale over which this occurs is the sound crossing 
time of the perturbation in the hot medium, $t_{\rm \lambda}=\lambda/\cb$, while for body modes it is 
the stream sound crossing time, $\tsc$ (M16, section 3.3). In \fig{lin_growth} we have normalized the 
perturbation amplitudes to unity at these times. Prior to this time the amplitude is not expected to 
grow appreciably, while afterwards it grows as ${\rm exp}(t/\tkh)$. The $\Mb=1$ simulation shows a good 
match to the analytic prediction, underestimating the growth rate by less than $20\%$, due to the smoothing 
in the initial conditions. The growth in the $\Mb=2$ case is oscillatory, likely due to the large smoothing 
in the unperturbed solution and to mode interactions, though during periods of growth the slope is a good 
match to the analytic prediction. Note that similar behavior was seen when examining the linear growth of 
body modes in 2d slab simulations (M16, figure 9).

\smallskip
The top row of \fig{lin_panel} shows the spatial form of the pressure perturbation in each simulation 
shortly after the fastest growing eigenmode has developed, $t\sim2.7\tkh\sim0.3\tsc$ for $\Mb=1$ and 
$t\sim6.9\tkh\sim 2.5\tsc$ for $\Mb=2$. For each simulation we show both an edge-on view (a slice in 
the $zy$ plane) and a face-on view (a slice in the $xy$ plane). The bottom row shows the analytic 
distribution of the pressure perturbation for the fastest growing eigenmode, computed following the 
procedure in section 2.4 of M16. The resemblance of the simulation results and the analytic prediction 
is striking, illustrating that the initial perturbation has evolved into eigenmodes and is dominated by 
the fastest growing mode. We conclude that our numerical scheme captures the linear evolution of the 
perturbations well.

%%%%%%%%%%%%%%%%%%%%%%%%%%%%%%%%%%%%%%%%%%%%%%%%  
\section{Convergence tests}
\label{sec:convergence} 

\smallskip
In this section we test convergence of our results to the number of cells per stream 
radius, the refinement structure of the grid, and the random seed used to generate the 
initial conditions. We also discuss how convergence may depend on the width of the 
smoothing layer between the stream and the background and on the dominant unstable mode. 
This section refers only to 3d cylinders, as the convergence of 2d slab simulations was 
discussed in P18.

\begin{figure*}
\begin{center}
\includegraphics[trim={-0.5cm 1.2cm 0.025cm 0}, clip, width =0.2823 \textwidth]{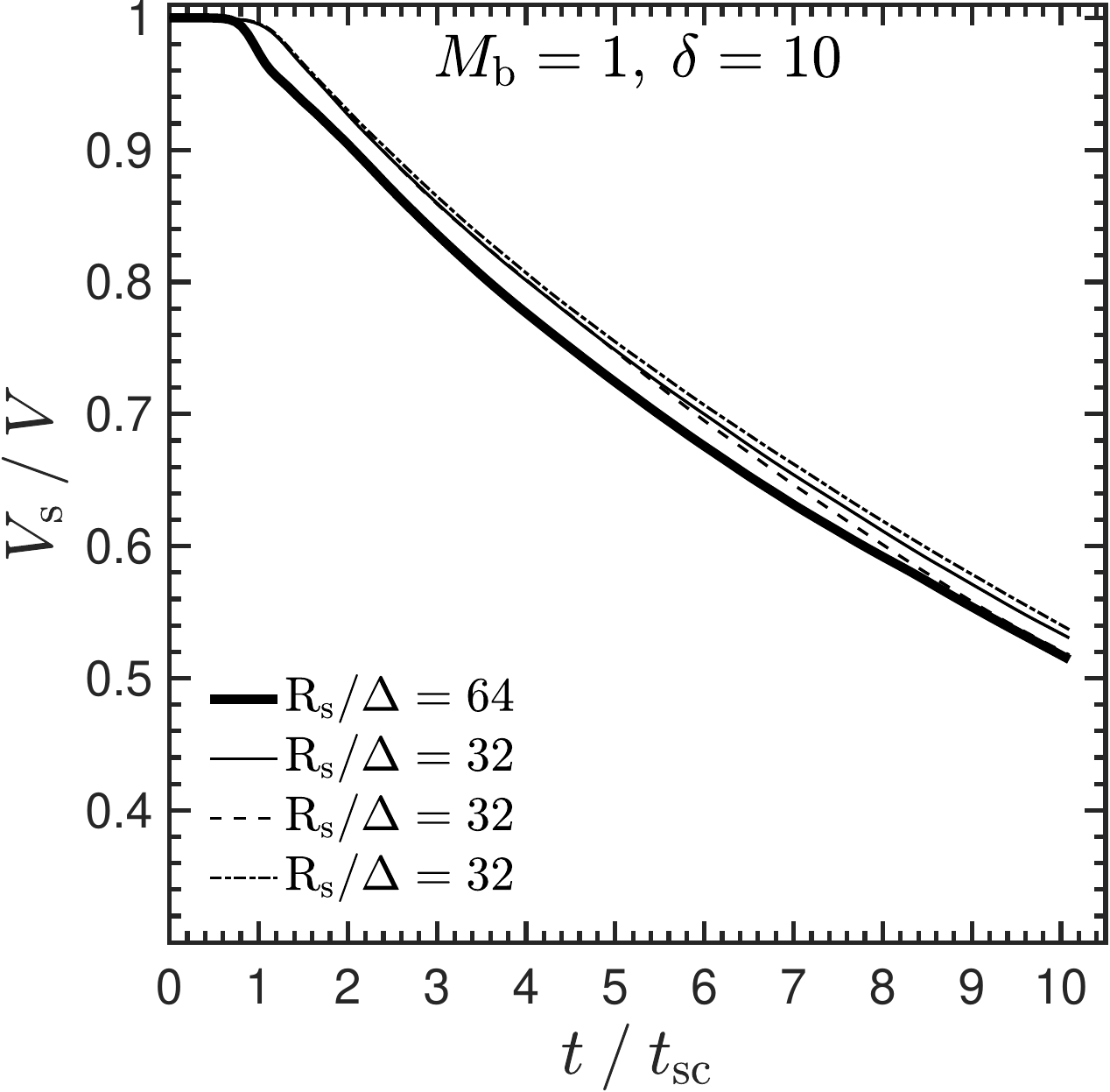}
\includegraphics[trim={1.81cm 1.2cm 0.025cm 0}, clip, width =0.233 \textwidth]{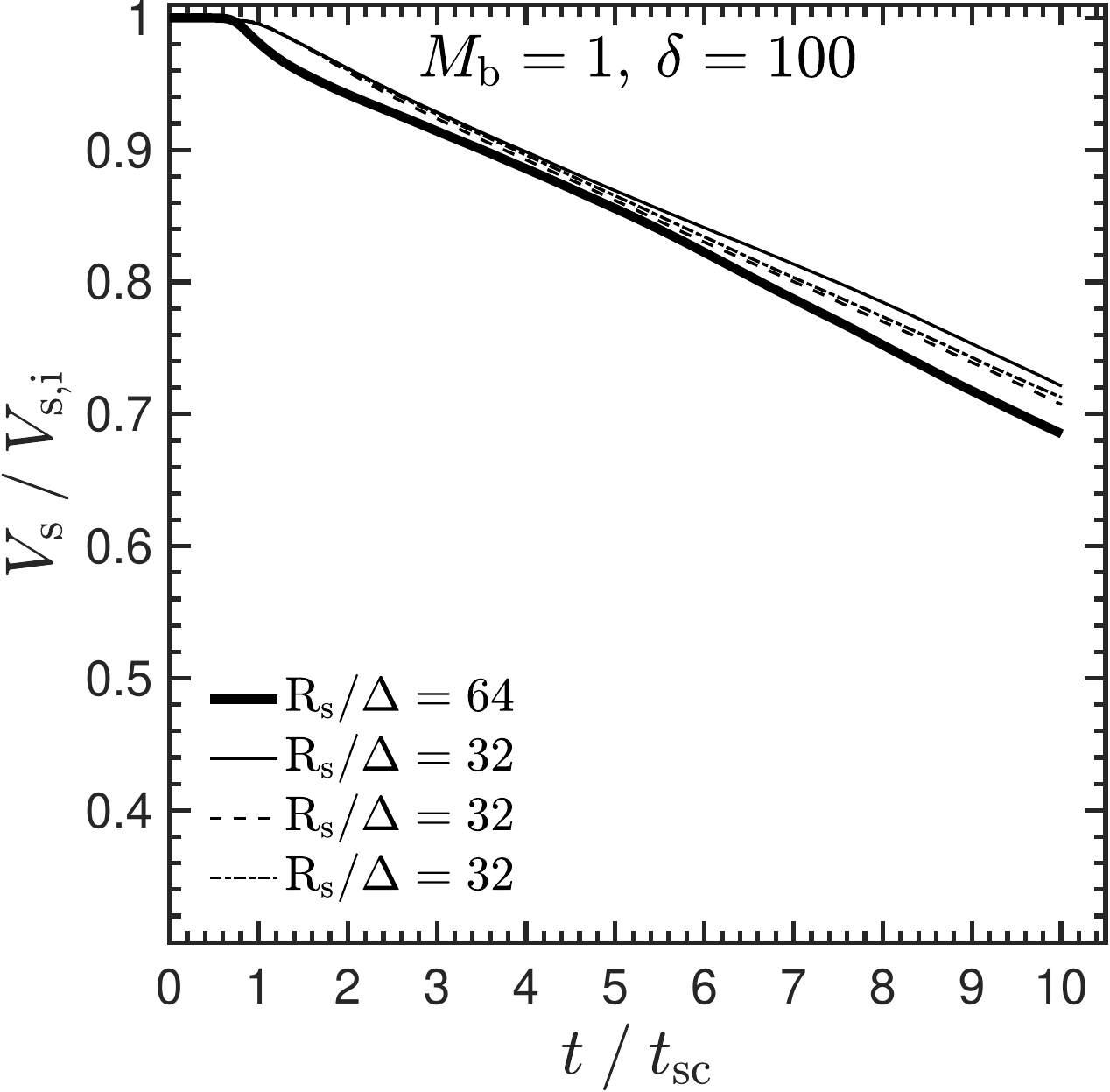}
\includegraphics[trim={1.81cm 1.2cm 0.025cm 0}, clip, width =0.233 \textwidth]{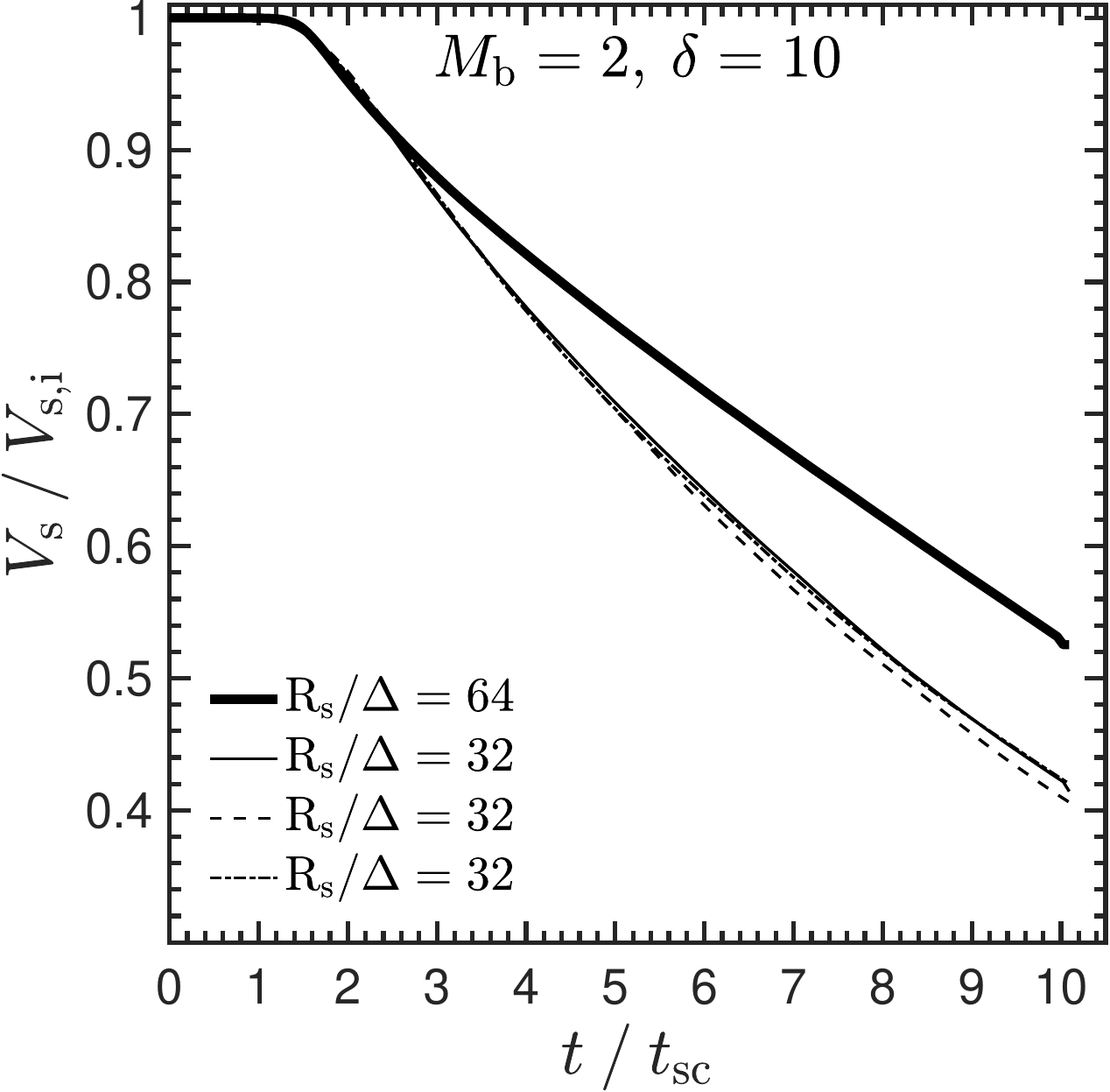}
\includegraphics[trim={1.81cm 1.2cm 0.025cm 0}, clip, width =0.233 \textwidth]{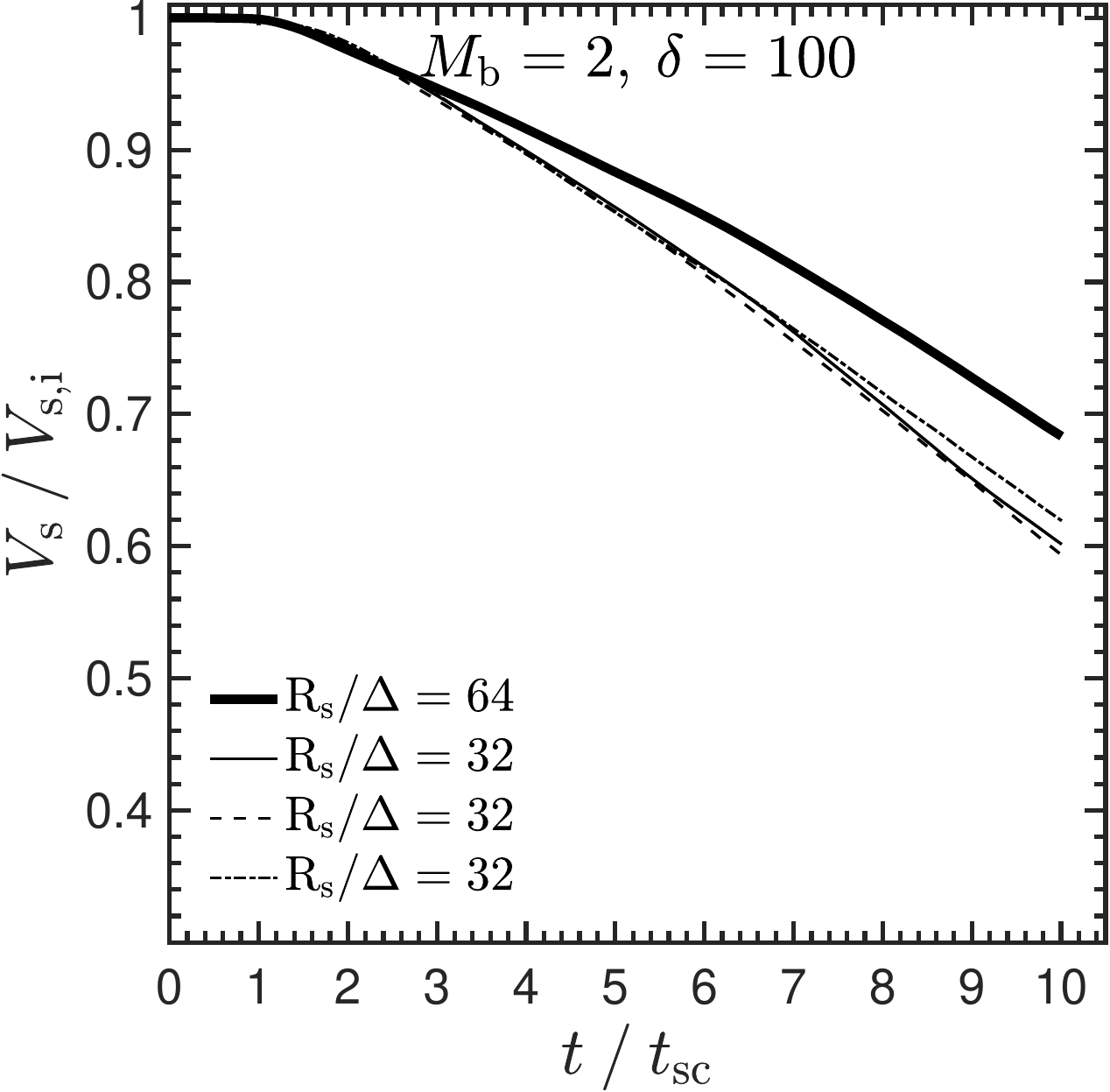}\\
\vspace{-0.22cm}
\includegraphics[trim={-0.5cm 1.2cm 0.025cm 0}, clip, width =0.2823 \textwidth]{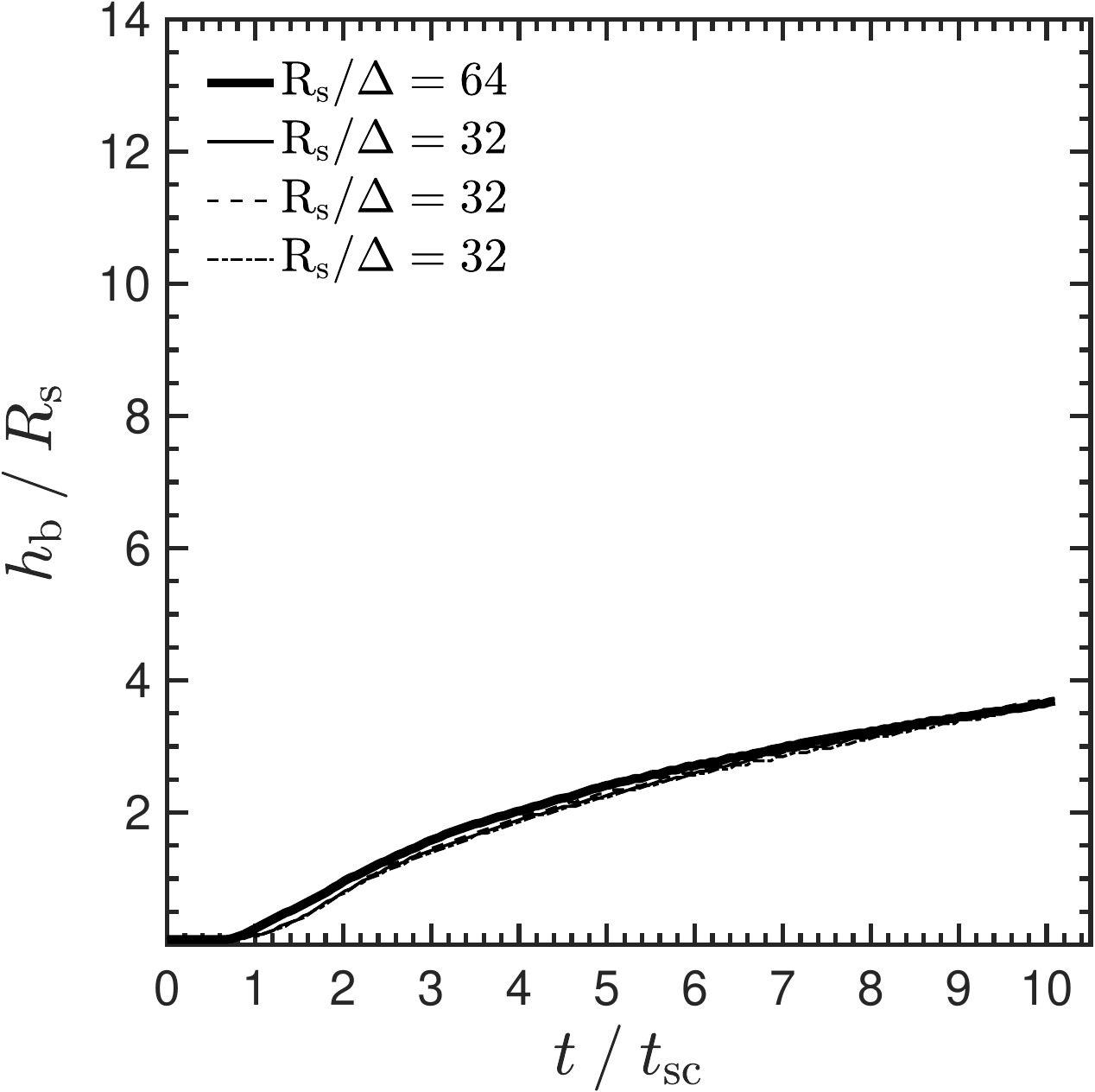}
\includegraphics[trim={1.81cm 1.2cm 0.025cm 0}, clip, width =0.233 \textwidth]{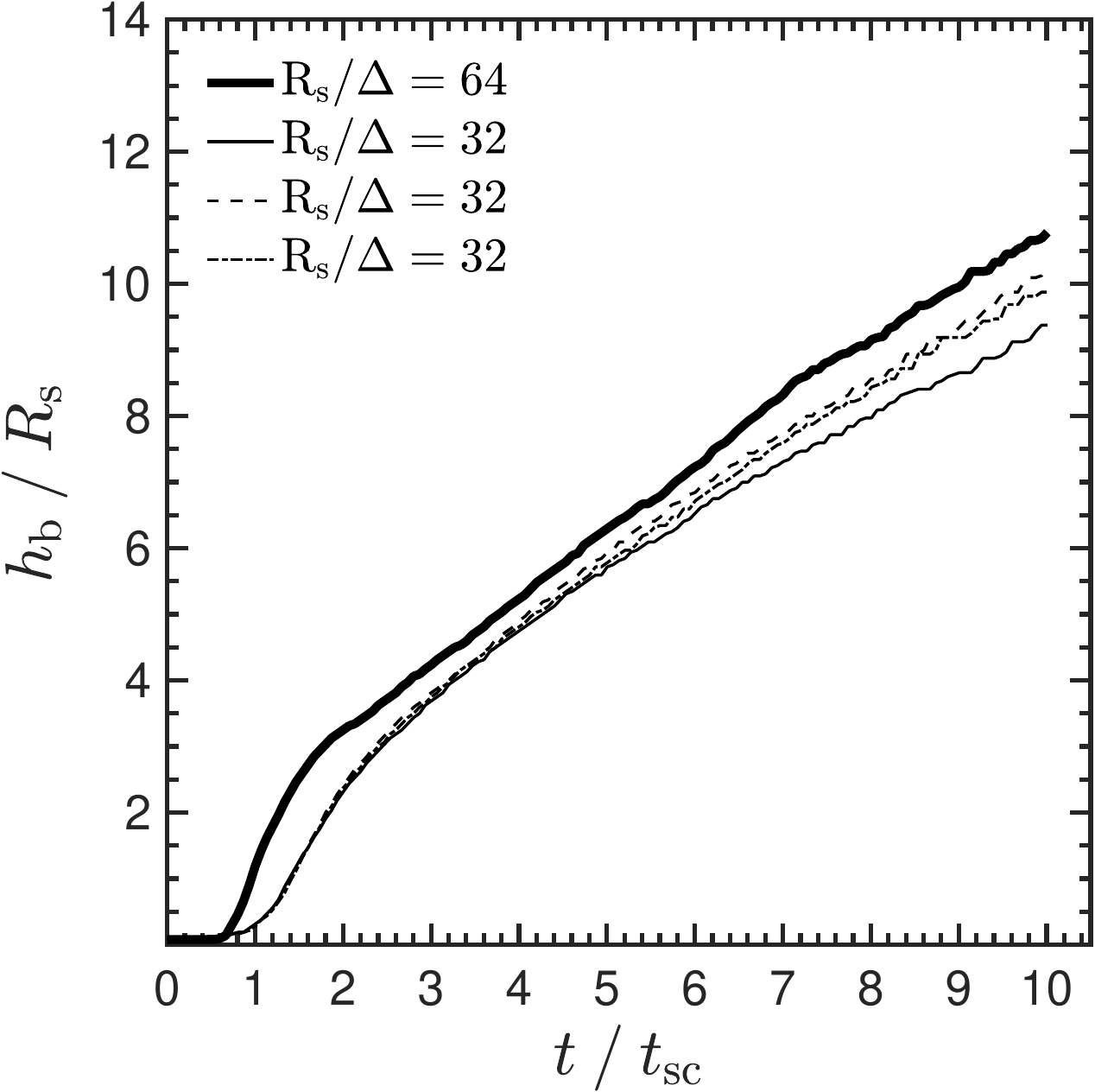}
\includegraphics[trim={1.81cm 1.2cm 0.025cm 0}, clip, width =0.233 \textwidth]{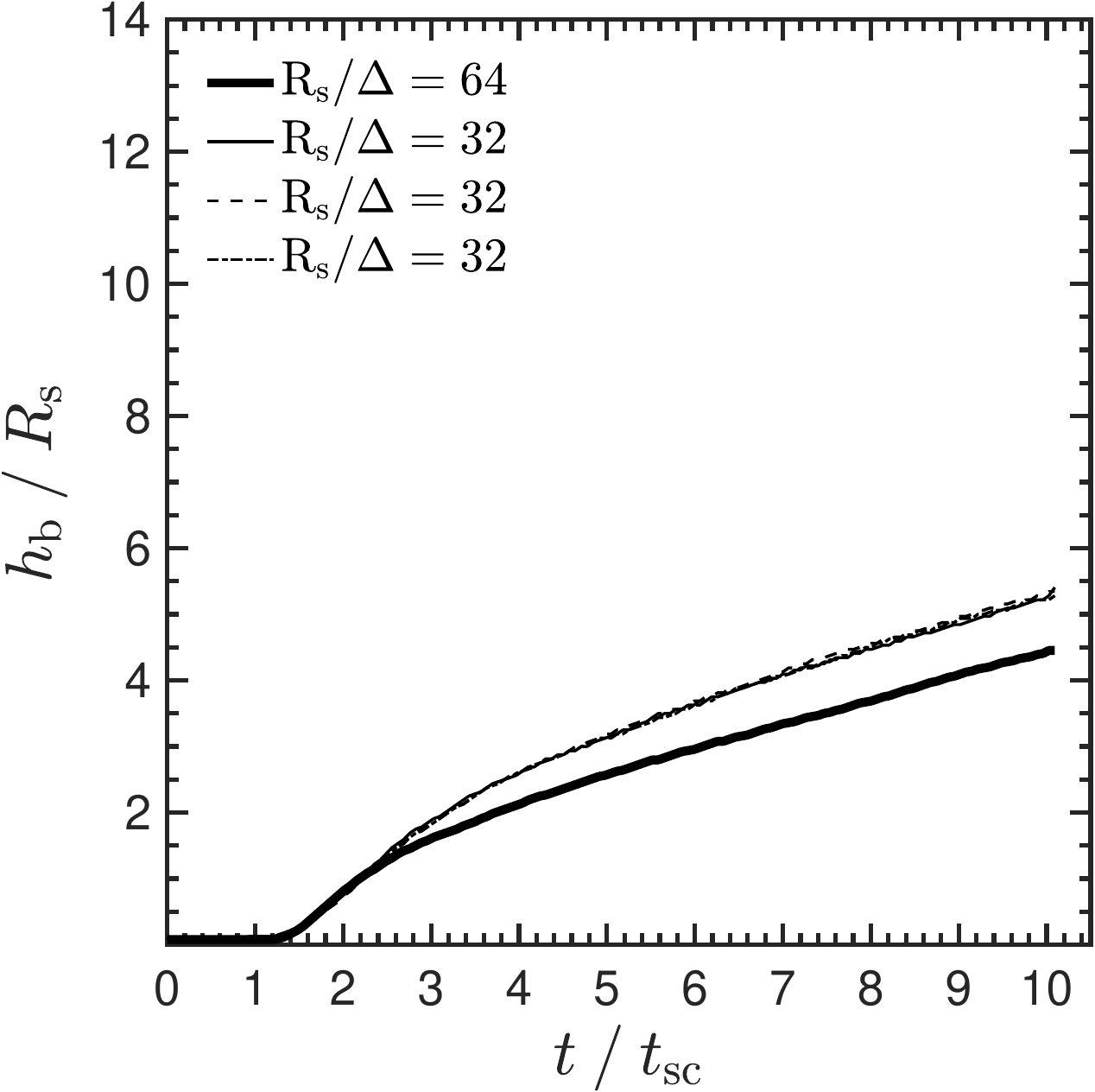}
\includegraphics[trim={1.81cm 1.2cm 0.025cm 0}, clip, width =0.233 \textwidth]{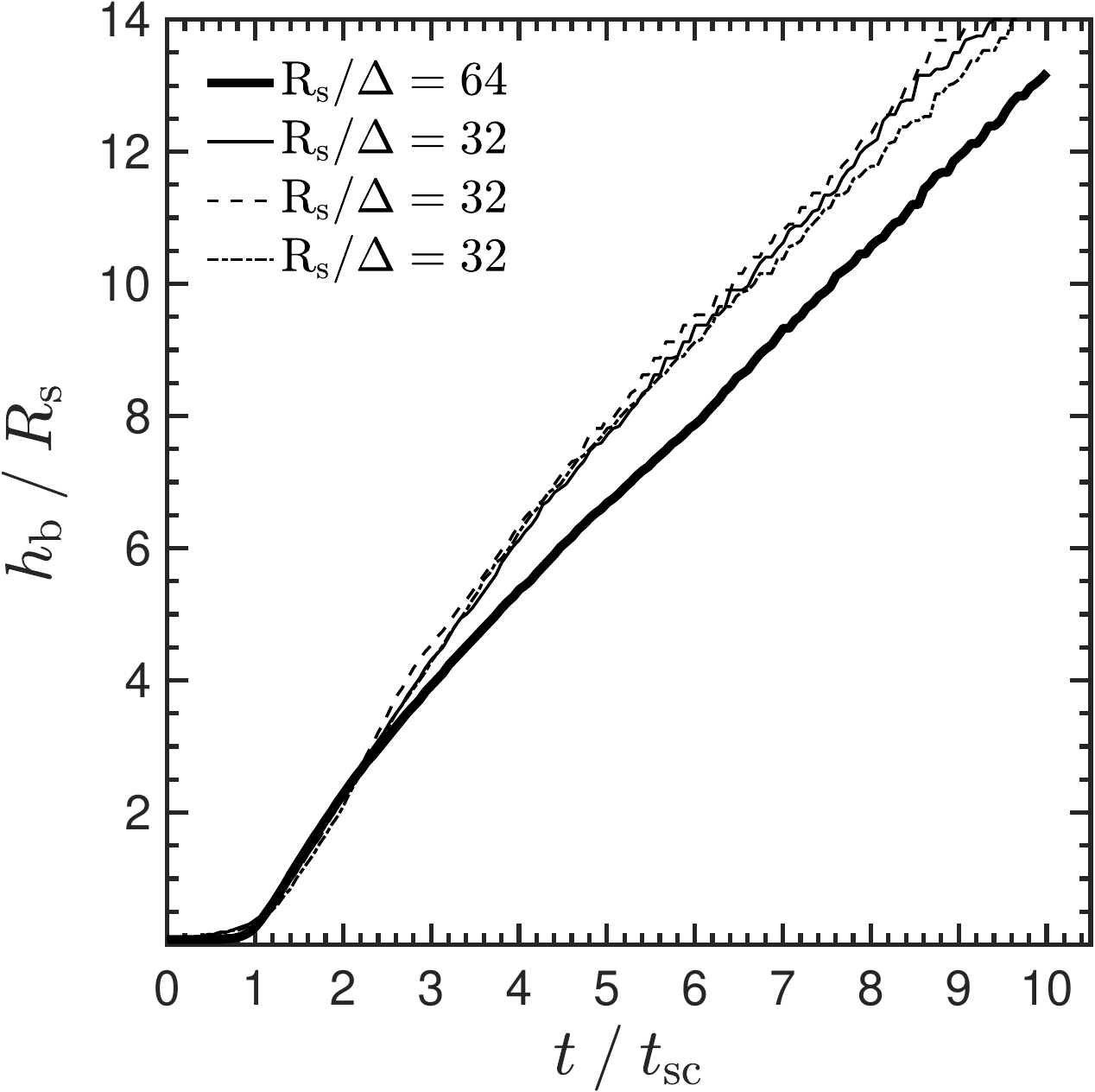}\\
\vspace{-0.2cm}
\includegraphics[trim={-0.5cm -0.3cm 0.025cm 0}, clip, width =0.2823 \textwidth]{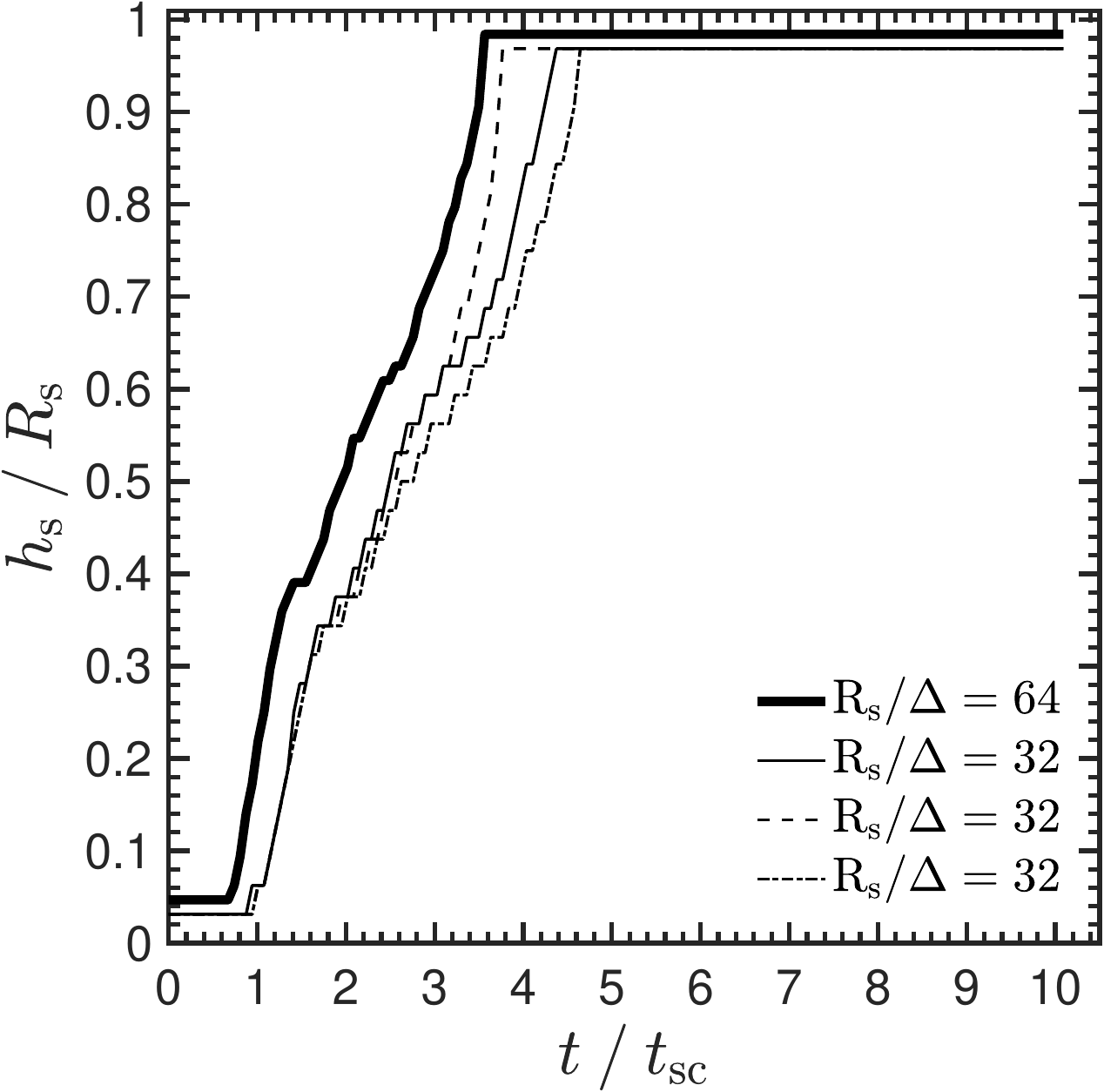}
\includegraphics[trim={1.81cm -0.3cm 0.025cm 0}, clip, width =0.233 \textwidth]{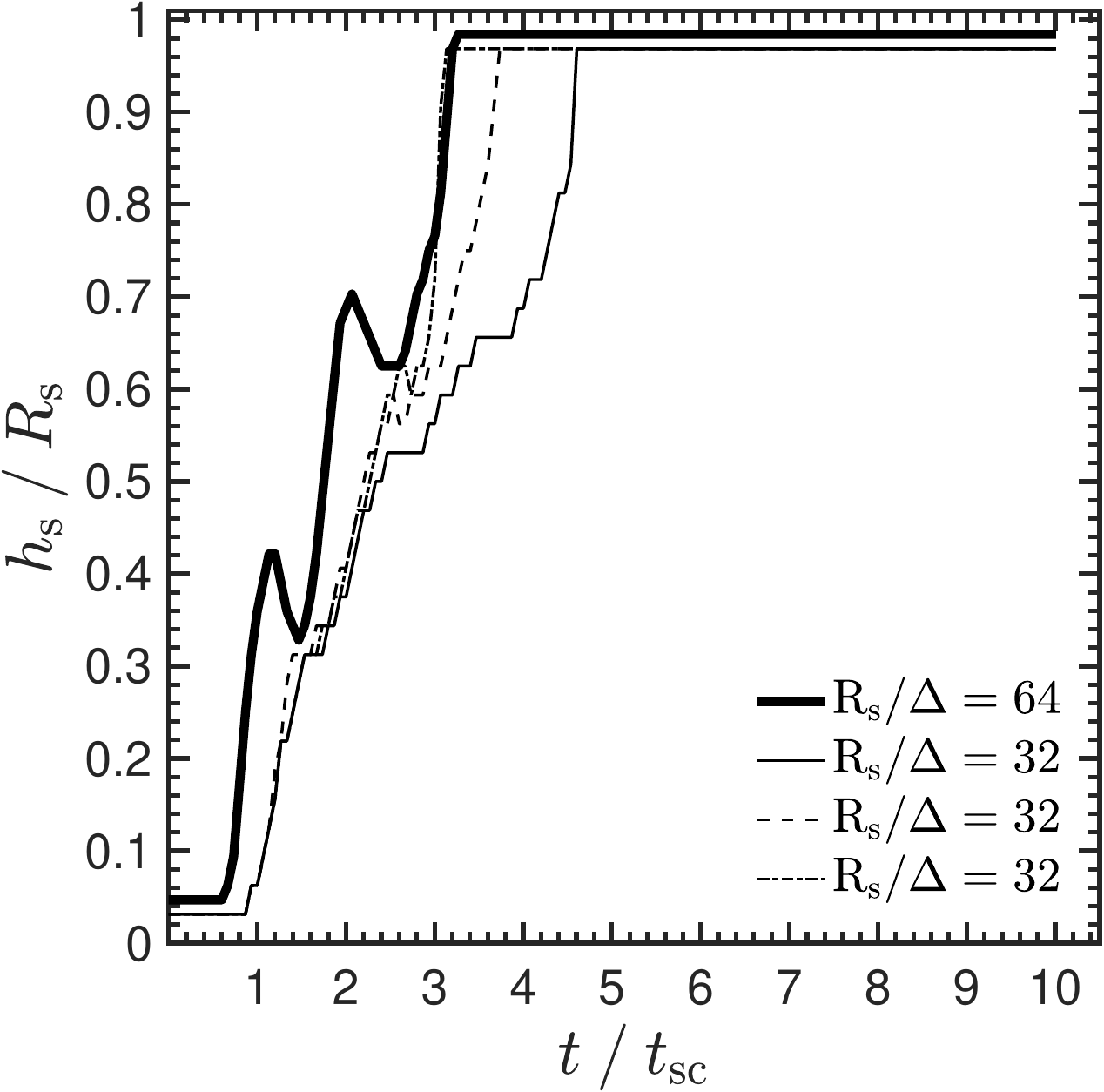}
\includegraphics[trim={1.81cm -0.3cm 0.025cm 0}, clip, width =0.233 \textwidth]{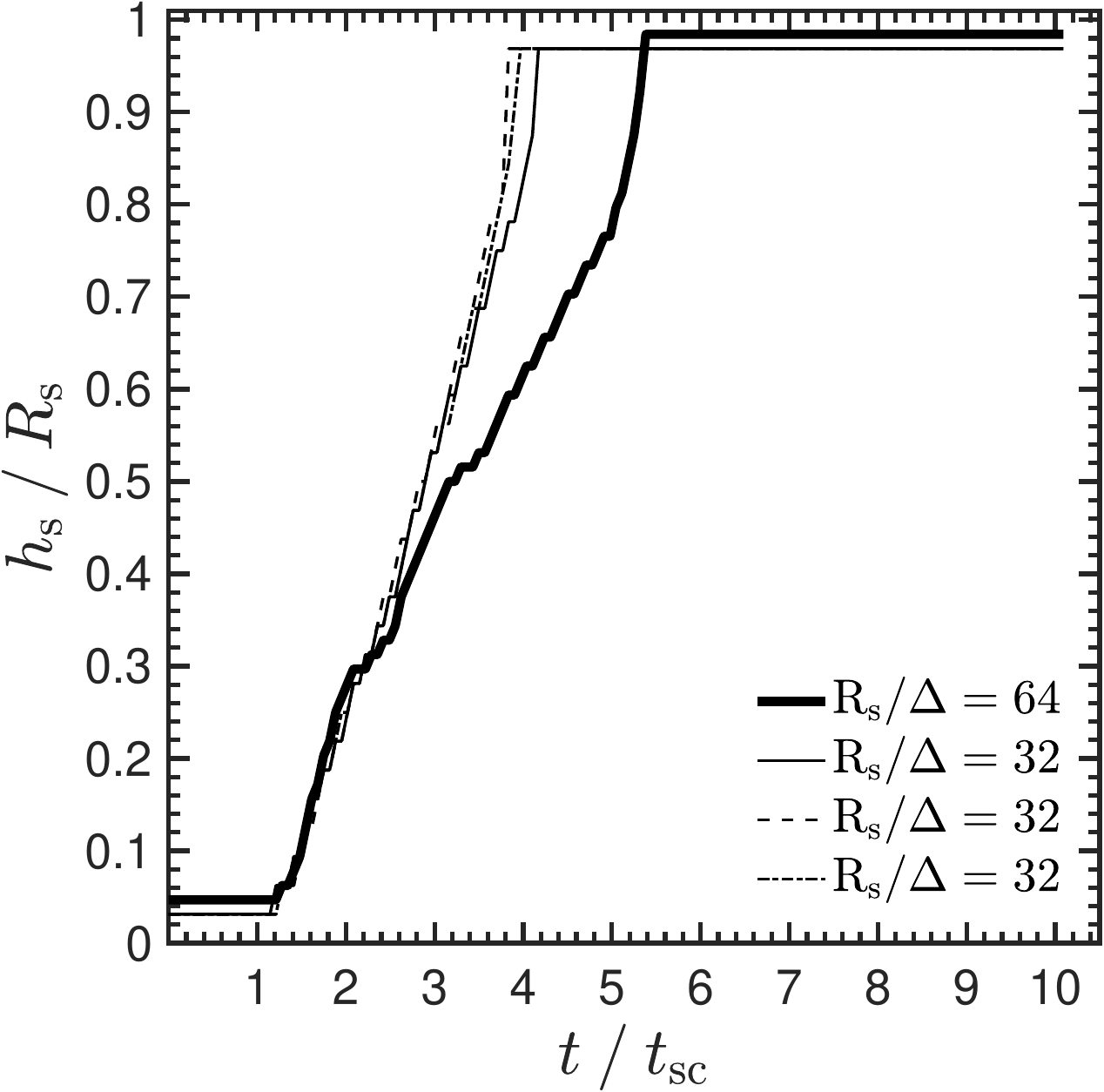}
\includegraphics[trim={1.81cm -0.3cm 0.025cm 0}, clip, width =0.233 \textwidth]{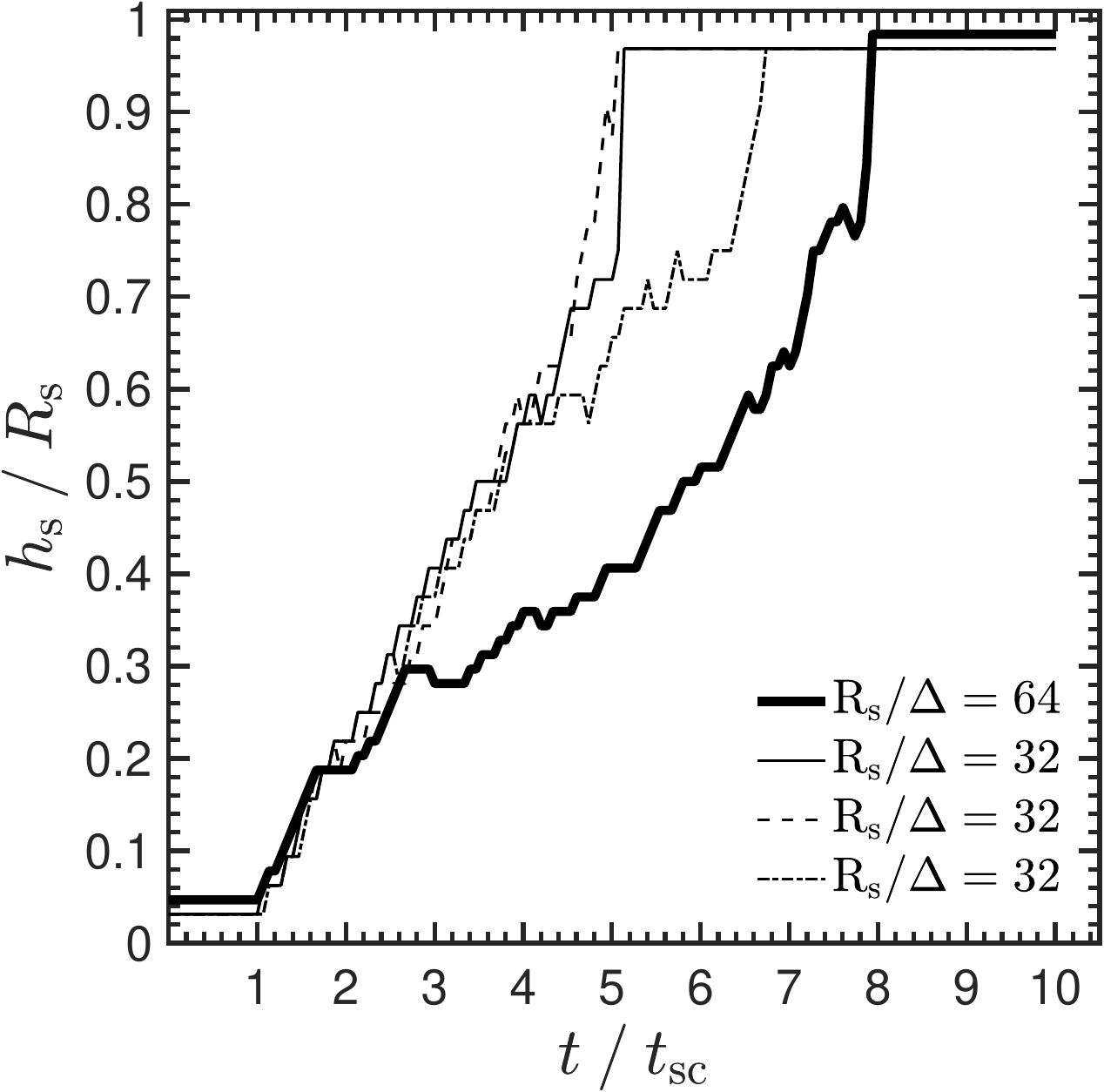}
\end{center}
\caption{Convergence of our results with resolution, and variability introduced by the random seed. 
We compare results of our fiducial simulations with 128 cells per stream diameter (thick solid lines), 
to simulations with 64 cells per diameter (thin solid, dashed, and dot-dashed lines) run with three 
different random realizations of the perturbation phases (the thin solid line represents the same 
phases as the thick solid line). All other numerical parameters have their fiducial values described 
in \se{methods}. We show the stream deceleration, $V_{\rm s}/V$ (top), and the shear layer growth into the 
background, $\hb$ (middle), and into the stream, $\hs$ (bottom). The left-two columns show surface mode 
simulations with $\Mb<M_{\rm crit}$ while the right-two columns show simulations with $\Mb>M_{\rm crit}$. 
In all cases, the variability associated with the random seed is extremely small. The largest variability 
is in the time when the shear layer consumes the entire stream, $\hs=\Rs$, as this is the most stochastic 
process. However, even here the variability about the mean is $\lsim 20\%$ in all cases. Surface mode 
simulations are converged at half our fiducial resolution. On the other hand, low resolution simulations 
with $\Mb>M_{\rm crit}$ are characterised by more rapid shear layer growth at $t\gsim 3\tsc$ when $\hb\gsim 2\Rs$, 
resulting in more rapid stream deceleration. This is due to the instability being dominated by high-order 
azimuthal surface modes, with azimuthal mode number $m\sim 12$ (\fig{smoothing_panel}). This leads to small 
initial eddies whose cascade to even smaller scales is unresolved, yielding more rapid shear layer growth 
as this process is dominated by the largest eddies. For high $\Mb$ simulations to converge, we require either 
higher resolution or a suppression of the high-$m$ modes.}
\label{fig:low_res_convergence} 
\end{figure*}

\subsection{Low Resolution and Random Seed}
\label{sec:convergence_1} 
\smallskip
\Fig{low_res_convergence} compares results of simulations with our fiducial resolution of 128 cells 
per stream diameter, $\Delta=\Rs/64$, to simulations at half the resolution, $\Delta=\Rs/32$. Furthermore, 
for the low resolution simulations we present results from three runs initiated with different random seeds. 
All runs were performed with our fiducial refinement strategy (\se{methods-grid}) and fiducial smoothing of 
the initial conditions (\se{methods-unperturbed}, $\sigma=\Rs/32$ in \equnp{ramp2}). We show results for 
$(\Mb,\delta)=(1,10)$, $(1,100)$, $(2,10)$, and $(2,100)$. The first two have $\Mb<M_{\rm crit}$ (\equnp{Mcrit}) 
and are representative of the simulations presented in \se{surface}, while the latter two have $\Mb>M_{\rm crit}$ 
and are representative of the simulations presented in \se{body}. For each simulation we show the stream deceleration 
(top, computed as in \fig{deceleration_surface}) and the growth of the shearing layer into the background (middle) 
and the stream (bottom, computed as in \fig{surface_h}). 

\smallskip
In all cases, the scatter induced by varying the random seed in very small, much smaller than the scatter 
induced in 2d slab simulations shown in P18. The smaller scatter likely results from the presence of many 
more eddies in 3d simulations than in 2d simulations due to the larger interface surface, resulting in less 
overall variability in the average merger times of eddies. Furthermore, the presence of a direct cascade in 
3d simulations leading to small scale turbulence reduces the macroscopic variability further. The largest 
variability is in the final dissintegration of the stream by the shearing layer, i.e. the time when $\hs=\Rs$. 
This is the most stochastic process, depending sensitively on when the largest eddies reach the stream axis. 
However, even here the scatter is less then $20\%$ about the mean.

\smallskip
The low $\Mb$ runs appear converged already at half our fiducial resolution. Stream deceleration and 
shear layer growth remain effectively identical at low resolution, modulo the aforementioned variability 
in the final stages of stream disruption. There is a slight systematic trend of shear layer growth begining 
slightly later in the low resolution simulations, by $<20\%$, resulting in a corresponding delay in the onset 
of deceleration. This is likely simply a result of the larger cell sizes requiring the interface to travel a 
larger distance before mixing is detected. Regardless, this has no effect on the rates of these processes.

\smallskip
The behaviour of the high $\Mb$ simulations is qualitatively different. At early times, the low resolution simulations 
behave identically to the fiducial runs. However, at $t\gsim 3\tsc$, when $\hb\gsim 2\Rs$, the growth rate of the 
shear layer becomes shallower in the fiducial runs than in the low resolution runs. This leads to a decrease in the 
deceleration rate, so the low resolution runs loose significantly more momentum by the end of the simulation. To explain 
these trends, we recall that these simulations are dominated by high-order azimuthal surface modes, with azimuthal mode 
number $m\sim 12$ (\fig{smoothing_panel}). As explained in \se{surface}, the shear layer growth rate of surface modes in 
3d cylinders decreases once $\hb\gsim 2\Rs$, since the energy cascade towards small scales results in less power at large 
scales to drive the growth. As the azimuthal mode number increases, the typical eddies become smaller and this cascade 
becomes harder to resolve. Thus, while it is converged for $m\sim 1$, it has not converged for $m\sim 12$, yielding a more 
rapid shear layer growth and stream deceleration as observed. We conclude that in order to properly resolve the late time 
behaviour of high-Mach number simulations, we must either go to higher resolution or else suppress the very high-$m$ modes. 
This is shown below.

\begin{figure*}
\begin{center}
\includegraphics[trim={-0.5cm 1.2cm 0.025cm 0}, clip, width =0.36 \textwidth]{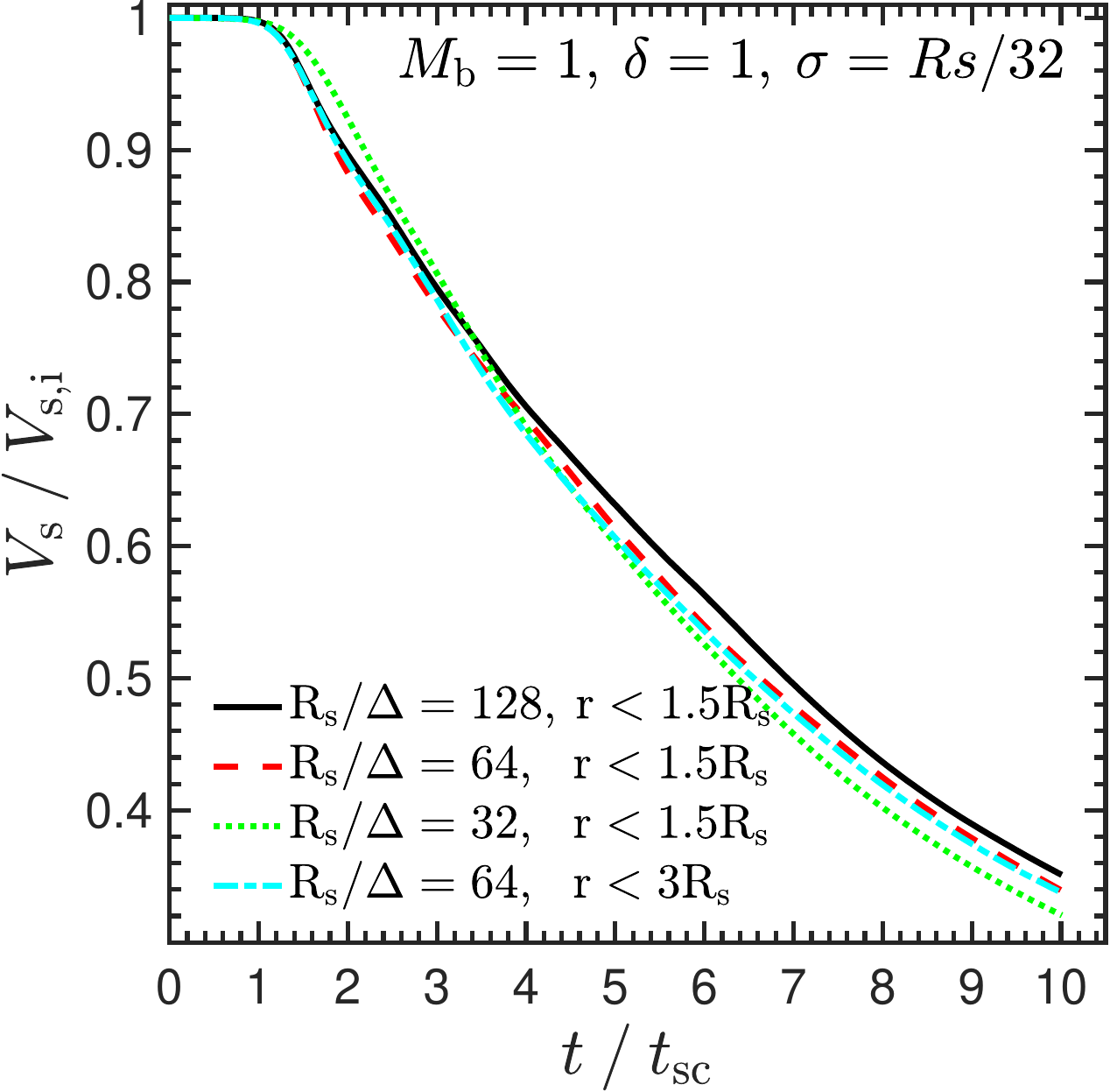}
\includegraphics[trim={1.81cm 1.2cm 0.025cm 0}, clip, width =0.297 \textwidth]{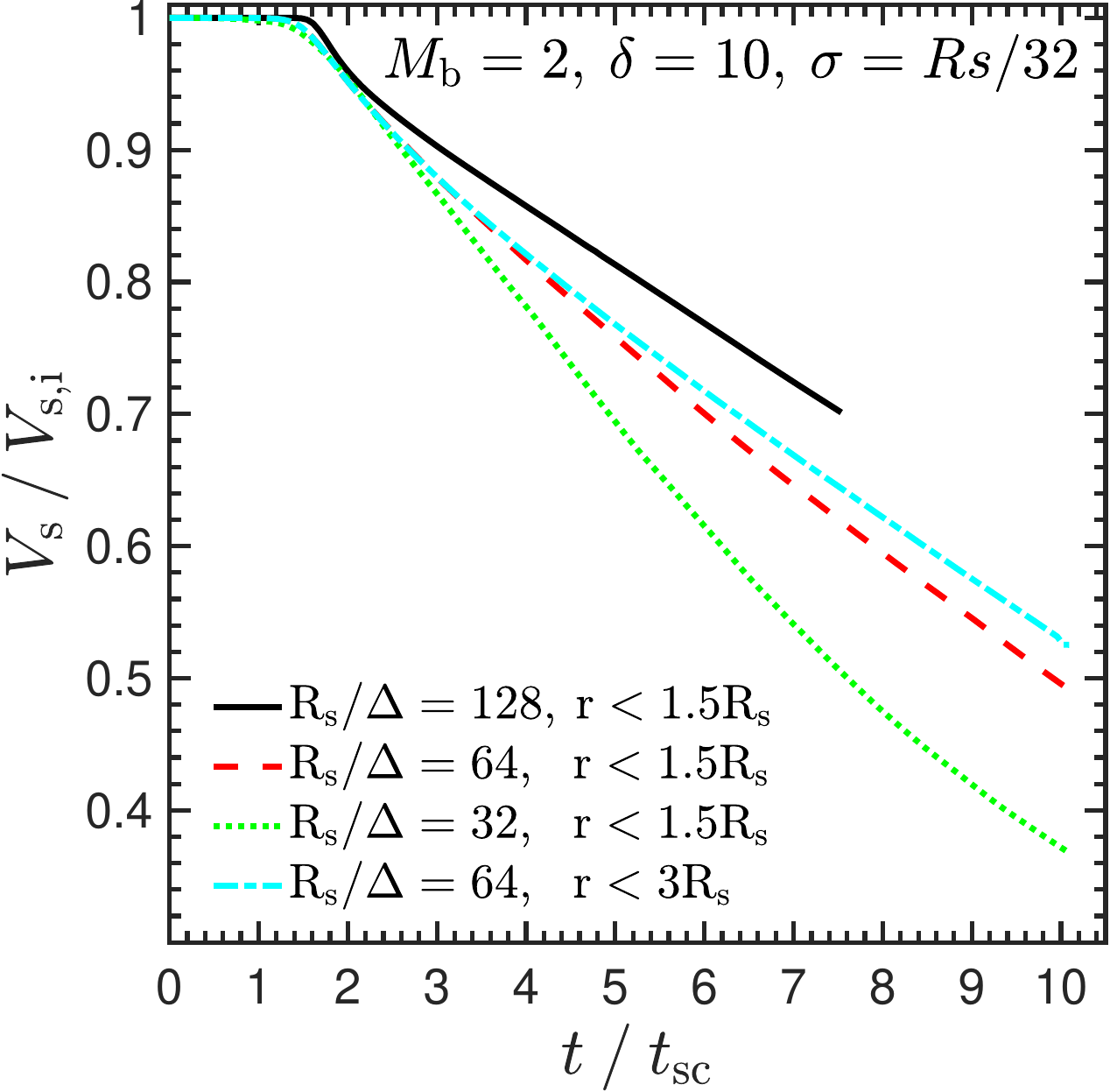}
\includegraphics[trim={1.81cm 1.2cm 0.025cm 0}, clip, width =0.297 \textwidth]{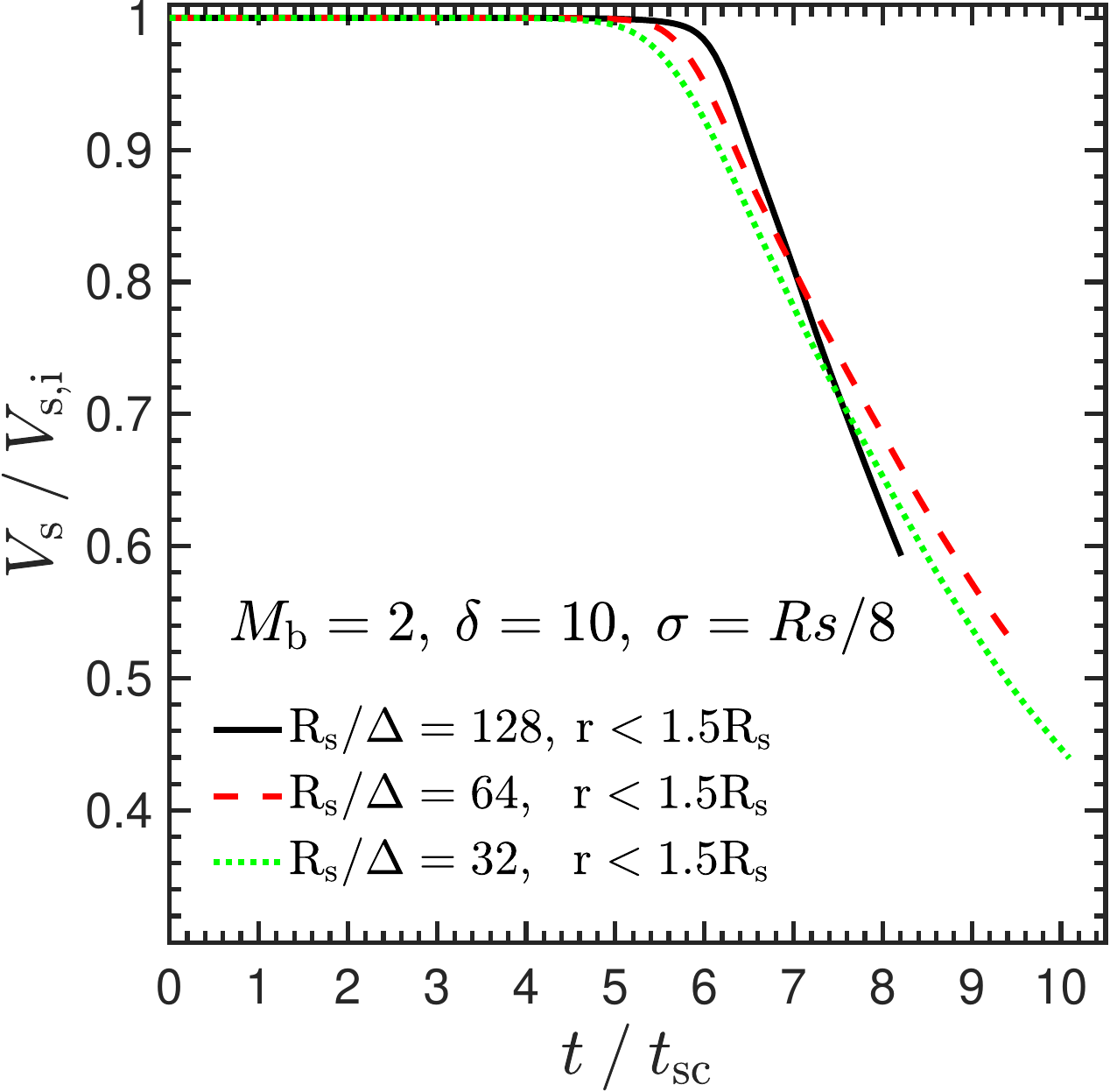}\\
\vspace{-0.15cm}
\includegraphics[trim={-0.5cm 1.2cm 0.025cm 0}, clip, width =0.36 \textwidth]{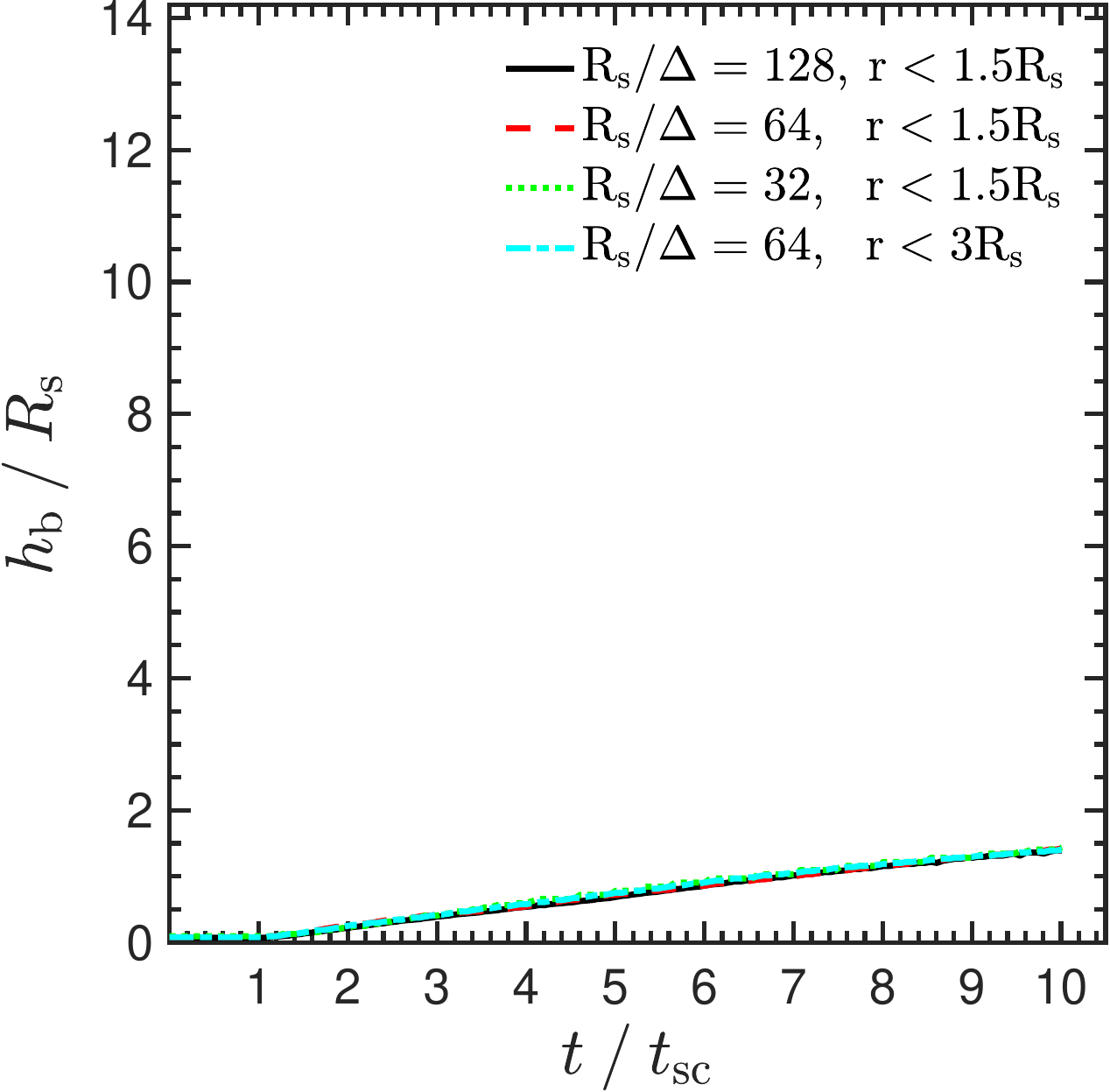}
\includegraphics[trim={1.81cm 1.2cm 0.025cm 0}, clip, width =0.297 \textwidth]{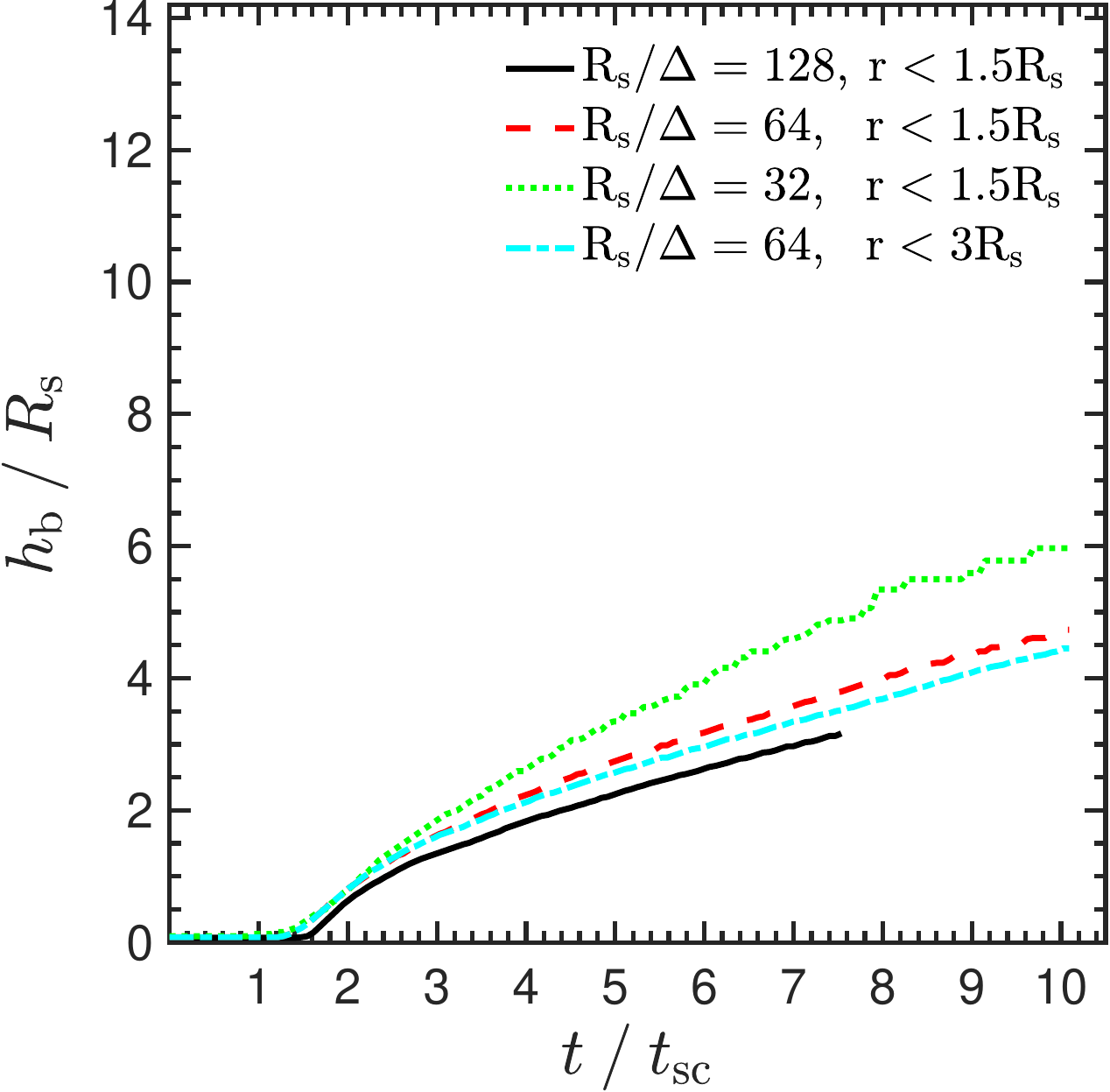}
\includegraphics[trim={1.81cm 1.2cm 0.025cm 0}, clip, width =0.297 \textwidth]{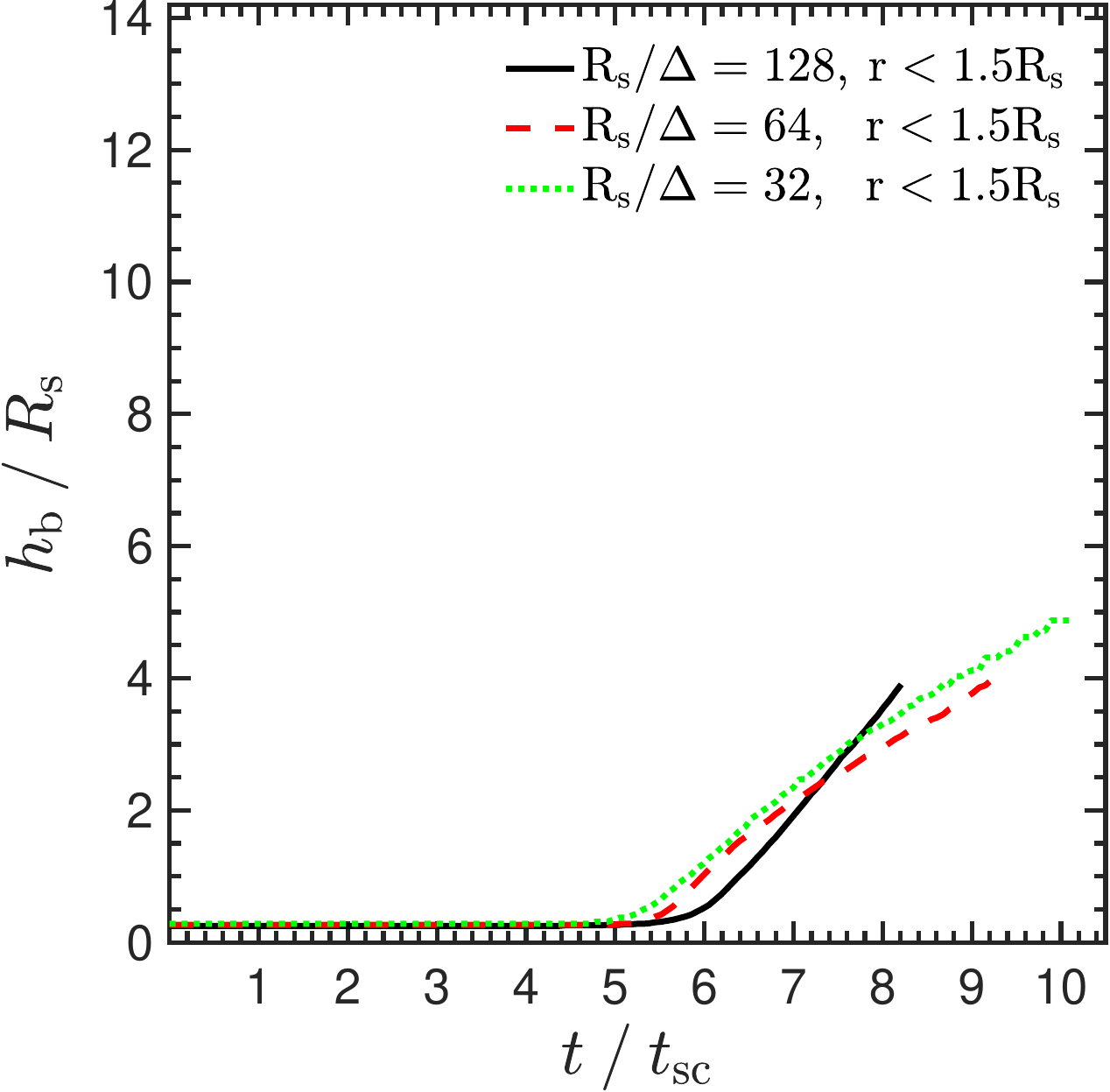}\\
\vspace{-0.15cm}
\includegraphics[trim={-0.5cm -0.3cm 0.025cm 0}, clip, width =0.36 \textwidth]{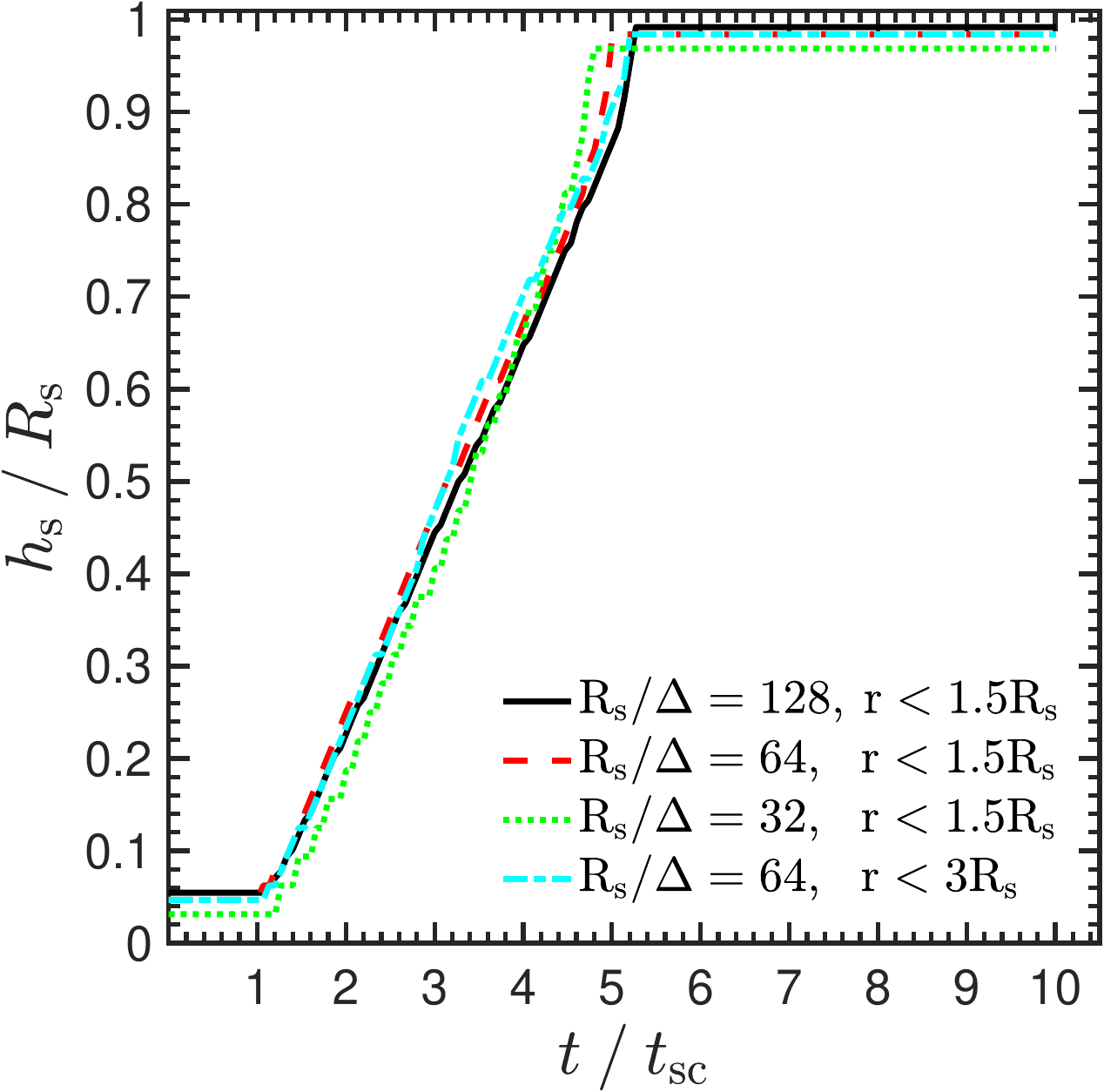}
\includegraphics[trim={1.81cm -0.3cm 0.025cm 0}, clip, width =0.297 \textwidth]{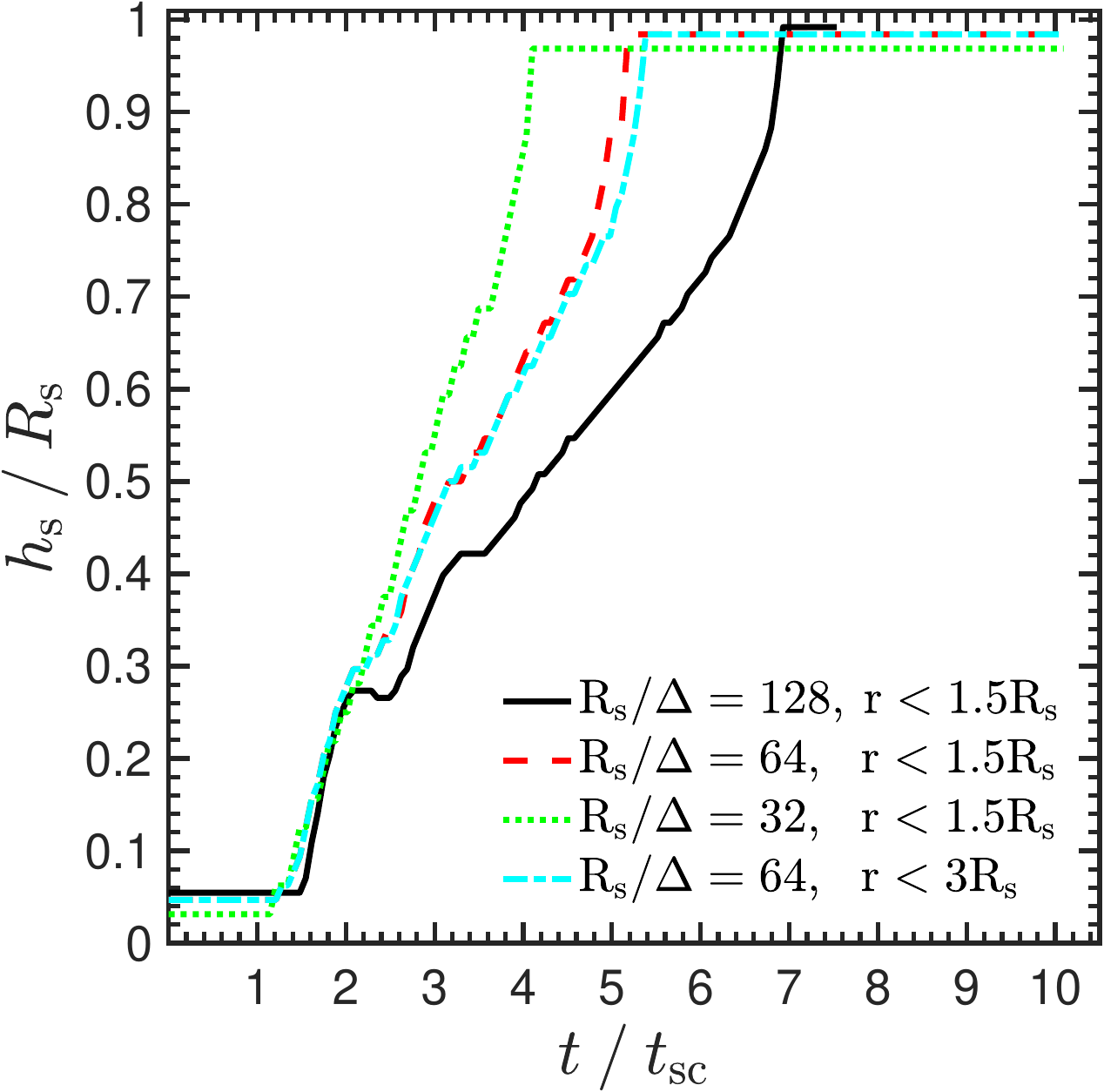}
\includegraphics[trim={1.81cm -0.3cm 0.025cm 0}, clip, width =0.297 \textwidth]{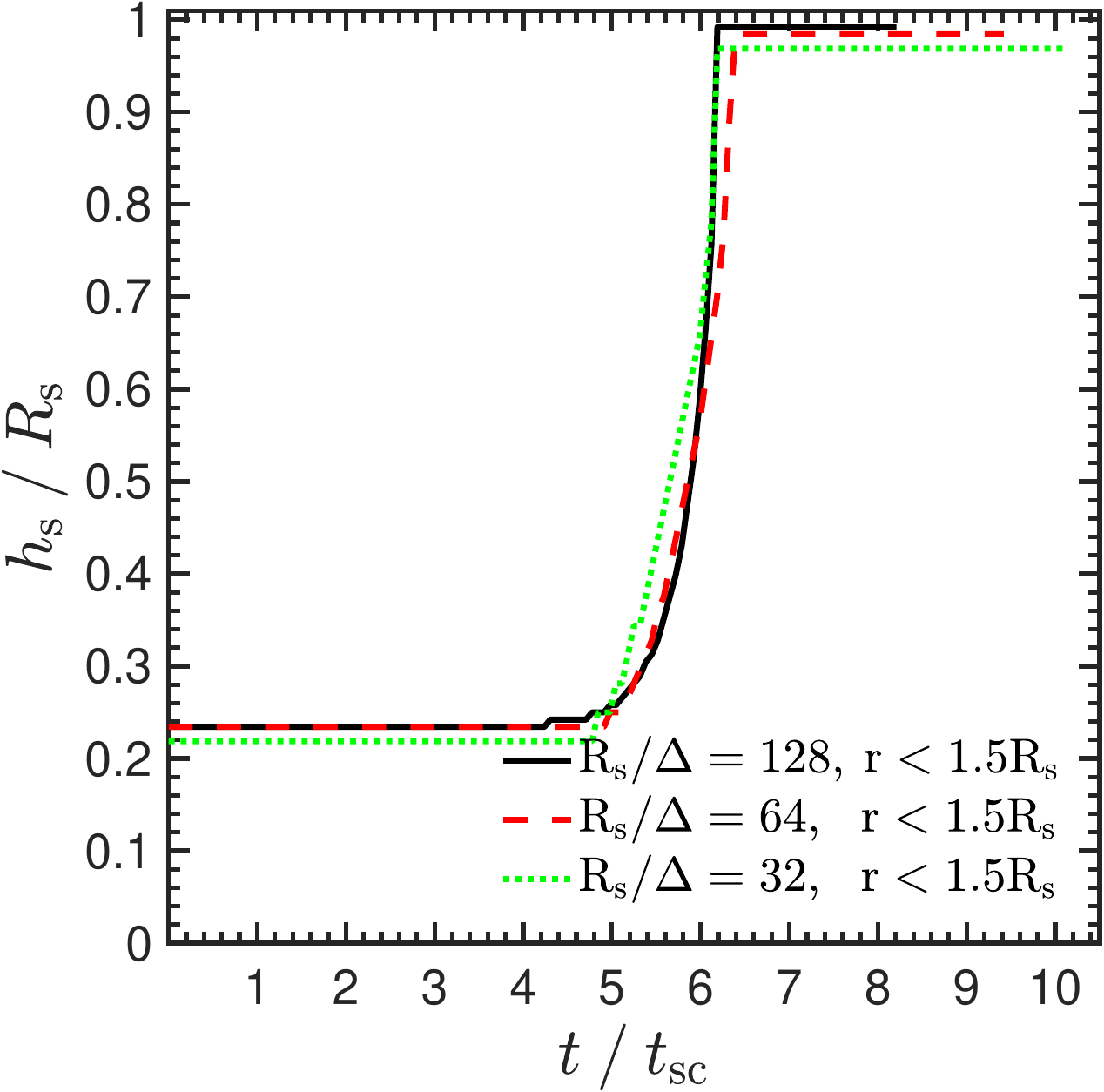}
\end{center}
\caption{Convergence of our results with resolution, and with the refinement scheme. We compare results 
of our fiducial simulations with 128 cells across the stream diameter and cell sizes increasing by a factor 
of 2 every $3\Rs$ in the $x$ and $y$-directions (red dashed lines, dubbed $r<3\Rs$ in the legend), to 
simulations with the same resolution but where the cell size is increased every $1.5\Rs$ in the $x$ and 
$y$-directions (cyan dash-dotted lines, dubbed $r<1.5\Rs$ in the legend). We also compare simulations at 
half and double the fiducial resolution with the $r<1.5\Rs$ refinement. As in \fig{low_res_convergence}, 
we show the stream deceleration, $V_{\rm s}/V$ (top), and the shear layer growth into the background, $\hb$ 
(middle), and into the stream, $\hs$ (bottom). We show a surface mode simulation with $(\Mb,\delta)=(1,1)$ 
on the left, and a body mode simulation with $(\Mb,\delta)=(2,10)$ in the middle and right columns. The middle 
column shows a case with narrow smoothing, $\sigma=\Rs/32$, as in \fig{low_res_convergence} so surface modes 
with $m\sim 12$ are unstable. The right-hand column shows a case with wide smoothing, $\sigma=\Rs/8$, so that 
modes with $m>4$ are stable. Note that this case does not show a simulation with the fiduial, $r<3\Rs$ refinement. 
In all cases shown, the difference between the two refinement schemes is extremely small, and we conclude 
that our results are converged with respect to this choice. The $(\Mb,\delta)=(1,1)$ case is converged with 
resolution at half our fiducial value, as inferred from \fig{low_res_convergence}. The $(\Mb,\delta)=(2,10)$ 
case with $\sigma=\Rs/32$ has not converged even at double our fiducial resolution. As inferred from 
\fig{low_res_convergence}, higher resolution runs evolve slower since they better resolve the breakup of the 
largest eddies of the $m\sim 12$ mode to smaller scales, leaving less power at large scales to drive the shear 
layer growth. However, when $\sigma=\Rs/8$, the $(\Mb,\delta)=(2,10)$ case has converged even at half our fiducial 
resolution. In this case, the instability is dominated by larger structures which are better resolved, associated 
with an $m=1$ body mode and an $m=4$ surface mode. We conclude that for body mode simulations, convergence can be 
acchieved by suppressing the growth of high-order azimuthal modes.}
\label{fig:high_res_convergence} 
\end{figure*}

\subsection{High Resolution and Refinement Scheme}
\label{sec:convergence_2} 
\smallskip
\Fig{high_res_convergence} compares results of simulations with our fiducial resolution of 128 
cells per stream diameter, $\Delta=\Rs/64$, to simulations with both half and double the resolution, 
$\Delta=\Rs/32$ and $\Rs/128$. For the fiducial resolution case, we also explore the effect of changing 
our refinement scheme from doubling the cell size every $3\Rs$ in the $x$ and $y-$directions to doubling 
the cell size every $1.5\Rs$. We show results for $(\Mb,\delta)=(1,1)$ and $(2,10)$, each with our fiducial 
smoothing of $\sigma=\Rs/32$ in \equ{ramp2}, and also for $(\Mb,\delta)=(2,10)$ with wider smoothing of 
$\sigma=\Rs/8$ as used in \se{body}. As in \fig{low_res_convergence}, we show the stream deceleration 
(top), and the shear layer growth in the background (middle) and the stream (bottom). 

\smallskip
In all cases, changing the refinement scheme has no affect on our results, as can be seen by comparing the 
red dashed lines to the cyan dot-dashed lines. The case $(\Mb,\delta)=(1,1)$ is converged. While the shear 
layer begins growing slightly later with low resolution compared to fiducial resolution, as seen in 
\fig{low_res_convergence}, this is not evident when comparing the fiducial resolution to higher resolution. 
The case $(\Mb,\delta)=(2,10)$ with $\sigma=\Rs/32$ is not converged even at our highest resolution. The trend 
is the same as described above based on \fig{low_res_convergence}, namely higher resolution runs evolve slower 
since they better resolve the turbulent cascade for $m\sim 12$ azimuthal modes, resulting in less power on 
large scales to drive the shear layer growth. However, when $\sigma=\Rs/8$ the $(\Mb,\delta)=(2,10)$ simulations 
have converged already at our low resolution. The instability in this case is dominated by the $m=4$ surface mode 
(\fig{smoothing_panel}) and the $m=1$ body mode, and thus involves larger structures which are easier to resolve. 
We conclude that our body mode simulations with wide smoothing layers presented in \se{body} are converged, while 
those with small smoothing layers evolve too quickly and have not converged.

%%%%%%%%%%%%%%%%%%%%%%%%%%%%%%%%%%%%%%%%%%%%%%%%  
\section{Evaluating Interface Distortions Corresponding to a Given Radial Velocity Amplitude}
\label{sec:ur_to_h}
When discussing \fig{h_body}, we wished to evaluate $t_{\rm NL}$ (\equnp{tNL}) for our simulations with 
$\sigma=\Rs/8$. However, this requires knowing the initial dispacement of the stream-background interface, $H_0$, 
while our fiducial simulations were initialized with perturbations to their radial velocity. We therefore wish 
to evaluate the amplitude of perturbations to the stream-background interface which correspond to a given amplitude 
of perturbations in radial velocity. In the linear regime, the relation between the interface distortion, $H_0$, and 
the radial velocity at the boundary inside the stream, $u_{\rm r}$, is (see equation 13 in M16 or equation C3 in P18) 
\be 
\label{eq:ur_to_h}
H_0 = u_{\rm r}\cdot {\rm REAL}\left(\frac{i}{Vk-\omega}\right),
\ee
{\no}where $V$ is the stream velocity, $k$ is the wavenumber, and $\omega$ is the complex frequency of the eigenmode.
$H_0$ and $u_{\rm r}$ are taken to be real, and represent the perturbation amplitudes. Writing $\omega=\omega_{\rm R}+i\omega_{\rm I}$ 
(where $\omega_{\rm I}=t_{\rm KH}^{-1}$) and $Vk=2\delta^{1/2}\Mb k\Rs \tsc^{-1}$ we obtain 
\be 
\label{eq:h_amp}
\frac{H_0}{\Rs}=\frac{u_{\rm r}}{\cs}\frac{2\omega_{\rm I}\tsc}{\left(\omega_{\rm I}\tsc\right)^2+\left(2\delta^{1/2}\Mb k\Rs-\omega_{\rm R}\tsc\right)^2}.
\ee

\smallskip
We wish to focus on the critical wavelength which leads to stream disruption via body modes. 
The values of $k\Rs$ and $\omega_{\rm I}\tsc$ for our five simulations with 
$(\Mb,\delta)=(5.0,1),\:(2.5,5),\:(2.0,10),\:(2.5,20)$, and $(2.0,100)$, can be inferred from \tab{body2}. 
In the order presented in \tab{body2} these are $k\Rs\sim 0.49,\:0.84,\:0.75,\:0.62,\:0.57$, and 
$\omega_{\rm I}\tsc\sim 1.33,\:1.82,\:1.85,\:1.88,\:1.96$. The corresponding values of $\omega_{\rm R}\tsc$ 
are $\sim 2.97,\:7.21,\:7.88,\:12.20,\:21.46$. Given $u_{\rm r}/\cs$, these can be used to evaluate $H_0/\Rs$ 
using \equ{h_amp}, which can then be used to evaluate $t_{\rm NL}$ using \equ{tNL}. Assuming that $u_{\rm r}/\cs$ 
at the critical wavelength is constant among all simulations\footnote{which is not in fact the case, as explained in 
\se{body} while discussing \fig{h_body}} and defining the value of $t_{\rm NL}$ for the case $(\Mb,\delta)=(5.0,1)$ as 
unity, we obtain for the the remaining four simulations $t_{\rm NL}\sim 0.73,\:0.68,\:0.67,\:0.62$.

%%%%%%%%%%%%%%%%%%%%%%%%%%%%%%%%%%%%%%%%%%%%%%%%  
\section{Stream Morphology for high-$\delta$}
\label{sec:add_figures} 

\begin{figure*}
\begin{center}
\includegraphics[trim={0.0cm 1.238cm 3.3cm 0}, clip, width =0.393 \textwidth]{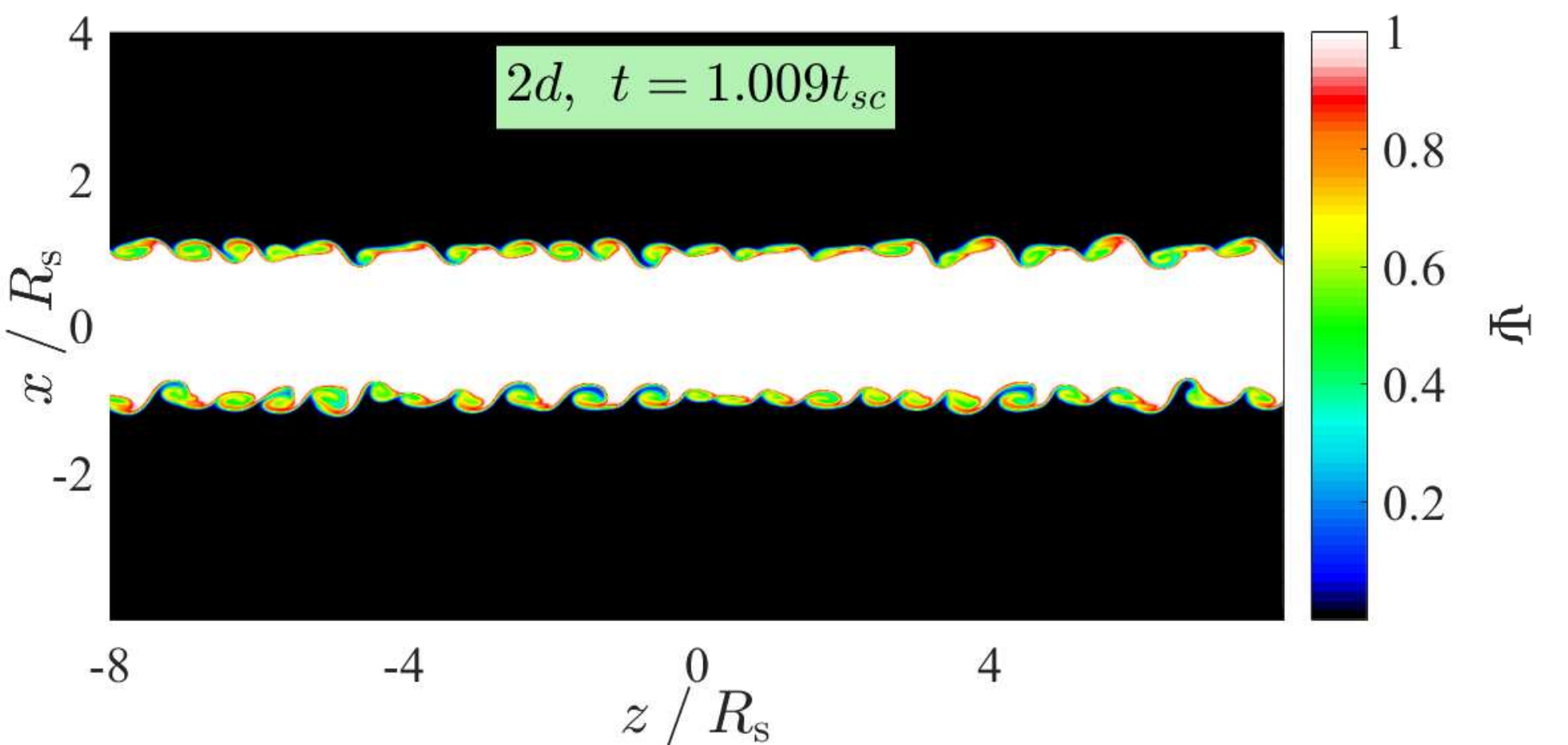}
\hspace{-0.3cm}
\includegraphics[trim={1.3cm 1.238cm 3.3cm 0}, clip, width =0.363 \textwidth]{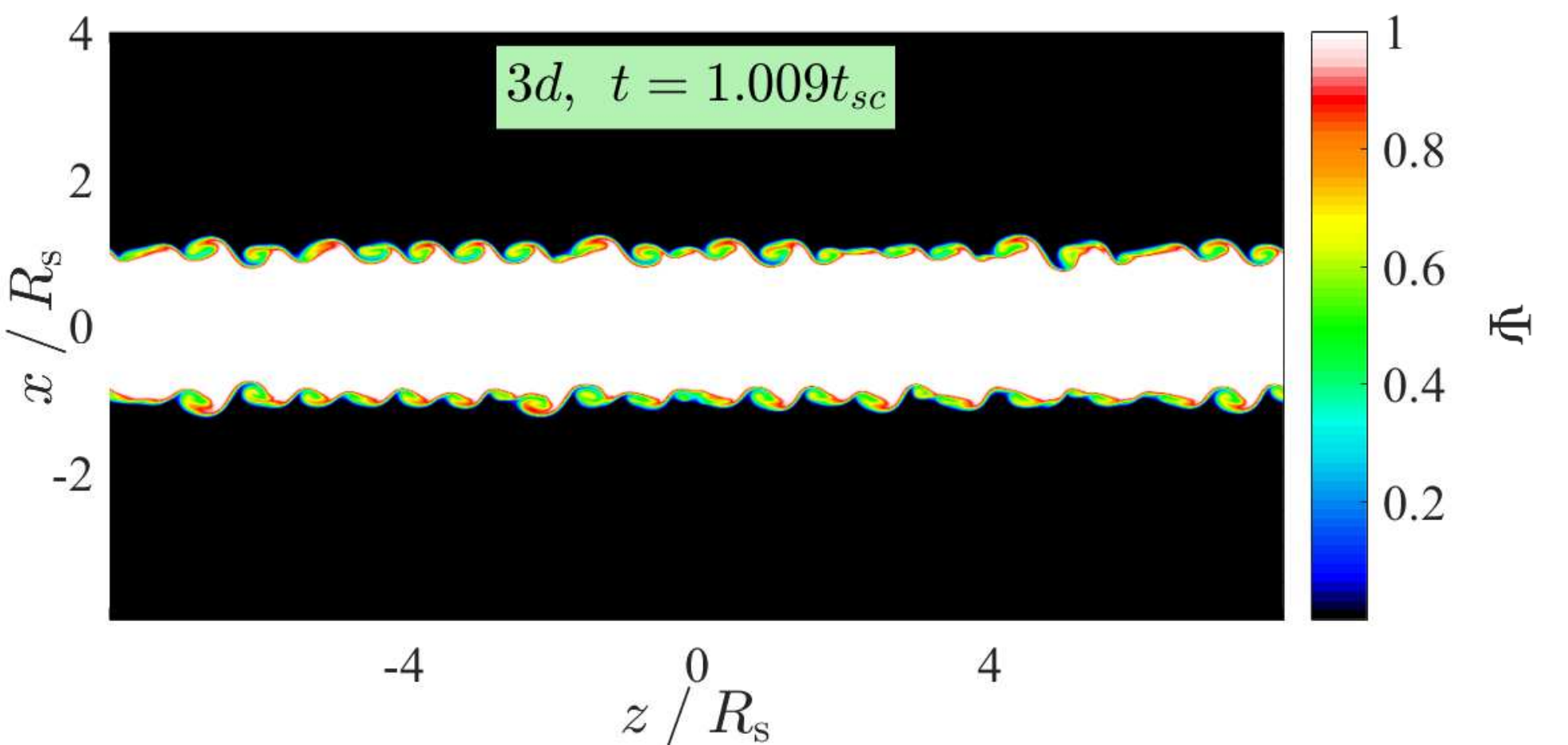}
\hspace{-0.29cm}
\includegraphics[trim={5.1cm 1.238cm 4.05cm 0}, clip, width =0.257 \textwidth]{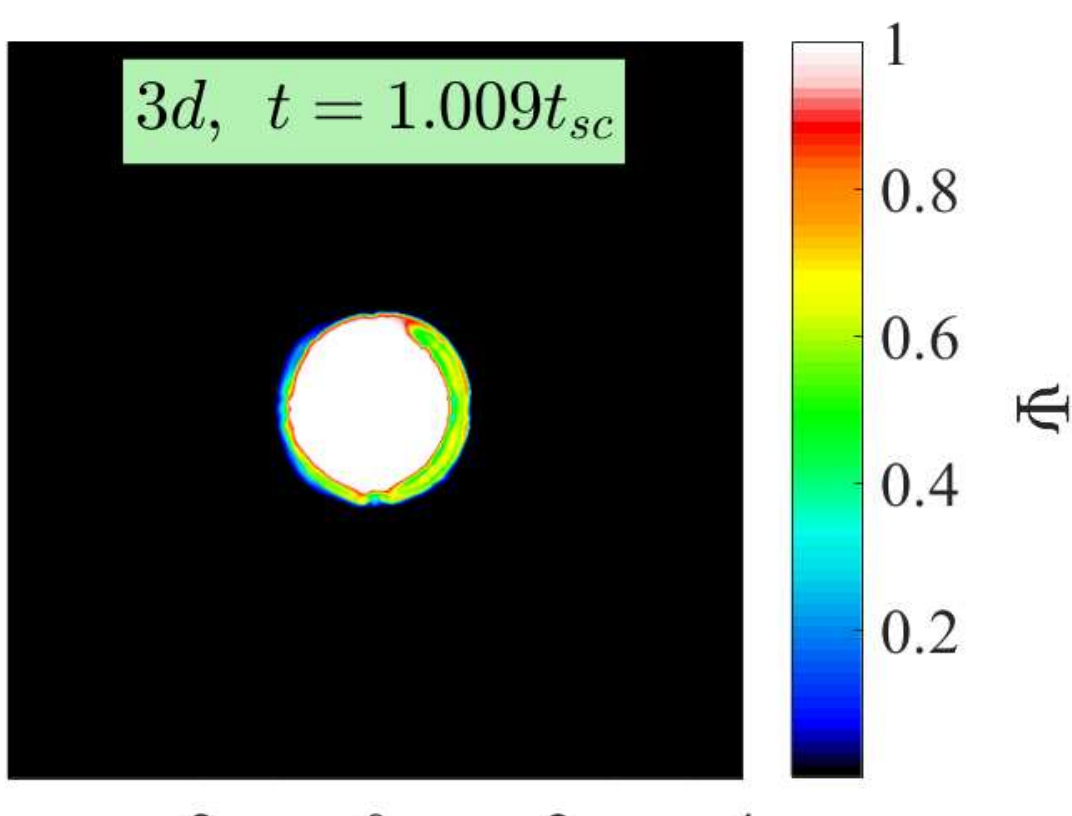}\\
\vspace{-0.09cm}
\includegraphics[trim={0.0cm 1.238cm 3.3cm 0.22cm}, clip, width =0.393 \textwidth]{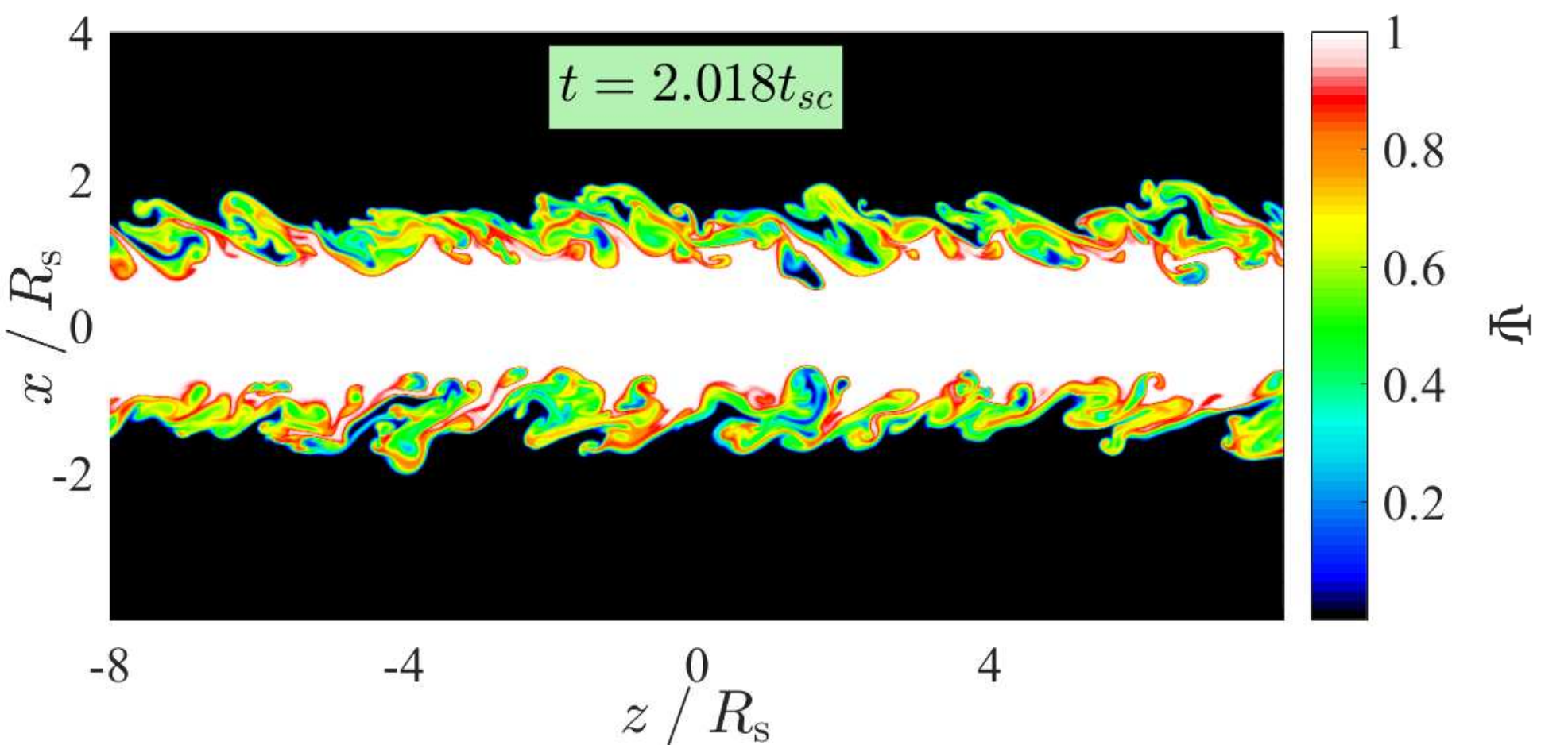}
\hspace{-0.3cm}
\includegraphics[trim={1.3cm 1.238cm 3.3cm 0.22cm}, clip, width =0.363 \textwidth]{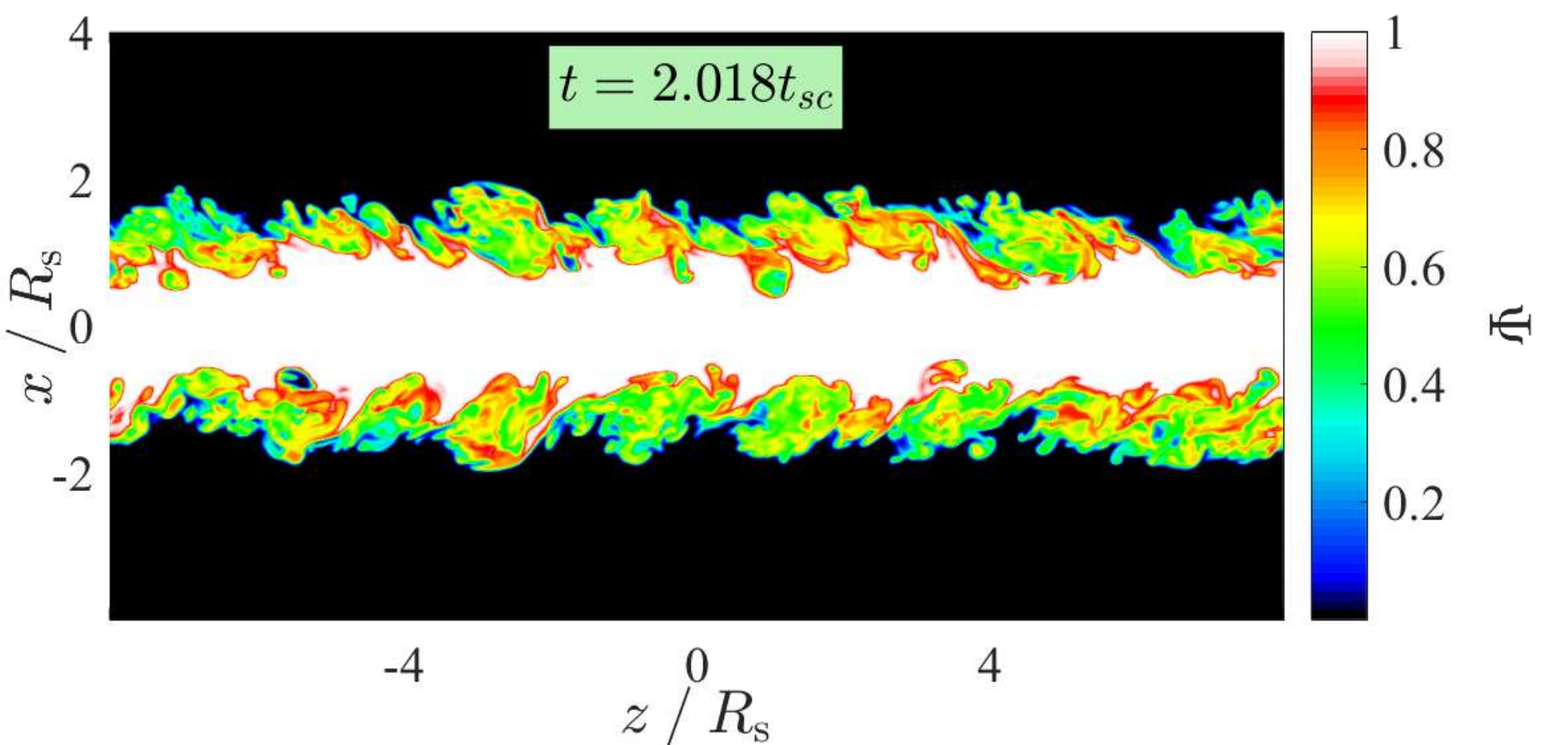}
\hspace{-0.29cm}
\includegraphics[trim={5.1cm 1.238cm 4.05cm 0.22cm}, clip, width =0.257 \textwidth]{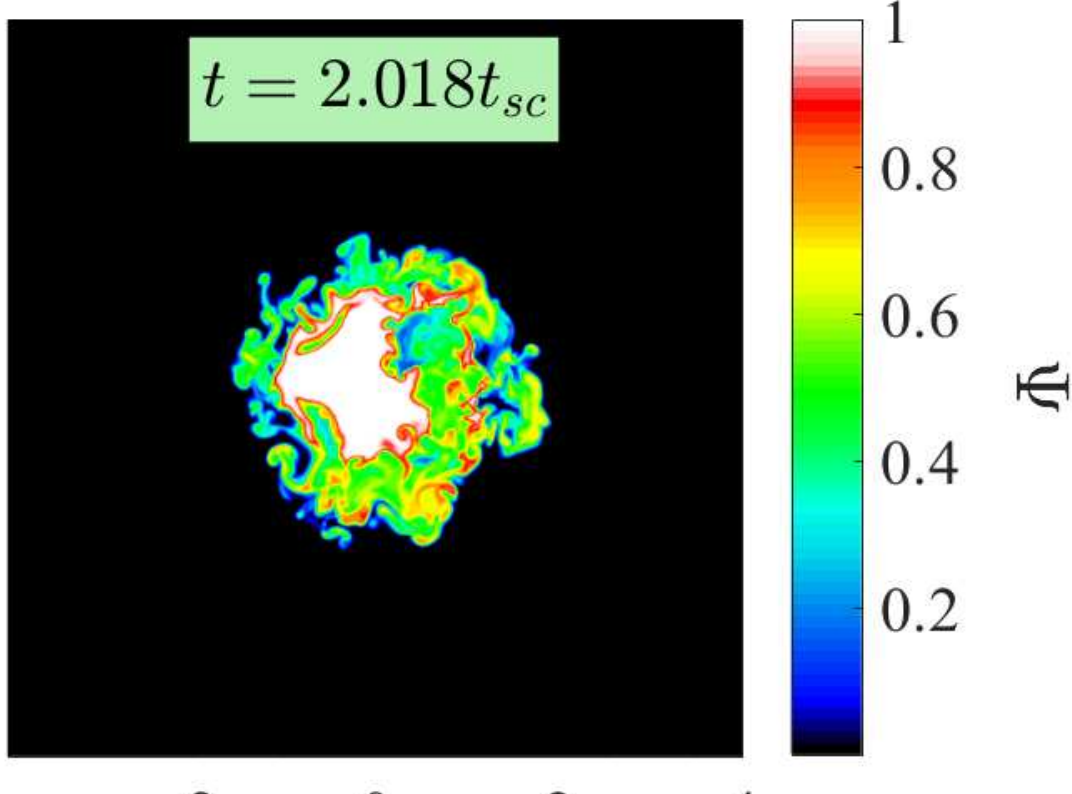}\\
\vspace{-0.09cm}
\includegraphics[trim={0.0cm 1.238cm 3.3cm 0.22cm}, clip, width =0.393 \textwidth]{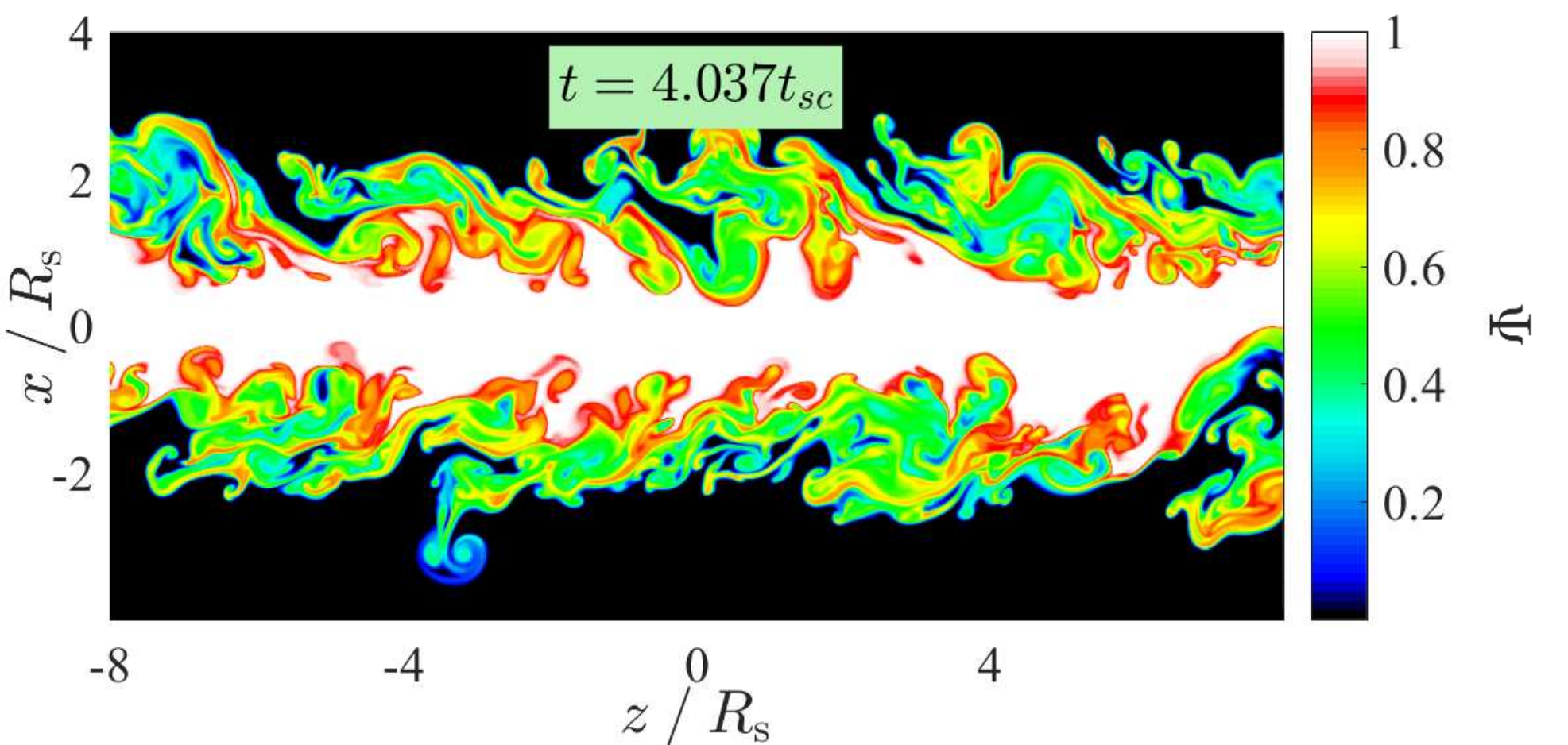}
\hspace{-0.3cm}
\includegraphics[trim={1.3cm 1.238cm 3.3cm 0.22cm}, clip, width =0.363 \textwidth]{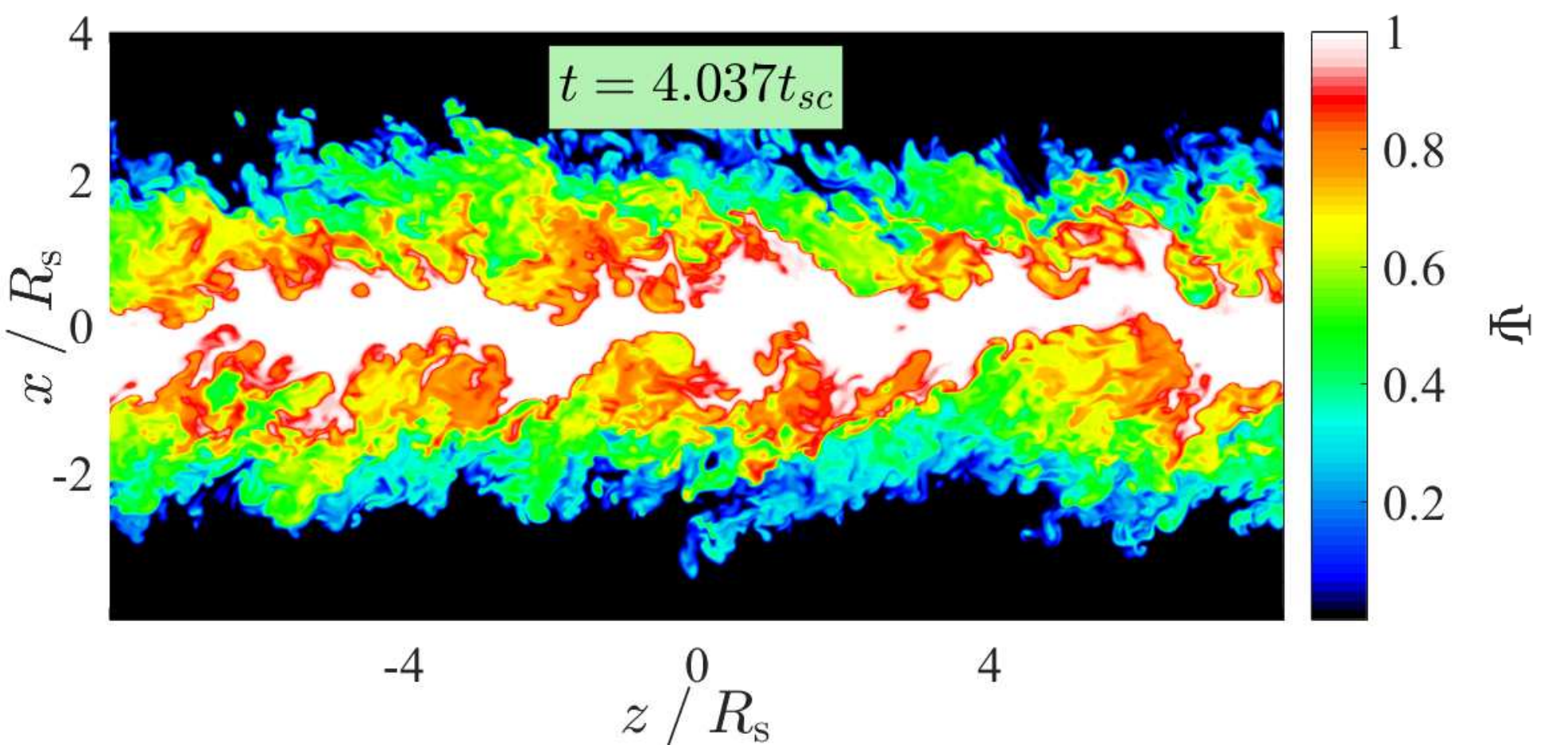}
\hspace{-0.29cm}
\includegraphics[trim={5.1cm 1.238cm 4.05cm 0.22cm}, clip, width =0.257 \textwidth]{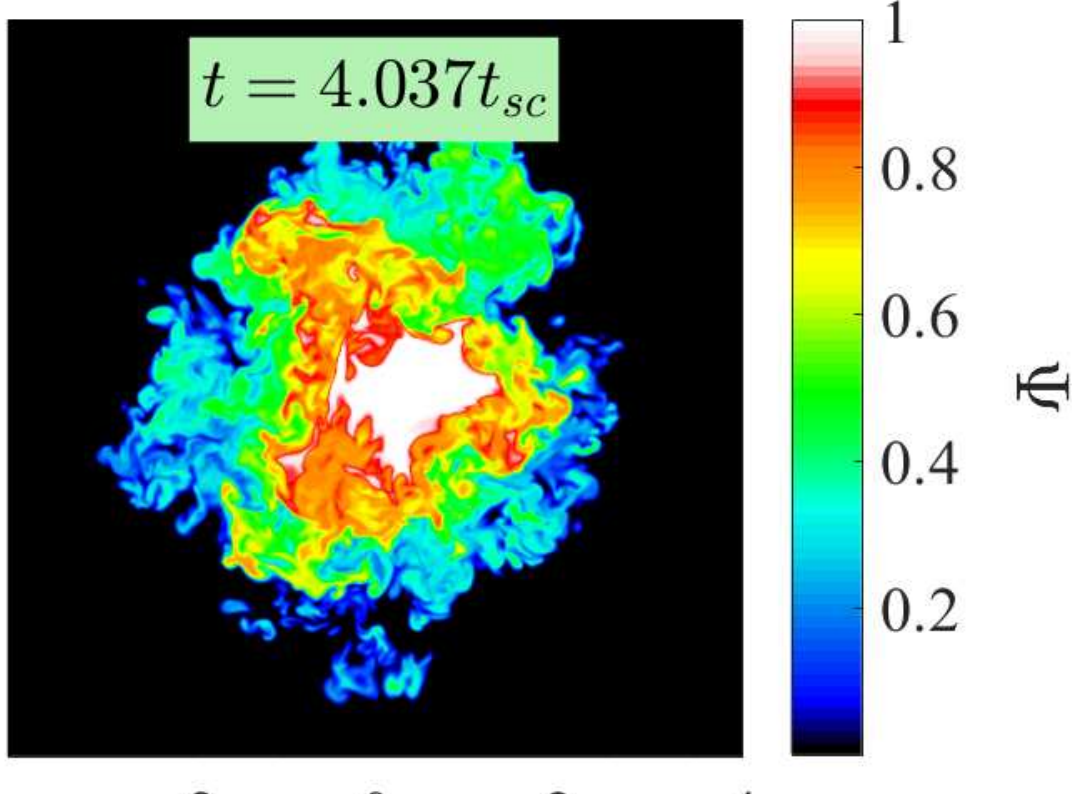}\\
\vspace{-0.09cm}
\includegraphics[trim={0.0cm 0.0cm 3.3cm 0.22cm}, clip, width =0.393 \textwidth]{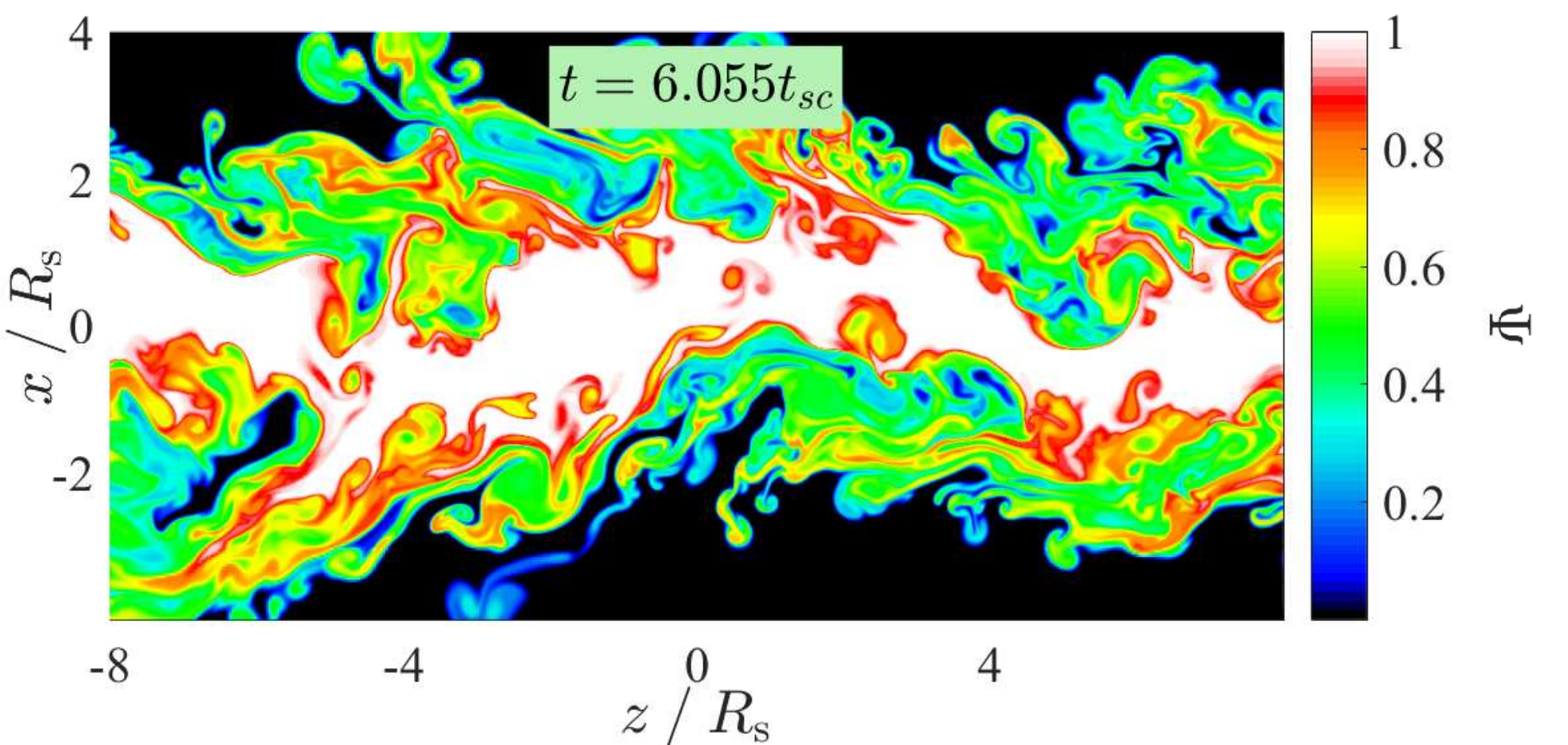}
\hspace{-0.3cm}
\includegraphics[trim={1.3cm 0.0cm 3.3cm 0.22cm}, clip, width =0.363 \textwidth]{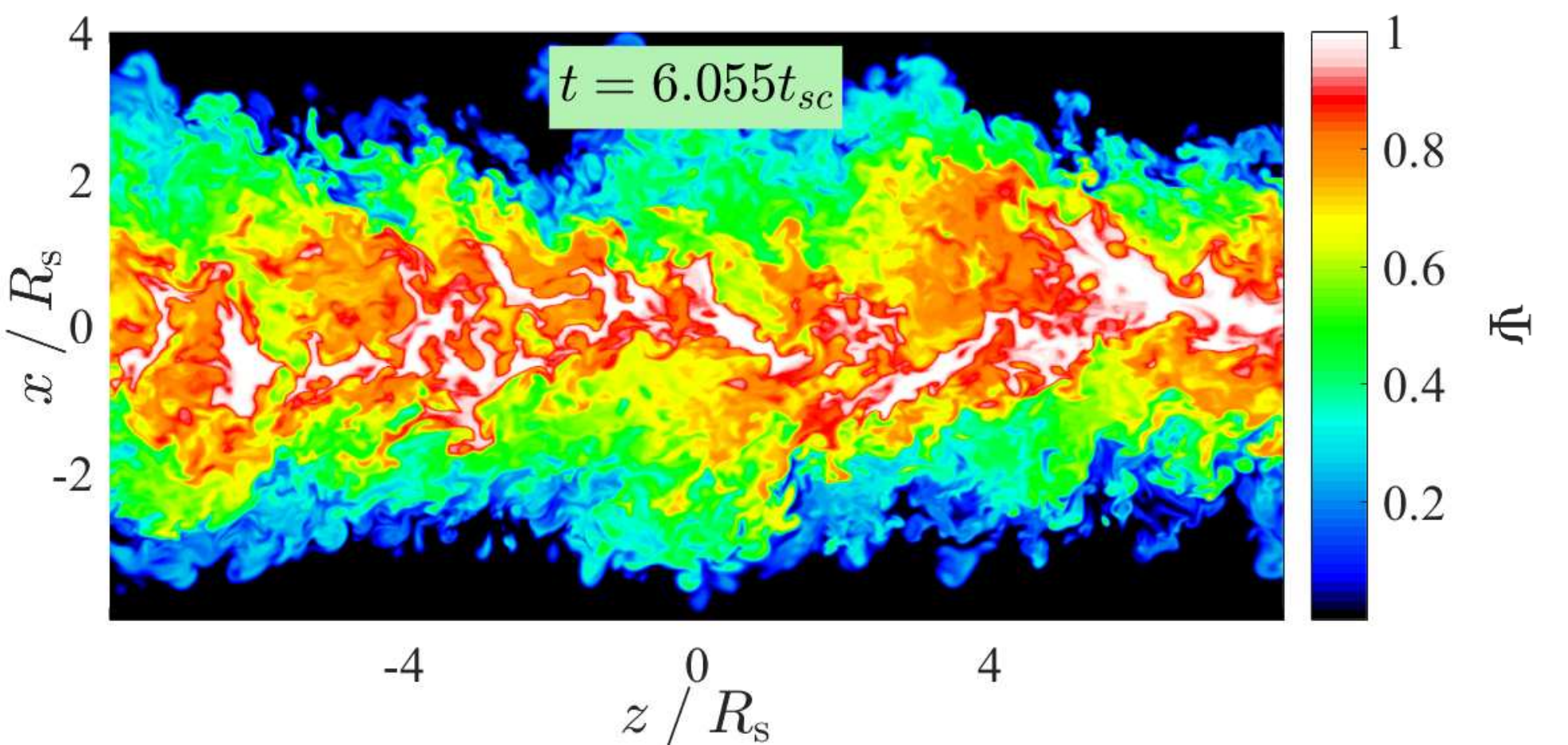}
\hspace{-0.29cm}
\includegraphics[trim={5.1cm 0.0cm 4.05cm 0.22cm}, clip, width =0.257 \textwidth]{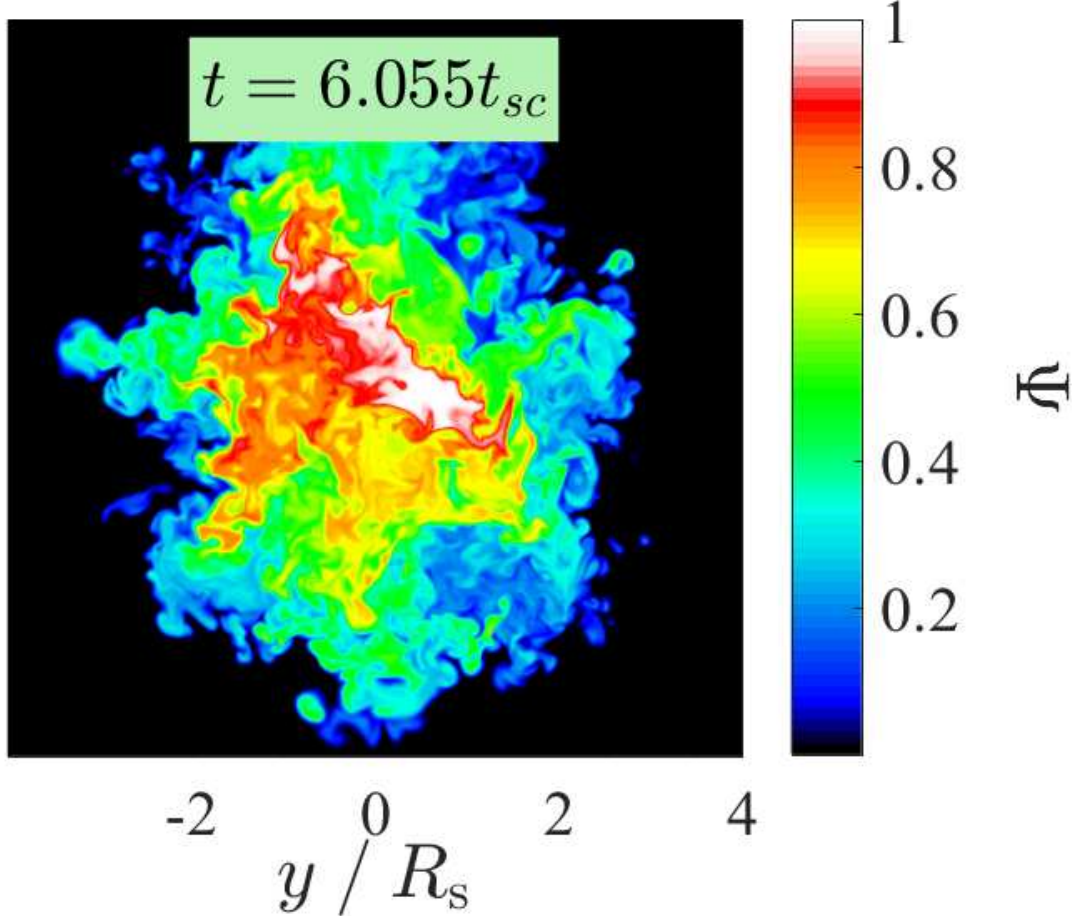}
\end{center}
\caption{Same as \fig{colour_panel_M1D1} but for a simulation with $(\Mb,\delta)=(1,10)$. 
We show snapshots at $t=1,\,2,\,4,$ and $6\tsc$. As for the case with $\delta=1$, at $t\sim \tsc$ 
the edge-on distribution of $\psi$ in the 3d cylinder appears nearly identical to its distribution 
in the 2d slab simulation, while the face-on distribution is dominated by symmetric, $m=0$, and 
antisymmetric, $m=1$, modes. At $t=2\tsc$, the face-on view of the cylinder is dominated by higher-order 
azimuthal modes, though the edge-on view remains qualitatively similar to the slab despite the appearance 
of small scale modes. At later times, the 2d slab is still dominated by large scale eddies, while in 3d 
cylinders the largest eddies have broken up and generated small-scale turbulence, leaving very little 
unmixed fluid in the stream. By $6\tsc$ there is very little unmixed fluid in the 3d stream, while the 2d 
slab still has a large unmixed core. The break up of the largest eddies and the appearance of small scale 
modes occurs after a smaller number of stream sound crossing times compared to the case with $\delta=1$ 
shown in \fig{colour_panel_M1D1}. This is because, as discussed in \se{surface}, this occurs when $\hb\sim 2\Rs$ 
which happens after a smaller number of stream sound crossing times for denser streams with longer sound 
crossing times.
}
\label{fig:colour_panel_M1D10} 
\end{figure*}

\begin{figure*}
\begin{center}
\includegraphics[trim={0.0cm 1.238cm 3.3cm 0}, clip, width =0.393 \textwidth]{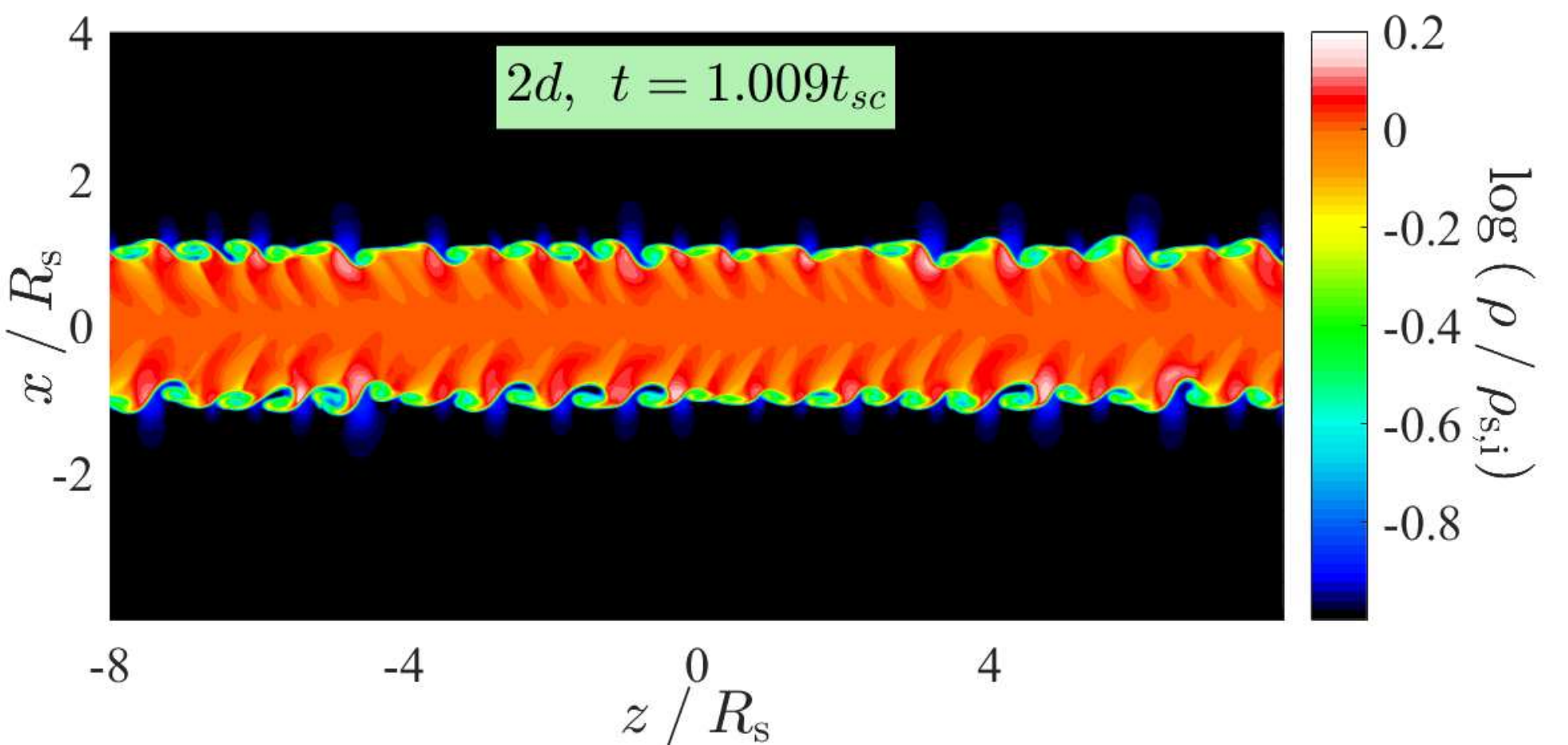}
\hspace{-0.3cm}
\includegraphics[trim={1.3cm 1.238cm 3.3cm 0}, clip, width =0.363 \textwidth]{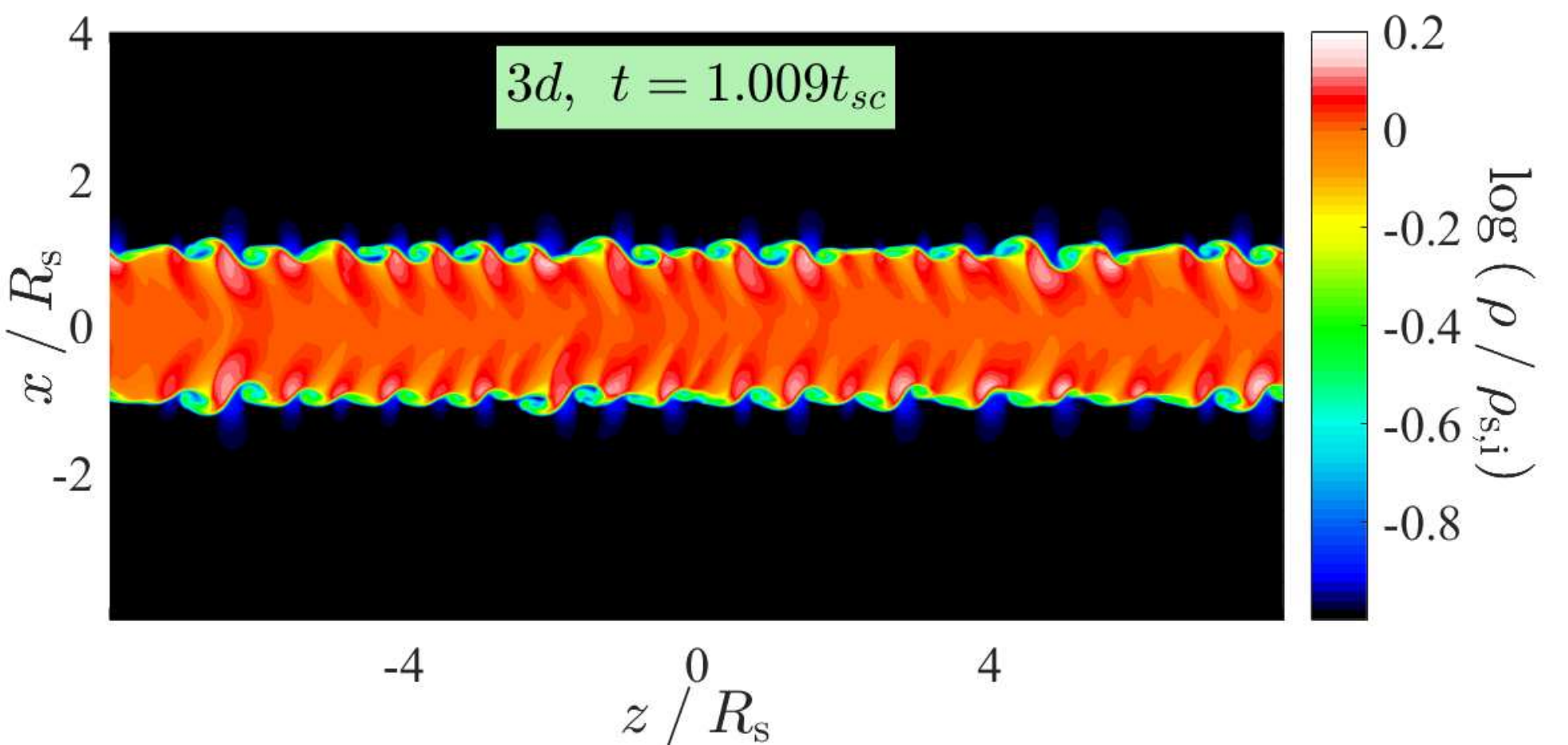}
\hspace{-0.29cm}
\includegraphics[trim={5.1cm 1.238cm 4.05cm 0}, clip, width =0.257 \textwidth]{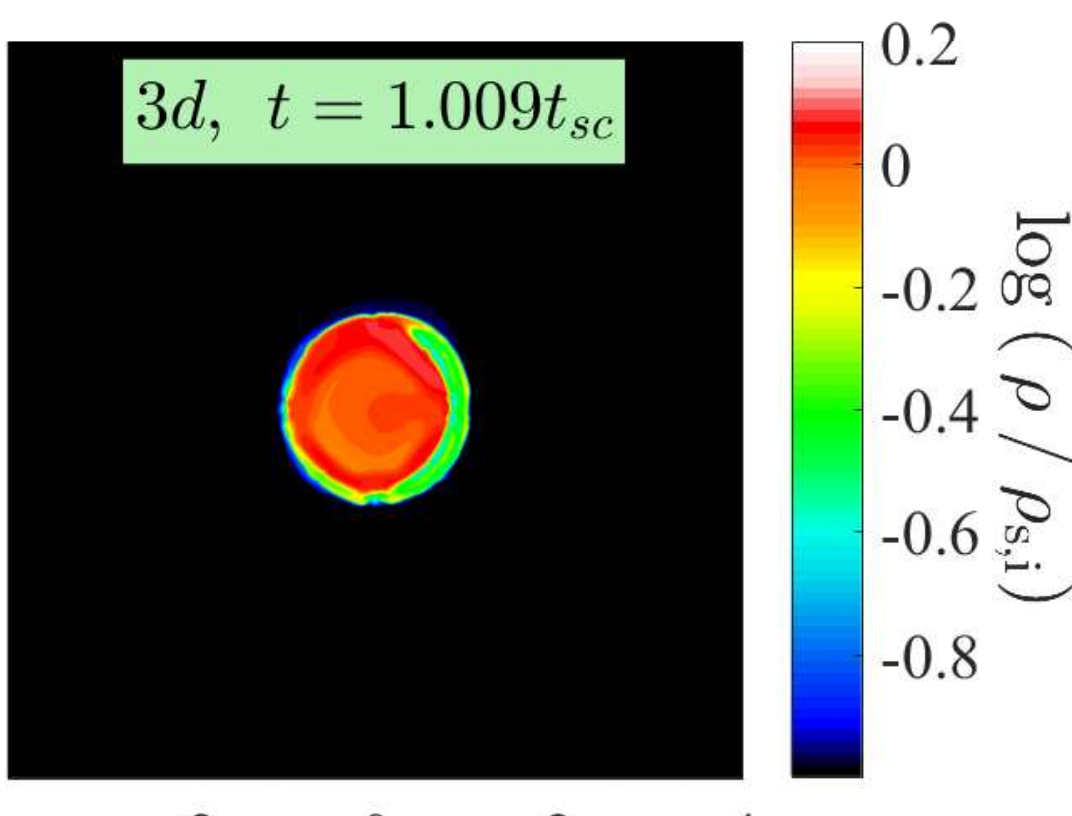}\\
\vspace{-0.09cm}
\includegraphics[trim={0.0cm 1.238cm 3.3cm 0.22cm}, clip, width =0.393 \textwidth]{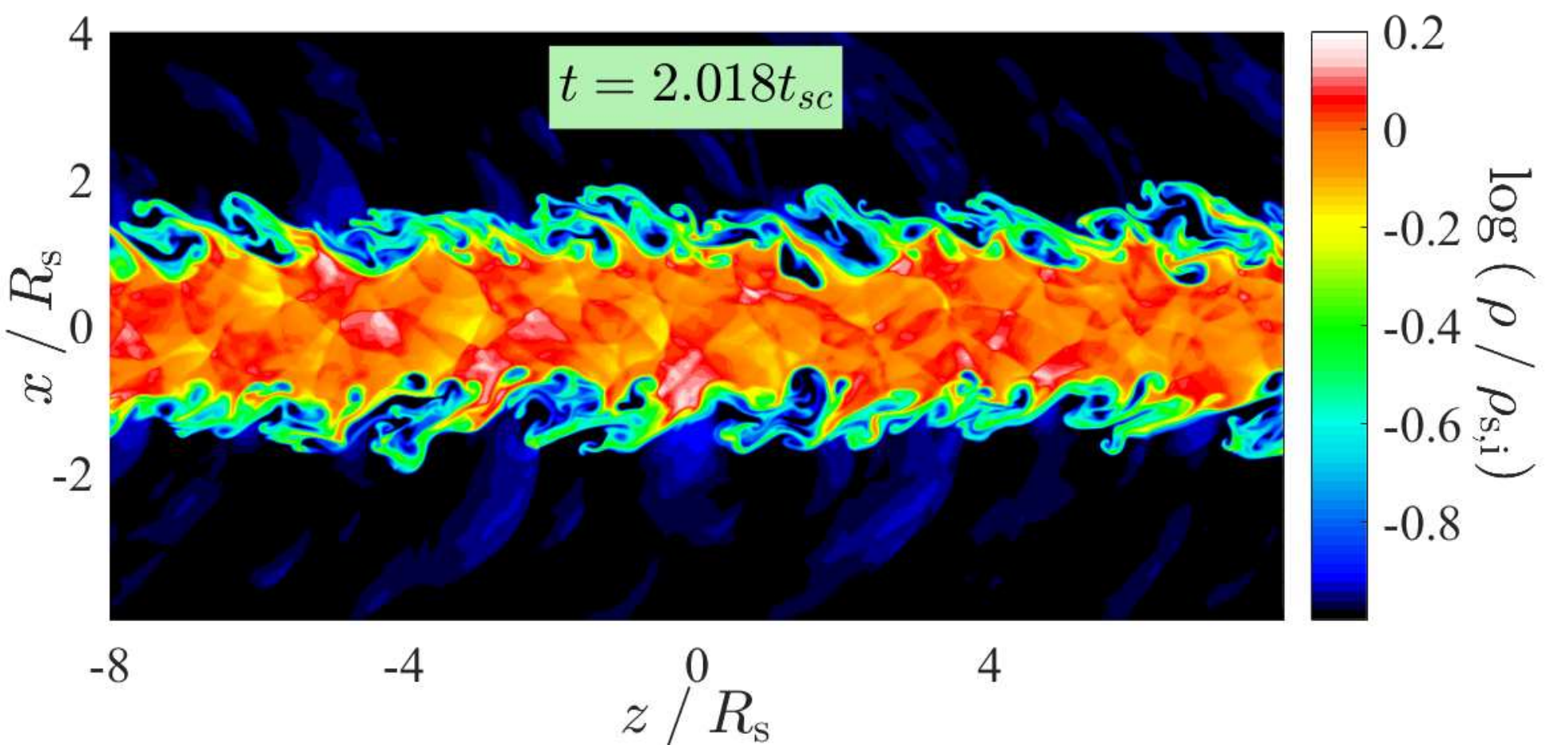}
\hspace{-0.3cm}
\includegraphics[trim={1.3cm 1.238cm 3.3cm 0.22cm}, clip, width =0.363 \textwidth]{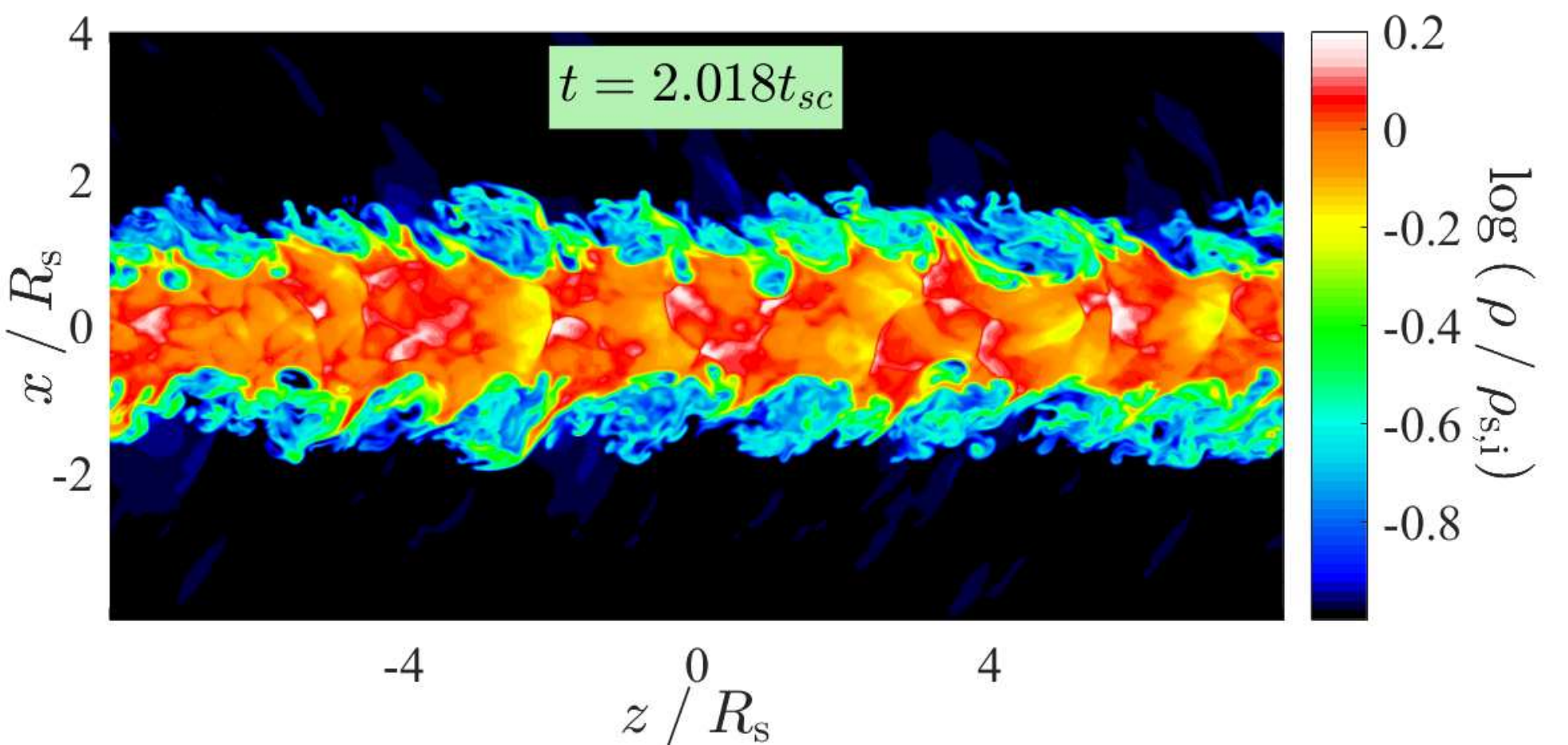}
\hspace{-0.29cm}
\includegraphics[trim={5.1cm 1.238cm 4.05cm 0.22cm}, clip, width =0.257 \textwidth]{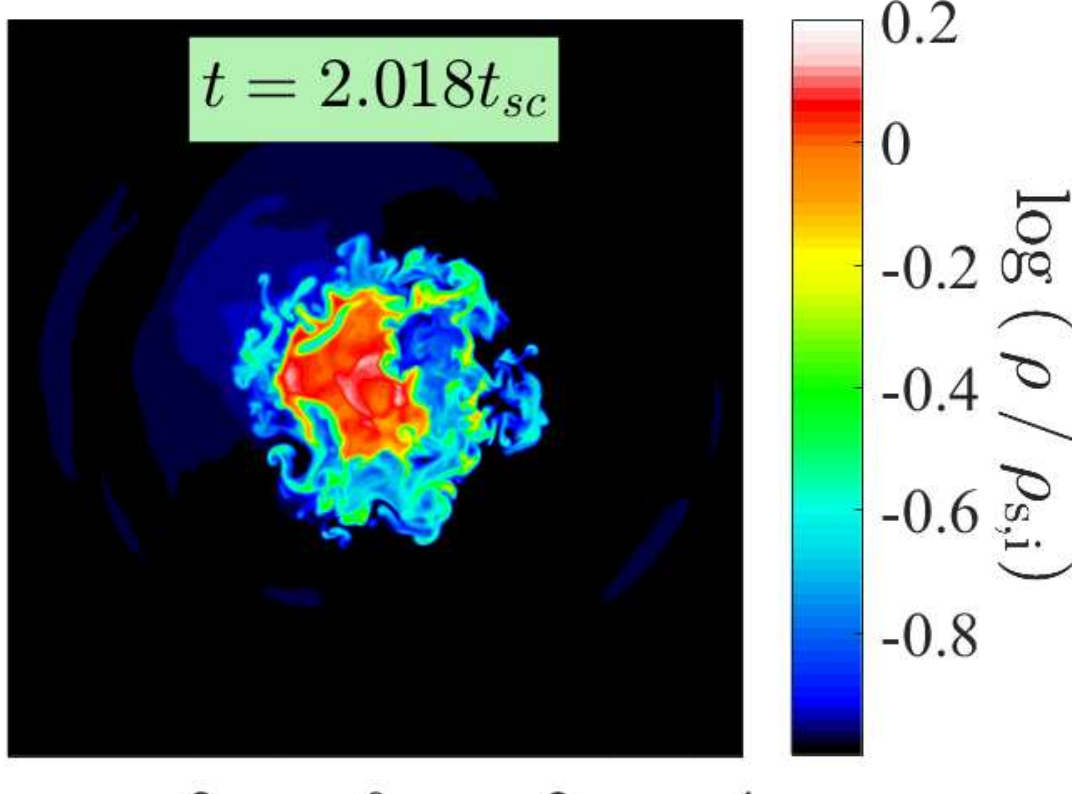}\\
\vspace{-0.09cm}
\includegraphics[trim={0.0cm 1.238cm 3.3cm 0.22cm}, clip, width =0.393 \textwidth]{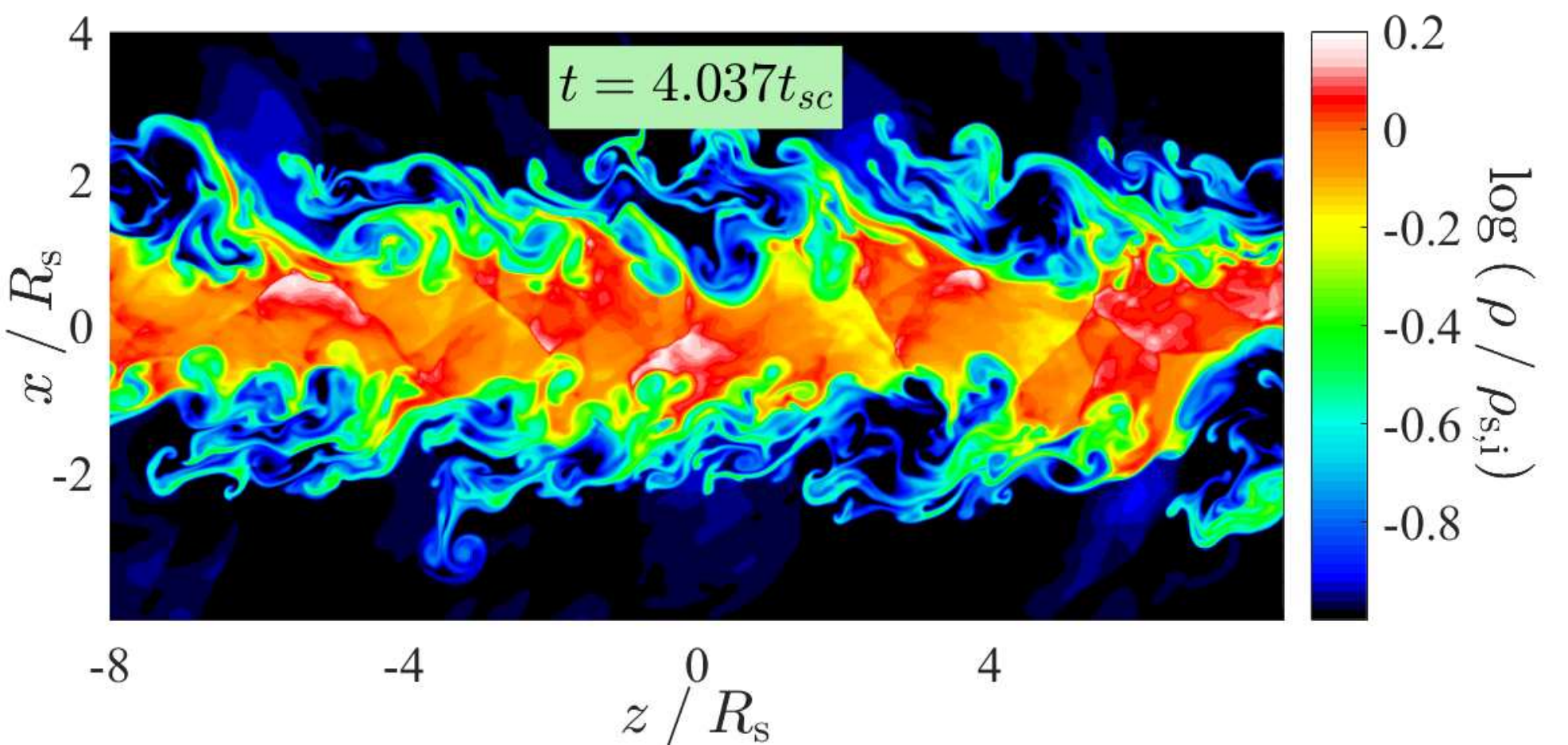}
\hspace{-0.3cm}
\includegraphics[trim={1.3cm 1.238cm 3.3cm 0.22cm}, clip, width =0.363 \textwidth]{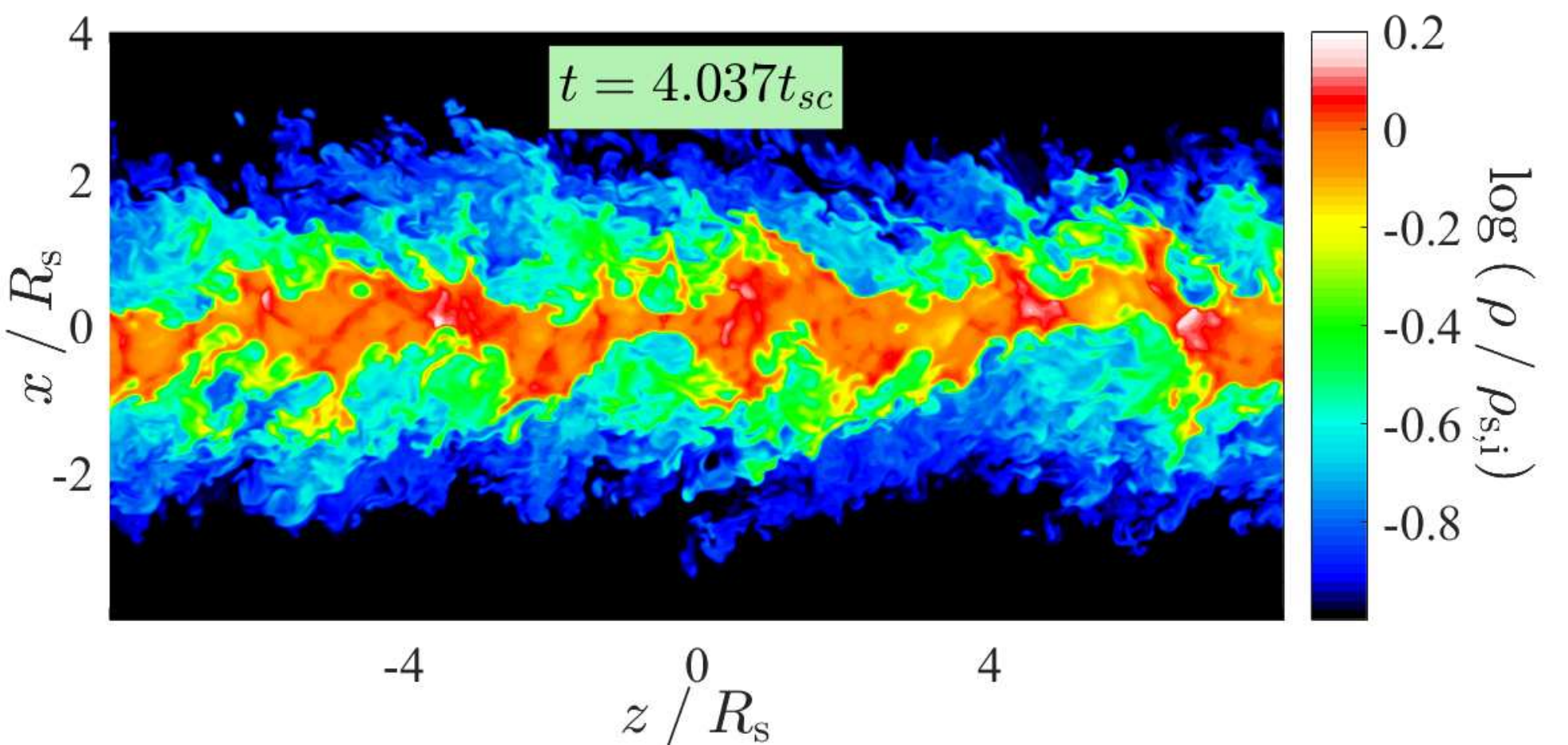}
\hspace{-0.29cm}
\includegraphics[trim={5.1cm 1.238cm 4.05cm 0.22cm}, clip, width =0.257 \textwidth]{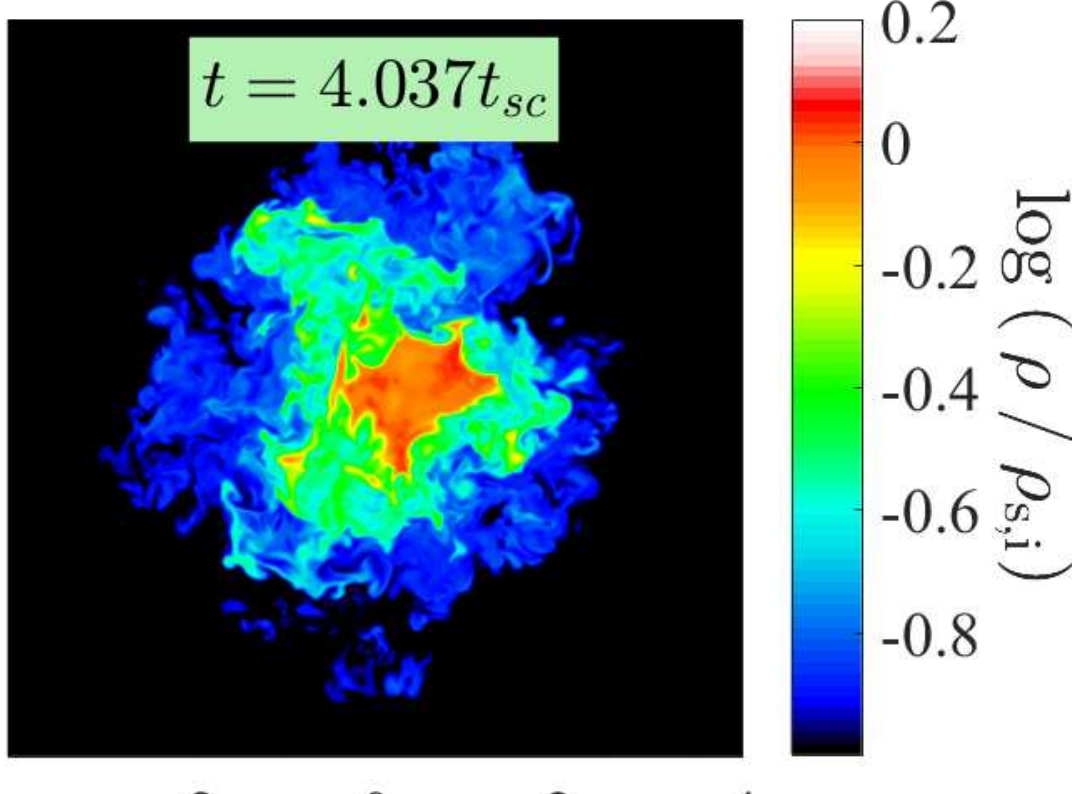}\\
\vspace{-0.09cm}
\includegraphics[trim={0.0cm 0.0cm 3.3cm 0.22cm}, clip, width =0.393 \textwidth]{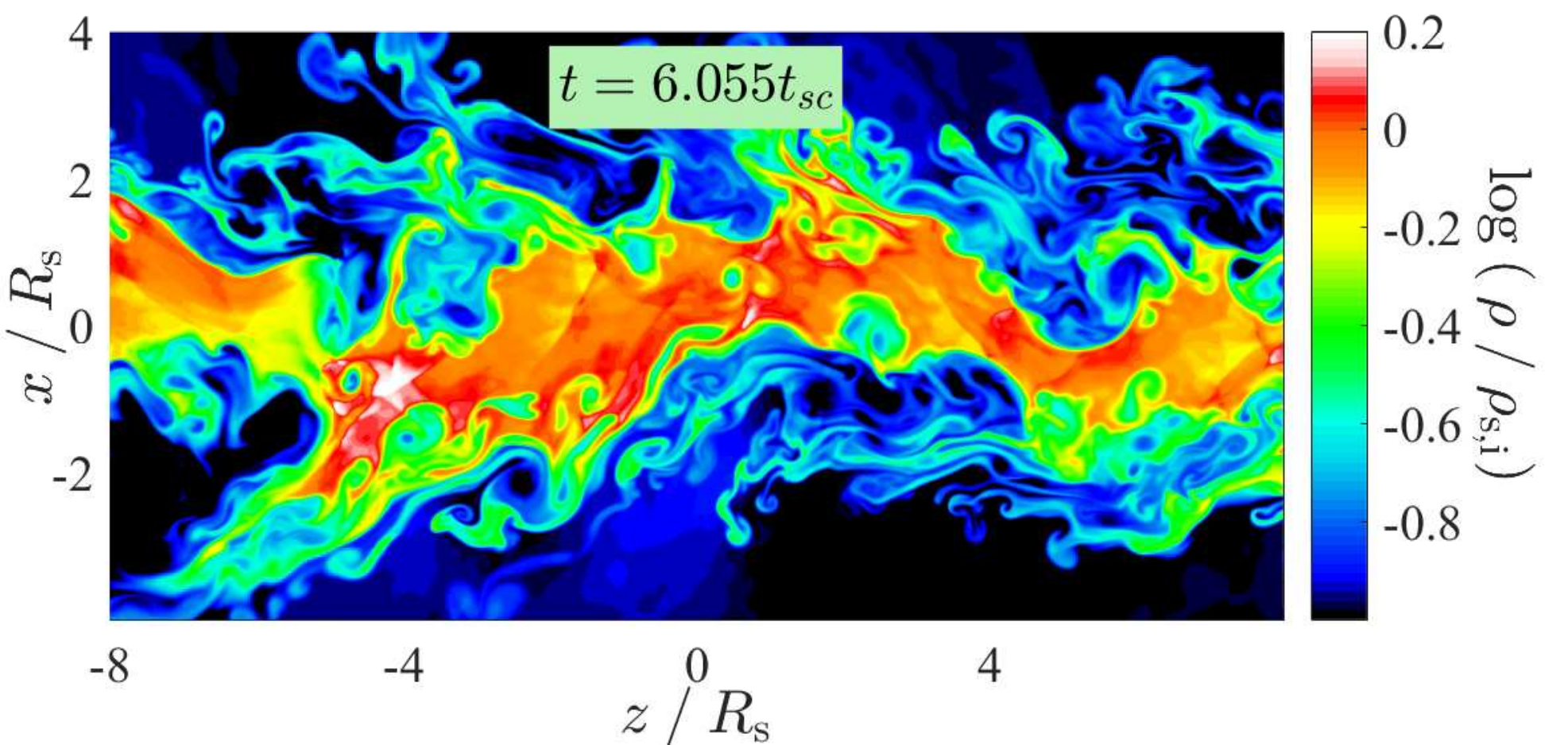}
\hspace{-0.3cm}
\includegraphics[trim={1.3cm 0.0cm 3.3cm 0.22cm}, clip, width =0.363 \textwidth]{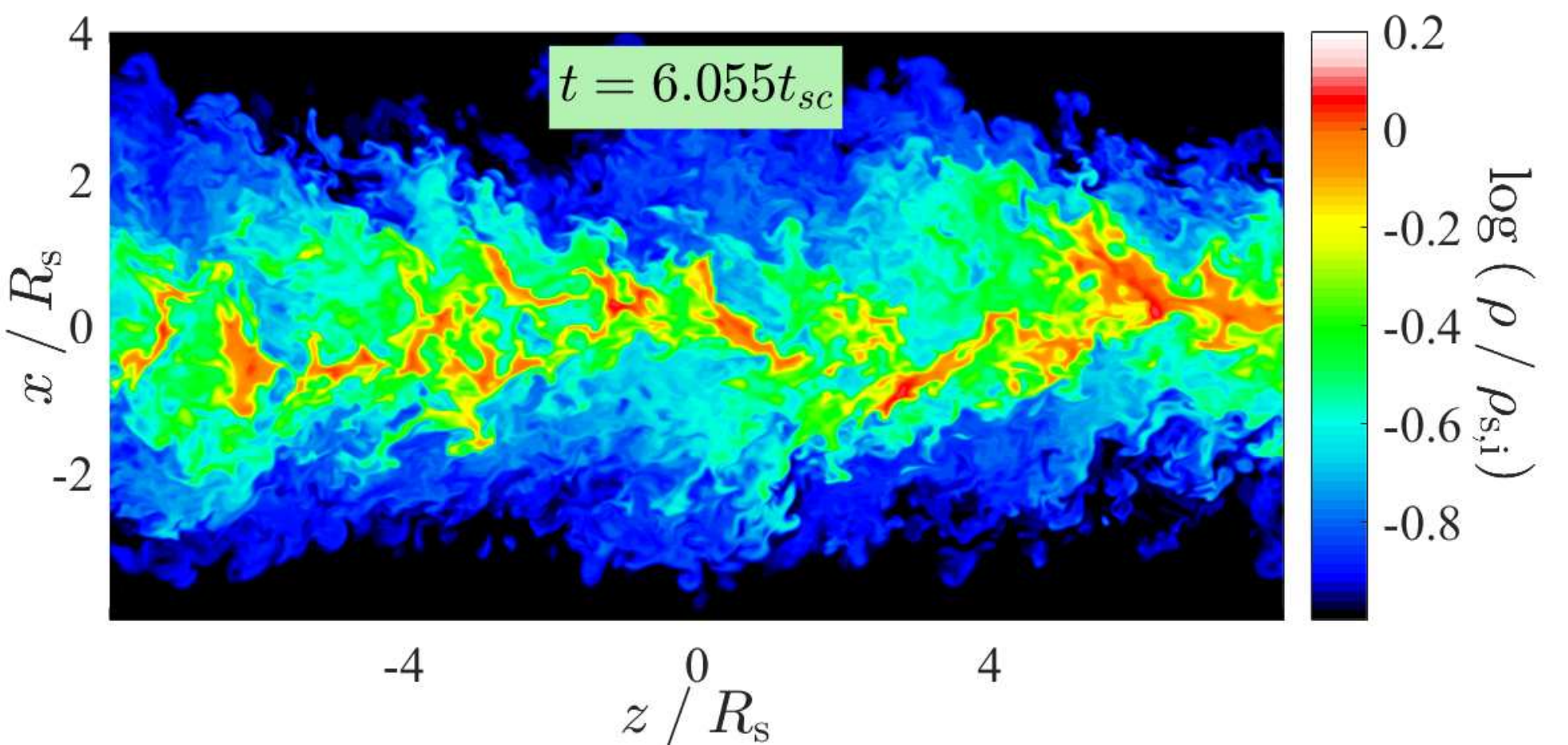}
\hspace{-0.29cm}
\includegraphics[trim={5.1cm 0.0cm 4.05cm 0.22cm}, clip, width =0.257 \textwidth]{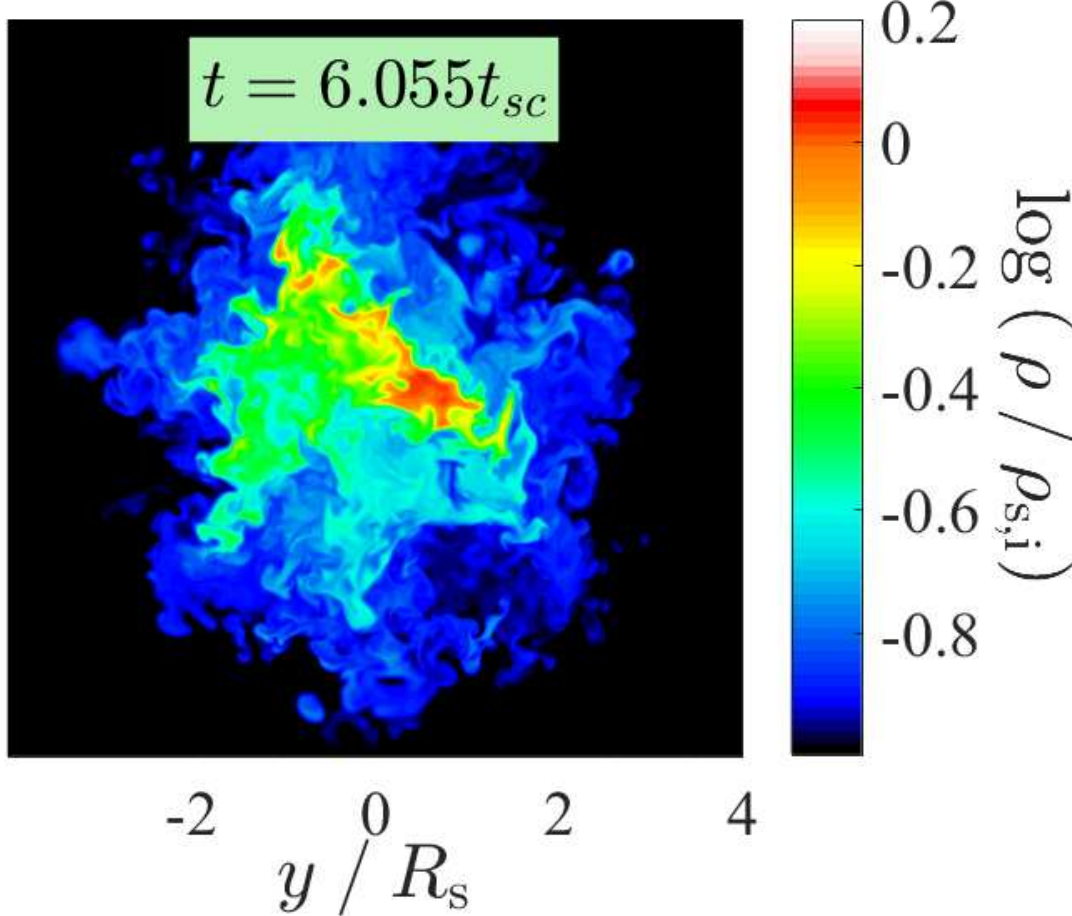}
\end{center}
\caption{Same as \fig{colour_panel_M1D10} but where the colour shows the density relative to the initial 
stream density, $\rho/\rhos$, rather than the passive scalar $\Psi$. At $t\sim \tsc$ the density perturbations 
concentrated near the stream-background interface, characteristic of surface modes. At later times, there are 
density perturbations in the stream interior as well, driven in part by weak shocks. In the 3d cylinder at late 
times, the dense regions in the stream interior correspond to the unmixed regions seen in \fig{colour_panel_M1D10}.
}
\label{fig:density_panel_M1D10} 
\end{figure*}

\begin{figure*}
\begin{center}
\includegraphics[trim={0.0cm 1.238cm 3.3cm 0}, clip, width =0.393 \textwidth]{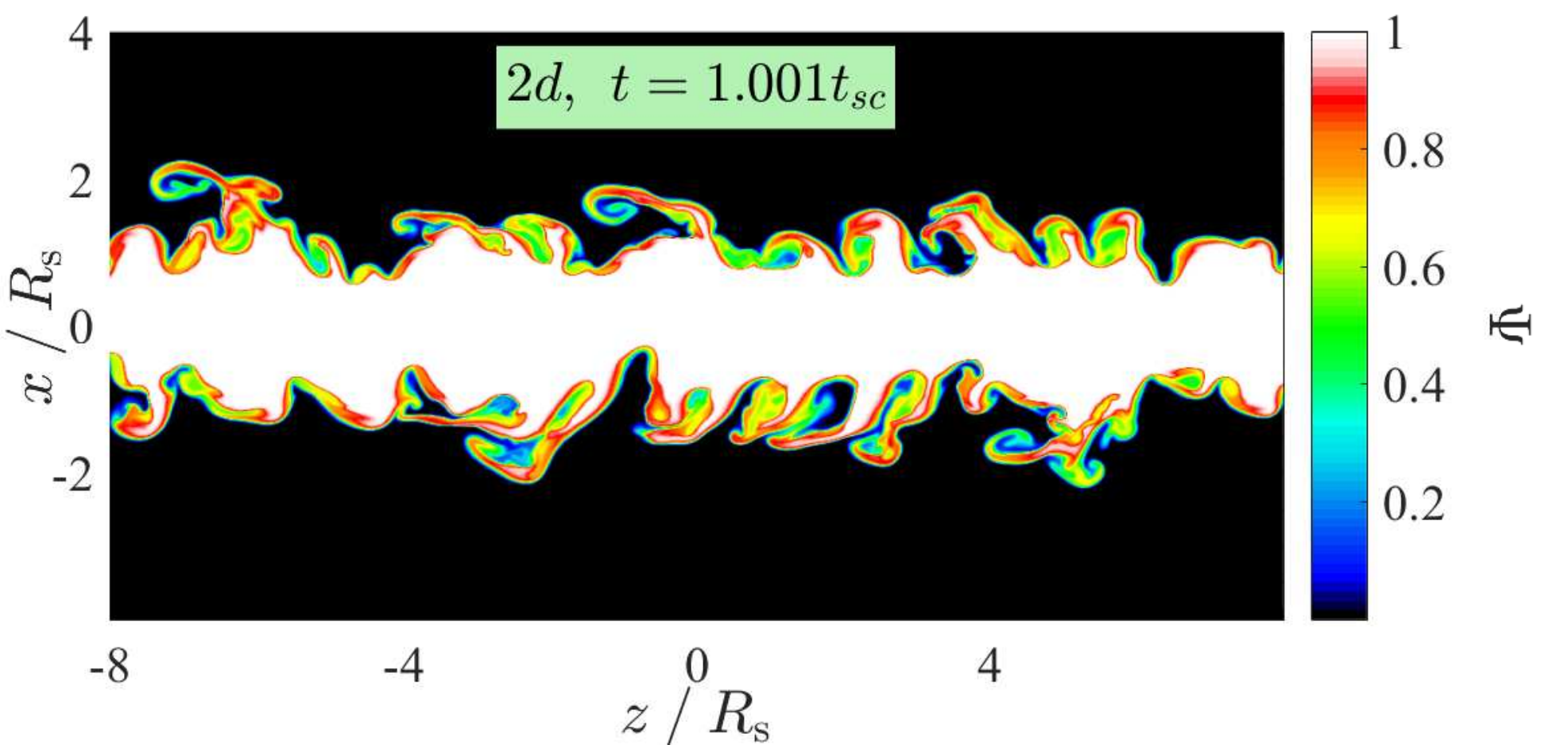}
\hspace{-0.3cm}
\includegraphics[trim={1.3cm 1.238cm 3.3cm 0}, clip, width =0.363 \textwidth]{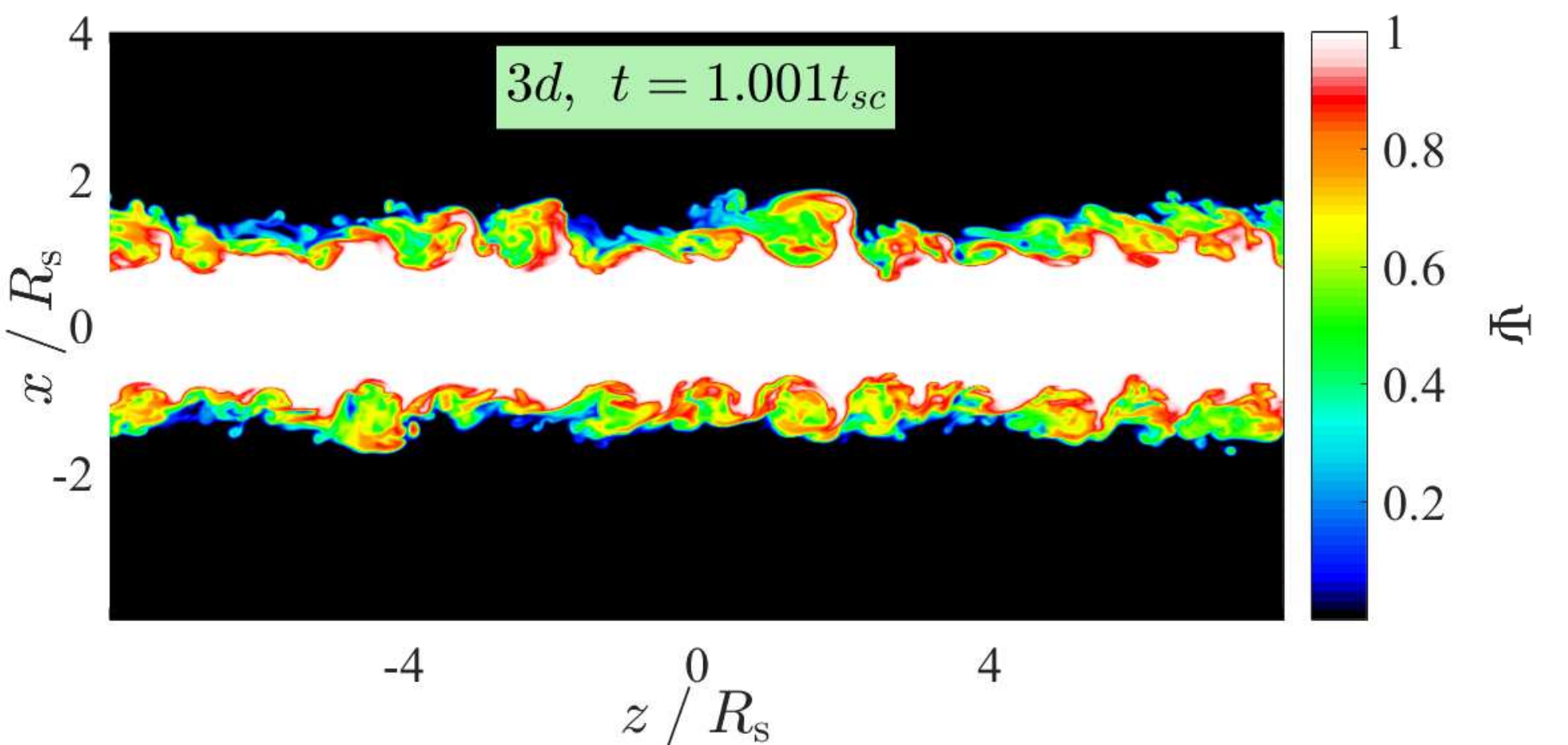}
\hspace{-0.29cm}
\includegraphics[trim={5.1cm 1.238cm 4.05cm 0}, clip, width =0.257 \textwidth]{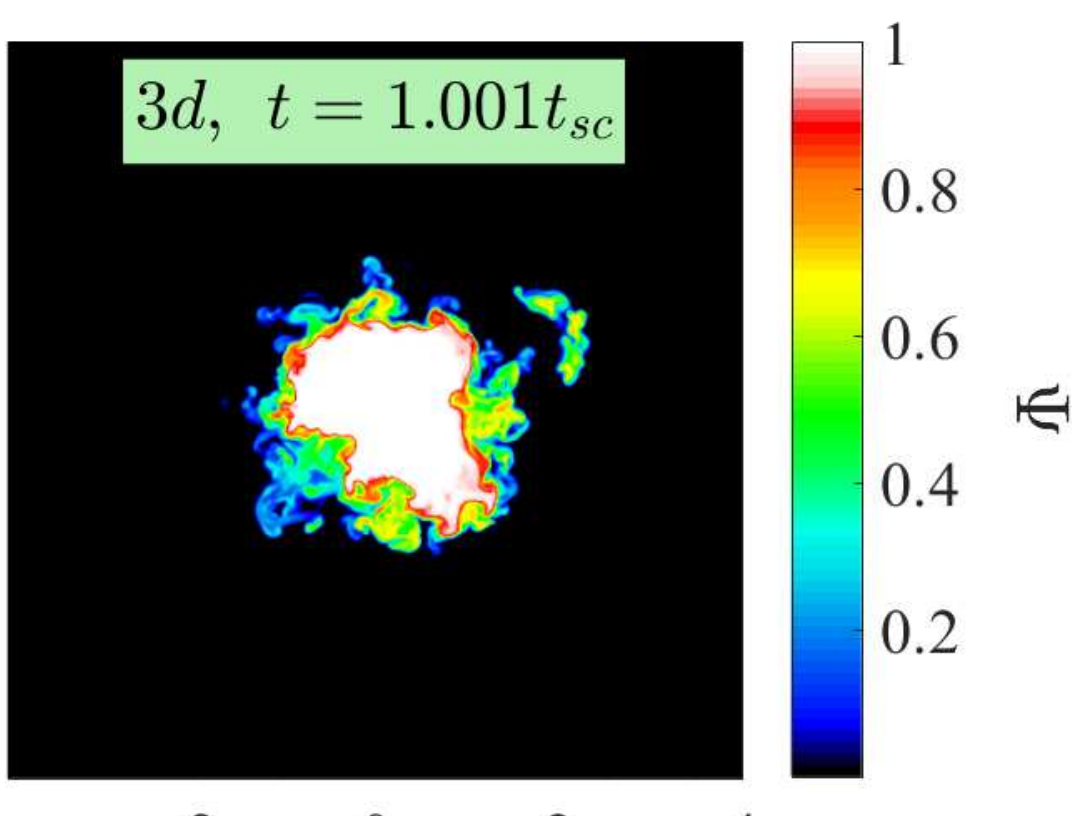}\\
\vspace{-0.09cm}
\includegraphics[trim={0.0cm 1.238cm 3.3cm 0.22cm}, clip, width =0.393 \textwidth]{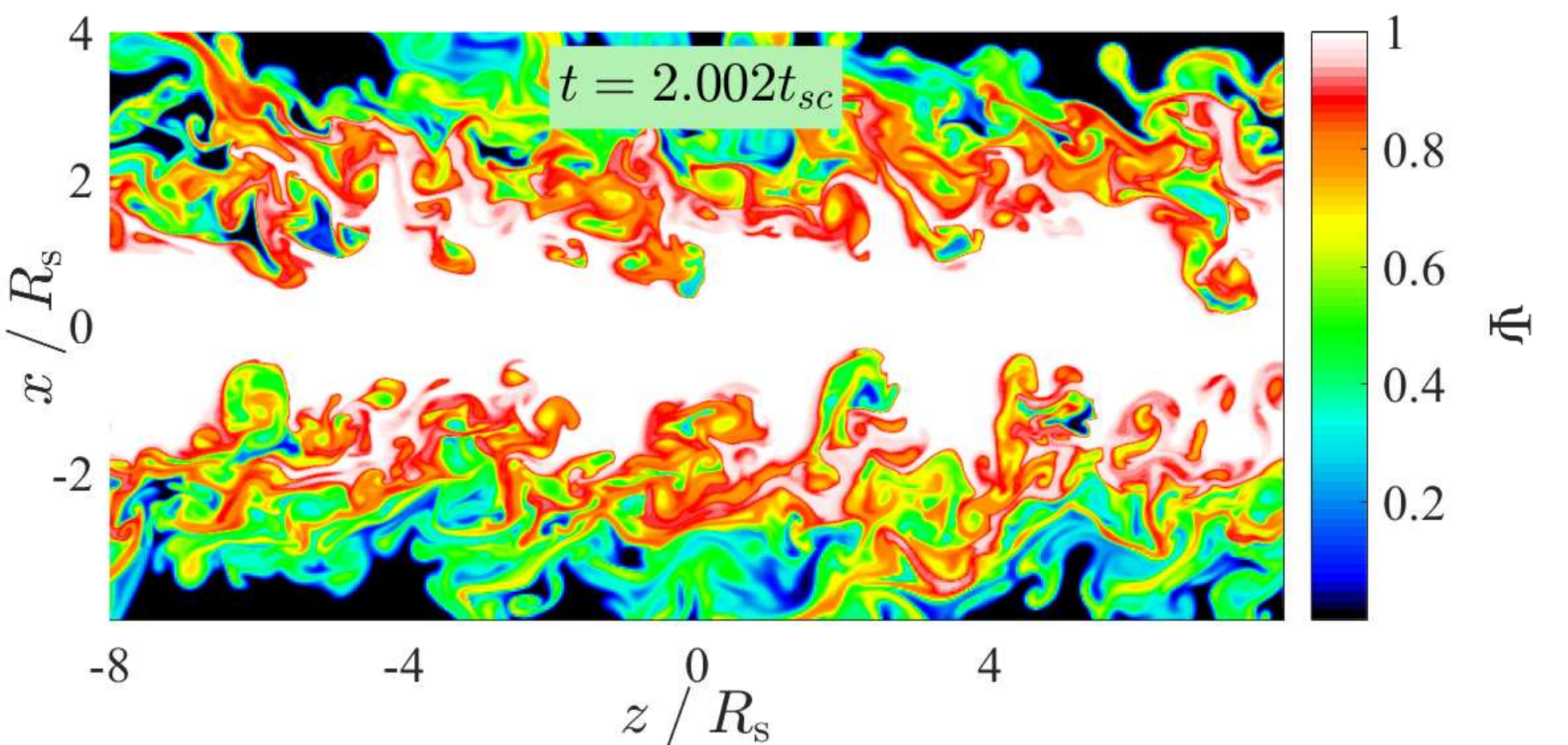}
\hspace{-0.3cm}
\includegraphics[trim={1.3cm 1.238cm 3.3cm 0.22cm}, clip, width =0.363 \textwidth]{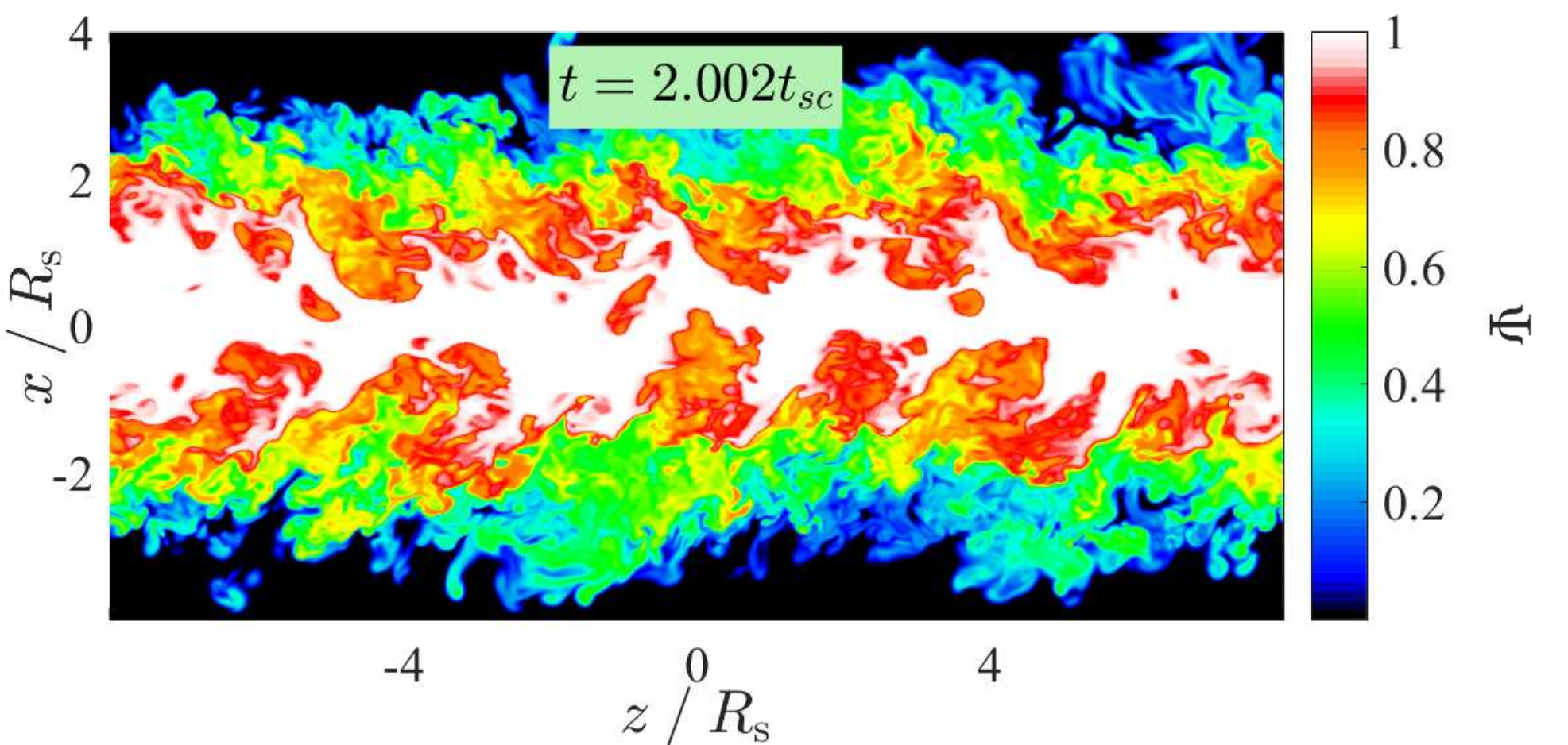}
\hspace{-0.29cm}
\includegraphics[trim={5.1cm 1.238cm 4.05cm 0.22cm}, clip, width =0.257 \textwidth]{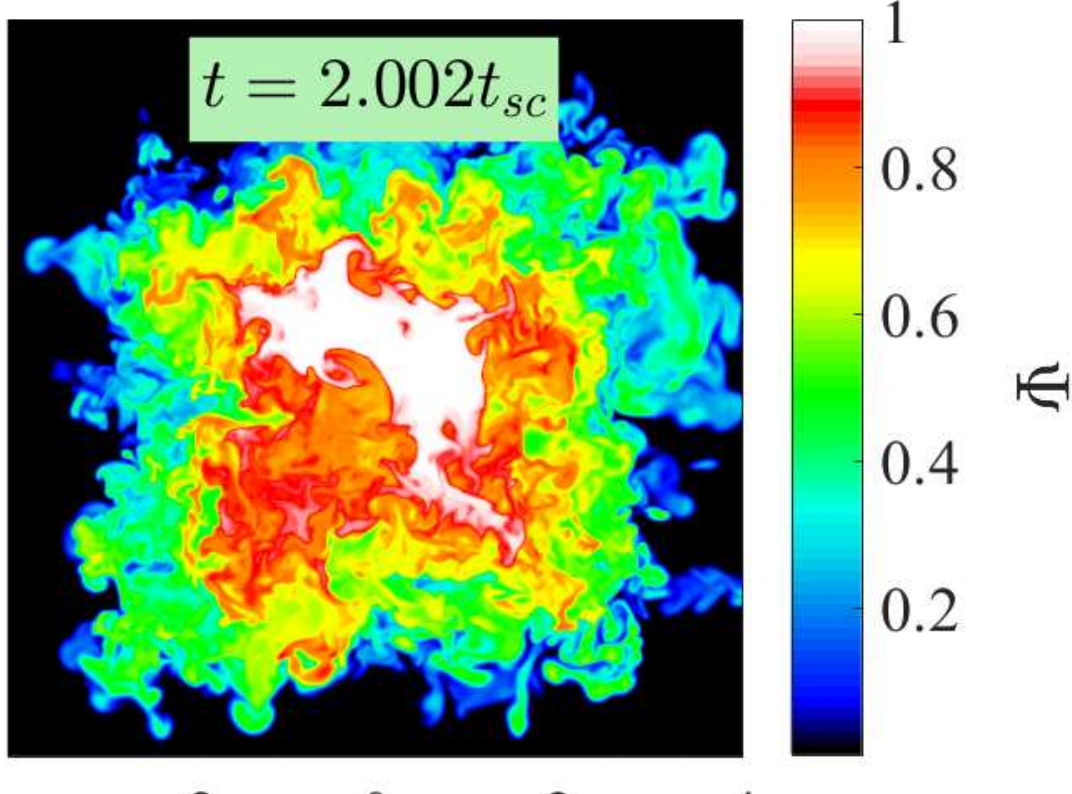}\\
\vspace{-0.09cm}
\includegraphics[trim={0.0cm 1.238cm 3.3cm 0.22cm}, clip, width =0.393 \textwidth]{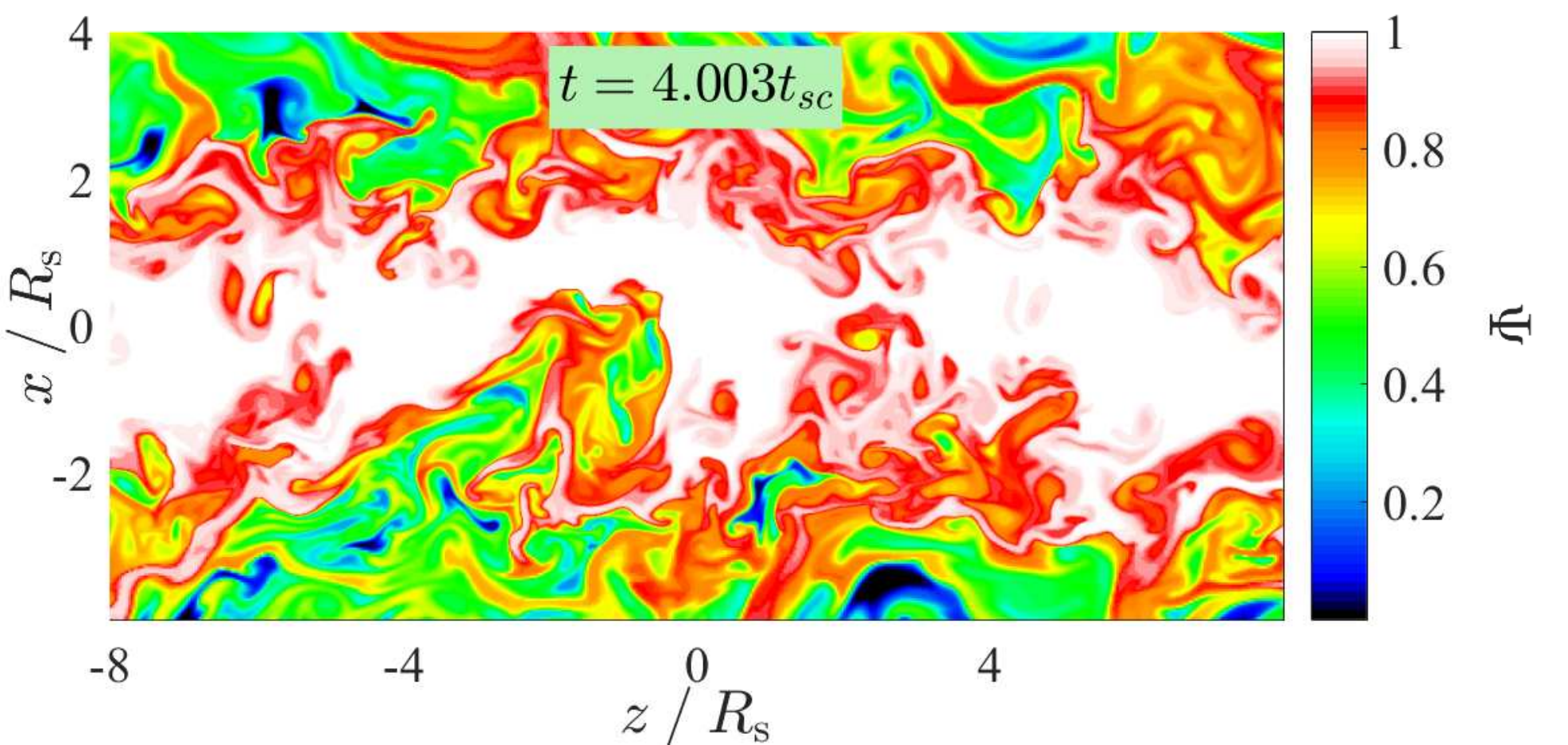}
\hspace{-0.3cm}
\includegraphics[trim={1.3cm 1.238cm 3.3cm 0.22cm}, clip, width =0.363 \textwidth]{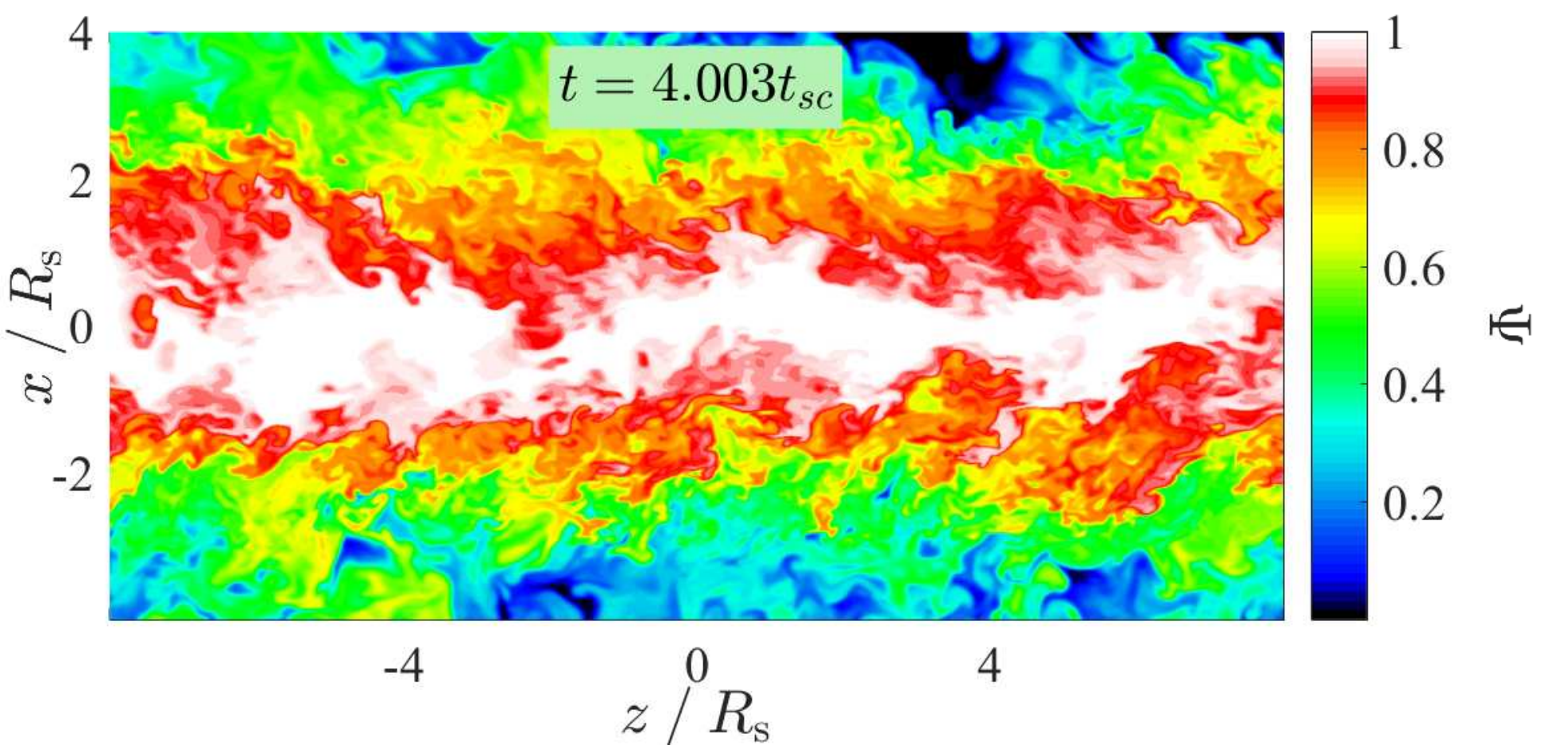}
\hspace{-0.29cm}
\includegraphics[trim={5.1cm 1.238cm 4.05cm 0.22cm}, clip, width =0.257 \textwidth]{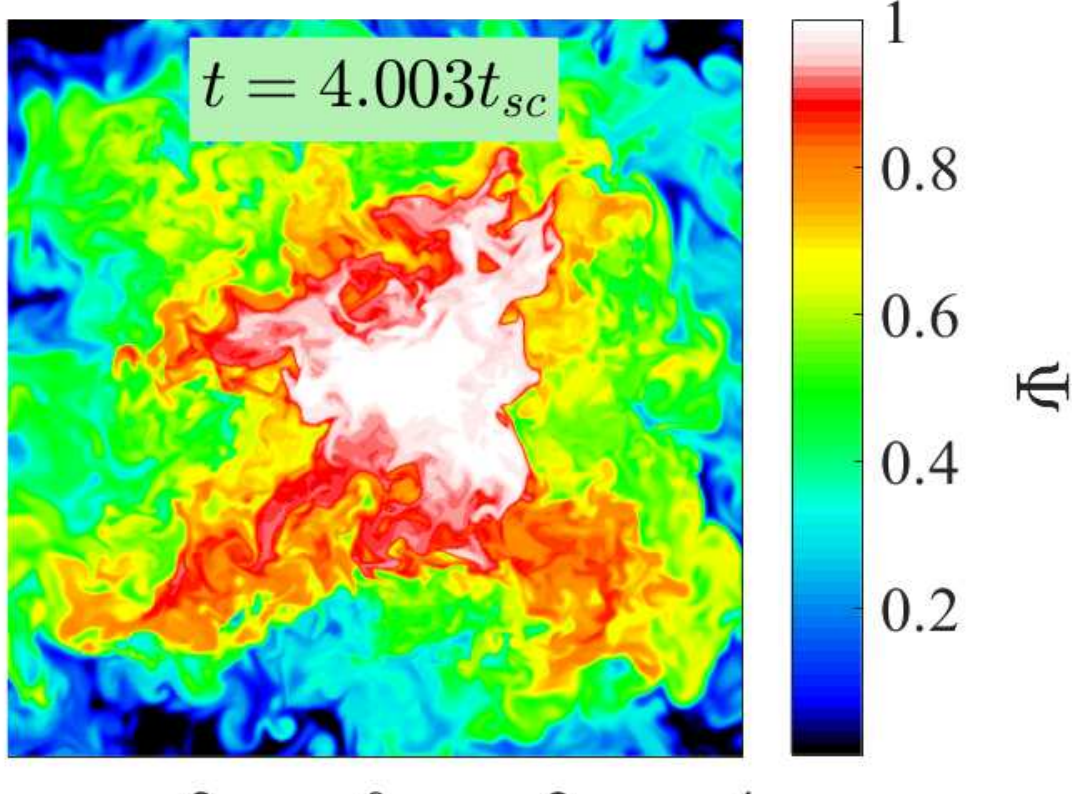}\\
\vspace{-0.09cm}
\includegraphics[trim={0.0cm 0.0cm 3.3cm 0.22cm}, clip, width =0.393 \textwidth]{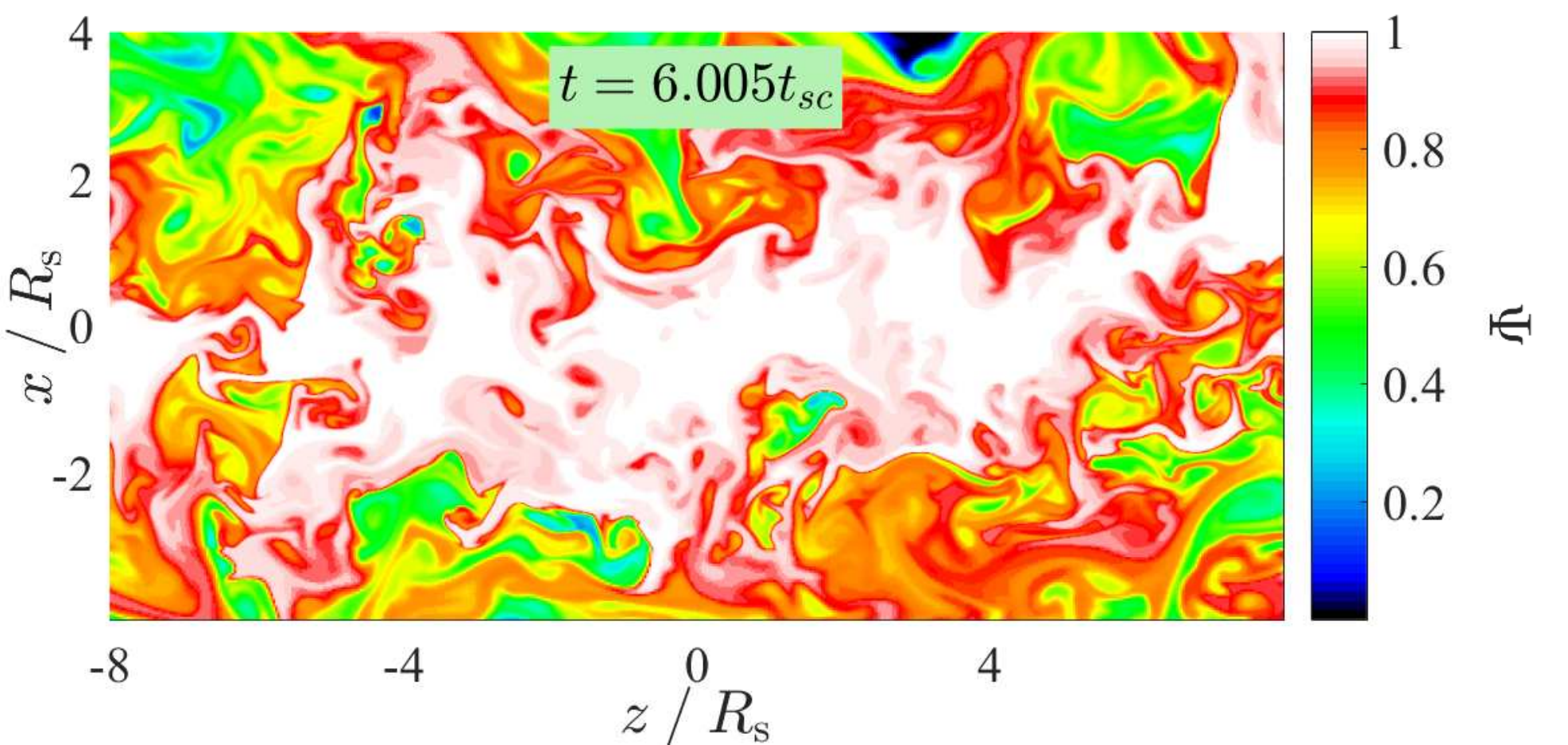}
\hspace{-0.3cm}
\includegraphics[trim={1.3cm 0.0cm 3.3cm 0.22cm}, clip, width =0.363 \textwidth]{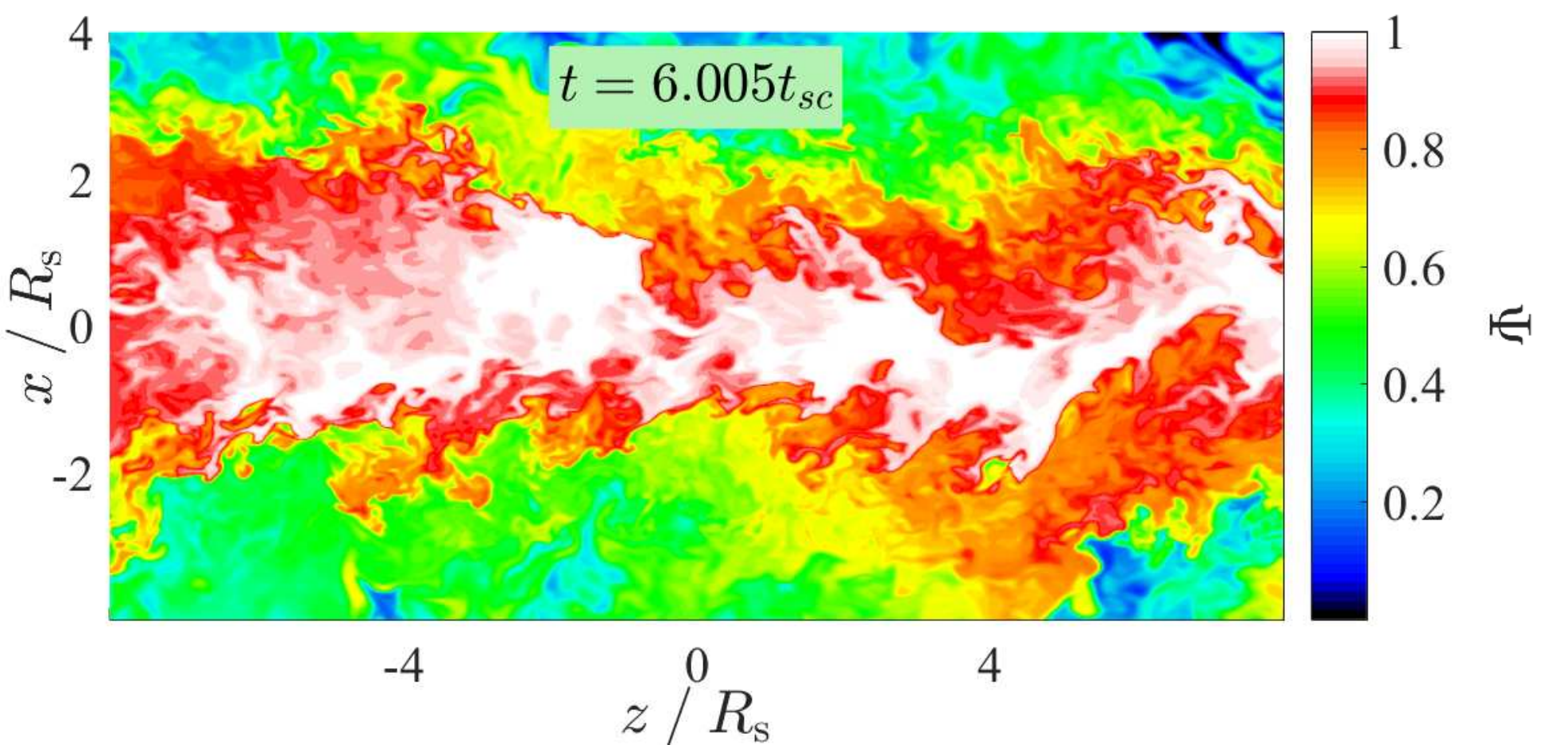}
\hspace{-0.29cm}
\includegraphics[trim={5.1cm 0.0cm 4.05cm 0.22cm}, clip, width =0.257 \textwidth]{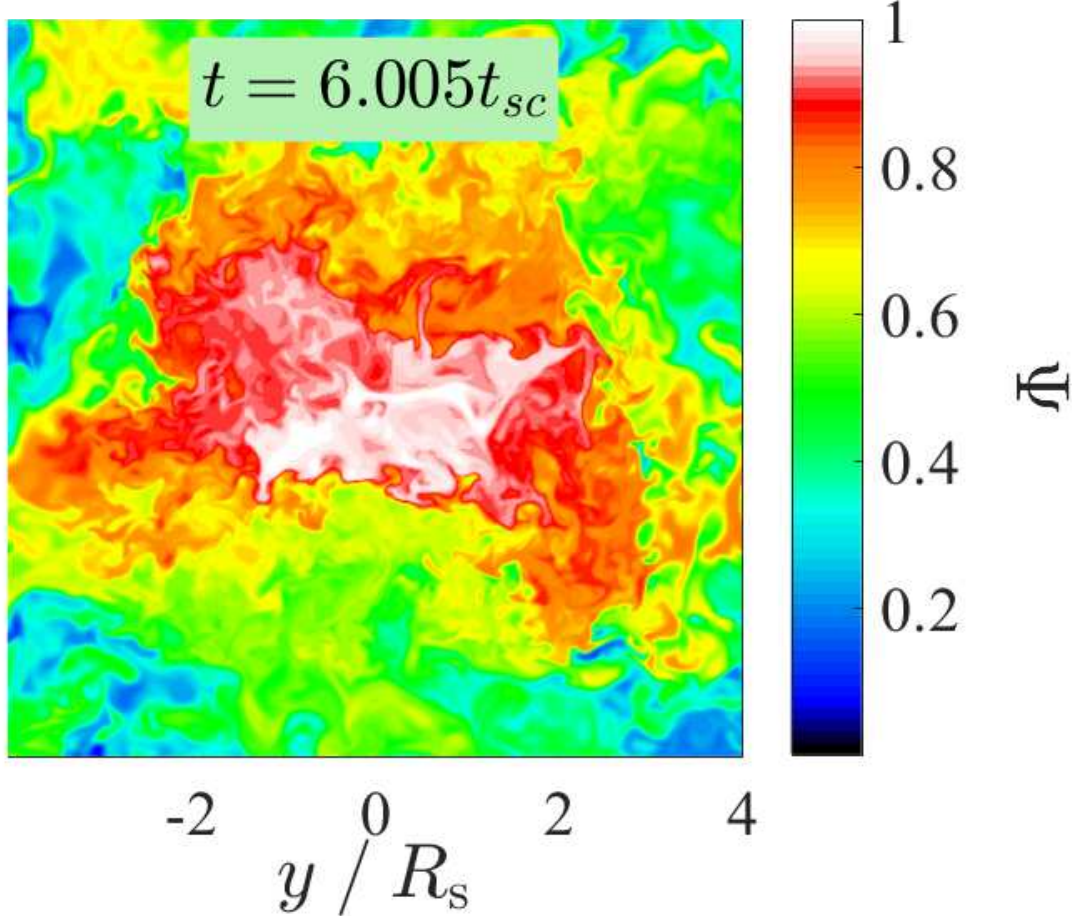}
\end{center}
\caption{Same as \fig{colour_panel_M1D10} but for a simulation with $(\Mb,\delta)=(1,100)$. 
As for the cases with $\delta=1$ and $10$, the edge-on distribution of $\psi$ in the 3d cylinder 
at $t\sim \tsc$ is reasonably similar to its distribution in the 2d slab simulation, though small 
scale modes are already visible in the 3d simulation since these appear once $\hb\sim 2\Rs$ which 
happens at $t\sim 0.9\tsc$. At later times, the stream expansion rate in the 3d simulation is 
noticeably slower than in the 2d slab. This is a manifestation of the phenomenon shown in the left 
panel of \fig{surface_h} for one case, and mentioned several times in the text of \se{results}, whereby 
the expansion rate of the shear layer into the background in 3d cylinders decreases to roughly half the 
value of 2d slabs once $\hb\sim 2\Rs$. Note that at $t\sim 6\tsc$ the 3d simulation still contains a lot 
of unmixed fluid in the stream, since $t_{\rm dis,\,surface}\sim 10\tsc$ for this case, according to 
\equ{tau_diss_2d}.
}
\label{fig:colour_panel_M1D100} 
\end{figure*}

\begin{figure*}
\begin{center}
\includegraphics[trim={0.0cm 1.238cm 3.3cm 0}, clip, width =0.393 \textwidth]{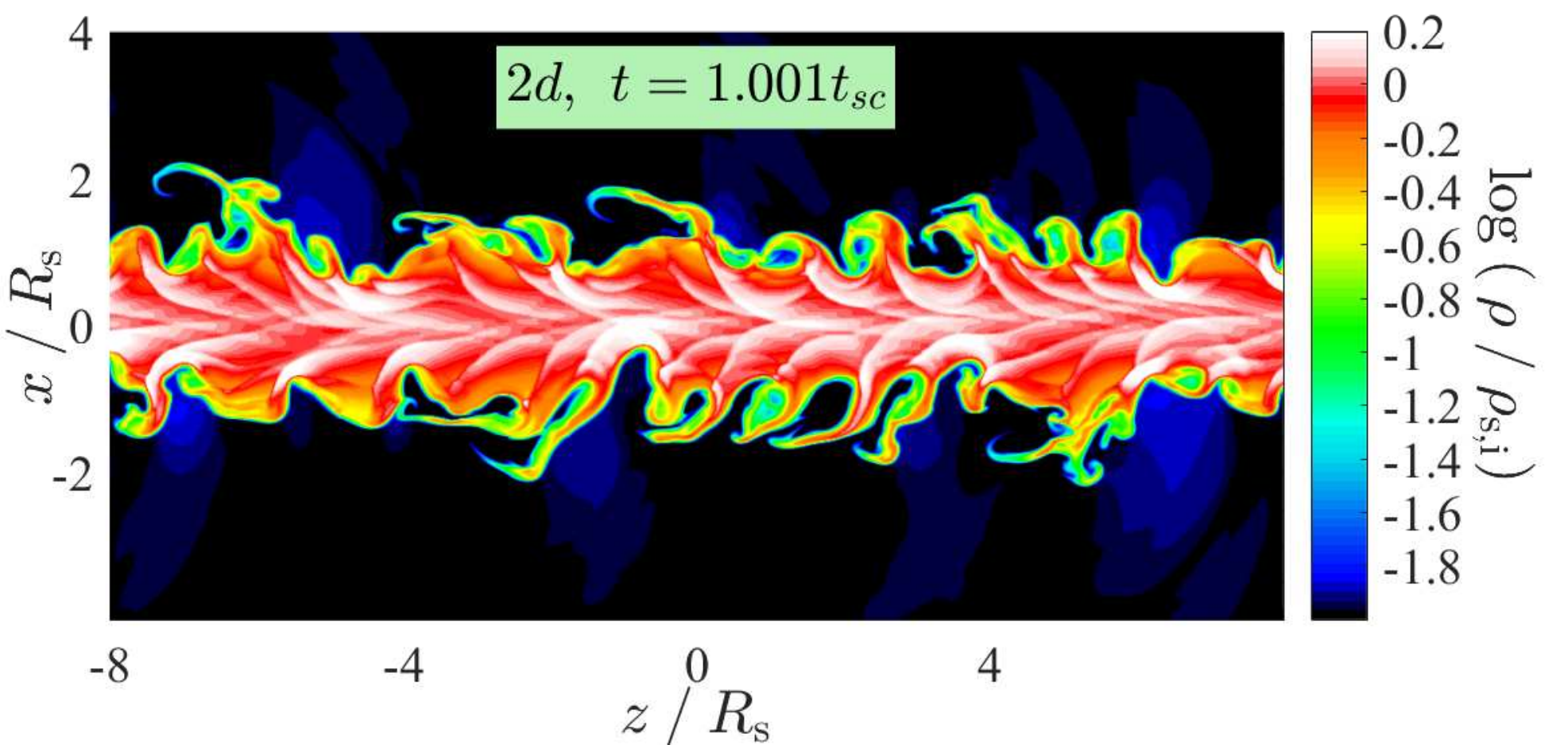}
\hspace{-0.3cm}
\includegraphics[trim={1.3cm 1.238cm 3.3cm 0}, clip, width =0.363 \textwidth]{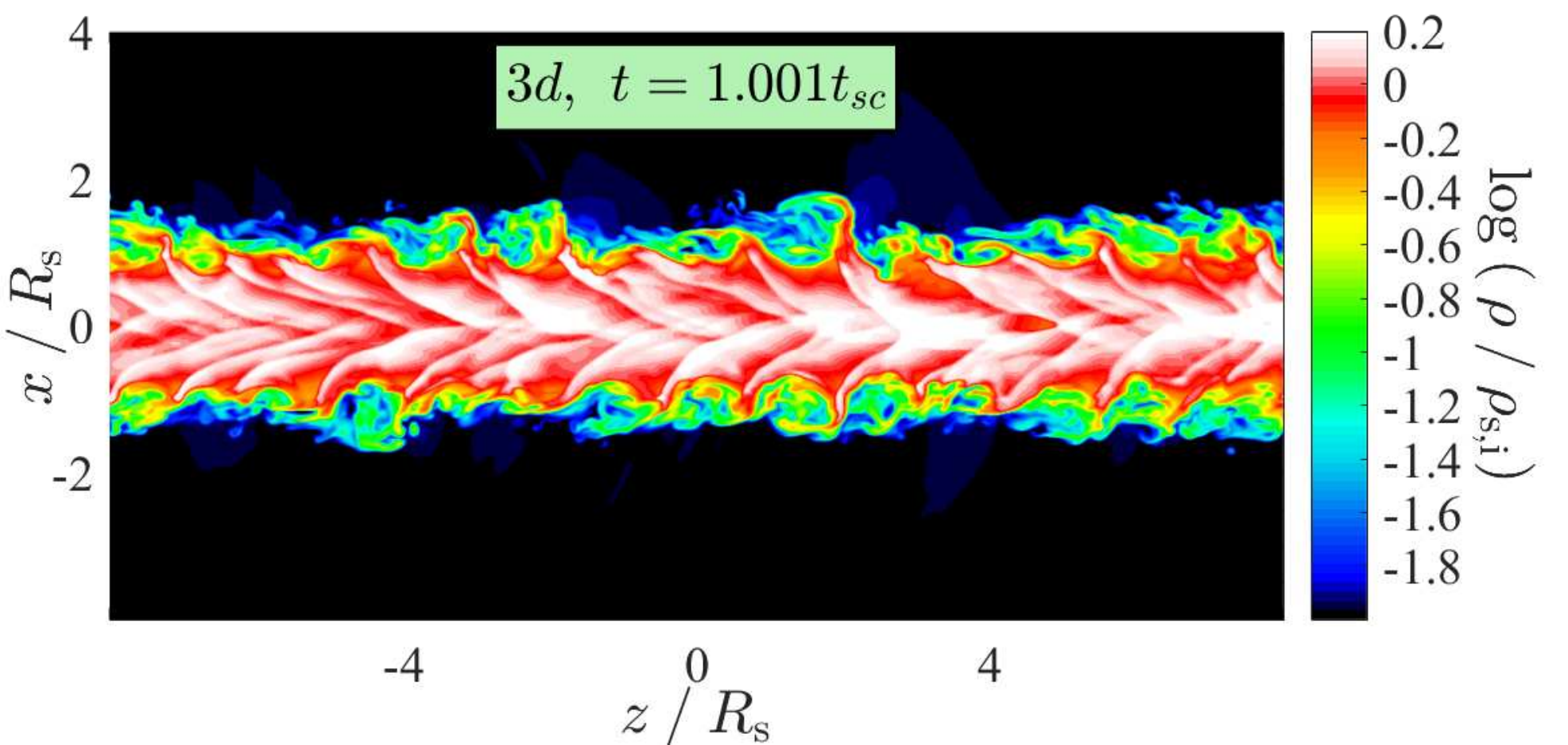}
\hspace{-0.29cm}
\includegraphics[trim={5.1cm 1.238cm 4.05cm 0}, clip, width =0.257 \textwidth]{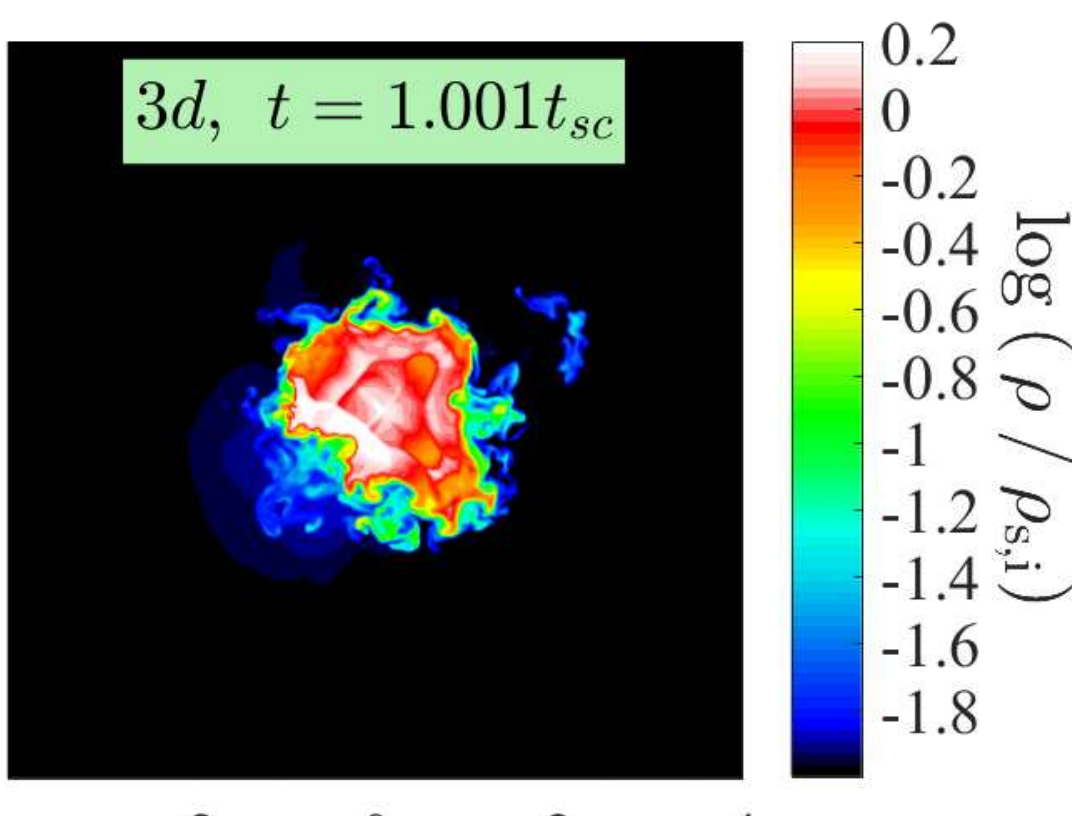}\\
\vspace{-0.09cm}
\includegraphics[trim={0.0cm 1.238cm 3.3cm 0.22cm}, clip, width =0.393 \textwidth]{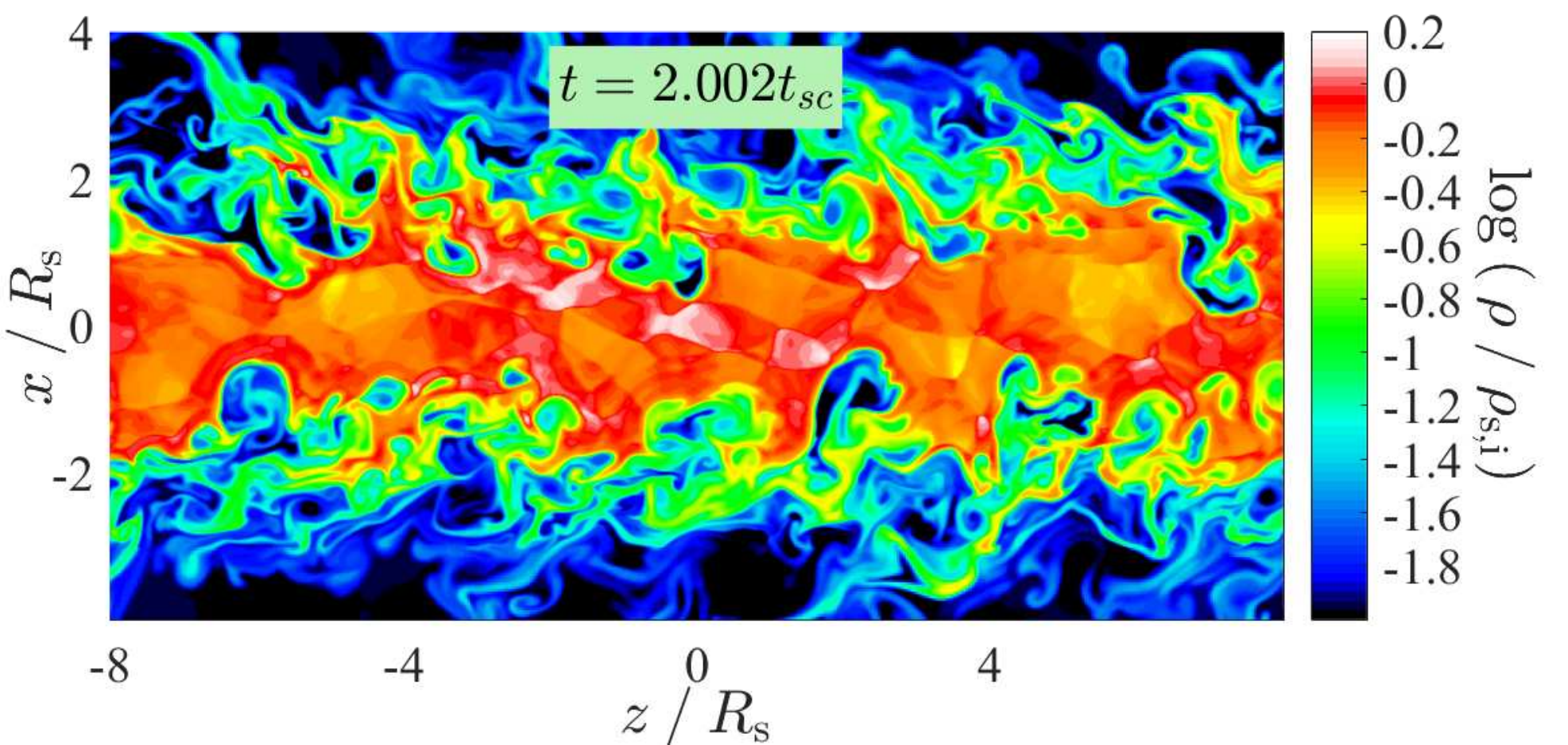}
\hspace{-0.3cm}
\includegraphics[trim={1.3cm 1.238cm 3.3cm 0.22cm}, clip, width =0.363 \textwidth]{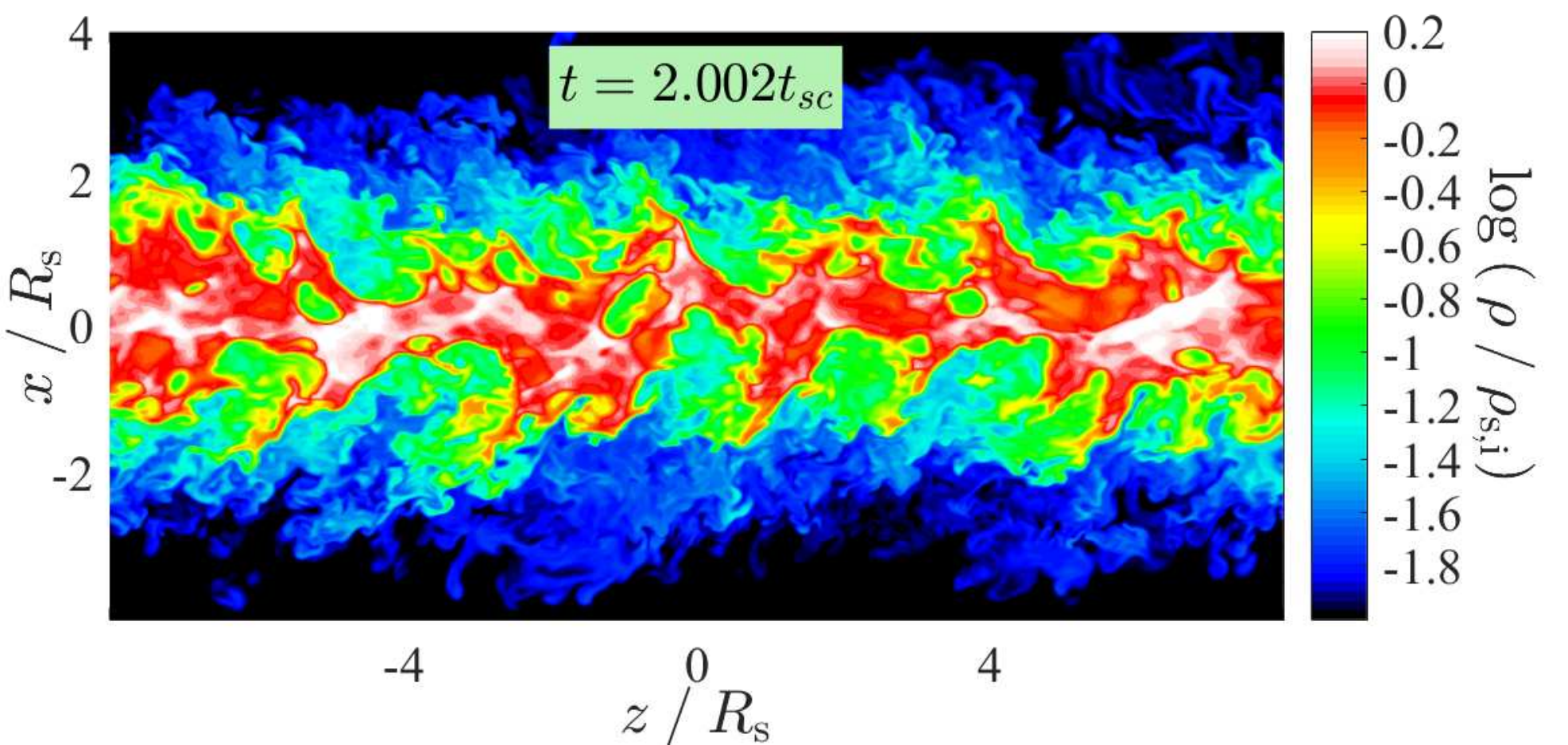}
\hspace{-0.29cm}
\includegraphics[trim={5.1cm 1.238cm 4.05cm 0.22cm}, clip, width =0.257 \textwidth]{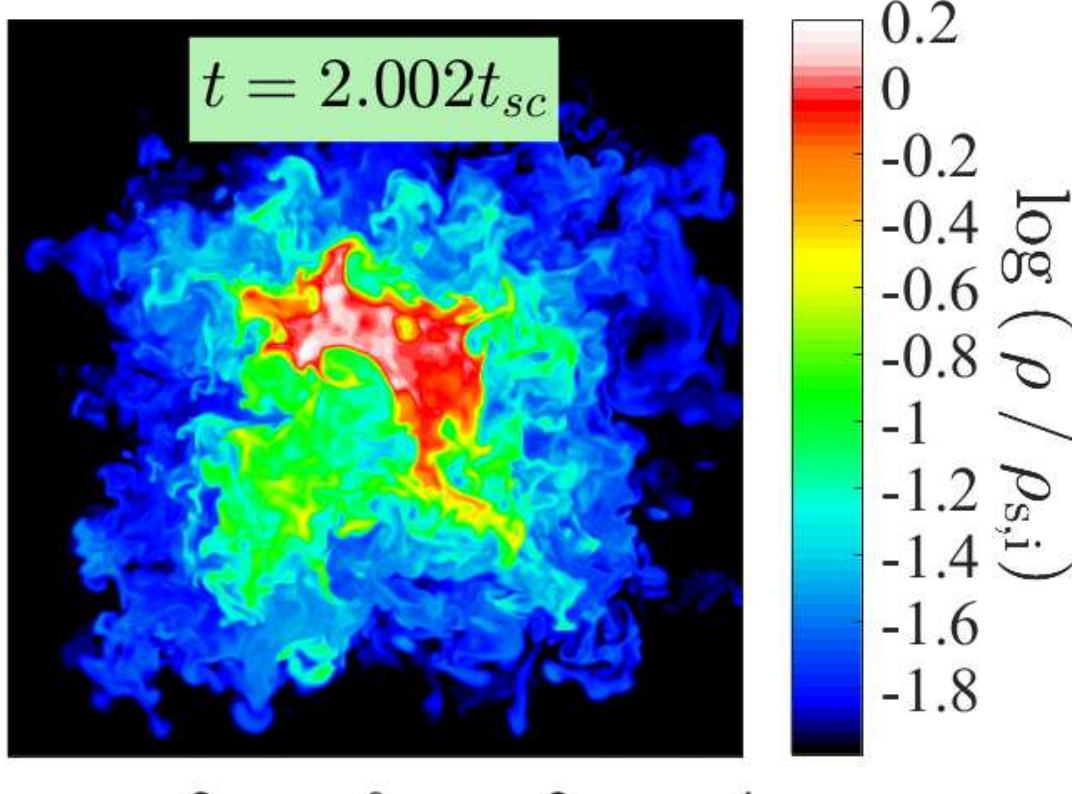}\\
\vspace{-0.09cm}
\includegraphics[trim={0.0cm 1.238cm 3.3cm 0.22cm}, clip, width =0.393 \textwidth]{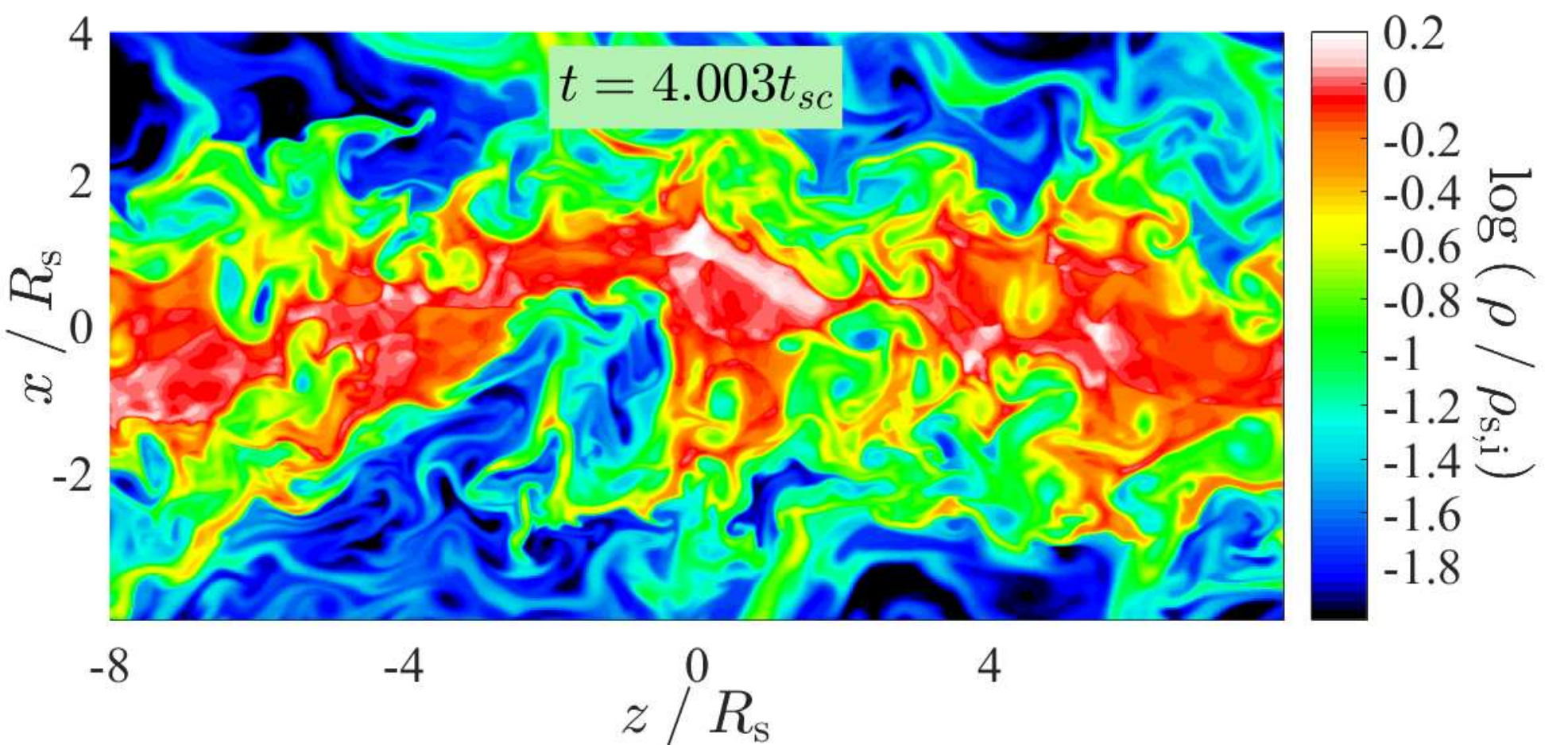}
\hspace{-0.3cm}
\includegraphics[trim={1.3cm 1.238cm 3.3cm 0.22cm}, clip, width =0.363 \textwidth]{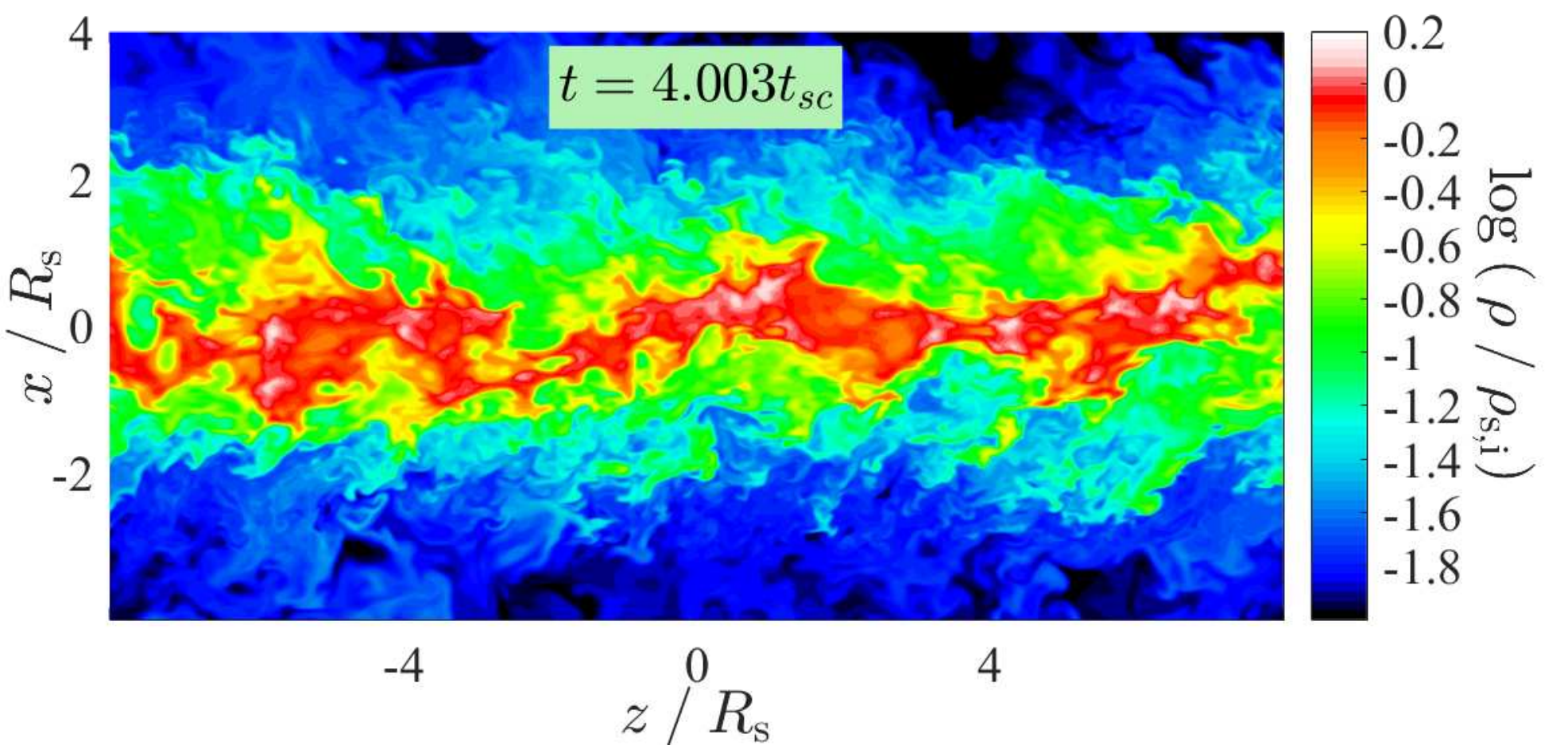}
\hspace{-0.29cm}
\includegraphics[trim={5.1cm 1.238cm 4.05cm 0.22cm}, clip, width =0.257 \textwidth]{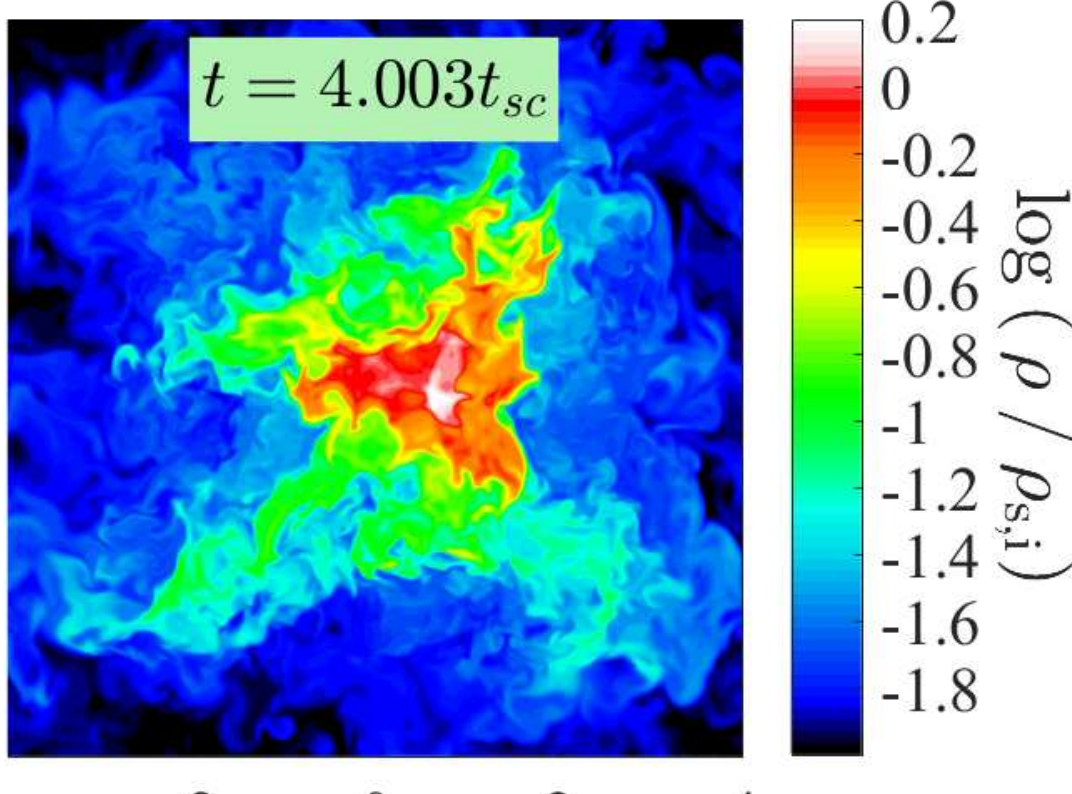}\\
\vspace{-0.09cm}
\includegraphics[trim={0.0cm 0.0cm 3.3cm 0.22cm}, clip, width =0.393 \textwidth]{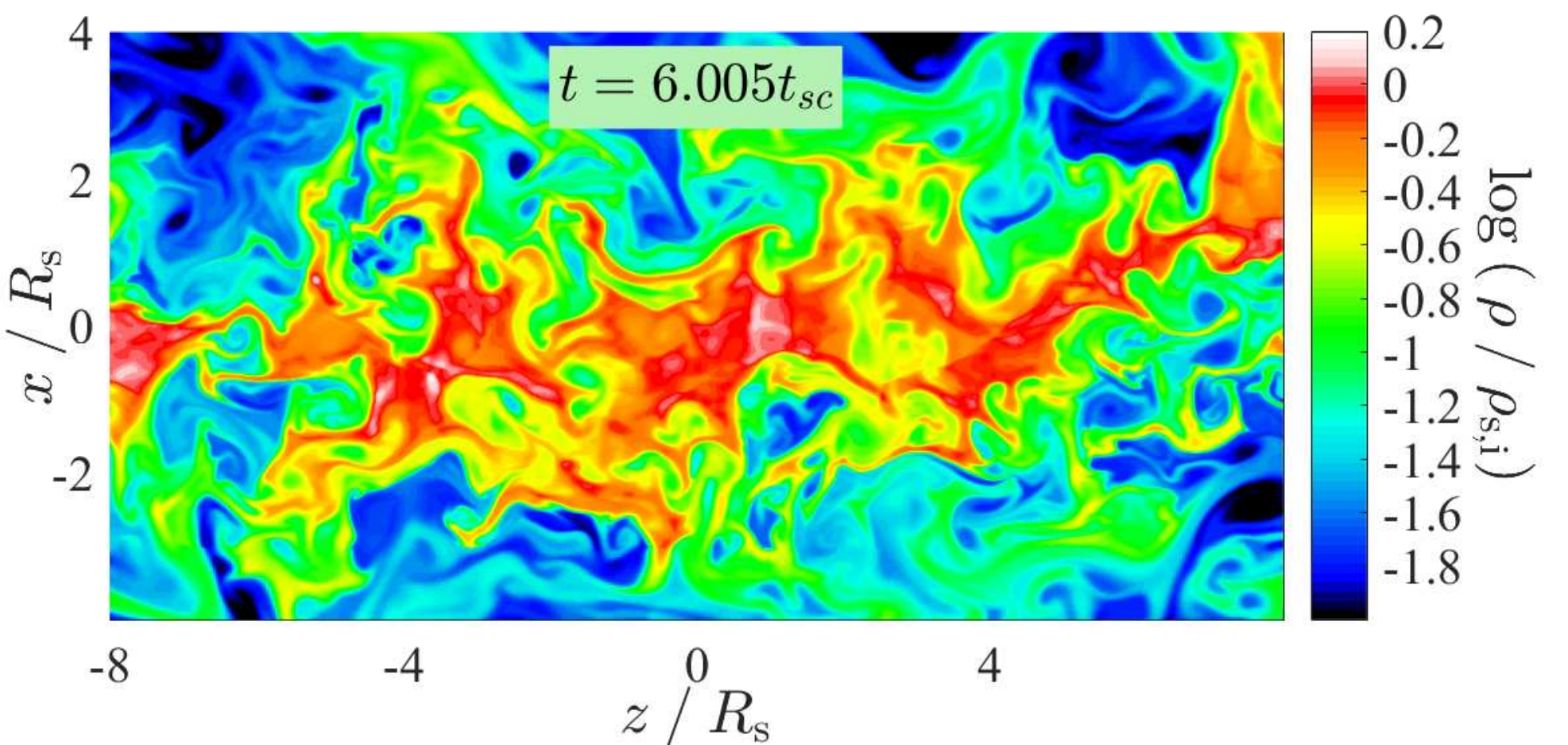}
\hspace{-0.3cm}
\includegraphics[trim={1.3cm 0.0cm 3.3cm 0.22cm}, clip, width =0.363 \textwidth]{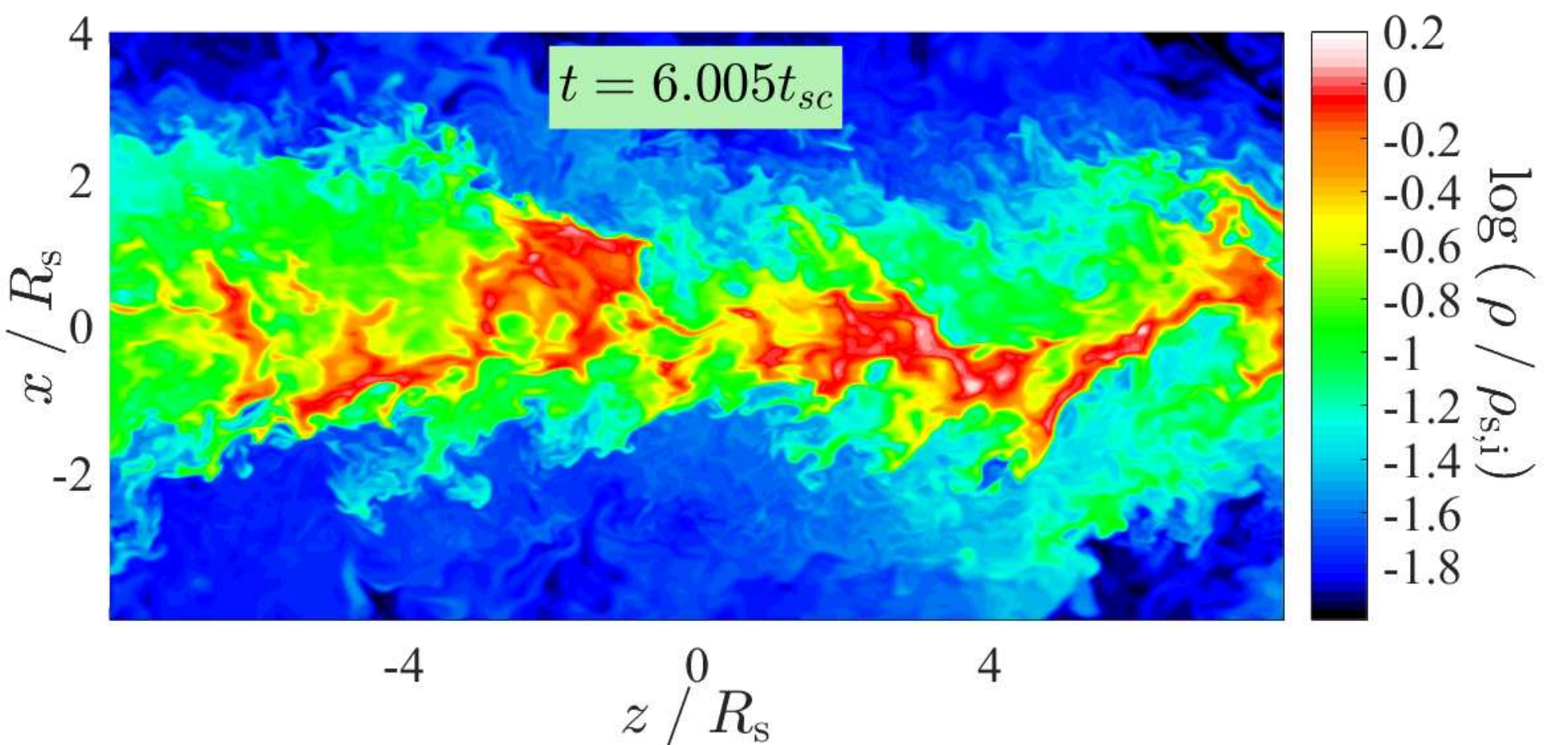}
\hspace{-0.29cm}
\includegraphics[trim={5.1cm 0.0cm 4.05cm 0.22cm}, clip, width =0.257 \textwidth]{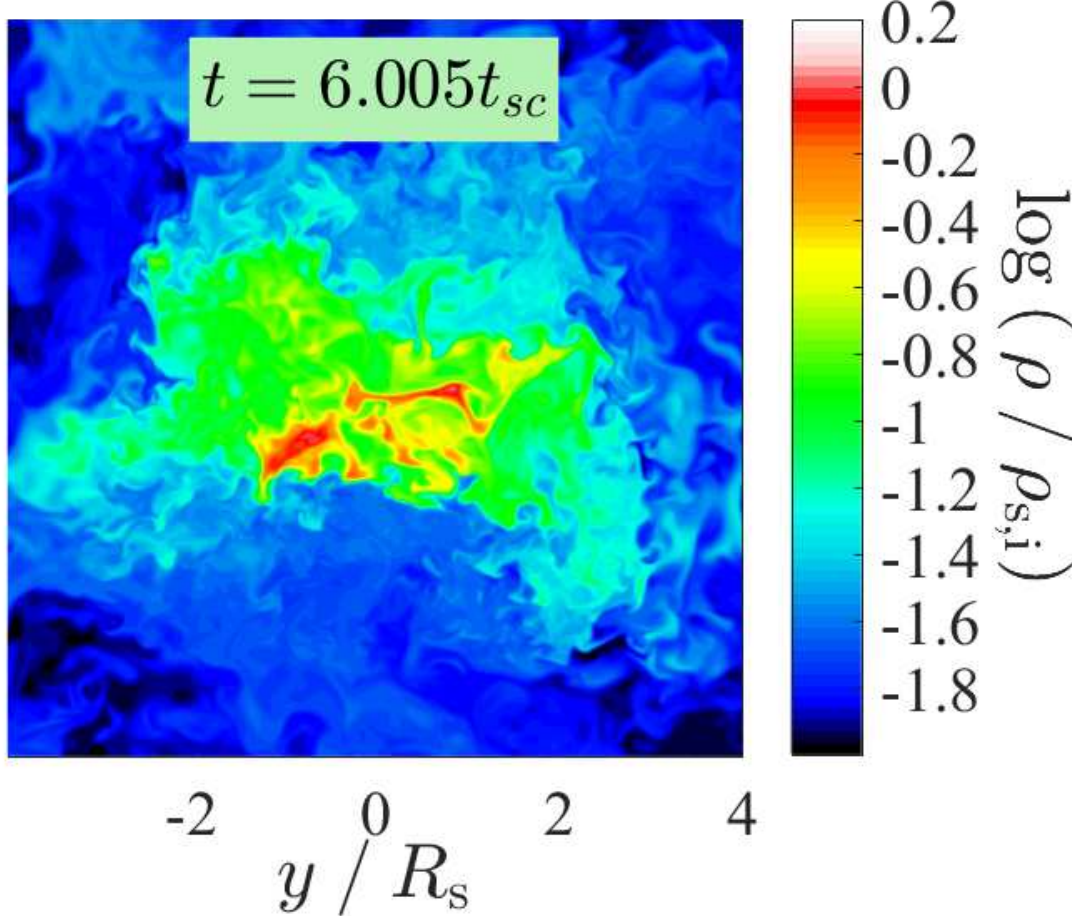}
\end{center}
\caption{Same as \fig{colour_panel_M1D100} but where the colour shows the density relative to the initial 
stream density, $\rho/\rhos$, rather than the passive scalar $\Psi$. Large density fluctuations are visible at 
$t\sim \tsc$, triggered by shocks. These are stronger in 3d than in 2d. $M_{\rm tot}\sim 0.91$ for this simulation, 
so while body modes are formally stable we are very close to the regime where they are unstable, and they may be 
triggered by numerical effects, as was also noted by M16, even if the instability is dominated by surface modes. 
By $t\sim 2\tsc$ these density fluctuaions have dissipated as the stream has begun to mix with the background. By 
$6\tsc$ the mean density in the stream has been diluted by a factor of $\sim 10$, with the densest remaining regions 
corresponding to the few unmixed regions with $\Psi\sim 1$. As also noted in \fig{colour_panel_M1D100}, the 3d 
cylinder seems to have expanded less than the 2d slab. 
}
\label{fig:density_panel_M1D100} 
\end{figure*}

\begin{figure*}
\begin{center}
\includegraphics[trim={0.0cm 1.238cmcm 3.3cm 0}, clip, width =0.45 \textwidth]{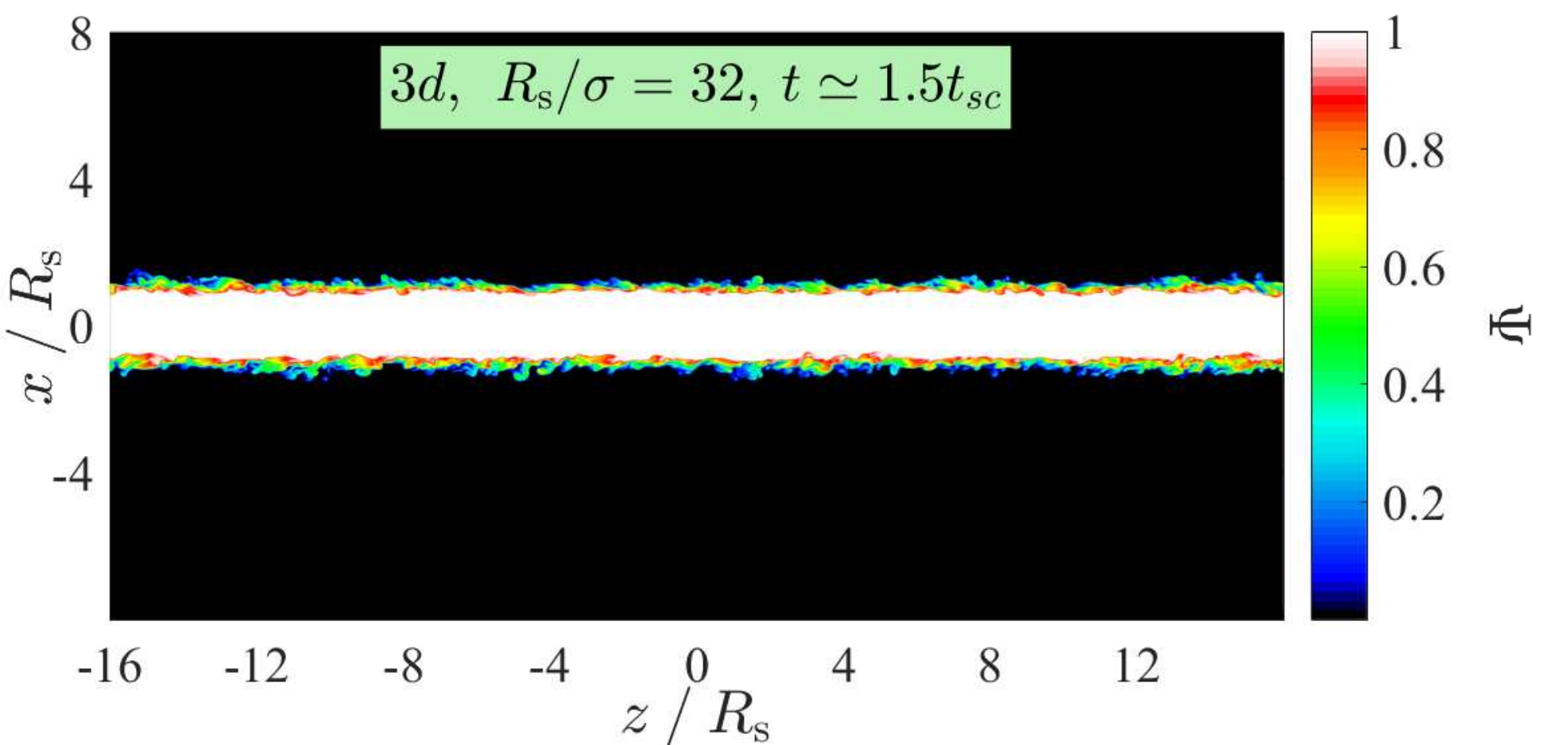}
\hspace{-0.3cm}
\includegraphics[trim={1.3cm 1.238cmm 0.1cm 0}, clip, width =0.501 \textwidth]{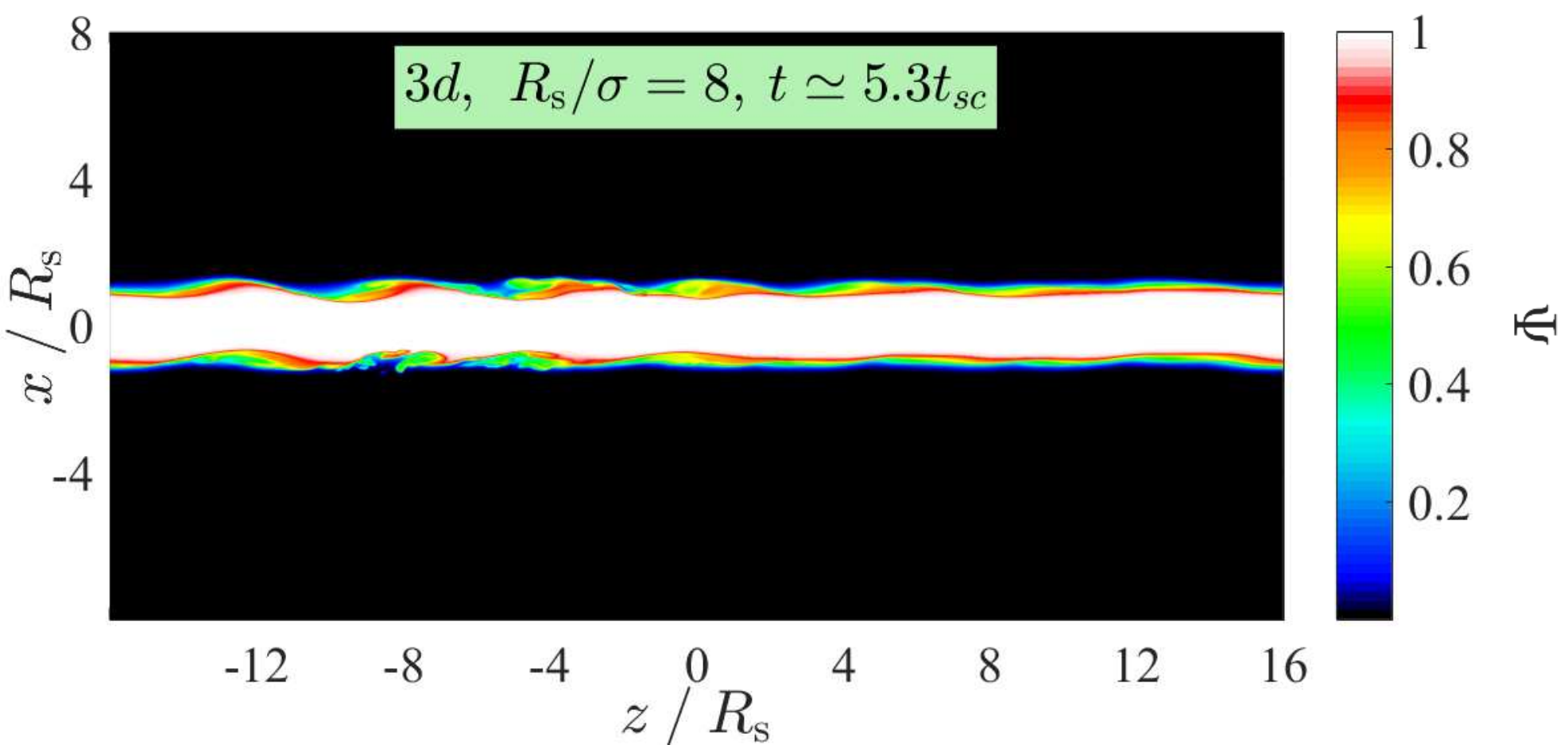}\\
\vspace{-0.09cm}
\includegraphics[trim={0.0cm 1.238cmcm 3.3cm 0}, clip, width =0.45 \textwidth]{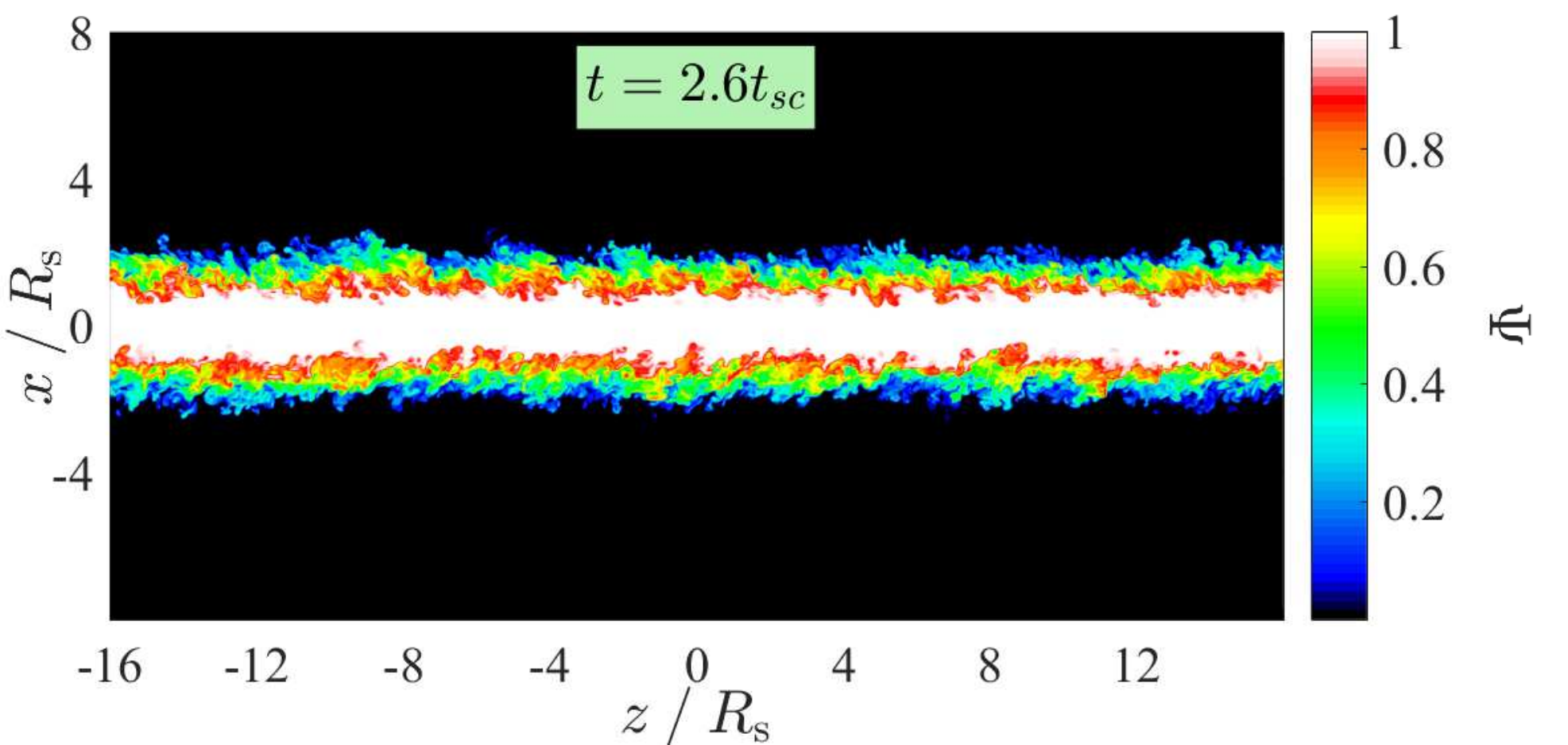}
\hspace{-0.3cm}
\includegraphics[trim={1.3cm 1.238cmcm 0.1cm 0}, clip, width =0.501 \textwidth]{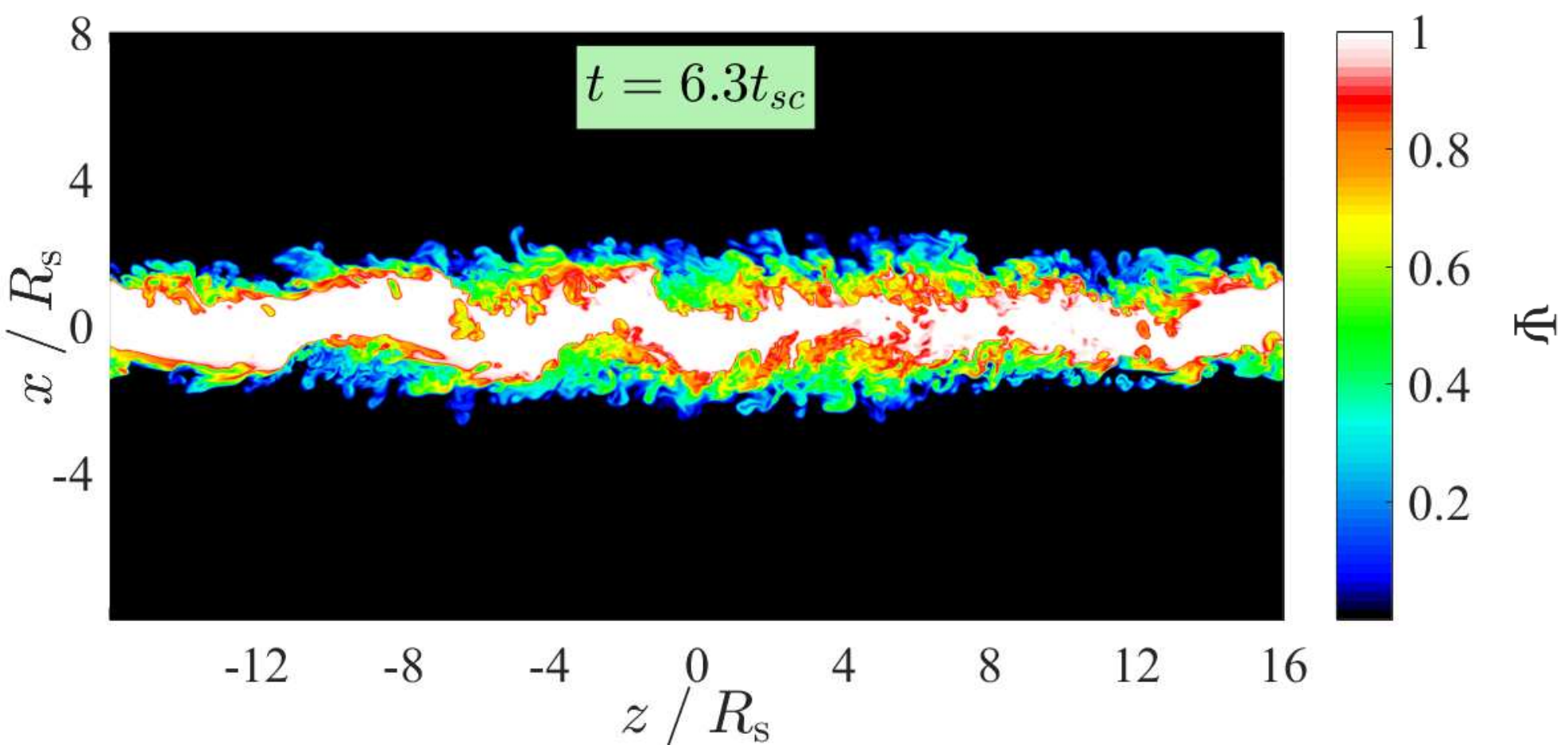}\\
\vspace{-0.09cm}
\includegraphics[trim={0.0cm 0.0cm 3.3cm 0}, clip, width =0.45 \textwidth]{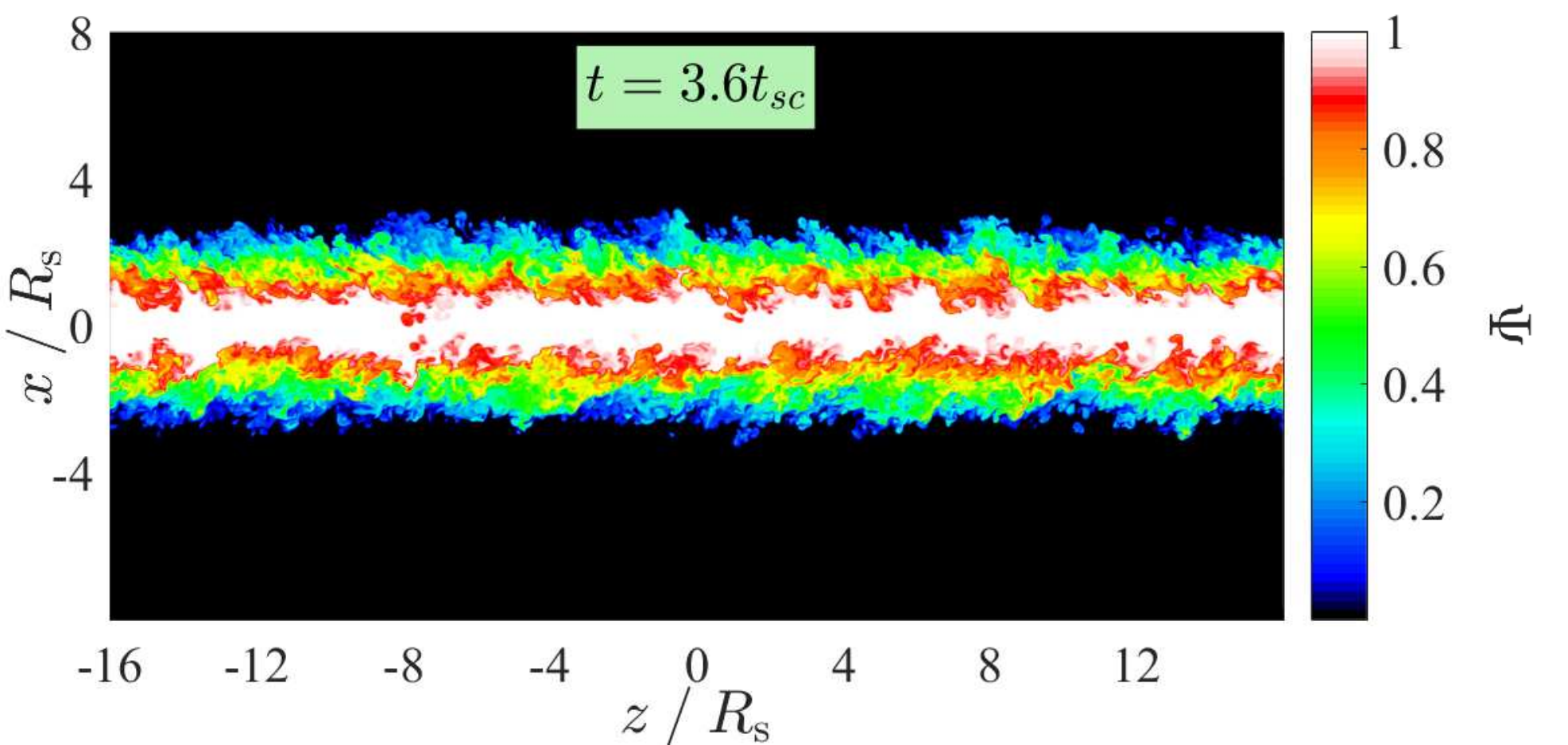}
\hspace{-0.3cm}
\includegraphics[trim={1.3cm 0.0cm 0.1cm 0}, clip, width =0.501 \textwidth]{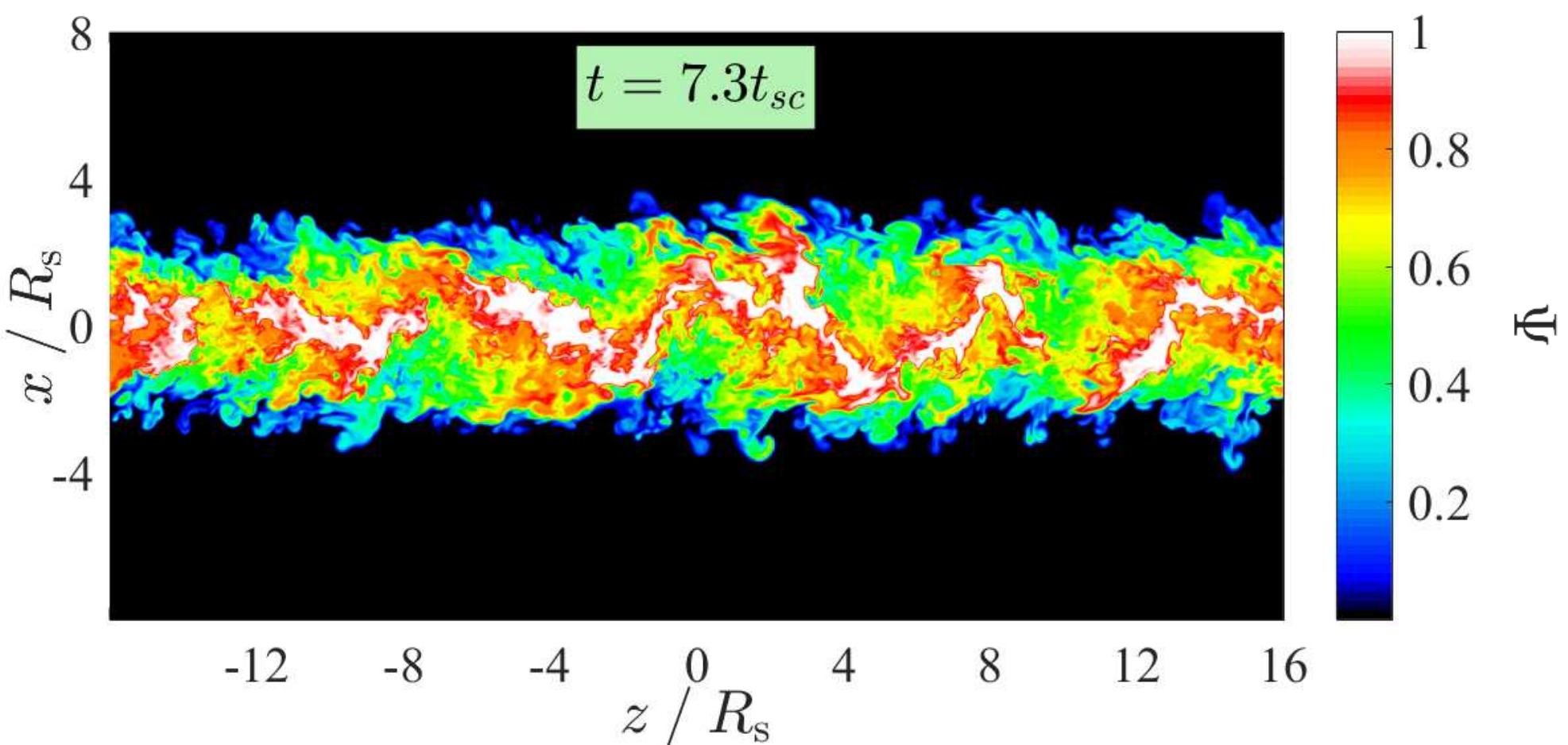}\\
\end{center}
\caption{Same as \fig{colour_panel_M5D1} but for a simulation with $(\Mb,\delta)=(2,10)$. Similar 
to \fig{colour_panel_M5D1}, the top row shows a snapshot just as $\hb$ begins to grow (\fig{h_body}), 
while the middle and bottom rows show snapshots 1 and 2 sound crossing times later, respectively. 
When $\sigma=\Rs/32$, the stream-background interface is dominated by small-scale structure when $\hb$ 
begins growing, at $t\sim 1.5\tsc$. The ensuing growth of $\hb$ is due to an expanding shear layer, as 
in the surface mode simulations described in \se{surface}. At $\sim 3.6\tsc$ there is still an unmixed 
core at the centre of the stream, which does not fully mix until $t\sim 6\tsc$. On the other hand, when 
$\sigma=\Rs/8$, the stream-background interface is dominated by a large-scale sinusoidal perturbation with 
relatively little small scale structure when $\hb$ begins growing, at $t\sim 5.3\tsc$. By $t\sim 6.3\tsc$ 
a sinusoidal structure with $\lambda\sim 6.5\Rs$ clearly dominates, in reasonable agreement with the predicted 
critical wavelength of $\sim 8.5\Rs$ (\tab{body2}). However, at the same time small scale structure has begun 
to develop, due to unstable surface modes with high-$m$. Within one additional sound crossing time these 
small-scale perturbations have efficiently mixed the stream and the background fluids, though the large scale 
sinusoidal mode is still visible. By $t\sim 7.5\tsc$ there is no unmixed fluid left in the stream.
}
\label{fig:colour_panel_M2D10} 
\end{figure*}

\begin{figure*}
\begin{center}
\includegraphics[trim={0.0cm 1.238cmcm 3.3cm 0}, clip, width =0.45 \textwidth]{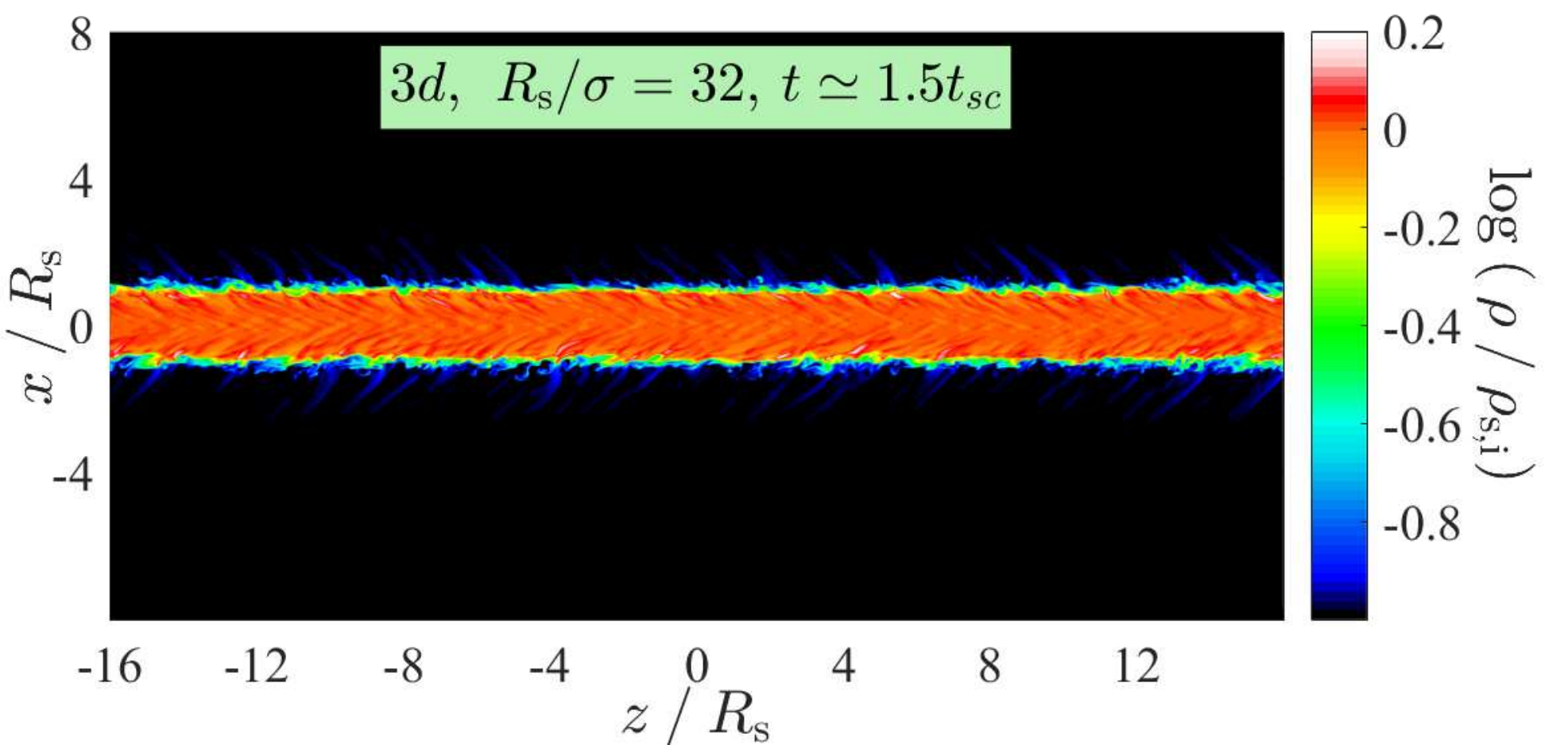}
\hspace{-0.3cm}
\includegraphics[trim={1.3cm 1.238cmm 0.1cm 0}, clip, width =0.501 \textwidth]{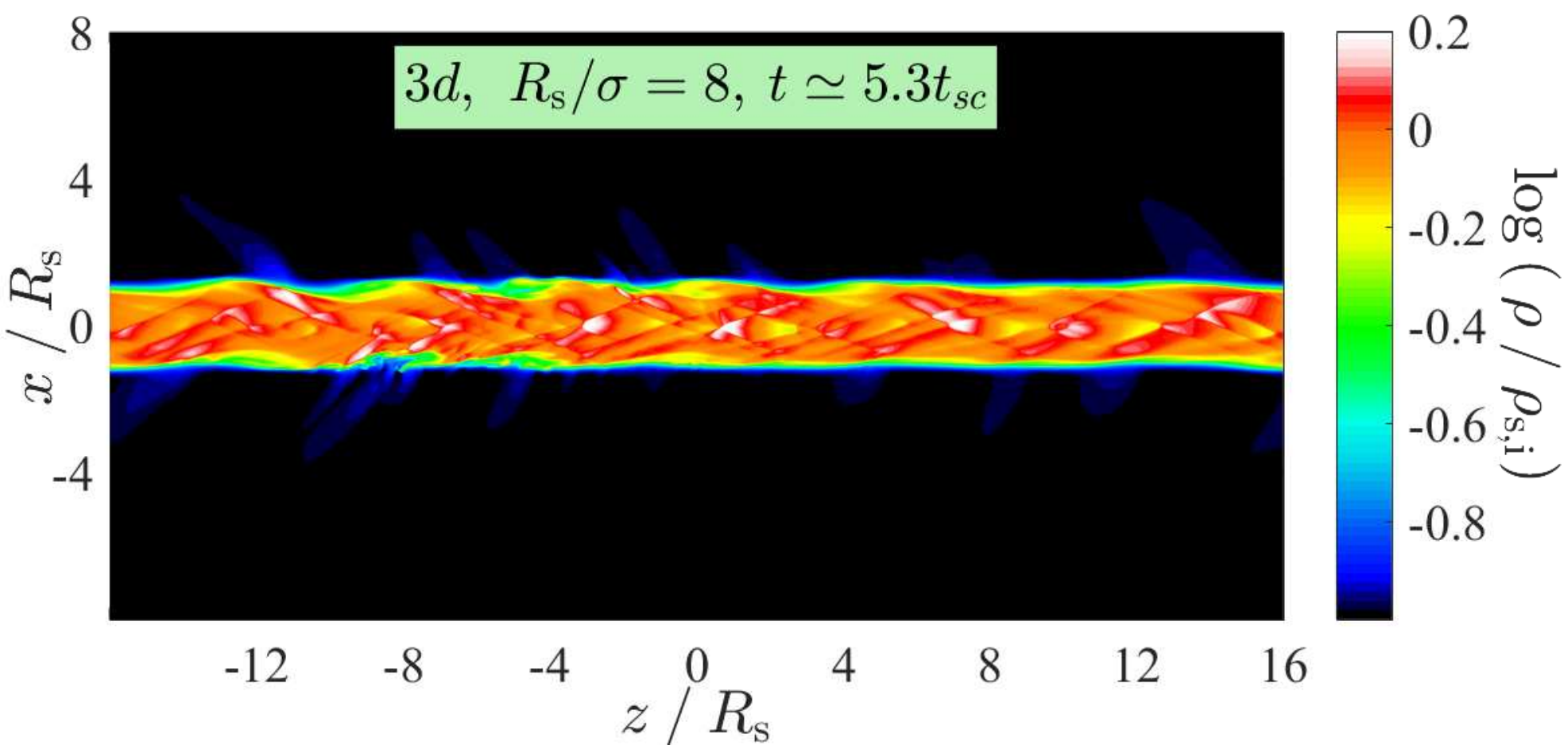}\\
\vspace{-0.09cm}
\includegraphics[trim={0.0cm 1.238cmcm 3.3cm 0}, clip, width =0.45 \textwidth]{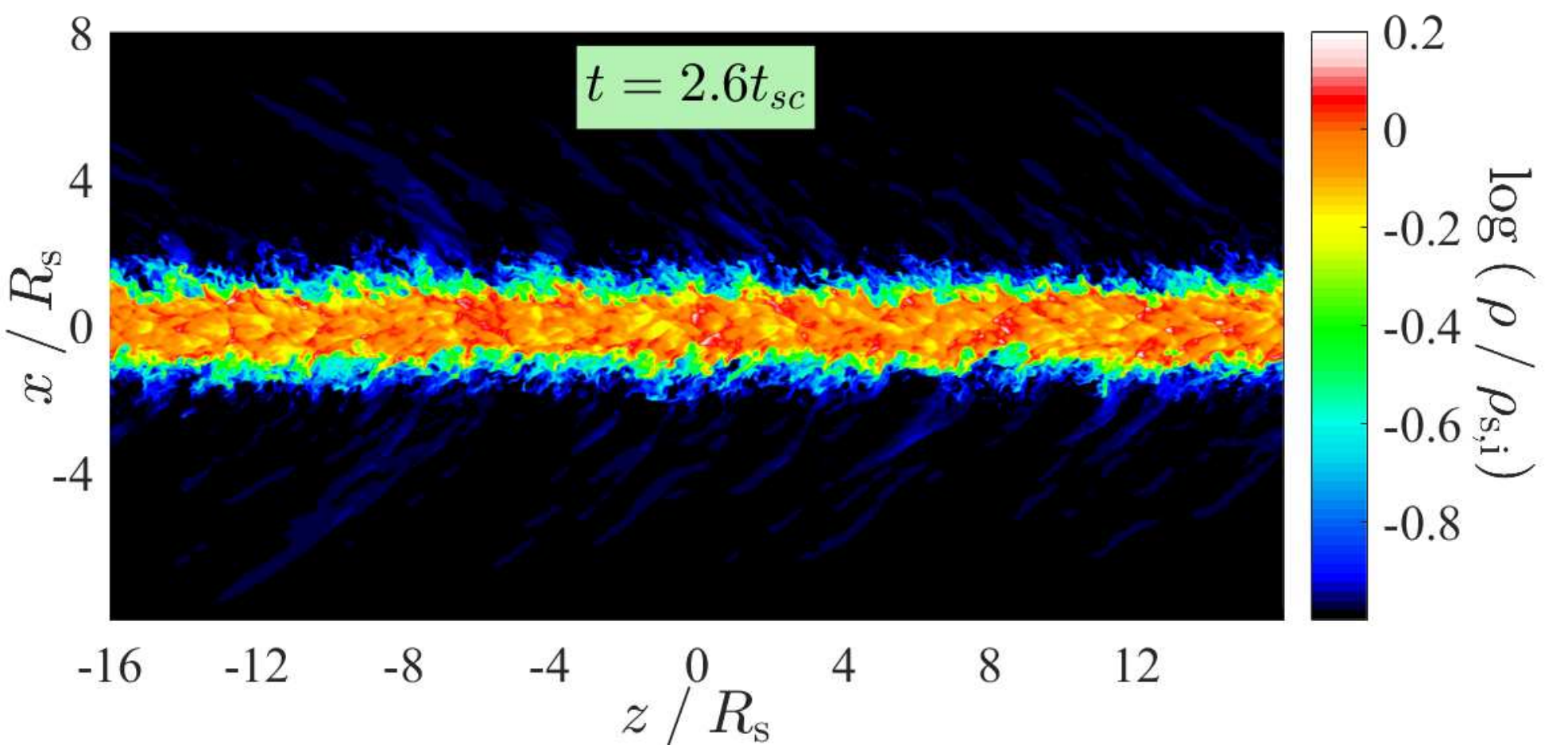}
\hspace{-0.3cm}
\includegraphics[trim={1.3cm 1.238cmcm 0.1cm 0}, clip, width =0.501 \textwidth]{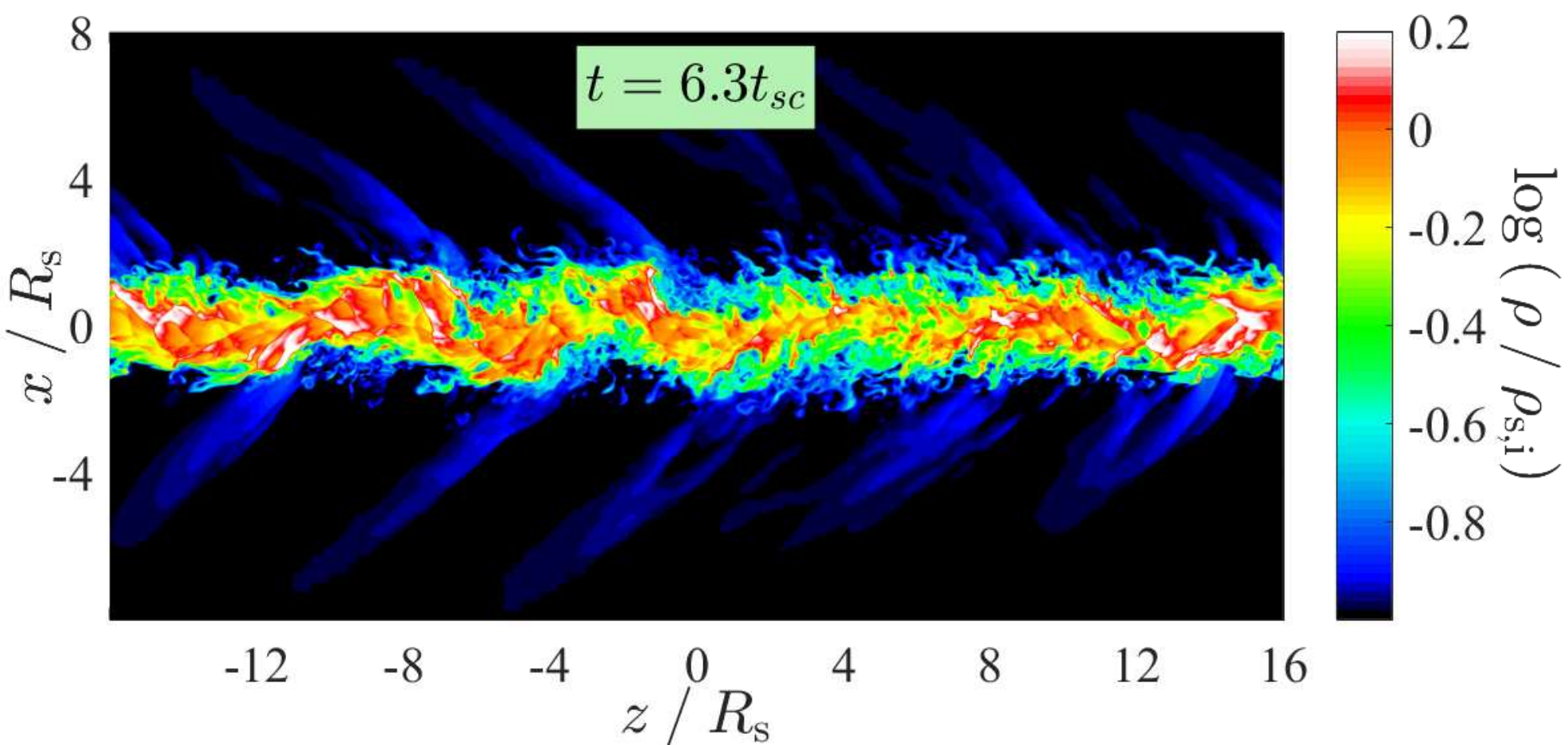}\\
\vspace{-0.09cm}
\includegraphics[trim={0.0cm 0.0cm 3.3cm 0}, clip, width =0.45 \textwidth]{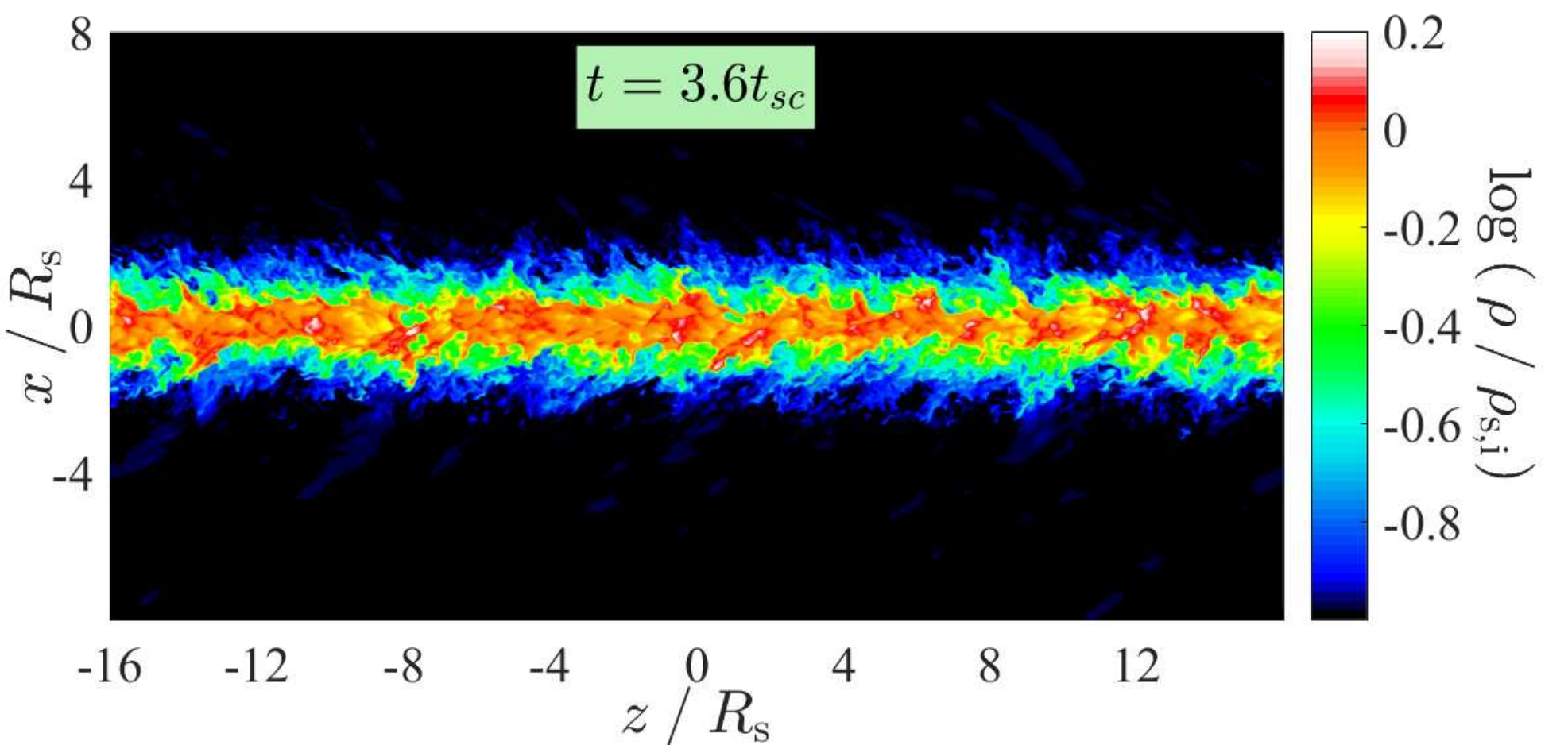}
\hspace{-0.3cm}
\includegraphics[trim={1.3cm 0.0cm 0.1cm 0}, clip, width =0.501 \textwidth]{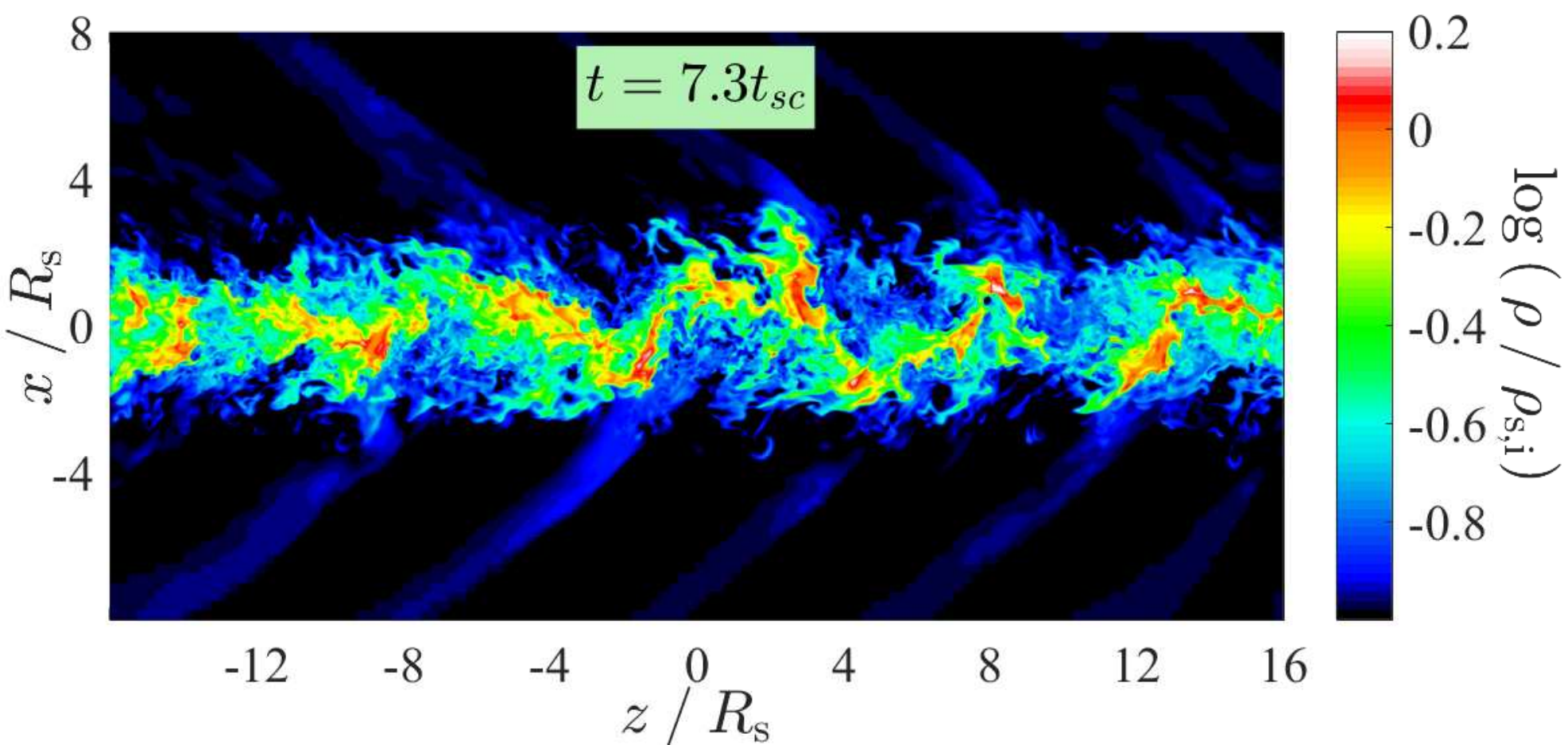}\\
\end{center}
\caption{Same as \fig{colour_panel_M2D10} but where the colour shows the density relative to the initial 
stream density, $\rho/\rhos$, rather than the passive scalar $\Psi$. The criss-cross patern of density perturbations 
in the stream interior at early times (top row) is typical of body modes. These are more pronounced in the simulation 
with $\sigma=\Rs/8$ because body modes have longer to develop in this case. Once the large scale perturbation 
has developed and the stream and background begin to mix, the density quickly becomes diluted. Note also the oblique 
shock waves propagating into the background, which facilitate the transfer of momentum to the background gas in the 
body mode regime.
}
\label{fig:density_panel_M2D10} 
\end{figure*}

\label{lastpage} 
 
\end{document}